\pdfoutput=1
\documentclass{svjour3} 
\smartqed  
\usepackage{graphicx}
\usepackage{cancel}
\usepackage{color}
\usepackage{float}
\usepackage{url}
\usepackage{latexsym}
\usepackage{multirow}
\usepackage{longtable}
\usepackage[draft]{minted}
\definecolor{mygray}{gray}{0.9}
\usepackage{soul} 
\sethlcolor{mygray}
\usepackage[title]{appendix}

\journalname{
Empirical Software Engineering}

\begin{document}

\title{Clones in Deep Learning Code: What, Where, and Why?}

\author{Hadhemi Jebnoun         \and
        Md Saidur Rahman \and
        Foutse Khomh \and
        Biruk Asmare Muse
}

\authorrunning{Jebnoun et al.}

\institute{Hadhemi Jebnoun
\and
Md Saidur Rahman \and Foutse Khomh \and Biruk Asmare Muse
\\\email{\{hadhemi.jebnoun,saidur.rahman,foutse.khomh, biruk-asmare.muse\}@polymtl.ca}
\at DGIGL, Polytechnique Montreal, QC, Canada,
}

\date{Received: date / Accepted: date}

\maketitle

\begin{abstract}
Deep Learning applications are becoming increasingly popular worldwide.
Developers of deep learning systems like in every other context of software development strive to write more efficient code in terms of performance, complexity, and maintenance.
The continuous evolution of deep learning systems imposing tighter development timelines and their increasing complexity may result in bad design decisions by the developers. Besides, due to the use of common frameworks and repetitive implementation of similar tasks, deep learning developers are likely to use the copy-paste practice leading to clones in deep learning code. Code clone is considered to be a bad software development practice since developers can inadvertently fail to properly propagate changes to all clones fragments during a maintenance activity. However, to the best of our knowledge, no study has investigated code cloning practices in deep learning development. The majority of research on deep learning systems mostly focusing on improving the dependability of the models. Given the negative impacts of clones on software quality reported in the studies on traditional systems and the inherent complexity of maintaining deep learning systems (e.g., bug fixing), it is very important to understand the characteristics and potential impacts of code clones on deep learning systems. 
This paper examines the frequency, distribution, and impacts of code clones and the code cloning practices in deep learning systems.  
To accomplish this, we use the NiCad clone detection tool to detect clones from 59 Python, 14 C\#, and 6 Java based deep learning systems and an equal number of traditional software systems. We then analyze the comparative frequency and distribution of code clones in deep learning systems and the traditional ones. 
Further, we study the distribution of the detected code clones by applying a location based taxonomy. In addition, we study the correlation between bugs and code clones to assess the impacts of clones on the quality of the studied systems. Finally, we introduce a code clone taxonomy related to deep learning programs based on 6 DL software systems (from 59 DL systems) and identify the deep learning system development phases in which cloning has the highest risk of faults.
Our results show that code cloning is a frequent practice in deep learning systems and that deep learning developers often clone code from files contain in distant repositories in the system. In addition, we found that code cloning occurs more frequently
during DL model construction, model training, and data pre-processing. And that hyperparameters setting is the phase of deep learning model construction during which cloning is the riskiest, since it often leads to faults.


\end{abstract}

\section{Introduction}
\label{intro}
Deep learning (DL) is a subset of machine learning (ML), which in turn is a subset of artificial intelligence (AI), the science of mimicking human capabilities by machines. DL is part of a broader family of ML methods inspired by the mechanism and structure of the human brain, a network of billions of neurons. This structure is simulated by networks of artificial neurons known as artificial neural networks (ANN). 
DL networks are artificial neural networks with more than three layers. DL enables computers to build concepts from data based on simpler concepts \cite{Goodfellow-et-al-2016}.  
The field of DL is currently revolutionizing almost every industry in countless ways and is having a gigantic impact on domains like healthcare, communication, transport, finance, etc. 
DL models have high learning capacity that allows them to capture increasingly complex patterns directly from data without the need of handcrafted feature engineering \cite{Goodfellow-et-al-2016}.

Traditionally, software systems are constructed deductively by writing down the rules that govern the behavior of the system as program code. However, with DL, these rules are inferred from training data and they are generated inductively. This consequently reduces the considerable manual work necessary to handcrafting features as required for classical machine learning approaches. DL models also have more sophisticated architecture than traditional ML models, capturing long-range dependencies and modeling them with data.

The main purpose of deep learning is to construct models with high performance that is able to learn patterns from the input data in order to make predictions for new data. 
To find the optimal model, deep learning practitioners used to implement quickly prototypes by experimenting with different configurations. They then compare the performance of the different models to identify the best configuration leading to the most efficient model. And as DL developers may have to follow the same or similar steps to build models  with or without some modifications, they often may end up writing poor code and duplicated functions or blocks known as clones. 
Also, code clones are often created through code reuse. Indeed, reusing code with or without modification by copying and pasting fragments from one location to another is a common practice during software development and maintenance activities, including deep learning development \cite{liu2020using}. For example, deep learning developers can clone models' architectures and model (hyper)parameters settings or initialization for similar model implementations. 

Code clones are a kind of code smell and code smells are violations of some fundamental design principles. 
The existence of code smells \cite{fowler1999refactoring} may affect the maintenance and the evolution of the software systems. 
In traditional systems, they negatively impact software quality \cite{suryanarayana2014refactoring} and tend to increase technical debts, and consequently incur additional development and maintenance costs. 
Some types of code smells may also increase the consumption of certain resources (processor, memory, etc.) \cite{gottschalk2012removing,vetro2013definition} and, consequently, hinder the deployment of efficient and sustainable solutions. Anawer et. al. \cite{Anwar2019SEAA} reported that duplicate code may be related to increased energy consumption by the software applications. For deep learning systems, concerns are growing regarding the high energy consumption by deep learning systems.
Given the known negative impacts of code clones on traditional software systems and the complexity of deep learning systems, it is reasonable to assume that code duplication is likely to pose challenges to the maintenance of deep learning projects. 
In fact, ML-based systems are expected to have the maintenance concerns of traditional systems as well as those specific to ML.

Earlier studies on traditional software systems show that about 7\%-23\% of code in the software repositories are cloned code \cite{RoyCK2010}. 
The intuitive benefit of clones is a short term productivity gain by code reuse. Some earlier studies reported positive impacts of clones \cite{kapser2008cloning} and showed that clones are not necessarily harmful \cite{hotta2010duplicate,krinke2011cloned}.  
However, there are substantial empirical evidences showing that clones can negatively impact software quality by affecting the stability of the code \cite{Lozano:2008,Mondal:2012sac,Rahman_SCAM2014} and making it bug-prone \cite{barbour2011late,Barbour_JSEP:2013,juergens2009code,Li2006cpminer,Li_ICSE_2012,Rahman_SCAM_2017,Wagner_SANER_2016}, and consequently adding complexities and costs to software maintenance. 
To leverage the benefits of code reuse while consciously avoiding possible issues, developers should be aware of clones and manage clones properly to ensure their consistent evolution \cite{RoyKeynote2014}. 

Although code cloning practices, as well as their impacts, have been widely investigated for traditional software systems, we know little about the coding practices in ML-based systems despite the recent upsurge in the development of machine learning; in particular DL-based systems. 
No study to date has examined the code quality of DL software systems by studying the distribution and impacts of code clones and by investigating the risks associated with their existence in the DL-based systems. We aim to fill in this gap by empirically investigating code cloning practices in DL systems. Specifically, we aim to 
understand the extent to which DL developers duplicate code, as well as the impact of code clones on the maintenance of DL systems.  
To the best of our knowledge, this paper presents the first empirical study on DL code clones, where one performs a comparative study of the distribution and bug-proneness of clones in deep learning and traditional software systems.
In this paper, we analyze clones in 59 Python, 14 C\#, and 6 Java based deep learning systems and an equal number of traditional software systems. We compare the distribution of clones from the perspectives of clone types and their locations. To gain further insights into the reasons behind developers' code cloning practices in the studied deep learning systems, we randomly select six deep learning projects to perform a manual analysis of their code clones. We build a taxonomy of DL code clones in which we assign each detected code clones to the corresponding deep learning phase. We further study the relation between bug-proneness and code cloning in the context of deep learning systems. 
Finally, we identify the phases of the development process of deep learning systems in which code cloning has the highest risk of bugs.
\newline
Our empirical study resulted in the following key findings:
\begin{itemize}
    \item Cloning is frequent in deep learning code. We found that code clones occurrences in deep learning code are higher than in traditional code. All three clone types (Type 1, Type 2, and Type 3) 
    are more prevalent in deep learning code than in traditional systems.
    \item Fragments of code clones in deep learning systems are dispersed. The majority of code clones in deep learning code are located in different files i.e, in the files in same or in different directories.
   
    \item Code clones in deep learning code are likely to be more defect-prone compared to non-cloned code. Type 3 clones (i.e., clones with differences because of added, deleted, or modified lines) are at higher risk of bugs compared to other clone types. 
    
    \item We classify code clones by deep learning phases and found that three main DL phases are more prone to code cloning: model construction (36.08\%), model training (18.56\%), and data preprocessing (18.56\%). 
   
    \item We found clones from the following phases to have the highest risk of bugs:
    model construction (50\%), model training (20\%), data collection (13.3\%), data preprocessing (10\%), data post-processing (3.3\%), and hyperparameter tuning (3.3\%). In brackets we provided the percentage of clones that experienced a bug fixing change. 

\end{itemize}

The rest of the paper is organized as follows. In Section \ref{sec:2}, we describe basic concepts of code clones, deep learning, and the bug-proneness of code clones. In Section \ref{sec:3}, we present our study design by introducing the research questions and methodology. Our findings and discussions are presented in Section \ref{sec:4}. Section \ref{sec:6} discusses the threats to validity. Section \ref{sec:7} presents the related work and Section \ref{sec:8} concludes the paper and introduces some future work.

\section{Background}
\label{sec:2}
In this section, we briefly discuss the concepts and terminology related to our empirical study. 
We give an overview of the characteristics of the deep learning systems and trends in the deep learning application development. We also provide brief descriptions of the phases of deep learning application development.
In addition, we give an overview of the taxonomy of code clones, common causes behind code cloning, and the impacts of clones on software systems based on existing literature.

\subsection{Deep Learning}
\label{sec:dl-phases}

In this section, we review the main concepts of DL and discuss the different phases of the life cycle of DL system development.

Deep Learning (DL) is a sub-domain of Machine Learning (ML) that involves stacking of multiple layers of neural networks to provide a powerful model with the ability to learn from data. Machine learning techniques other than deep learning rely heavily on feature engineering. Whereas deep learning is capable to learn representation from raw data which makes deep learning a powerful ML technique. Thanks to the increase in the amount of data available and the advancement in computer infrastructure both hardware and software, deep learning has earned growing popularity recently and has dealt with complex applications with increasing accuracy over time \cite{Goodfellow-et-al-2016}. Deep learning has been applied in various fields and in several activities and services in daily life including transportation, health, and finance \cite{heaton2017deep,miotto2018deep,nguyen2018deep}. Deep learning has contributed to several research fields such as computer vision \cite{farabet2012learning,tompson2014joint}, speech recognition \cite{hinton2012deep,sainath2013deep}, machine translation \cite{bordes2014question,collobert2011natural}, software engineering \cite{gupta2017deepfix,li2018deep,white2016deep}, etc.

\begin{figure}[htpb]
\centering
\includegraphics[scale=0.40]{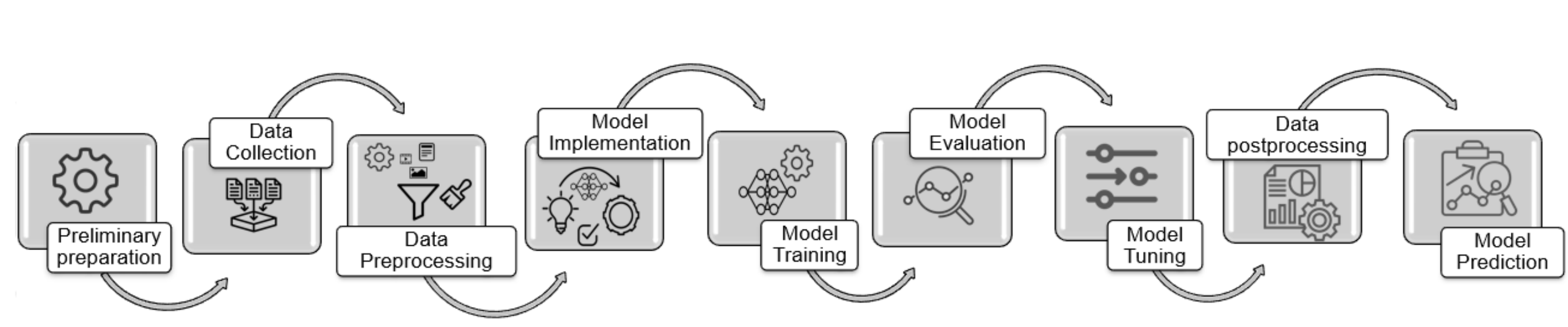}
\caption{Deep Learning Workflow}
\label{fig:DLWorkflow}
\end{figure}

There is an abundant literature on the software quality assurance of traditional software systems \cite{al2017empirical,buckley1984software,hamdan2015quality,kumlander2010towards,rochimah2015non}. However, very few studies have investigated quality issues in deep learning software systems. Deep learning software development may face all the development and maintenance challenges of a traditional software system, in addition to DL/ML specific challenges related to their dependence to data, inductive nature, and the complexity to understand their behavior \cite{samek2017explainable}. 
To help developers create reliable deep learning systems efficiently, it is important to understand their design and implementation practices and how these practices impact the quality of DL software systems. 
In the following, we describe the steps followed to construct a deep learning model. We adopted this development phases from the workflow described by Han et al. \cite{han2020programmers}. This workflow was inferred from analyzing various deep learning frameworks (e.g, Tensorflow, Pytorch, and Theano) Fig. \ref{fig:DLWorkflow} shows the steps of this workflow that we detail in the following. We present 9 steps represented in a linear diagram, however, deep learning workflows are non-linear and may contain several feedback loops \cite{amershi2019software}. 
We add the data post-processing and the data collection phases to this workflow to get a more accurate classification of code clones regarding development activities. 
We provide our conjectures of why deep learning practitioners may duplicate code to perform each step of this workflow. Regarding other code smells, we assume that they may exist in every step of the deep learning workflow, since they correspond to poor coding practices on lines and functions or classes, regarding their complexity and length.

\textbf{Preliminary preparation:} It is a coding-based phase in which developers prepare the environment, resolve installation issues, frameworks/libraries versions, configuring hardware requirements (CPU, GPU management), etc. 
This initial step may be prone to code clones since if deep learning developers use different models, each model requires a specific configuration. These configurations are likely to be same or similar for closely similar models. Consequently, we may find exact duplication and near-miss clones in the code used to perform this task. 
In addition, to parallelize data preparation, setting the number of cores, and the number of threads per core, developers may duplicate code with different values. Hence, the spread of code duplication to perform this DL phase. 

\textbf{Data collection:} Deep learning practitioners start by gathering data required to model the target business problem. The collection could be from available open source or internal dataset. This step could be done by reading file(s) from the disk, or by calling a REST or Web service API, or by using data collector functionalities provided by the deep learning frameworks. 
Whenever deep learning practitioners need to collect data, they will use the same or similar call or logic possibly by modifying only the source paths. Thus, they may end up duplicating code for data acquisition or data streaming. 

\textbf{Data preprocessing:} Once the dataset is selected and collected, data should be prepared as input to the chosen architecture of the models. Each model requires input data of specific characteristics regarding size, shape, format, and data type. This phase is done before model training and it depends on the required input for each model. It is common for each model and the developers may opt for code reuse by copy-paste to implement common functions leading to duplicate code. Even if those are not exact clones, they will be near-miss (closely similar) clones as they have the same or  similar algorithmic logic. After preprocessing the data to be suitable to model learning, the dataset is split into three different subsets as follows: training data, validation data, and test data. 
As for splitting data, it is a common practice used by DL practitioners. 
The majority of deep learning frameworks provide ready to use functions to perform those pre-processing data tasks.
Calling functions to collect data with or without modifications for each specific need is likely to introduce clones in deep learning code. 

\textbf{Model implementation:} In this step, deep learning developers construct and configure the DL model with the chosen architecture. Then comes the hyperparameters set up and the selection of activation functions, loss functions, and model optimizers. Another option is to use pretrained models that are available from online sources or load them from disk storage. This practice is used to speed up model construction and training steps. This phase is considered as the most crucial phase in deep learning development since it is dedicated to the issues related to the model itself: the choice of the model \cite{han2020programmers}. The setup of deep learning models has common steps. These steps are blocks of code that are common between models, and each performs a sequence of calls to DL routines. Due to the use of the same frameworks and libraries, the DL code can have duplicated blocks of code between functions or cloned functions.

\textbf{Model training:} Once the model implementation and data preparation are complete, the model is ready to be trained. The training process updates the parameters iteratively to minimize the loss, i.e., the prediction error. At the end of the training, the model is generated with better accuracy and performance \cite{han2020programmers}. Since many DL models may share the same algorithmic logic and share some computational functions (e.g., loss function, activation function), the deep learning code may have duplicated code fragments that are either exact or near-miss clones.  

\textbf{Model evaluation:} At this level, the trained model is ready to be evaluated on its performance. Thus, deep learning practitioners need the validation dataset, that was hidden from the model during training, for evaluation. The evaluation of the model is frequently done by visualizing the performance metrics of the trained model; assessing 
the changes to the loss function, model accuracy, etc \cite{han2020programmers}. Koenzen et al. \cite{koenzen2020code} have shown that code snippets related to visualization tend to have a duplication rate of up to 21\% in Jupyter notebook. 
Evaluating deep learning models is an integral part of the DL development process. It allows developers to find the best model's configuration. This step gives a better idea of how well the model may perform on unseen data.
This is an essential phase in the model construction process, hence one can find the code for model evaluation in each deep learning project.
Their logic is similar if not exact. Hence, deep learning code may contain duplicated functions or blocks of code used to perform model evaluations. 

\textbf{Model tuning:} To optimize the performance of the model, hyper\-/parameters are tuned.
Using an unsuitable loss function, incorrectly initializing the weights, or choosing an inappropriate learning rate will negatively affect the model's performance. This step is empirical, it is usually done through a trial and error process that aims to compute the optimum values of models hyper-parameters (e.g, grid search technique \cite{bergstra2012random}). Hyper-parameter tuning is an essential phase to improve the model's performance. The implementation of this technique is either done by deep learning practitioners or by invoking ready to use optimization functions from a framework. Because these functions have the same logic, calling the same or similar set of framework routines to perform this task will likely result in duplicated code blocks or functions in the deep leaning code.  

\textbf{Data postprocessing:} After an inductive process of learning from raw data, the output of the model may not be well-suited to represent the prediction results in an application specific and user-interpretable from. Hence, prediction results should be post-processed to be more meaningful and informative to end-users. This phase is frequently used in object detection, where the code interprets output by assigning the class with a higher probability to each object or by drawing the resulted bounding boxes on an image. If models have a common objective, such as detecting objects, they may induce code duplication, during data post-processing.

\textbf{Model prediction:} After training the model or using a pretrained model, the model is ready to make a prediction for new given data. The model prediction is implemented frequently as a function named \textit{predict} and a call to this function. When using the same logic, steps of implementation, and renaming strategy, the deep learning model prediction code may have duplicated code blocks or functions.

\subsection{Code clones}
Code clones are exact or similar copies of code fragments usually created by copying and pasting code fragments for code reuse. It can be similar code fragments, with renamed or added lines. The code fragments are usually identified by their file names, start line number, and end line number. Code clones could be detected by pair or by class.

\textbf{Clone pair}: Clone detection result can be represented by pairs of fragments. Two fragments that are clones to each other form a clone pair.

\textbf{Clone class}: Code clones can also be presented as clone classes. Each clone class contains a set of fragments that are clones to each other.

\subsubsection{Clone Taxonomies} \label{sec:cloneTypes}
In our study, we are interested in exploring two kinds of clone taxonomies, similarity-based and location-based clone taxonomy. We will explain both of them in the next two subsections.
\paragraph{\textbf{Similarity-based Clone Taxonomy}}
Basically, there are two kinds of similarities between the two code fragments: functional (semantic) and textual (syntactic). 

\textbf{Textual similarity} is when a copied fragment is used with or without minor modification.
There are three types of syntactically similar clones: \begin{itemize}
\item \textbf{\textit{Type 1:}} Identical code clones except for differences in white-spaces, layouts and comments. It is known as exact clones. Table \ref{tab:Type1Example} presents an example of two fragments of code clones where the difference between them is the comment highlighted in grey. The pair of code fragments are exact copies of each other. Hence, they are clones of Type 1. 
\item \textbf{\textit{Type 2:}} Syntactically identical code clones except for differences in identifiers name, data types, whitespace, layouts, and comments are Type 2 clones. As shown in Table \ref{tab:Type2Example}, the two fragments will be exact when we ignore the naming differences (function name, name of input variables). These two code fragments are Type 2 clones of each other.
\item \textit{\textbf{Type 3:}} Code clones with some modification, addition or deletion of lines in addition to a difference in identifiers, data types, whitespaces, and comments. Examples of two Type 3 code fragments are showed in Table \ref{tab:Type3Example}. These two code fragments are different in the function name and the addition of 2 lines for another condition in the second code fragment.
\end{itemize}

\begin{table}[htpb]
\caption{Type 1 Clones }

\label{tab:Type1Example}
    \centering
    \begin{tabular}{|c|c|}
    \hline

\begin{minipage}{0.46\textwidth}
\vspace{0.1cm}
\begin{minted}[fontsize=\scriptsize,highlightlines={2},highlightcolor=mygray]{python}
def forward_activation(self, X):
    #compute post activation value of X
    if self.activation_fct == "sigmoid":      
        return 1.0/(1.0 + np.exp(-X))    
    elif self.activation_fct == "tanh":      
        return np.tanh(X)    
\end{minted}
\vspace{0.05cm}
\end{minipage}
& 
\begin{minipage}{0.46\textwidth}
\vspace{0.1cm}

\begin{minted}[fontsize=\scriptsize,highlightlines={},highlightcolor=mygray]{python}
def forward_activation(self, X):     
    if self.activation_fct == "sigmoid":      
        return 1.0/(1.0 + np.exp(-X))    
    elif self.activation_fct == "tanh":      
        return np.tanh(X)    
  \end{minted}
  \vspace{0.05cm}
  \end{minipage} \\
    \hline
    \end{tabular}
\end{table}

\begin{table}[htpb]
\caption{Type 2 Clones }

\label{tab:Type2Example}
    \centering
    
    \begin{tabular}{|c|c|}
    \hline

\begin{minipage}{0.47\textwidth}
\vspace{0.1cm}
\begin{minted}[fontsize=\scriptsize,highlightlines={1,3,5,},highlightcolor=mygray]{python}
def forward_activation_fct(self, X):
    if self.activation_fct == "sigmoid":      
        return 1.0/(1.0 + np.exp(-X))    
    elif self.activation_fct == "tanh":      
        return np.tanh(X)    
\end{minted}
\vspace{0.05cm}
\end{minipage}
& 
\begin{minipage}{0.47\textwidth}
\vspace{0.1cm}

\begin{minted}[fontsize=\scriptsize,highlightlines={1,3,5},highlightcolor=mygray]{python}
def forward_activation(self, input):     
    if self.activation_fct == "sigmoid":      
        return 1.0/(1.0 + np.exp(-input))    
    elif self.activation_fct == "tanh":      
        return np.tanh(input)    
  \end{minted}
  \vspace{0.05cm}
  \end{minipage} \\
    \hline
   
    \end{tabular}
    
\end{table}

\begin{table}[htpb]
\caption{Type 3 Clones }

\label{tab:Type3Example}
    \centering
    
    \begin{tabular}{|c|c|}
    \hline

\begin{minipage}{0.45\textwidth}
\vspace{0.1cm}
\begin{minted}[fontsize=\scriptsize,highlightlines={1,8,9},highlightcolor=mygray]{python}
def forward_activation_fct(self, X):
    if self.activation_fct == "sigmoid":      
        return 1.0/(1.0 + np.exp(-X))    
    elif self.activation_fct == "tanh":      
        return np.tanh(X)    

\end{minted}
\vspace{0.05cm}
\end{minipage}
& 
\begin{minipage}{0.49\textwidth}
\vspace{0.1cm}

\begin{minted}[fontsize=\scriptsize,highlightlines={1,6,7},highlightcolor=mygray]{python}
def forward_activation(self, x):     
    if self.activation_fct == "sigmoid":      
        return 1.0/(1.0 + np.exp(-x))    
    elif self.activation_fct == "tanh":      
        return np.tanh(x)    
    elif self.activation_fct == "relu":      
        return np.maximum(0,x) 

  \end{minted}
  \vspace{0.05cm}
  \end{minipage} \\
    \hline
   
    \end{tabular}
    
\end{table}

\begin{table}[htpb]
\caption{Type 4 Clones }

\label{tab:Type4Example}
    \centering
    \begin{tabular}{|c|c|}
    \hline

\begin{minipage}{0.43\textwidth}
\vspace{0.1cm}
\begin{minted}[fontsize=\scriptsize,highlightlines={},highlightcolor=mygray]{python}
def forward_activation(self, X):
    if self.activation_fct == "sigmoid":      
        return 1.0/(1.0 + np.exp(-X))    
    elif self.activation_fct == "tanh":      
        return np.tanh(X)    
\end{minted}
\vspace{0.05cm}
\end{minipage}
& 
\begin{minipage}{0.49\textwidth}
\vspace{0.1cm}

\begin{minted}[fontsize=\scriptsize,highlightlines={},highlightcolor=mygray]{python}
def forward_activation(self, x):
    vals = { "sigmoid" : 1.0/(1.0+np.exp(-x)),
             "tanh" : np.tanh(x) }
    return vals[self.activation_fct]   
  \end{minted}
  \vspace{0.05cm}
  \end{minipage} \\
    \hline
    \end{tabular}
\end{table}

\textbf{Functional similarity} is when two pieces of code are similar in functionality without being written in a textually identical/similar way. This kind of similarity is called semantic clones and referred to as \textbf{\textit{Type 4}}. Table \ref{tab:Type4Example} shows an example of Type 4 clone. The two functions differ syntactically, but they achieve the same result, which is to compute the post-activation of X with respect to the activation function.

In our study, we are interested in detecting both exact (Type 1) and near-miss clones (Type 2 and Type 3).
Thus, we use the most recent version of Nicad (NiCad-5.2) \cite{Nicad5.2}, at the date of launching clone detection on our subject systems. We use NiCad because it was found to achieve higher precision and recall in near-miss clone detection \cite{svajlenko2014evaluating}.

\paragraph{\textbf{Location-based Clone Taxonomy}}

Clone taxonomies are categorized based on three attributes: similarities, location, and refactoring opportunities as introduced in the survey by Roy and Cordy \cite{roy2007survey}.
In this paper, we follow the location-based taxonomy proposed by Kapser and Godfrey \cite{kapser2003toward}. They introduced a categorization scheme for code clones, and applied their taxonomy in a case study performed on the file system of the Linux operating system. They provide a hierarchical classification of clones using attributes such as locations and functionality. Their taxonomy mainly consists of three partitions of the physical locations of clones in the source code as follows:

\begin{itemize}
\item \textbf{Same file} when clones reside in different locations of the same file.
\item \textbf{Same directory} when clones belong to different files but within the same directory. 
\item \textbf{Different directories} when clones are detected in different files and different directories.
\end{itemize}
They further sub-classified the clones by the type of the region in which they are located (i.e., function, loop, function ending, etc).
In our study, we apply the same location-based classification and propose a sub-classification of clones based on the functionalities related to deep learning. We perform this classification by manual analysis of the clones and using the location-based taxonomy. We use the different steps of the
deep learning workflow presented above (Figure \ref{fig:DLWorkflow})
to label the detected clones. 

\subsubsection{Bug-proneness of Code Clones}
Code cloning facilitates code reuse and thus intuitively increases productivity. However, this productivity gain may be outweighed by the negative impacts of clones on software maintenance as suggested by empirical evidences from different studies \cite{Lozano:2008,Mondal:2012sac,Rahman_SCAM2014}.
For example, code-duplication increases both size (code bloating) and complexity of the software system. Because of these confounding factors, software maintainability may become increasingly complicated. 
One of the key challenges posed by code clones is ensuring the consistent evolution of clones during software maintenance; meaning that all cloned copies should be updated with necessary changes. This is because inconsistent changes to clones or missing change propagation are likely to introduce bugs 
\cite{Barbour_JSEP:2013,barbour2018investigation,juergens2009code,gode2011frequency}. As consistent changes to clones is important, missing changes, once identified, should also be propagated (late propagation) accordingly. However, late propagation of changes to clones have also been found to be prone to bugs introduction \cite{aversano2007clones,barbour2011late,Barbour_JSEP:2013}. 

Again, as the cloned copies of code fragments are expected to evolve consistently, cloned code are likely to experience frequent changes, and thus negatively affect the stability of the software systems \cite{Lozano:2008,lozano2010tracking,Mondal:2012sac,Rahman_SCAM2014,gode2011frequency}. The instability of clones, in turn, has also been found to be related to bugs \cite{Rahman_SCAM_2017}. The bug-proneness of clones may also vary based on the types of clones \cite{Mondal_ICSME_2015}. Several other prior research works have also investigated the bug-proneness of code clones \cite{Mondal_ICSME_2019,Li2006cpminer,Li_ICSE_2012}. These multiple studies on the impacts of clones, from different perspectives, show that the bug-proneness of clones is an important concern. 
Given the complexity and lack of explainability or the `black-box' nature of the deep learning models, testing deep learning based systems and thus fixing bugs are quite challenging \cite{Braiek2019,BRAIEK2020}. Thus, it will be of interest to study the relationship between software bug-proneness and code clones in the deep learning applications context. Since, duplicating code is a common practice in the deep learning development process, as reported by deep learning practitioners in a recent survey \cite{liu2020using}. In this paper, we aim to empirically analyze deep learning systems to understand the extent and impacts of clones. We aim to raise the awareness of deep learning developers on the impact of code cloning, since it is likely to add more complexity and cost to the development and maintenance of deep learning systems.
\vspace{-0.5cm}
\section{Study Design} 
\label{sec:3}
In this section, we first present our research questions by highlighting the motivation and research objectives. Then we describe our research methodologies. 
\subsection{Study Objectives}
In our empirical study, we first examine the distribution of code clones in deep learning code in terms of clone type and clone location. Then, we compare them to the distribution of code clones in traditional code. Second, we investigate the relationship between code clones and bug-proneness in deep learning code. Third, we examine the reasons behind code duplication in deep learning code and build a taxonomy of code clones occurring in different phases of the development process of deep learning systems.
Finally, we determine the riskiest phases or activities of deep learning application development by analyzing the bug-proneness of clones in each phase. To achieve these above-mentioned research objectives, we empirically investigate the following five research questions:

\textbf{RQ$1$: Are code clones more prevalent in deep learning code than traditional source code?} 

Code reuse by code cloning is a common practice in software development. Despite the intuitive productivity gain that one can expect from reusing code through code cloning, there are evidences showing that clones can negatively impact software quality; increasing complexity and maintenance costs. \cite{juergens2009code} 
Although code clones have been widely investigated for traditional software systems \cite{RoyKeynote2014}, the impact of code cloning in deep learning systems is still unknown. 
The widespread use of common open-source libraries and frameworks and the use of code examples from crowd-source question-answering web sites (e.g., Stack Overflow) may introduce duplicate code in DL systems. Moreover, given the code reuse from open-source repositories and the repetitive use of similar development phases or tasks (e.g., data preprocessing, model training) during deep learning system development, it is reasonable to expect that code clones would exist in deep learning systems. Since we know from the studies on traditional software systems, that the impacts of clones vary based on the types and frequency of clone occurrences, it is therefore important to examine the prevalence and distribution of clones in deep learning systems. A comparative analysis of code cloning in traditional and deep learning based systems is important to understand if deep learning code is more prone to clones than traditional programs, and therefore deserve a special attention, from the research community and tools builders to 
help practitioners manage clones 
in deep learning systems efficiently.

\textbf{RQ$2$: How are code clones distributed in deep learning code in comparison to traditional source code?}

Code clones location impacts the refactoring cost. Navigating into distant duplicated code fragments adds comprehension overhead. Respectively, dispersed code clones can be hard to manage and may incur an increased cost of maintenance.
To understand where deep learning practitioners duplicate code, we study the distribution of code clones in deep learning and traditional codes. We use the taxonomy proposed by Kapser and Godfrey \cite{kapser2003toward} to categorize the detected code clones by their locations (i.e, same file, same directory, and different directories). 

\textbf{RQ$3$: Do cloned and non-cloned code suffer similarly from bug-proneness in deep learning projects?} \\
Studies of code clones in traditional software systems suggest that clones can have an adverse impact on the maintainability of the system; increasing the risk of fault \cite{barbour2011late,Barbour_JSEP:2013}. However, it is important but yet to know the impacts of cloning in deep learning systems especially on the bug-proneness. 
This research question investigates the relationship between bug-proneness and code clones occurrences in deep learning code by performing Mann Whitney statistical test and analyzing the effect size. We examine the impacts of different types of clones on the bug-proneness of deep learning code and assess the effort required to fix bugs in cloned and non-cloned code (by computing the time to fix of each bug).  

\textbf{RQ$4$: Why do deep learning developers clone code?}\\
Given the complexity of deep learning code, constructing an efficient DL model can be a tedious job. DL developers should be experienced in the problem domains and should also have a sufficient understanding of deep learning techniques. They also need to have coding skills with deep learning frameworks as well as the ability to manage the computing resources. When faced with a new task, to mitigate the risk of writing erroneous code, DL developers may often duplicate the code of an existing 
tested model with the same or similar logic with or without modifications, depending on the requirements of their task. 
To understand activities in the development of deep learning applications that are more prone to code duplication, we conduct a manual analysis of
code clone classes. We categorize clones based on the development phases in which they occurred.
This analysis allows us to identify activities (i.e., phases) in the deep learning development process where code cloning occurs frequently. 

\textbf{RQ$5$: In which phases of deep learning development code cloning is more prone to faults? }\\
Since previous works on traditional systems \cite{barbour2018investigation} have shown that the risk of faults in cloned code vary depending on the types of code that is cloned, we are interested in investigating whether clones occurring at certain stages of the development process of deep learning systems or in certain functions of the deep learning code are more bug-prone than others. Therefore, for this research question, we compare the risk of faults of clones found in the different functions of the deep learning code. 

\subsection{Study Overview}
\begin{figure}[ht]
\centering
\includegraphics[width=1\textwidth]{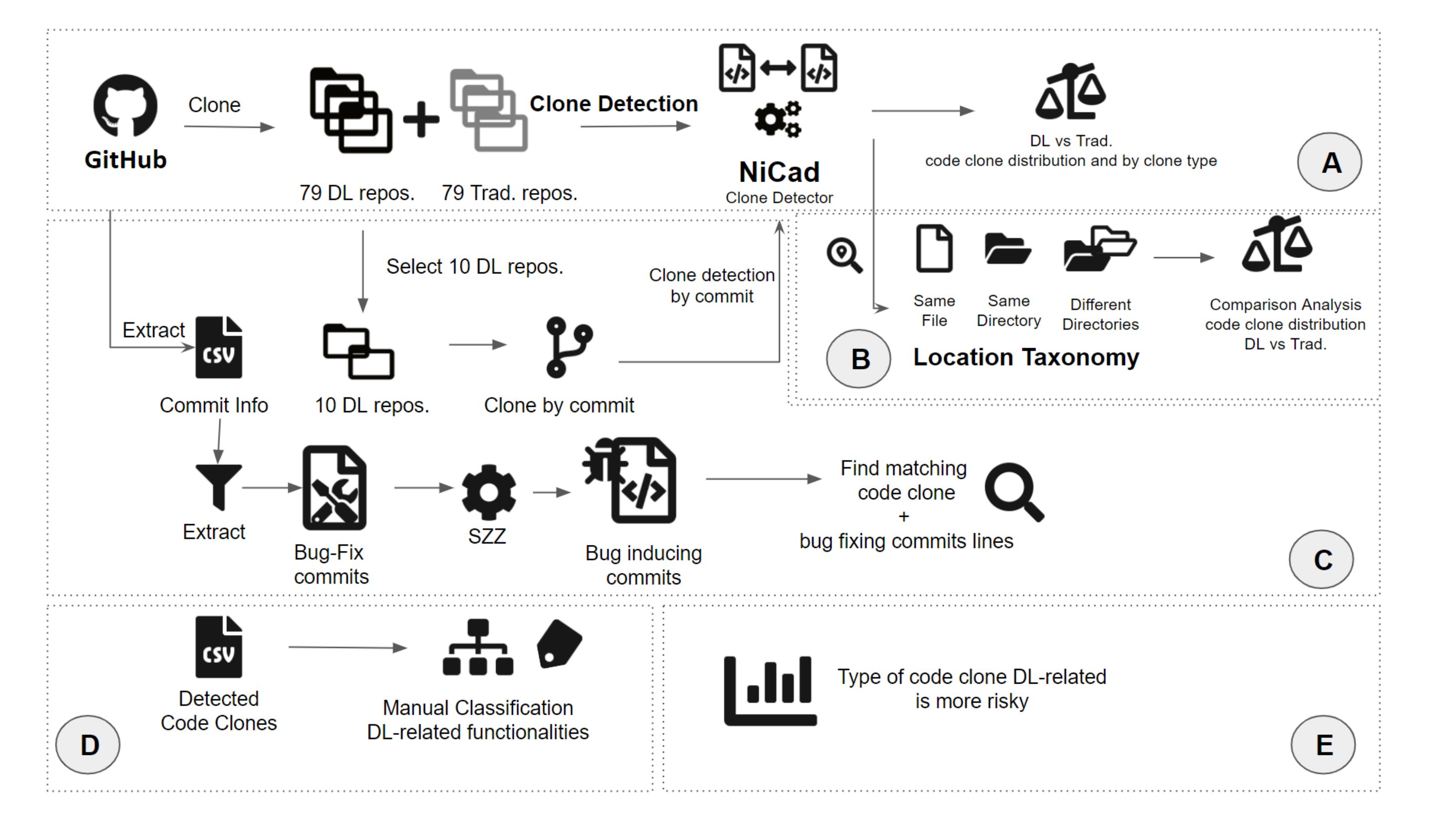}
\caption{Study Overview A- Detecting Code Clones in Deep Learning and Traditional Repositories, B- Applying Code Clone Location Taxonomy, C- Studying the Relationship between Bug-proneness and Code Clones, D-Classifying Code Clones Manually Based on DL-related Functionalities, E-Exploring riskier DL-related Code Clones}
\label{fig:methodology}
\end{figure}

In this section, we present our study design as shown in Figure \ref{fig:methodology}. We divide our methodology into five main steps.  
\begin{description}
\item[Fig \ref{fig:methodology}-A] We first clone 79 DL and 79 traditional repositories from GitHub, the detailed methodology is described in the Subsection \ref{subsubsec:subjectsystems}. We then detect code clones for both DL and traditional open-source projects using the NiCad clone detector (details are presented in section \ref{subsubsec:clonedetection}). We finally compare the clone distribution between both type of systems (i.e., DL and traditional systems) in terms of lines of code and clone types.  
\item[Fig \ref{fig:methodology}-B] We analyze the distribution of code clone by applying the location taxonomy. As shown in Fig \ref{fig:methodology}-B, we have three locations where there might be code clones classes: same file, same directory, and different directories (details are presented in Section \ref{subsubsec:LocationTaxonomyMethod}). An arrow towards a balance as shown in the figure \ref{fig:methodology}-B represents the comparative analysis of the distribution of code clones between deep learning and traditional systems in terms of localization.
\item[Fig \ref{fig:methodology}-C] The third part involves studying the relationship between code clones occurrences and bug-proneness. We detect code clones for each commit of a set of six deep learning repositories (discussed in Section \ref{subsubsec:preprocessing}). Then, we identify bug-fixing commits relying on the commit history. Next, we extract bug-inducing commits by applying the SZZ algorithm. Once we have bug-inducing commits with their corresponding changed lines, we match these buggy lines with lines that are cloned to find the relationship between bug and clones (details are shown in Section \ref{subsubsec:cloneBugMethod}).
\item[Fig \ref{fig:methodology}-D] We manually classify code clones that are DL-related to construct a taxonomy of code clones in DL code (see Section \ref{subsubsec:DLtaxonomy}). 
\item[Fig \ref{fig:methodology}-E] Finally, we examine the risk of bugs of code clones occurring in specific phases of the DL development.
\end{description}

\subsubsection{Subject Systems}
\label{subsubsec:subjectsystems}
We mined 138 open-source projects (59 Python, 6 Java and 14 C\# deep learning projects with equal number of traditional projects in each programming language) from GitHub. We used the same Python projects investigated in Jebnoun et al. \cite{jebnounscent} study on detecting code smells in DL code. Our empirical study is primarily focused on the set of projects \cite{JebnounScentGithub} in Python, as it is the most widely used language in the machine learning field \cite{braiek2018open}. However, we also analyzed Java and C\# projects for the generalizability of the findings. Deep learning projects were selected by first searching repositories using DL-related keywords (e.g., deep learning, deep neural network, convolutional neural network) \cite{jebnounscent} and manually filtering out 
tutorials and some projects with a low number of releases (to obtain our 59 Python, 6 Java and 14 C\# DL repositories). To obtain our dataset of traditional systems (i.e., non deep-learning code), we used a benchmark from an existing study by Chen et al. \cite{chen2018understanding} (similarly to  Jebnoun et al. \cite{jebnounscent} study). This benchmark consists of 106 GitHub repositories with at least 1k stars each from which we randomly selected a subset of 59 traditional Python projects.  
We use the same set of 59 traditional Python projects studied by Jebnoun et al. \cite{jebnounscent}, for the comparative analysis in this study. We apply the same selection criteria for selecting 6 Java and 14 C\# traditional projects.

\subsubsection{Preprocessing of source code repositories}
\label{subsubsec:preprocessing}
\paragraph{\textbf{Selection of a Subset of Subject Systems:}}
Our research questions (RQ1, RQ2) are based on the analysis of 59 Python, 6 Java, and 14 C\# deep learning systems and an equal number of traditional systems. For RQ3, we analyzed 59 Python, 2 Java and 2 C\# systems for each of the DL and traditional categories. However, we randomly select 6 Python deep learning projects out of the 59 projects to analyze the distribution and impacts of clones associated with different phases or activities in deep learning system development by manual investigation (RQ4 and RQ5). We select a subset of systems to keep the cost and effort of manual analysis in a feasible range. We provide more details about the 6 DL repositories : name of the repository, number of commits, total lines of code, url and size in appendix \ref{appendixStudyDesign}.
We use this subset of systems to study the relationship between bug-proneness and code clones in deep learning code because detection of code clones by NiCad can be very time-consuming, especially for the detection of clones at every commit as we did for RQ4 and RQ5. So, we use the six selected systems to perform manual analysis for extracting the taxonomy of code clones in deep learning code regarding development steps. 
Our selected systems are from diverse application domains with varying (small to medium) size (SLOC) and lengths (in number of commits) of evolution history.

\paragraph{\textbf{Computing Source Code Lines:}}
We use the SLOCCount \cite{sloccount} tool to compute the total source lines of code (SLOC) of Python, Java and C\# code in each project from corresponding languages. SLOCCount is a software metrics open-source tool developed by David A. Wheeler supporting several programming languages. It is robust to handle variations in many languages such as the use of string constants as comments in Python code. Since one project could contain different programming languages, SLOCCount results list the available languages in the system. We consider only the size of the Python, Java and C00\# code based on the type of the systems. We execute the SLOCCount tool for both DL and traditional repositories. We also measure the SLOC for each commit of each DL repository from the selected set of six deep learning repositories for further analysis. We normalize the detected lines of code clones by the size of the project (SLOC) to have a more accurate comparison of deep learning and traditional systems regarding the frequency and distribution of clones. We divide the total lines of cloned code for each project by the total lines of code of the corresponding project. This gives a normalized (per LOC) representation of the frequency of clones for comparison. 

\subsubsection{Code Clone Detection}
\label{subsubsec:clonedetection}
For detecting code clones, we use the \textit{NiCad} tool \cite{Nicad5.2}. NiCad \cite{cordy2011nicad} can detect both exact and near-miss clones with high precision and recall \cite{roy2009mutation} with respect to blocks and functions granularity.

\paragraph{\textbf{NiCad Settings:}}  Table \ref{tab:nicadSettings} shows the setup for detecting the three types of clones. We detect code clones with a minimum size of five lines of code. We detect both exact clones (Type 1) and near-miss (Type 2 and Type 3) clones. 
We use the blind renaming option of NiCad for the detection of Type 2 and Type 3 clones. 
Type 3 clones are detected with a dissimilarity threshold of 30\%. 

\begin{table}[htpb]
\caption{NiCad Settings}
\label{tab:nicadSettings}
\begin{tabular}{|c|c|c|c|c|}
\hline
\textbf{\begin{tabular}[c]{@{}l@{}}Clone Type\end{tabular}} &
  \textbf{\begin{tabular}[c]{@{}l@{}}Identifier Renaming\end{tabular}} &
  \textbf{\begin{tabular}[c]{@{}l@{}}Dissimilarity  Threshold\end{tabular}} &
  \textbf{Size (LOC)} 
  \\ \hline
\textbf{Type 1} & none  & 0\%  & {[}5-2500{]} \\ \hline
\textbf{Type 2} & blind & 0\%  & {[}5-2500{]}  \\ \hline
\textbf{Type 3} & blind & 30\% & {[}5-2500{]} \\ \hline
\end{tabular}
\end{table}

\paragraph{\textbf{Clone Detection: }}
We detect code clones for a particular snapshot and for each commit of each repository.
To perform a comparative analysis of the frequency and distribution of clones in deep learning and traditional code, we detect three types of code clones (Type 1, Type 2, and Type 3) in both deep learning and traditional code using the NiCad with settings detailed in Table \ref{tab:nicadSettings}. 
We detect clones on a particular snapshot, which in our case is the last available version of the project on GitHub at the time of cloning the repository \cite{sha1DLRepos}. For RQ1, we use both granularities (function and block) to detect code clones. For the rest of the research questions (RQ2-RQ5), we analyze code clones at function granularity.
To study the relation between the bug-proneness and code clones, we use the commit history to extract commits information of each repository from the set of six deep learning repositories and detect code clones for every commit of the repositories.

\paragraph{\textbf{Results Cleaning: }}To perform clone-type based investigation, we need to separate each type of clones. This is because by definition (as in Section \ref{sec:cloneTypes}), Type 2 clones include Type 1 clones and Type 3 clones include both Type 1 and Type 2 clones. To separate Type 2 clones, we exclude matching clone classes between Type 1 and Type 2 from the outputs of Type 2 clone detection results from NiCad. As NiCad results for Type 2 contain clone classes from Type 1 since the set of Type 1 clones (exact copies) is a subset of the set of Type 2 clones (e.g. with differences in identifier names). Hence, we remove matching clone classes based on the clone fragment specifications (file path, start and end line number, etc). 
Thus, if the same clone class (i.e., containing the same set of clone fragments specifications) exists in both Type 1 and Type 2 clone detection results, we remove such clone classes from Type 2 results as they are Type 1 clones. So, the filtered Type 2 results contain exclusively Type 2 clones without any Type 1 clone fragment in it. 
Similarly, we exclude the matching Type 1 and Type 2 classes from Type 3 clone detection results to have Type 3 clones exclusively. This is because Type 3 clone results by definition contain Type 1 and Type 2 clone fragments which need to be removed to perform the clone-type centric analysis correctly. 

\subsubsection{Location taxonomy based labeling of clones}
\label{subsubsec:LocationTaxonomyMethod}
To answer RQ2 that investigates the distribution of code clones in deep learning and traditional codes in terms of location, we apply the taxonomy used by Kapser and Godfrey \cite{kapser2003toward}. It categorizes the detected clones based on their relative locations in the file system structure for both types of projects. 
We label a clone class by \textbf{\textit{`Same file'}} when all the fragments in the detected clone class are from the same file. We label a clone class by \textbf{\textit{`Same directory'}} when all the files associated with the detected clone class belong to the same directory. And finally, we assign a clone class to the \textbf{\textit{`Different directories'}} label, when clone fragments are from different files located in different directories. 

We then calculate the proportion (in percentage) of both clones fragments and cloned LOCs distributed over the location types defined by the taxonomy. 
We perform Mann-Whitney statistical test to compare whether the distribution of code clones in DL and traditional systems are significantly different. We further extend our comparative analysis by considering individual clone types.

\subsubsection{Labeling clones based on the taxonomy of deep learning tasks}
\label{subsubsec:DLtaxonomy}
Since common functional steps are followed to build deep learning models and the same deep learning libraries are used, deep learning practitioners may often copy-paste ready to use functionalities with or without modification. This aims to gain productivity and to reduce the risk of writing erroneous new code when tested implementations are available. We expect code duplication not only in all deep learning phases (i.e., building model and training) but also in model testing. 

Therefore, we categorize different types of cloning practices by DL phases and label them via a manual inspection of the code clones found in the selected six DL repositories. We found 595 clone fragments in these six deep learning repositories. We manually analyzed each of these clone fragments.  
We used a bottom-up approach, where we first assigned each clone fragment with a label corresponding to the DL task or functionality. We further grouped the clone fragments in each subcategory and mapped them to the different phases of the development workflow of deep learning models, as discussed in Section \ref{sec:dl-phases}. We ensure that the relation between a subcategory and a category is: ``to perform". For example, \textit{initialize weights} \textit{to perform} a \textit{model construction}. Here, the sub-category is `initializing weights' and category is `model construction'. We also categorize and label functions that are not related to deep learning such as logging, test, etc, as \textit{`others'}. The manual classification for generating this DL taxonomy was done by at least two authors, all with academic and industry backgrounds. The resulting taxonomy was then cross-validated, and disagreements were resolved by group discussion. 

Tables \ref{tab:examplecodeclonesTaxonomyModelTraining}, \ref{tab:examplecodeclonesTaxonomyDataCollection}, and  \ref{tab:examplecodeclonesTaxonomyDataPreprocessing} present some real-world examples from the detected clones. In case of exact (Type 1) clones, we show only one code fragment from the clone class since all the fragments are identical except for formatting differences. For near-miss clones, we show both fragments in a clone pair. In Tables \ref{tab:examplecodeclonesTaxonomyModelTraining}, \ref{tab:examplecodeclonesTaxonomyDataCollection}, and \ref{tab:examplecodeclonesTaxonomyDataPreprocessing}, we present the clone types of the example fragments, and the sub-category and category of DL-related phases to which the clone class or fragments belong to. Our objective is to first assign a sub-category (DL sub-task) for each clone class and then group them into top-level categories (DL phases). 

The first example in Table \ref{tab:examplecodeclonesTaxonomyModelTraining} represents a Type 1 clone fragment. Here, the function \texttt{iou(box1, box2)} computes Intersection Over Union (IOU) between the predicted box and any ground truth box. It is a metric that computes the accuracy of YOLO (You Only Look Once) \cite{redmon2016you}, a real-time object detection system based on Convolutional Neural Network (CNN). Therefore, we designate this clone class to the sub-category `Measure model accuracy'. As this computation function is performed to train the model, we label it as the \textbf{`Model training'} category. 

\begin{table}[htpb]
\caption{Example of Clone Fragment related to 'Model Training' }
\label{tab:examplecodeclonesTaxonomyModelTraining}
    \centering

    \begin{tabular}{|c|c|c|c|}
    \hline
Code Fragment & Clone Type & Sub-category & Category \\
\hline
\begin{minipage}{0.41\textwidth}
\vspace{0.1cm}
\begin{minted}[fontsize=\scriptsize]{python}
def iou(box1, box2):
    tb = (min(box1[0] + 0.5 * box1[2],
    box2[0] + 0.5 * box2[2]) -
    max(box1[0] - 0.5 * box1[2],
    box2[0] - 0.5 * box2[2]))
    lr = (min(box1[1] + 0.5 * box1[3],
    box2[1] + 0.5 * box2[3]) -
    max(box1[1] - 0.5 * box1[3], 
    box2[1] - 0.5 * box2[3]))
    if tb < 0 or lr < 0:
        intersection = 0
    else:
        intersection =  tb*lr
    return intersection / (box1[2] 
    * box1[3] + box2[2] * box2[3] 
    - intersection)

\end{minted}
\vspace{0.05cm}
\end{minipage}
& 
 Type 1
  &
  \begin{tabular}[c]{@{}l@{}}Measure\\ model\\ accuracy\end{tabular} & Model training\\
    \hline
\end{tabular}\\

\end{table}

The second example contains two fragments of Type 3 clones as shown in Table \ref{tab:examplecodeclonesTaxonomyDataCollection}. The purpose of the first function (\texttt{read\_images\_from\_file}) is to read multiple images from the given file. Whereas the second function (\texttt{read\_images\_from\_url}) reads input images from a given URL. Thus, the function of this clone class is to load data. Since load data is performed to collect data, we assign this clones class to the\textbf{ `Data collection' } phase of deep learning. 

\begin{table}[htpb]
\caption{Example of Clone Fragment related to 'Data Collection'}

\label{tab:examplecodeclonesTaxonomyDataCollection}
    \centering

    \begin{tabular}{|c|c|c|c|}
    \hline
Code Fragment & Clone Type & Sub-category & Category \\
\hline
\begin{minipage}{0.41\textwidth}
\vspace{0.1cm}
\begin{minted}[fontsize=\scriptsize]{python}
def read_images_from_file(filename, 
rows, cols, num_images, depth=1):
    """Reads multiple images 
    from a single file."""
    ...
    _check_describes_image_geometry
    (rows, cols, depth)
    with open(filename, 'rb') 
    as bytestream:
        return images_from_bytestream
        (bytestream, rows, cols,
        num_images, depth)
\end{minted}
\vspace{0.05cm}
\end{minipage}
& \multirow{2}{*}{Type 3} & \multirow{2}{*}{Load data} & \multirow{2}{*}{Data collection} \\ \cline{1-1}
\begin{minipage}{0.41\textwidth}
\vspace{0.1cm}
\begin{minted}[fontsize=\scriptsize]{python}

def read_images_from_url(url,
rows, cols, num_images, depth=1):
    """Reads multiple images 
    from a single URL."""
    ...
    _check_describes_image_geometry
    (rows,cols, depth)
    with urllib.request.urlopen(url)
    as bytestream:
        return images_from_bytestream
        (bytestream, rows, cols,
        num_images, depth)

\end{minted}
\vspace{0.05cm}
\end{minipage}
  &                                  &                            &                                  \\ \cline{1-2}
    \hline 

\end{tabular}\\

\end{table}

The last example in Table \ref{tab:examplecodeclonesTaxonomyDataPreprocessing} is a Type 1 clone fragment. This code fragment (\textit{process\_inceptionv3\_input}) prepares a given image for the model input requirements. In this case, the model is Inception v3 \cite{szegedy2016rethinking}, which is a Deep Convolutional Neural Network with 48 layers. We assign this clone class to the `Resize image' sub-category and we label it as belonging to the \textbf{`Data preprocessing'} category.

\begin{table}[htpb]
\caption{Example of Clone Fragment related to 'Data pre-processing'}
\label{tab:examplecodeclonesTaxonomyDataPreprocessing}
    \centering

    \begin{tabular}{|c|c|c|c|}
    \hline
Code Fragment & Clone Type & Sub-category & Category \\
    \hline 
  \begin{minipage}{0.41\textwidth}
\vspace{0.1cm}
\begin{minted}[fontsize=\scriptsize]{python}
def process_inceptionv3_input(img):
    image_size = 299
    mean = 128
    std = 1.0/128
    dx, dy, dz = img.shape
    delta = float(abs(dy - dx))
    if dx > dy:  #crop the x dimension
        img = img[int(0.5*delta):dx
        -int(0.5*delta), 0:dy]
    else:
        img = img[0:dx, 
        int(0.5*delta):dy
        -int(0.5*delta)]
    img = cv2.resize(img,
    (image_size, image_size))
    img = cv2.cvtColor(img,
    cv2.COLOR_BGR2RGB)
    for i in range(3):
        img[:, :, i] = (img[:, :, i] 
        - mean) * std
    return img

\end{minted}
\vspace{0.05cm}
\end{minipage}
& 
 Type 1
  &
  Resize image & \begin{tabular}[c]{@{}l@{}}Data\\ preprocessing\end{tabular}\\
    \hline
    \end{tabular}
\end{table}

\subsubsection{Code Clone and Bugs}
\label{subsubsec:cloneBugMethod}
Several studies have been focused on investigating the relationship between clones and bug-proneness. Some studies based on traditional software systems have shown that code clones may introduce bugs and negatively impact software maintainability \cite{barbour2011late,Barbour_JSEP:2013,juergens2009code,Li2006cpminer,Li_ICSE_2012,Lozano:2008,Mondal:2012sac,Rahman_SCAM2014,Rahman_SCAM_2017,Wagner_SANER_2016}.
Therefore, it is important to study whether and to what extent this relationship holds in the context of deep learning systems.

\paragraph{\textbf{Detecting bug fixing commits:}} We extract all the commits information from each of the six selected repositories. We leverage a keyword-based approach to classify commits relying on keywords occurrence such as \textit{`bug', `fix', `solve', `problem'} in the commit messages. We use the set of keywords used by Rosen et al. \cite{rosen2015commit}. At least one of the keywords from this set should be in the commit message to consider it as a bug-fixing commit.

\paragraph{\textbf{Bug Inducing commits:}} We use PyDriller \cite{PyDriller} to extract bug-inducing commits for each bug-fix commit. PyDriller is a python-based framework that supports mining information from Git repositories. We used PyDriller to extract information such as commits and diffs from each of the selected repositories. We mainly use this framework to get bug-inducing commits given a bug-fix commit. It returns the set of commits that last modified the lines that are changed in the files modified by the bug-fix commit by applying the SZZ algorithm. 

\paragraph{\textbf{Bug Proneness of code clone:}} We define a bug to be related to code clone if the lines changed in bug-fix commits are in between the start line and end line of the detected code clones. We further analyze to identify riskier DL-related functionalities that are likely to introduce bugs. We similarly match lines changed in bug-fix commits with the corresponding lines of the cloned DL-related functions. We then extract the most frequently occurring cloned DL-related functions related to bugs in DL projects. Therefore, we obtain which tasks of DL code are more likely to introduce bugs than others when cloned. We perform MWW tests for the distributions of code clones and non-clone code in the DL bug-fix commits along with effect size analyses. To determine if there is a statistically significant difference between the number of commits that fix bugs on cloned lines and the number of commits that fix bugs on non-cloned lines.

\paragraph{\textbf{Time to fix bugs when it is related to code clones:}}
We investigate whether or not clones have impacts on the time required to fix a bug. 
The objective is to know whether clones hinder bug fixing; making bugs long-lived in the deep learning systems. We thus study the comparative time it takes to fix bugs when it is related and not related to clones respectively. We compute the difference in time between bug fixing commit and their related bug inducing commit as introduced by Kim and Whitehead \cite{kim2006long}. To do so, we extract from the commit history the time of each bug-introducing commit as well as the time of their corresponding bug-fix commit. We calculate the bug-fix time by taking the difference between bug-fix commit and bug inducing commit. Once we have the time difference (in seconds) we carry out a non-parametric Man-Whitney test to find if there is any significant difference in the time required to fix bugs related and not related to clones.
\paragraph{\textbf{Testing statistical significance and the effect size:}}
To answer the research questions, we computed different metrics
based on the quantitative analysis of clones in deep learning and traditional systems, respectively. To verify whether the differences between the corresponding metrics for DL and traditional code are statistically significant, we performed Mann-Whiteny-Wilcoxon (MWW) \cite{Neuhauser2011} test (two-tailed, p-value significant at $<$0.5). We apply MWW test because this is a non-parametric test i.e., it does not require data to be normally distributed. Moreover, this test can be applied on small data set. However, MWW test only confirms whether the observed differences between two sets of values are by chance. So, the observed differences to be statistically significant, we also need to measure the effect size. We apply Cliff's delta \cite{macbeth2011cliff} effect size along with the MWW test. This also a non-parametric measure of effect size. 
The range of the values for cliffs delta effect size is [-1, +1]. The value of effect size is generally interpreated as small ($\sim0.20$), medium ($\sim0.5$), and large ($\sim0.8$). Larger effect size refers to stronger relationships between the set of observations tested. Here, +1 and -1 values for the effect size indicate the absence of overlap while value 0 indicates a lot of overlap between two set of samples. 
We used \texttt{mannwhitneyu()} functions for the Python package \texttt{scipy.stats} for MWW test and calculated effect size by  \texttt{cliffsDelat()} function as implemented in \cite{NeilErnstGithub}.

\section{Study Findings and Discussions}
\label{sec:4}
In this section, we present the results of our study in details, and
answer five research questions as follows. 
\subsection{\textbf{RQ\(1\) : Are code clones more prevalent in deep learning code than traditional source code?}}

Due to the complexity, the lack of explainability of deep learning code, and the excessive use of ready to use routines from popular deep learning frameworks and libraries, deep learning code is likely to have duplicated code fragments i.e., code clones. Although many studies investigated the distribution, evolution, and impacts of clones regarding traditional software, none of the existing studies, has investigated the code cloning practices in deep learning code. Thus, in this research question, we study the distribution of different types of code clones in both deep learning and traditional systems, to understand and compare the prevalence of clones in these two types of software systems. 

To answer this research question, we detect code clones in 59 deep learning systems and 59 traditional software systems developed in Python. We calculate the number of occurrences of code clones across different dimensions (e.g., project type, clone type, clone granularity). Then we compare the density of clones in DL-based systems with that of the traditional systems. To compare we need normalized values of the clone densities in different systems as the systems varies in code size. For normalization, first we calculate the total number of lines of code clones in each project and then divide that by the total number of source code lines (i.e., SLOC) for normalized representation of the clone density. We count SLOC using the tool SLOCCount as discussed in  section \ref{subsubsec:preprocessing}.  
We then perform the Mann-Whitney Wilcoxon (MWW) test \cite{Neuhauser2011} to compare the distribution of clones in deep learning and traditional systems by testing if there exists any statistically significant differences in clone densities in these two types of systems. To have deeper insights, we also compare the clone densities with respect to the three clone types (Type 1, Type 2, and Type 3). 
However, MWW test only verifies whether the difference between two set of observations is by chance and does not express the effect or magnitude of the differences. Thus, we calculate the effect size to determine the magnitude of the differences between each two distributions. We measure Cliff's Delta \cite{macbeth2011cliff}, which is a non-parametric estimate of effect size and does not require data to be normally distributed. 
When Cliff's Delta is beyond of 0.2 and below 0.5, then the effect size is low. When it is beyond 0.5 and below 0.8, the effect size is medium. Beyond 0.8, the effect size is large. In this research question, we present the findings for both clone granularities: function and block. 

\begin{figure}[htpb]
\centering
\includegraphics[width=.48\textwidth]{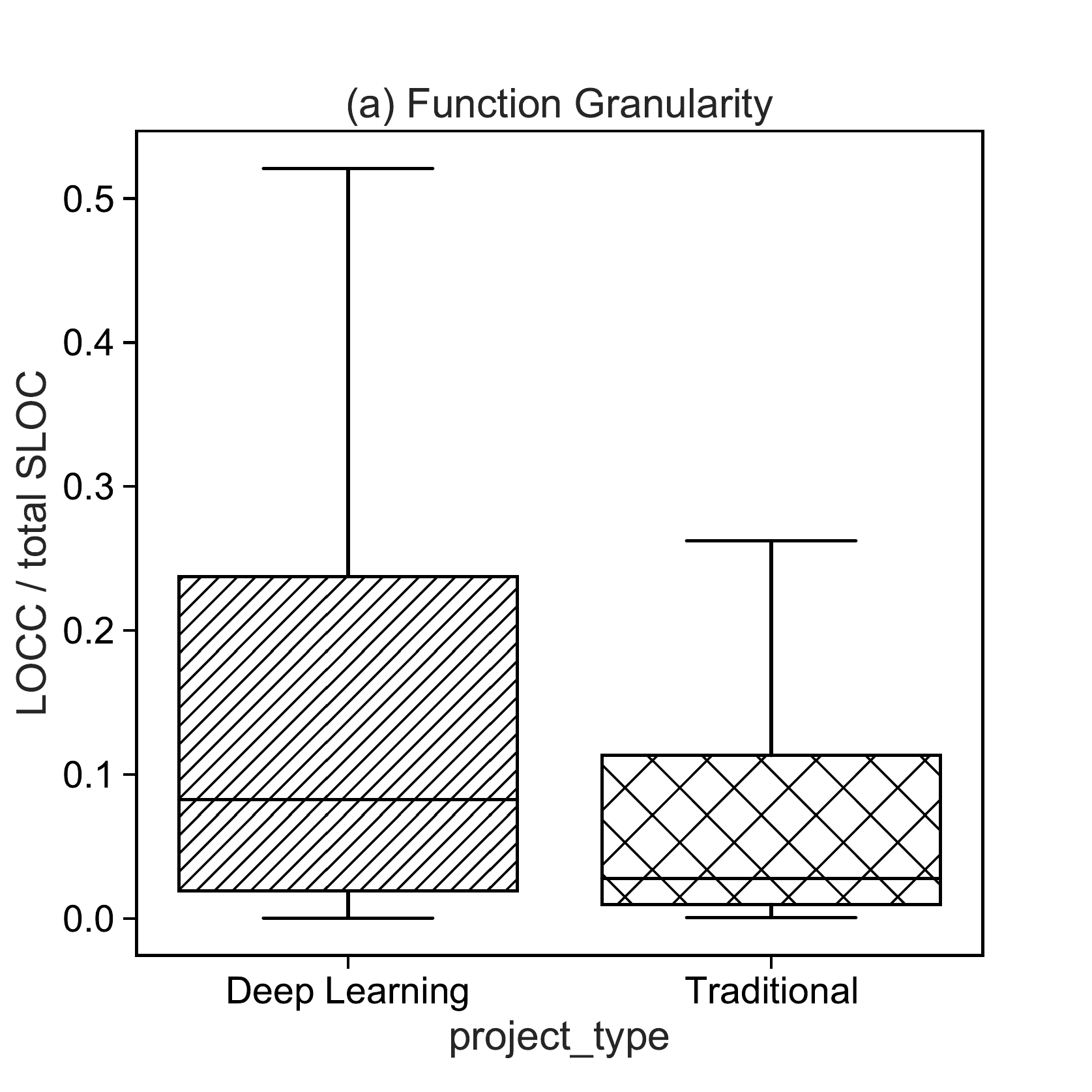}
\includegraphics[width=.48\textwidth]{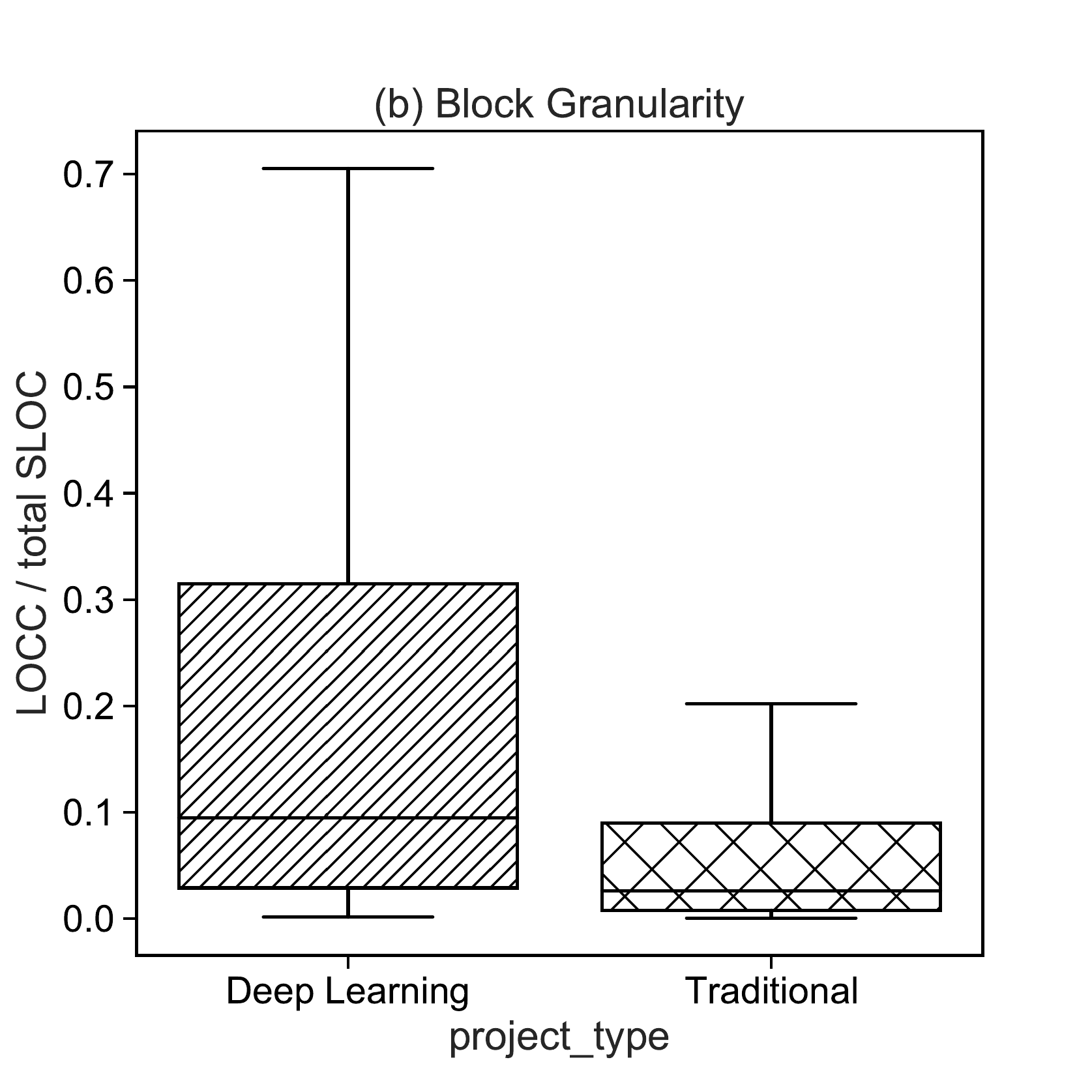}
\caption{Code Clones Occurrences in DL and Traditional Python Projects for Both Code Clones Granularities: (a) Function, (b) Block. \textbf{\textit{LOCC}}: Lines Of Code Clones, \textbf{\textit{SLOC}}: Source Lines Of Code.}
\label{fig:dl_trad_python}
\end{figure}

\paragraph{\textbf{Clone Occurrences by Project Type:}} 
Fig. \ref{fig:dl_trad_python} shows the comparison of clone occurrences between DL-based and traditional systems considering all clone types for both function and block granularity. The box plots represent \textit{clone density} defined by the normalized lines of code clones per lines of source code for both deep learning and traditional software systems. This metric computes the density of clones as a ratio of the total lines of cloned code and the total lines of source code in the corresponding systems. In Fig. \ref{fig:dl_trad_python}, we observe that the median of the clone density for deep learning systems is comparatively higher than that of traditional systems. We observe the similar differences for both function and block granularity. This suggests that deep learning systems tend to have higher proportions of cloned code compared to traditional systems. 

To investigate whether the observed differences that DL systems having higher density of clones compared to traditional systems are statistically significant i.e, the difference is not by chance, we perform Mann-Whitney Wilcoxon (MWW) tests \cite{Neuhauser2011} (two-tailed, significance at $<0.05$). We chose the MWW test because it is a non-parametric test and thus does not assume data to be normally distributed. Also, it can be applied on small sample sizes. We present our statistical test results in Table \ref{tab:p-value_types_python}. The column `All' in Table \ref{tab:p-value_types_python} shows the p-values for MWW test for all clone types. The p-values for function and block granularity are 1.78e-07 and 3.01e-10 respectively which are $< 0.05$ indicating that our observation is statistically significant. Thus, DL systems have higher density of clones compared to traditional systems. 

Now to observe the magnitude of the differences in clone density of DL and traditional systems, we analyze Cliff's delta effect sizes. As shown in Table \ref{tab:p-value_types_python}, the effect size in column `Total' (which includes all clone types) for function granularity belongs to the large category as it is equal to 0.8. Thus, we have an 80\% chance that deep learning code will have higher density of function clones than traditional code. Whereas for block granularity, we have an equal likelihood (0.53 $\sim$ 0.5) of having higher density of block clones in deep learning code in comparison to traditional code. 
When we consider all clone types, the observed higher clone density in deep learning systems in comparison to traditional systems is statistically significant with medium to large effect size.

\begin{figure}[htpb]
\centering
\includegraphics[width=.48\textwidth]{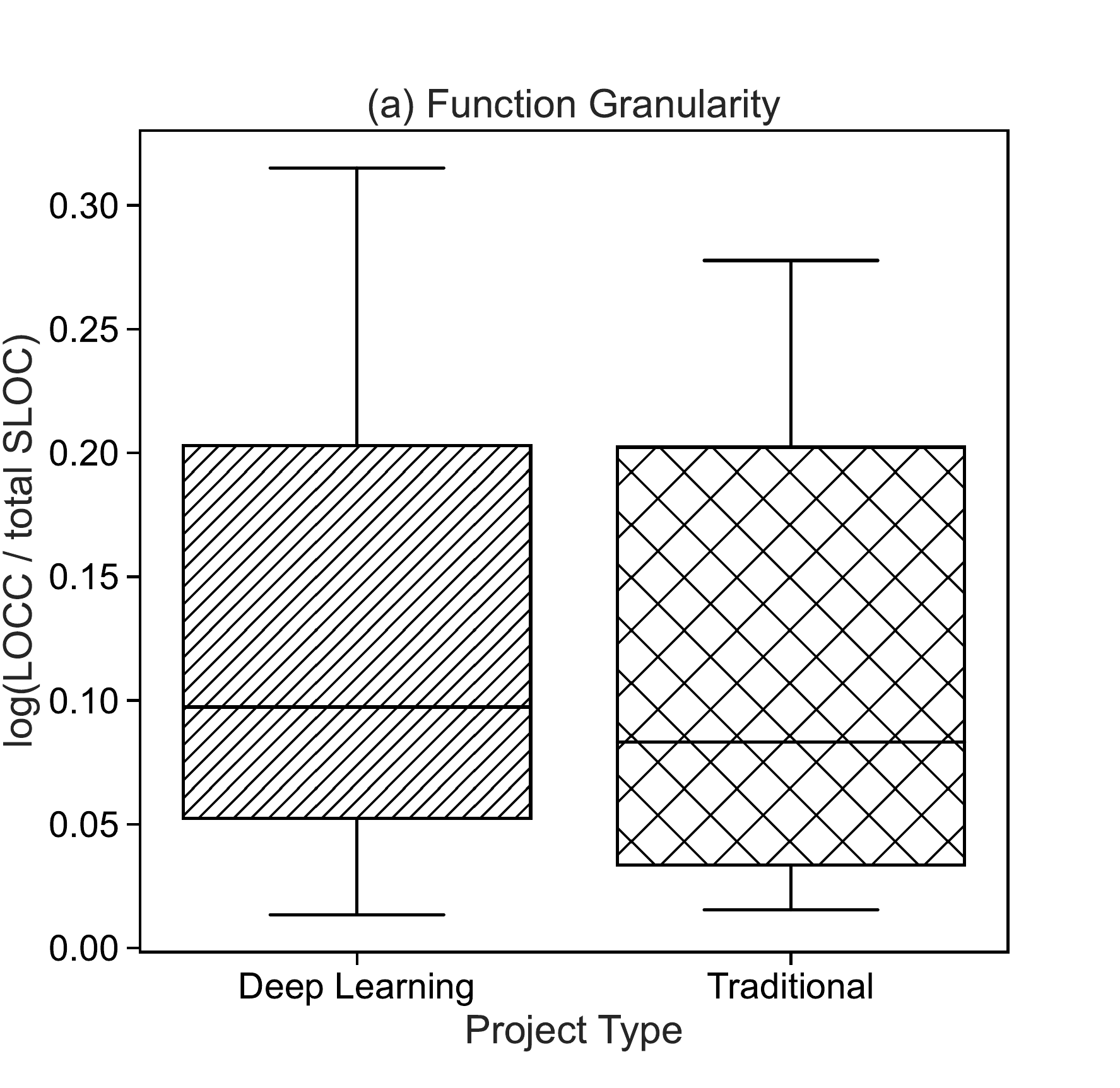}
\includegraphics[width=.48\textwidth]{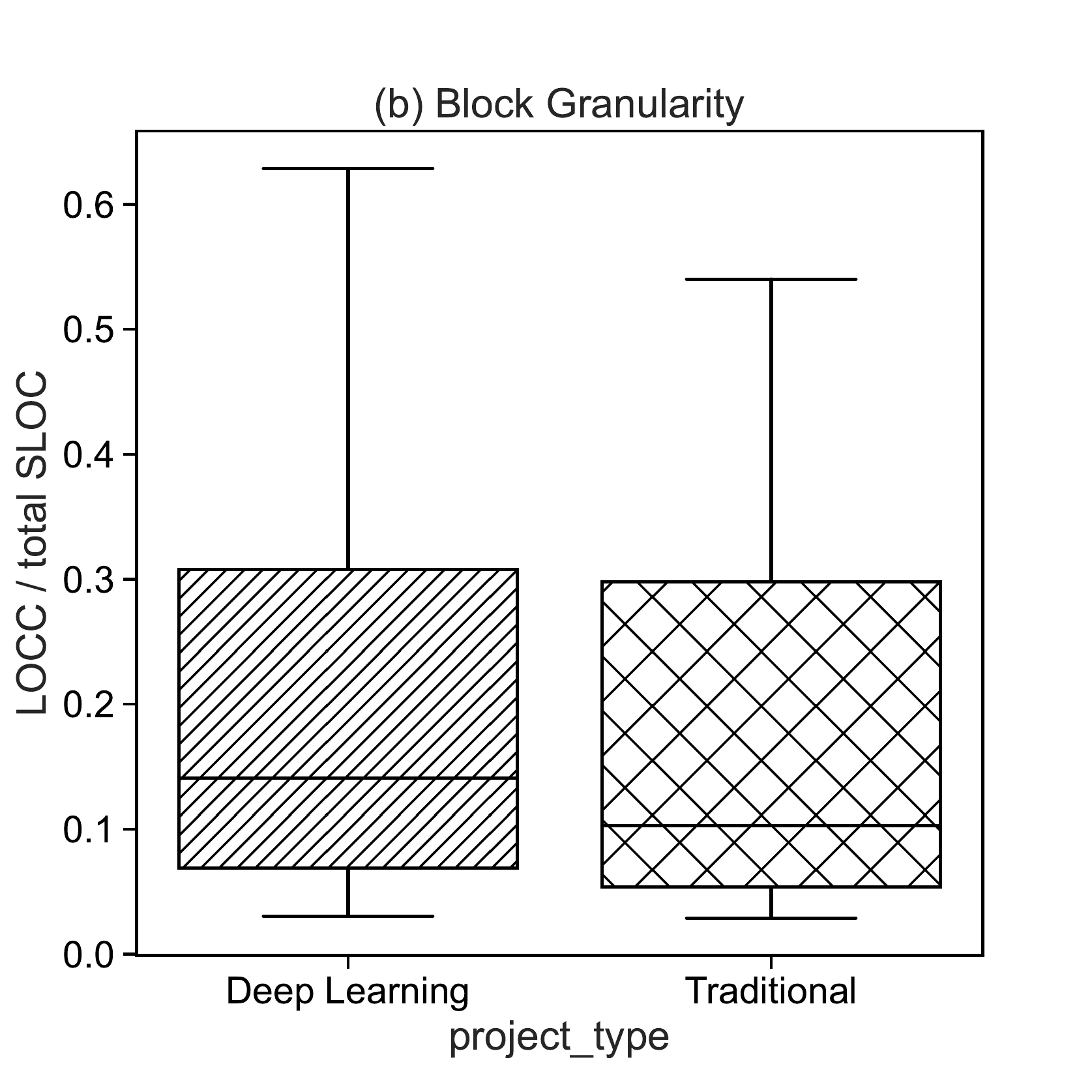}
\caption{Code Clones Occurrences in DL and Traditional Java Projects for Both Code Clones Granularities: (a) Function, (b) Block. \textbf{\textit{LOCC}}: Lines Of Code Clones, \textbf{\textit{SLOC}}: Source Lines Of Code.}
\label{fig:dl_trad_java}
\end{figure}

\begin{figure}[htpb]
\centering
\includegraphics[width=.48\textwidth]{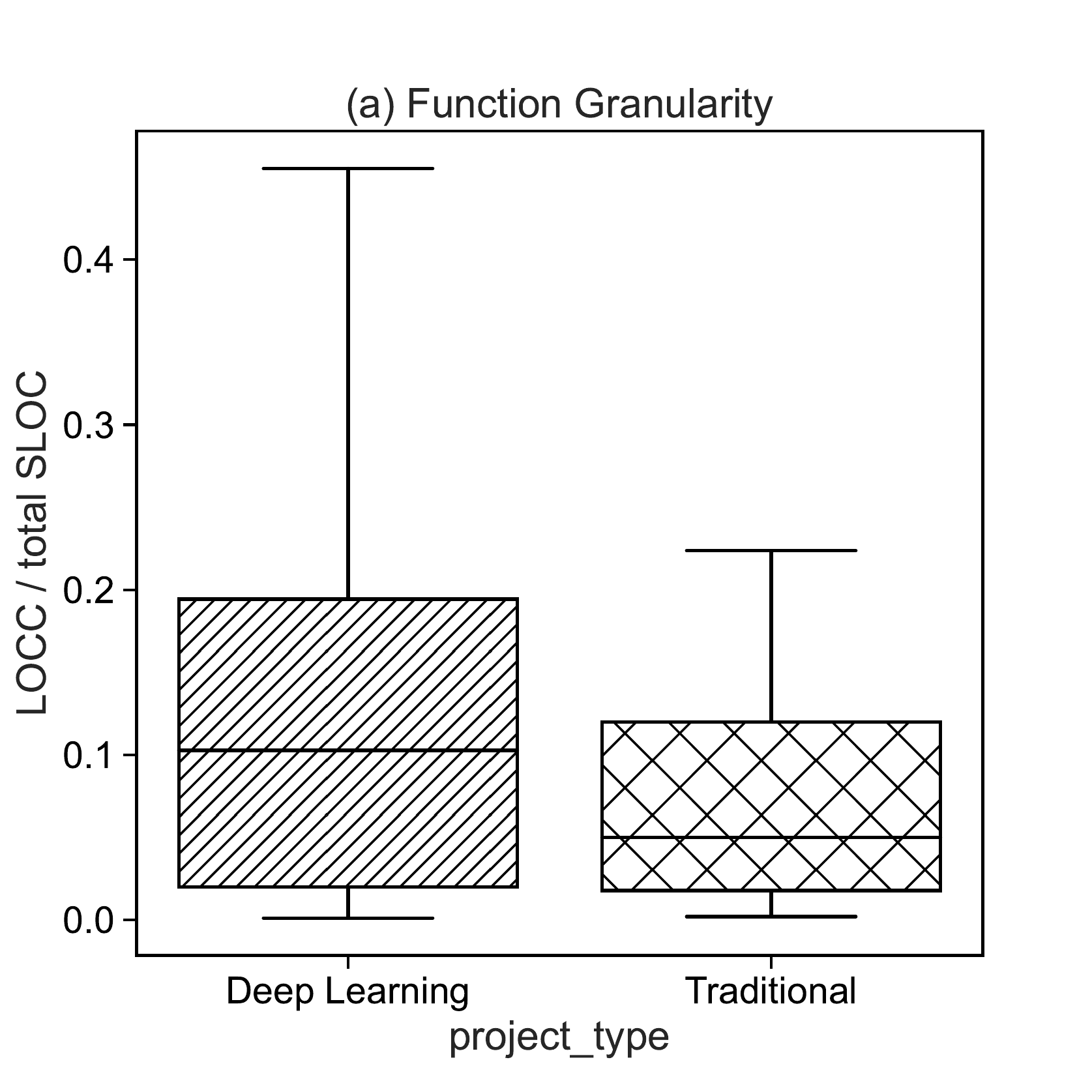}
\includegraphics[width=.48\textwidth]{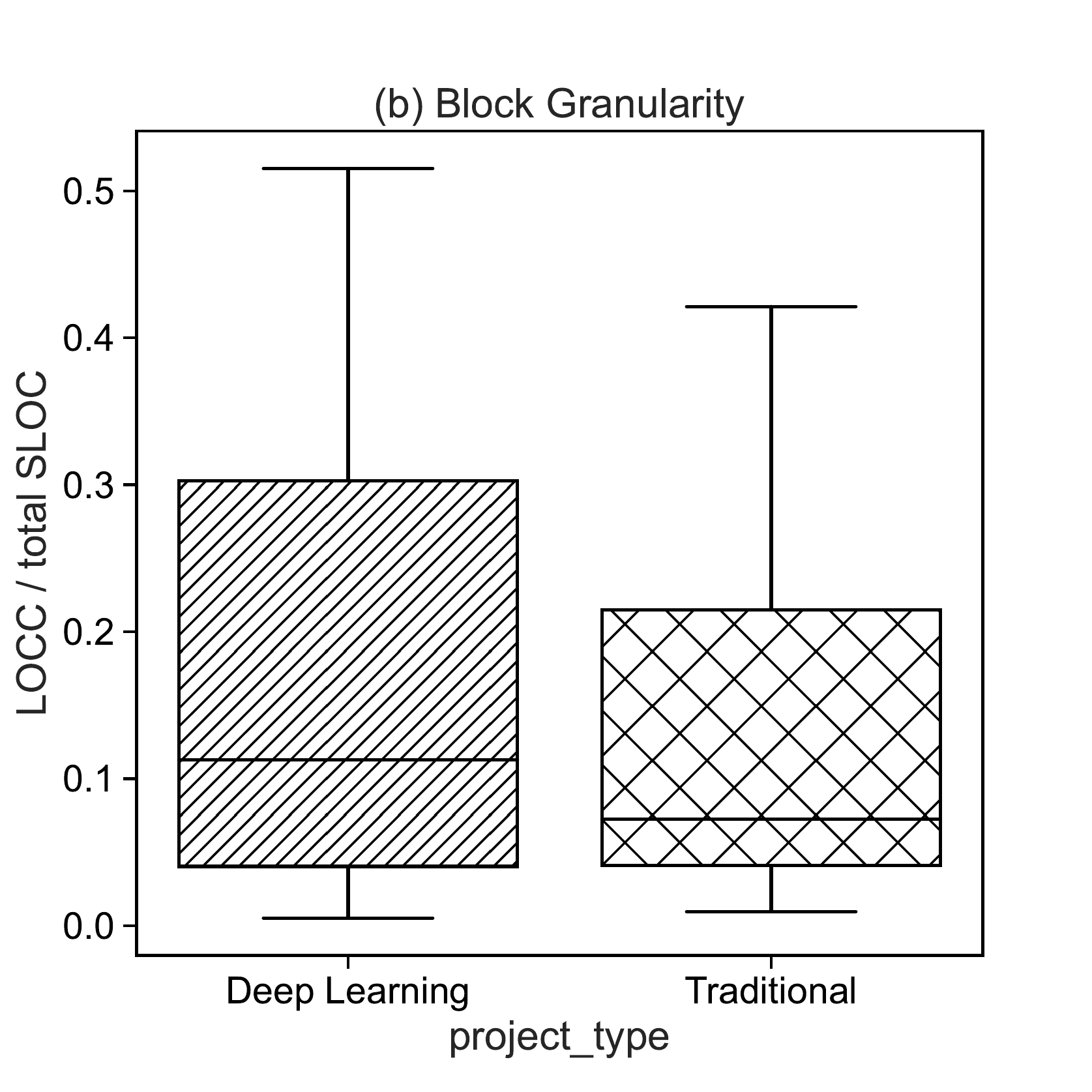}
\caption{Code Clones Occurrences in DL and Traditional C\# Projects for Both Code Clones Granularities: (a) Function, (b) Block. \textbf{\textit{LOCC}}: Lines Of Code Clones, \textbf{\textit{SLOC}}: Source Lines Of Code.}
\label{fig:dl_trad_csharp}
\end{figure}

We also studied 6 Java and 14 C\# systems to investigate whether the clone distributions observed in Python-based systems  generalize to other programming languages. 
Fig. \ref{fig:dl_trad_java} and Fig. \ref{fig:dl_trad_csharp} show the comparison of clone densities between DL-based and traditional Java and C\# systems respectively, considering all clone types for both function and block granularity. 
In both Fig. \ref{fig:dl_trad_java} and Fig. \ref{fig:dl_trad_csharp}, we observe that the median of the clone density for deep learning systems is comparatively higher than that of traditional systems. We observe similar differences for both function and block granularity. These observations for Java and C\# systems are similar to the observation from Python based systems. This suggests that deep learning systems tend to have higher proportions of cloned code compared to traditional systems. Table \ref{tab:p-value_types_java} and Table \ref{tab:p-value_types_csharp} show the results for statistical significance test (two-tailed MWW test, significant at $<$0.05) results. We do not observe statistically significant differences between the density of cloned code in Deep Learning and Traditional Java and C\# systems costrasting our results based on Python based systems. These differences between clone densities in Python system from that of Java and C\# systems could be explained by the differences in syntactic structures, compactness of code, availability and usage patterns of libraries, differences in object-oriented design, etc. 


We therefore conclude that deep learning systems tend to have higher proportions of cloned code compared to traditional systems. However, this may depend on confounding factors, i.e, programming languages, using the same libraries/frameworks, having the same decision logic across deep learning models construction, as explained by the difference in effect size for function (high) and block (medium) granularity, despite the differences in clone density being statistically significant for both granularities.

\begin{table}[htpb]
\caption{Mann-Whitney Test and Cliff's Delta Results between DL and Traditional Python Projects} 
\label{tab:p-value_types_python}
\resizebox{\textwidth}{!}{%
\begin{tabular}{|l|l|l|l|l|l|l|l|l|}
\hline
\multicolumn{1}{|c|}{\textbf{Clone Types}} &
  \multicolumn{2}{c|}{\textbf{Type 1}} &
  \multicolumn{2}{c|}{\textbf{Type 2}} &
  \multicolumn{2}{c|}{\textbf{Type 3}} &
  \multicolumn{2}{c|}{\textbf{All}} \\ \hline
\multicolumn{1}{|c|}{\textbf{Granularity}} &
  \multicolumn{1}{c|}{\textbf{Function}} &
  \multicolumn{1}{c|}{\textbf{Block}} &
  \multicolumn{1}{c|}{\textbf{Function}} &
  \multicolumn{1}{c|}{\textbf{Block}} &
  \multicolumn{1}{c|}{\textbf{Function}} &
  \multicolumn{1}{c|}{\textbf{Block}} &
  \multicolumn{1}{c|}{\textbf{Function}} &
  \multicolumn{1}{c|}{\textbf{Block}} \\ \hline
\textbf{p-value} &
  3.13e-06 &
  7.61e-08 &
  3.38e-05 &
  4.17e-06 &
  1.77e-03 &
  2.87e-04 &
  1.78e-07 &
  3.01e-10 \\ \hline
\textbf{Cliff's Delta} &
  0.58 &
  0.62 &
  0.46 &
  0.60 &
  0.32 &
  0.37 &
  0.8 &
  0.53 \\ \hline
\end{tabular}%
}
\end{table}

\begin{table}[htpb]
\caption{Mann-Whitney Test and Cliff's Delta Results between DL and Traditional Java Projects} 
\label{tab:p-value_types_java}
\resizebox{\textwidth}{!}{%
\begin{tabular}{|l|l|l|l|l|l|l|l|l|}
\hline
\multicolumn{1}{|c|}{\textbf{Clone Types}} &
  \multicolumn{2}{c|}{\textbf{Type 1}} &
  \multicolumn{2}{c|}{\textbf{Type 2}} &
  \multicolumn{2}{c|}{\textbf{Type 3}} &
  \multicolumn{2}{c|}{\textbf{All}} \\ \hline
\multicolumn{1}{|c|}{\textbf{Granularity}} &
  \multicolumn{1}{c|}{\textbf{Function}} &
  \multicolumn{1}{c|}{\textbf{Block}} &
  \multicolumn{1}{c|}{\textbf{Function}} &
  \multicolumn{1}{c|}{\textbf{Block}} &
  \multicolumn{1}{c|}{\textbf{Function}} &
  \multicolumn{1}{c|}{\textbf{Block}} &
  \multicolumn{1}{c|}{\textbf{Function}} &
  \multicolumn{1}{c|}{\textbf{Block}} \\ \hline
\textbf{p-value} &
  0.93 &
  0.92 &
  0.57 &
  0.64 &
  0.81 &
  0.64 &
  0.73 &
  0.60 \\ \hline
\textbf{Cliff's Delta} &
 0.055  &
 0.066  &
 0.222  &
 0.2  &
 0.111  &
 0.2  &
  0.018 &
  0.111\\ \hline
\end{tabular}%
}
\end{table}

\begin{table}[htpb]
\caption{Mann-Whitney Test and Cliff's Delta Results between DL and Traditional C\# Projects} 
\label{tab:p-value_types_csharp}
\resizebox{\textwidth}{!}{%
\begin{tabular}{|l|l|l|l|l|l|l|l|l|}
\hline
\multicolumn{1}{|c|}{\textbf{Clone Types}} &
  \multicolumn{2}{c|}{\textbf{Type 1}} &
  \multicolumn{2}{c|}{\textbf{Type 2}} &
  \multicolumn{2}{c|}{\textbf{Type 3}} &
  \multicolumn{2}{c|}{\textbf{All}} \\ \hline
\multicolumn{1}{|c|}{\textbf{Granularity}} &
  \multicolumn{1}{c|}{\textbf{Function}} &
  \multicolumn{1}{c|}{\textbf{Block}} &
  \multicolumn{1}{c|}{\textbf{Function}} &
  \multicolumn{1}{c|}{\textbf{Block}} &
  \multicolumn{1}{c|}{\textbf{Function}} &
  \multicolumn{1}{c|}{\textbf{Block}} &
  \multicolumn{1}{c|}{\textbf{Function}} &
  \multicolumn{1}{c|}{\textbf{Block}} \\ \hline
\textbf{p-value} &
  0.64 &
  0.39 &
  0.62 &
  0.42 &
  0.05 &
  0.22 &
  0.22 &
  0.31 \\ \hline
\textbf{Cliff's Delta} & 
  0.115 &
  0.197 &
  0.119 &
  0.183 &
 0.438  &
 0.510  &
  0.159 &
  0.127 \\ \hline
\end{tabular}%
}
\end{table}

\begin{figure}[htpb]
\centering
\includegraphics[width=.45\textwidth]{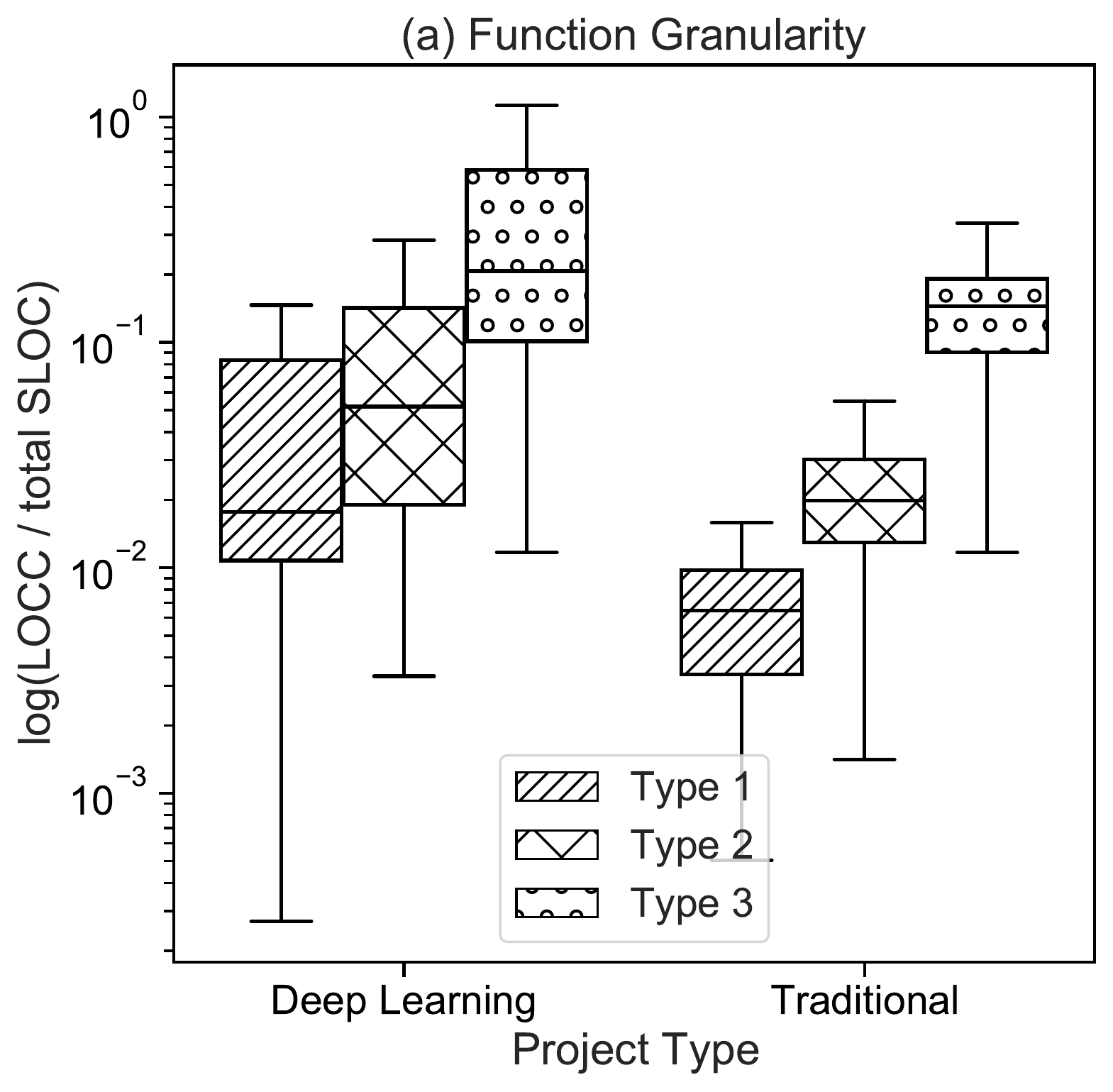}
\includegraphics[width=.45\textwidth]{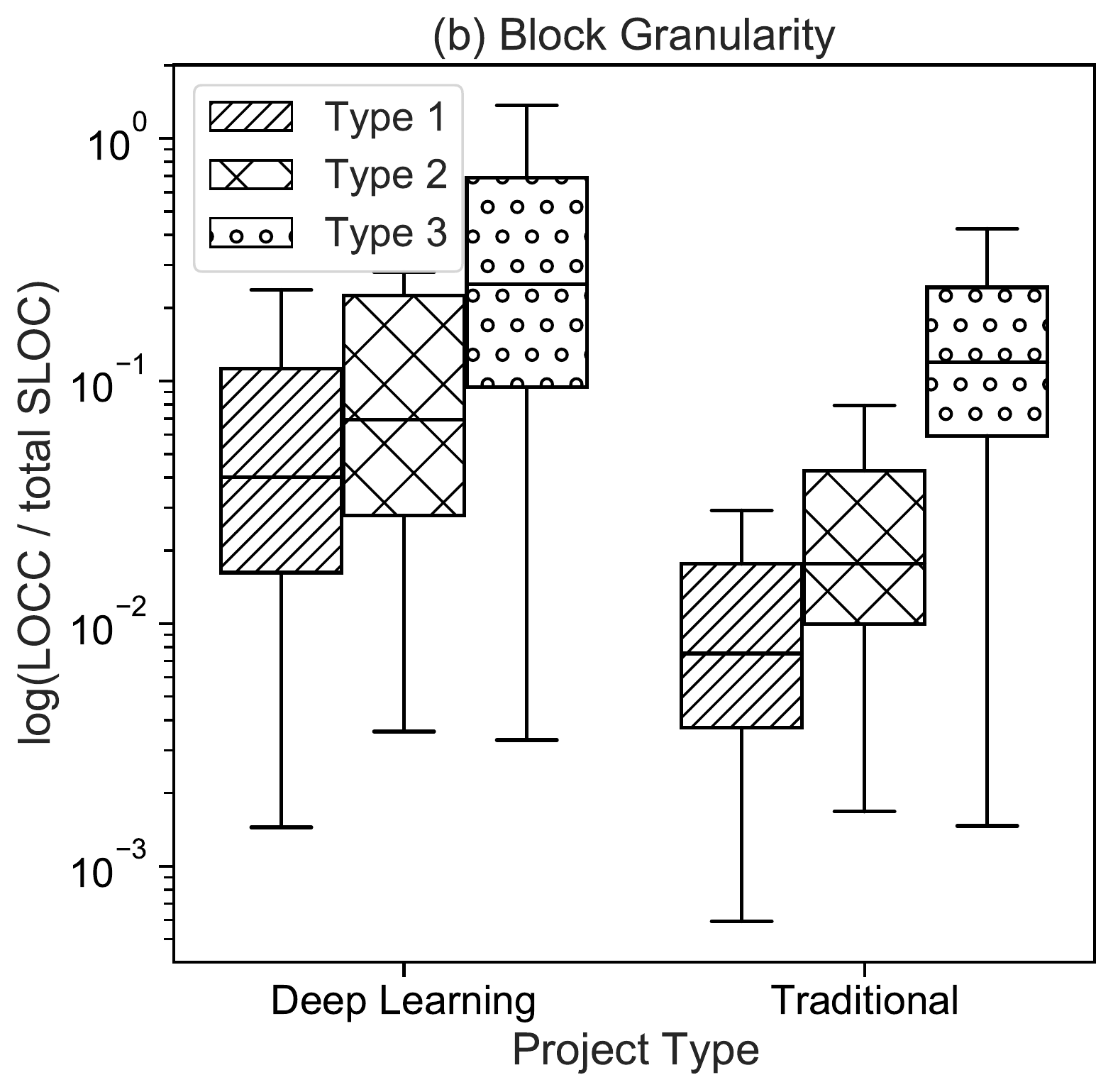}
\caption{Clone Density in DL and Traditional Python Projects for Clone Types and Granularity \textit{\textbf{LOCC: } Lines Of Code Clones}}
\label{fig:dl_trad_clone_type_python}
\end{figure}

\begin{figure}[htpb]
\centering
\includegraphics[width=.45\textwidth]{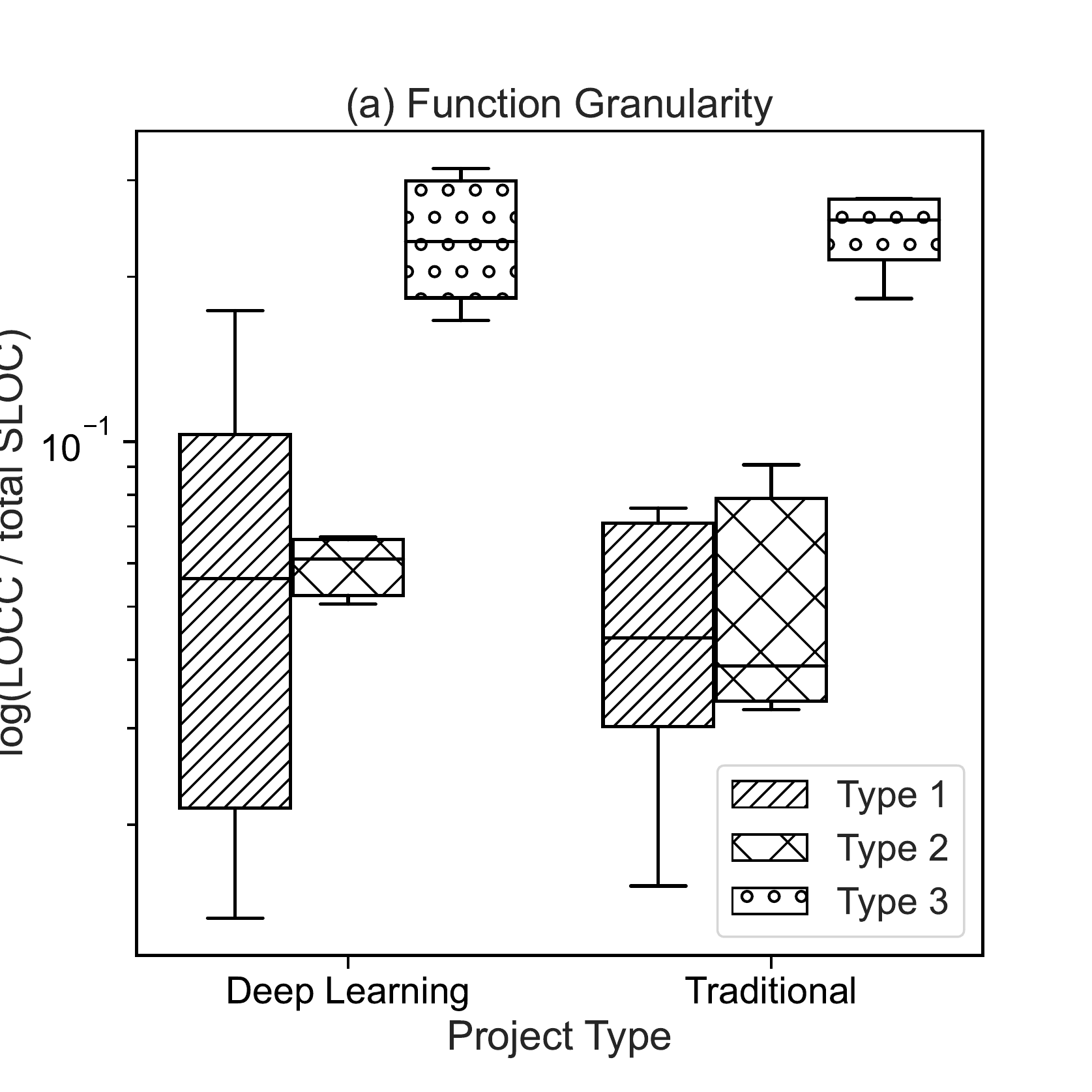}
\includegraphics[width=.45\textwidth]{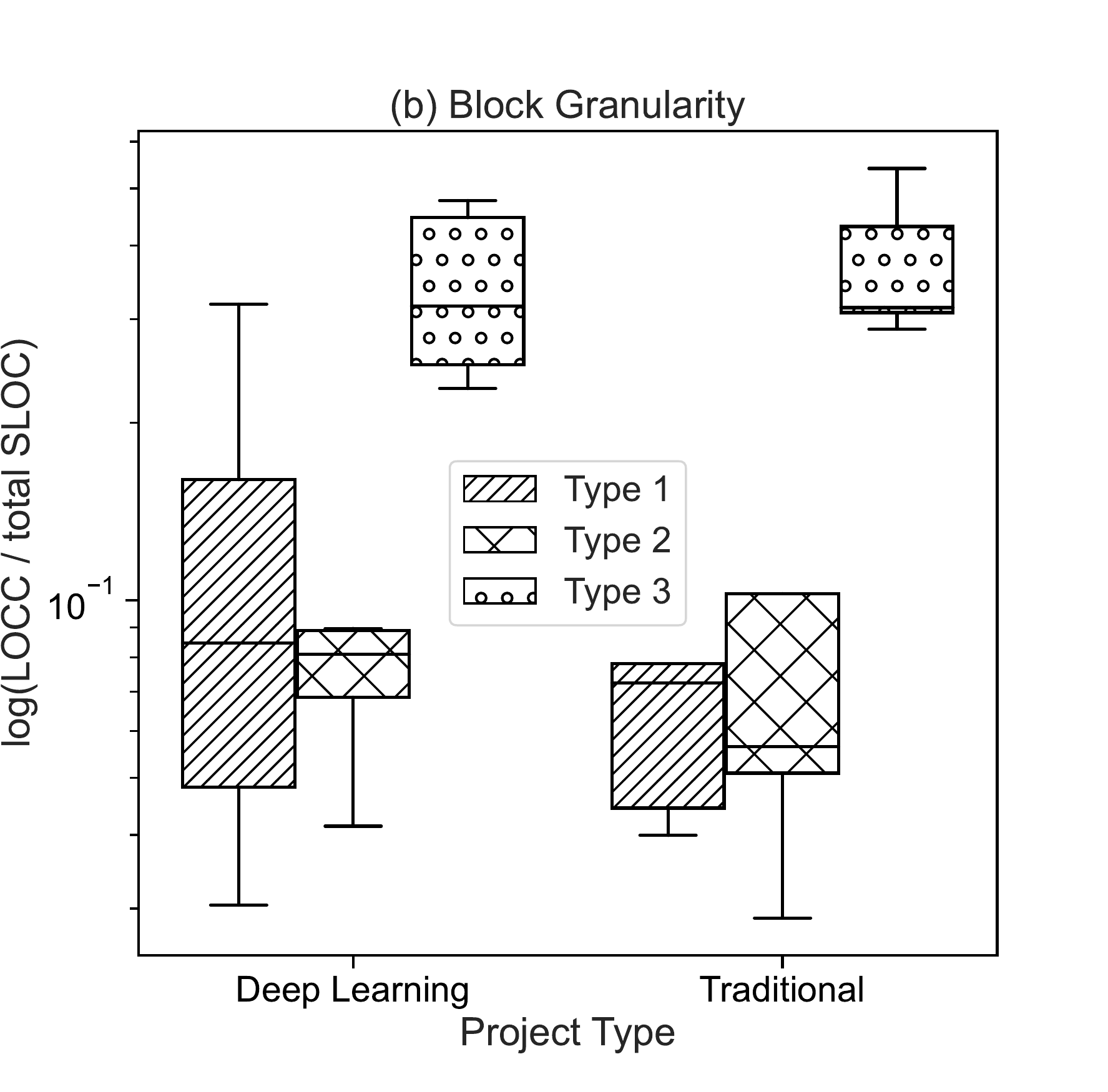}
\caption{Clone Density in DL and Traditional Java Projects for Clone Types and Granularity \textit{\textbf{LOCC: } Lines Of Code Clones}}
\label{fig:dl_trad_clone_type_java}
\end{figure}

\begin{figure}[htpb]
\centering
\includegraphics[width=.45\textwidth]{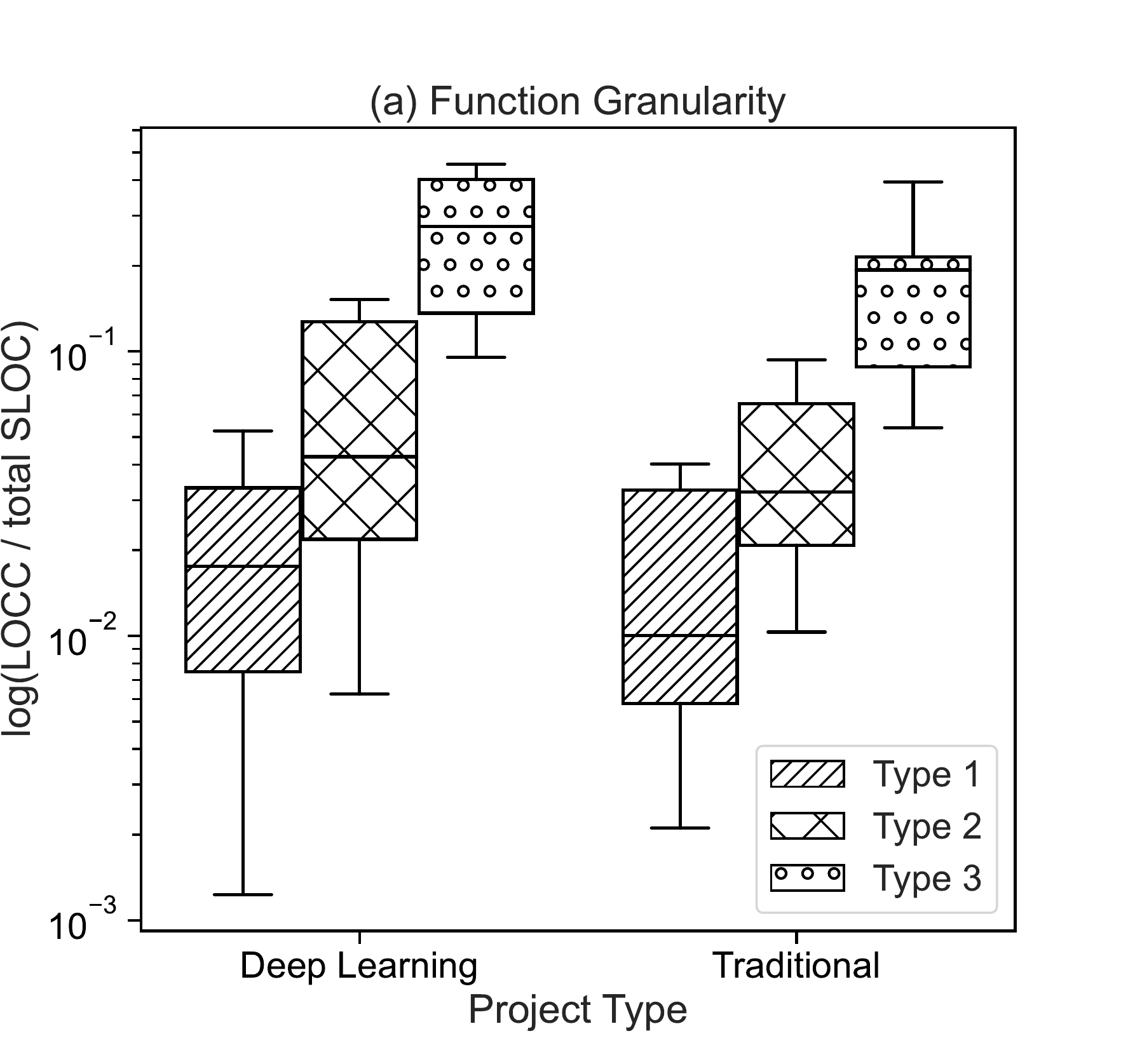}
\includegraphics[width=.45\textwidth]{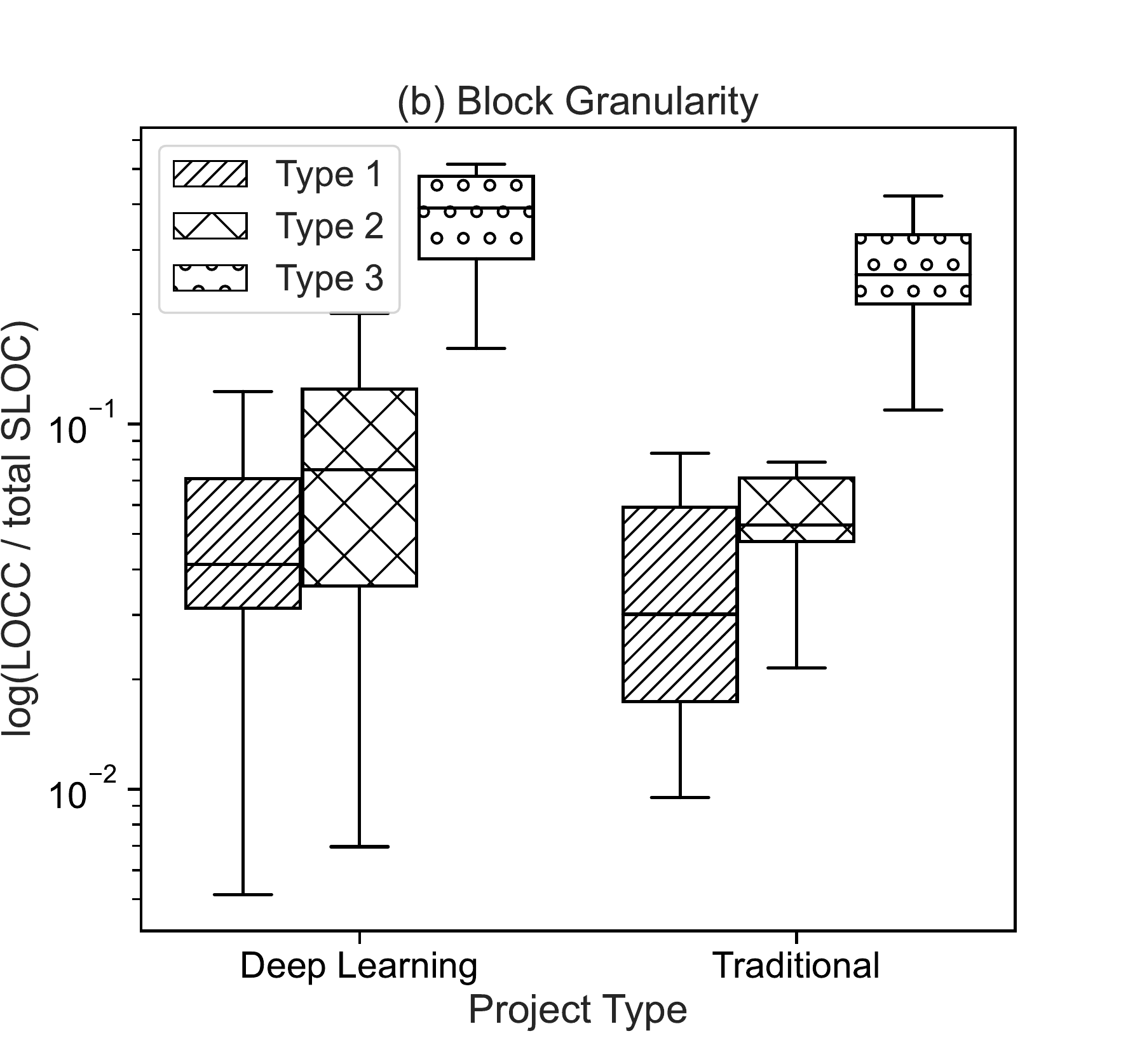}
\caption{Clone Density in DL and Traditional C\# Projects for Clone Types and Granularity \textit{\textbf{LOCC: } Lines Of Code Clones}}
\label{fig:dl_trad_clone_type_csharp}
\end{figure}

\paragraph{\textbf{Code Clone occurrences by Clone Type:}} 
In Fig. \ref{fig:dl_trad_clone_type_python}, we compare clone densities for deep learning and traditional systems regarding individual clone types (Type 1, Type 2, and Type 3) for both function and block granularity. To calculate clone density for individual clone types, we divide the total lines of cloned code for a particular clone type by the total number of source code lines in a given system. We calculate clone density for both function and block granularity. We use a log scale for the y-axis to make the results more clear.  
As shown in Fig. \ref{fig:dl_trad_clone_type_python}, for deep learning systems, the density of Type 3 clones is the highest followed by Type 2 and Type 1 clones respectively. We observe the same trends in comparative densities for all types of clones in traditional systems, and for both function and block granularity in both types of systems. 
Now, when we compare the clone densities in deep learning and traditional systems based on the box-plots in Fig. \ref{fig:dl_trad_clone_type_python}, we observe that for all types of clones, deep learning systems have higher median for clone densities compared to that of traditional systems. The same overall trend holds for both function and block granularity. This suggests that deep learning systems tend to have higher density of cloned code compared to traditional systems regarding all three clone types. 

To verify whether these differences of deep learning systems having higher densities of clones compared to traditional systems are statistically significant, we perform MWW tests (two-tailed, significance at 0.05) on the corresponding clone densities for all individual clone types, for both function and block granularity. We also compute the Cliff's delta effect size to determine the magnitude of the differences in clone densities for DL and traditional systems. 
Table \ref{tab:p-value_types_python} shows the p-values of the MWW tests along with the values for the corresponding effect size. 
We observe that there is a statistically significant difference between clone densities in deep learning code and traditional code with respect to Type 1 clones with p-values of 3.13e-06 and 7.61e-08, which are $<$ 0.05 for both function and block granularities, respectively. The values of Cliff's Delta are 0.58 and 0.62 which belong to the `medium' category. 
We also found a statistically significant difference in clone densities for Type 2 clones (p-values of 3.38e-05, 4.17e-06). The values of effect size for Type 2 for both function and block granularities are medium with values of 0.46 and 0.60. Similarly, we also obtained a statistically significant difference in clone densities between deep learning code and traditional code regarding Type 3 clones. The effect size for Type 3 clones is small with values of 0.32 for function granularity and 0.37 for the block granularity.  
Overall, the results of our clone-type based analysis show that Type 3 clones comprise the highest proportion of clones in deep learning code. Moreover, for all clone types, deep learning code has higher density of clones than traditional systems, and these differences are statistically significant. This indicates an overall trend of deep learning systems having more clones than traditional systems, although the magnitude of the differences may not always be `large'.

Fig. \ref{fig:dl_trad_clone_type_java} and Fig. \ref{fig:dl_trad_clone_type_csharp} show the distributions of clone densities in DL and traditional Java and C\# systems respectively. We observe that the median density of clones in DL systems tend to be higher compared to traditional systems for clone types in Java and C\# systems except for Type 3 clones in Java systems (Fig. \ref{fig:dl_trad_clone_type_java}). Our clone-type centric analysis show that densities of cloned code tend to be higher in DL systems compared to traditional systems. However, we do not observe these differences to be statistically significant.

As the dissimilarity threshold for the clone detection tool is known to be important confounding factor for the empirical analysis of clones, we also perform the analysis with 20\% dissimilarity threshold for NiCad. We presented our results for 20\% threshold in Appendix \ref{appRQ1} (Fig. \ref{fig:dl_trad_java_20}, Fig. \ref{fig:dl_trad_clone_type_java_20}, Fig. \ref{fig:dl_trad_csharp_20}, and Fig. \ref{fig:dl_trad_clone_type_csharp_20}). Our results show that the overall trends in the prevalence of clones in deep learning and traditional systems for 20\% threshold do not change from what we observed for 30\% threshold.

These results support our hypothesis about the existence of a higher proportion of code clones in deep learning code compared to traditional systems. The observed prevalence of code clones in deep learning systems underscores the importance of investigating the reasons for such cloning practices and their impacts on the quality of deep learning systems, which we address in the remaining research questions. However, further investigation is necessary to generalize the findings and to identify the factors that influence the prevalence of clones in DL and traditional software systems.

\vspace{4mm}
\fbox{\begin{minipage}{30em}
\textit{\textbf{Summary of findings (RQ1):}} As shown by our experimental results, the density of code clones in DL-based systems tend to be higher than that of traditional systems, and the difference is statistically significant for Python systems. Regarding clone types, all three clone types (Type 1, Type 2, and Type 3) are more prevalent in DL-based systems than in traditional software systems. The comparative higher prevalence of clones in DL systems compared to traditional systems are likely to be programming language dependent.

\end{minipage}}

\subsection{\textbf{RQ\(2\): How are code clones distributed in deep learning code in comparison to traditional source code?}}

As we observed in RQ1, the density of code clones in deep learning code tends to be higher compared to traditional systems. We aim to further explore the distribution of code clones in terms of their locations. Since, distant code clones may hinder the maintenance process by adding navigation and code comprehension overhead, it is of interest to study how the code clones are distributed in the deep learning system in terms of their locations. 

We categorize the detected code clones classes by their location based on the taxonomy proposed by Kapser and Godfrey \cite{kapser2003toward}. If a clone class contains code fragments that are all from the same file, we label them as belonging to the `Same file' category. When the clone fragments are from the same directory but from different files, we assign them to the `Same directory' category. Otherwise, if the clone fragments of a clone class are from files from different directories, we categorize them as belonging to the `Different directories' category (further details about this classification are provided in section \ref{subsubsec:LocationTaxonomyMethod}). We then count the proportion of lines of code clones of different location categories for deep learning and traditional systems. We also perform the same analysis for individual types of clones (Type 1, Type 2, and Type 3) to have insights on their comparative location-based distributions in DL and traditional systems.The analysis based on the proportion (in percentage) of the lines of code clones, shows the distribution of clones from a relative code size point of view. A few larger clone fragments in closer proximity are likely to have less navigation overhead than a higher number of small clone fragments scattered in distant location, even if they have equal total LOC of the larger clone fragments. So, fragment-based analysis of the distribution of clones is also important to have insights on the potential impacts of clones on the maintenance of the systems. So, we also analyze the location-based distribution of the percentages of clone fragments for different types of clones. 
In addition, we manually investigate samples of cloned fragments from each location category to gain insights about the characteristics of the clones and to better understand the intents and potential impacts of the proximity of clones (in the code base) and their relative distribution, and impacts on software quality. 

\begin{figure}[htpb]
\centering
\includegraphics[scale=0.7]{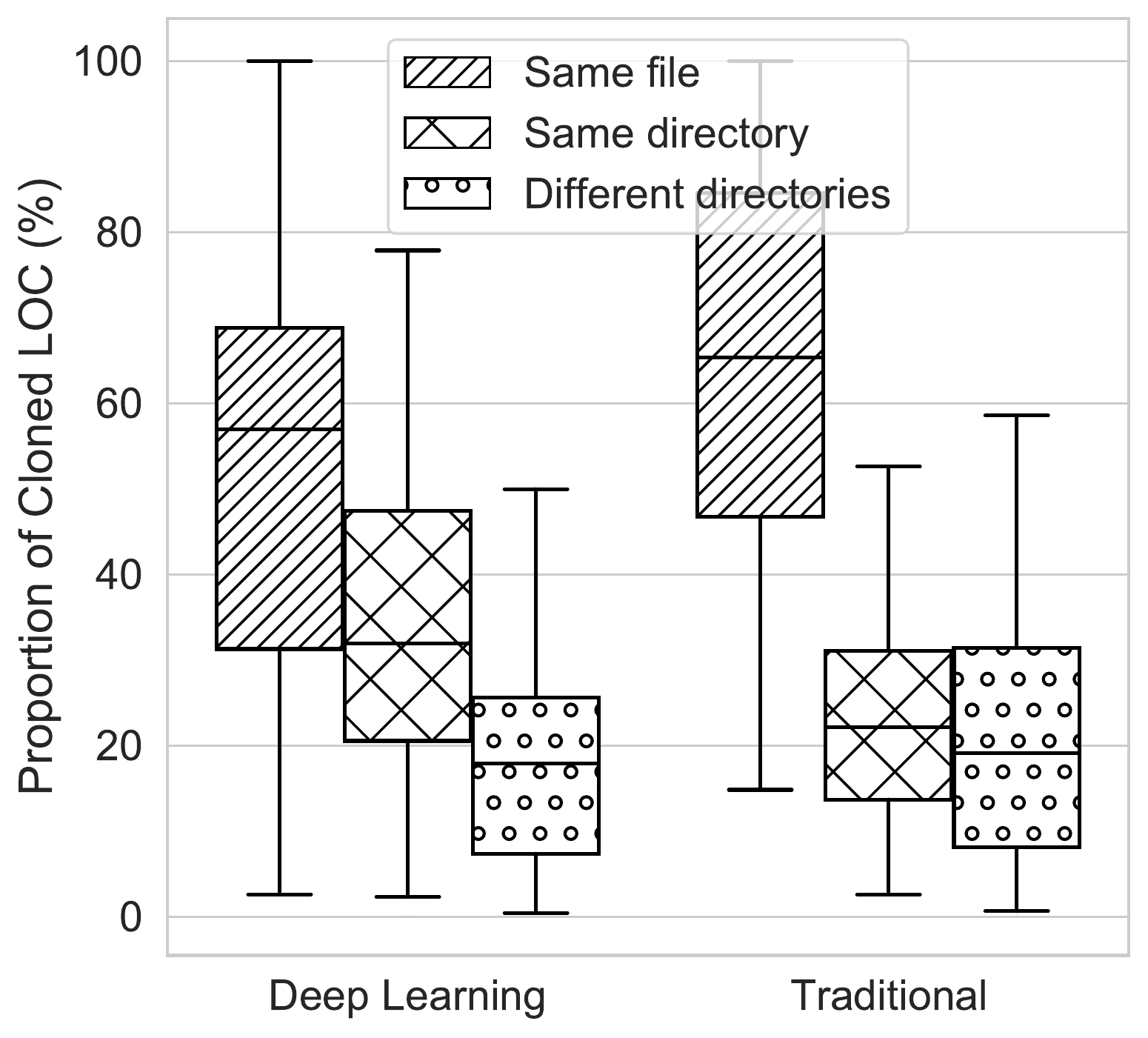}
\caption{Code Clones Distribution by Location in DL and Traditional Python Systems Regarding Percentage of Lines of Code Clones (LOCC). i.e, (LOCC/total LOCC)x 100} 
\label{fig:CloneLocation_python}
\end{figure}

\begin{figure}[htpb]
\centering
\includegraphics[scale=0.7]{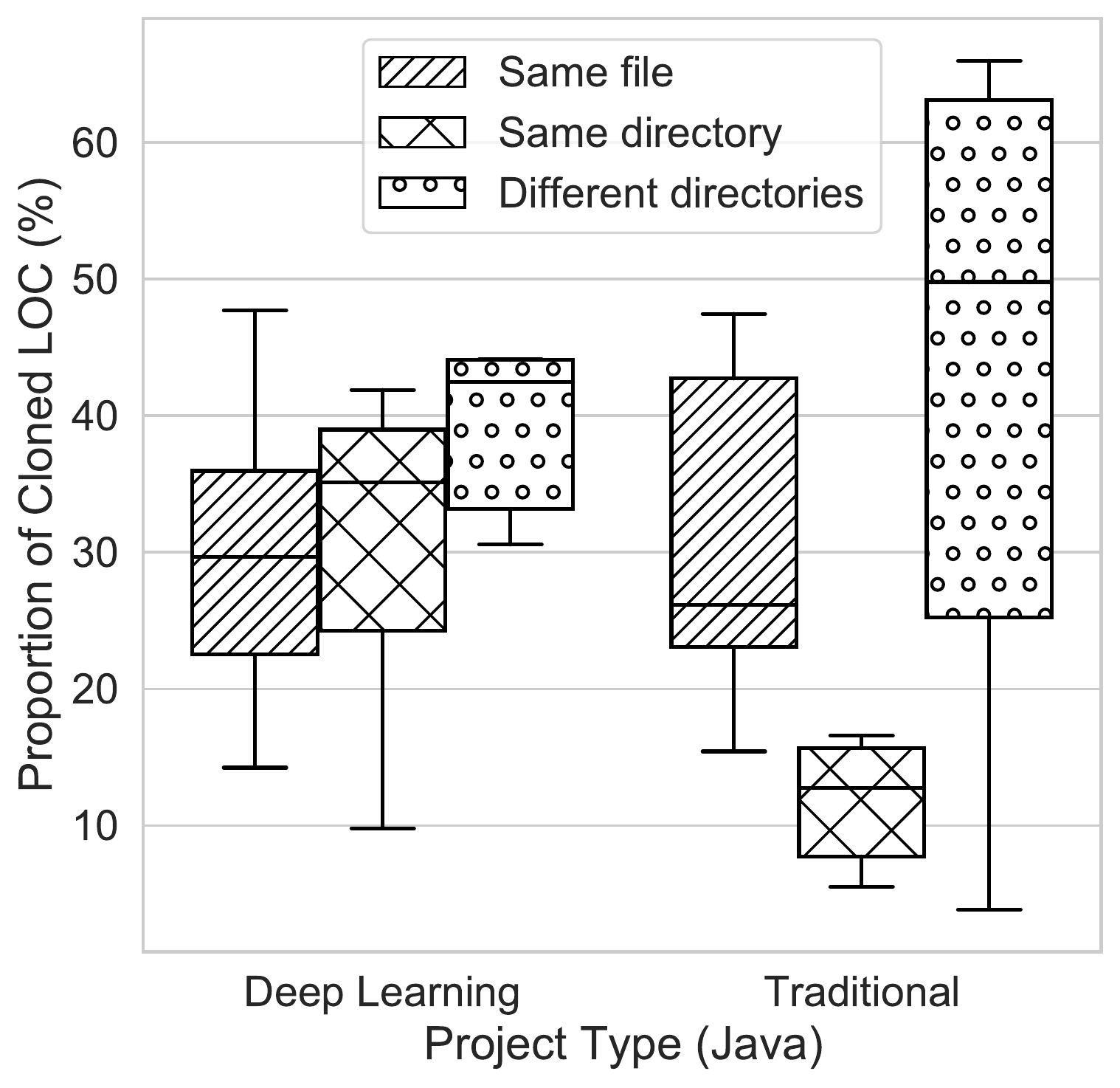}
\caption{Code Clones Distribution by Location in DL and Traditional Java Systems Regarding Percentage of Lines of Code Clones (LOCC). i.e, (LOCC/total LOCC)x 100} 
\label{fig:CloneLocation_java}
\end{figure}

\begin{figure}[htpb]
\centering
\includegraphics[scale=0.65]{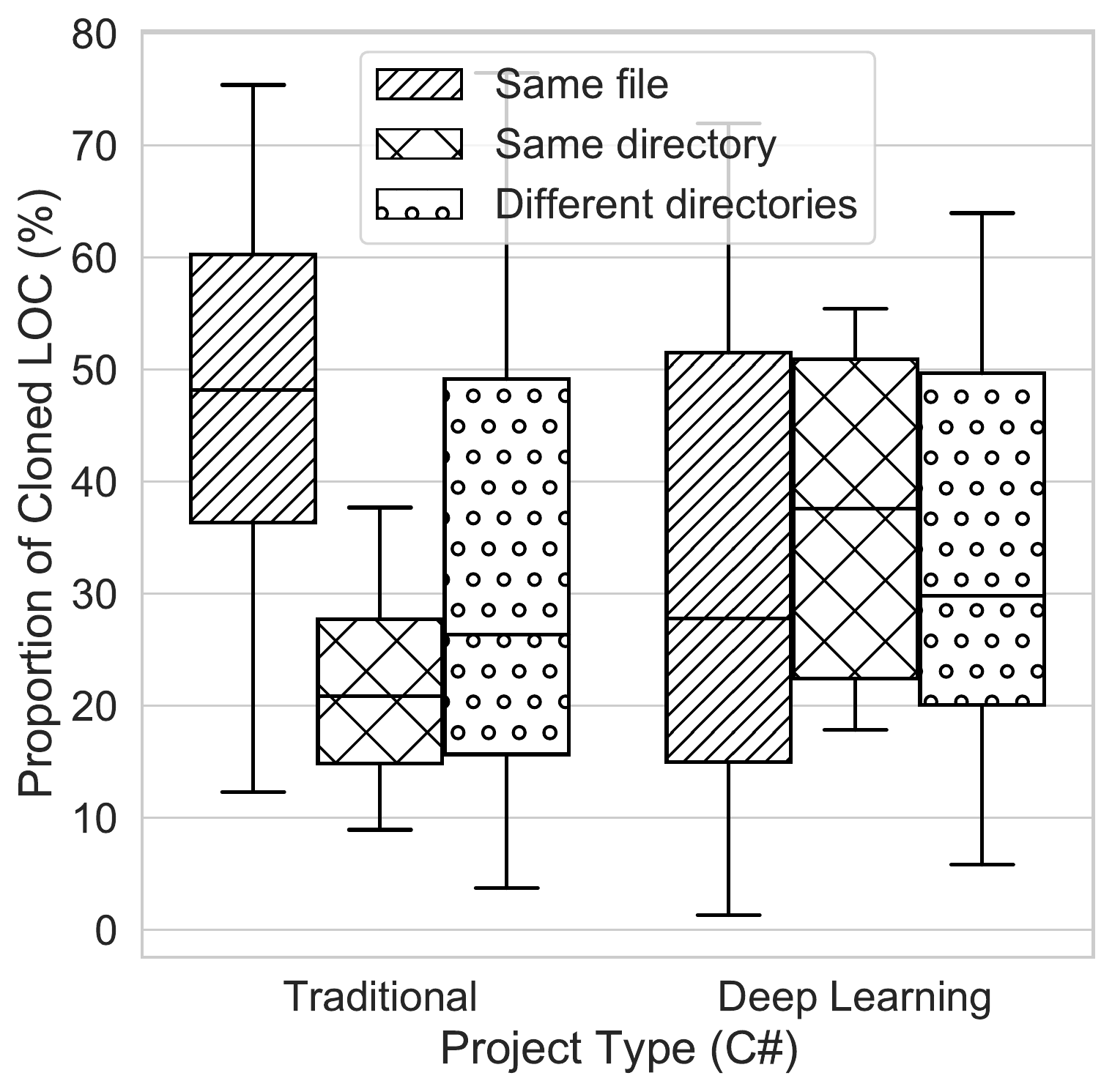}
\caption{Code Clones Distribution by Location in DL and Traditional C\# Systems Regarding Percentage of Lines of Code Clones (LOCC). i.e, (LOCC/total LOCC)x 100} 
\label{fig:CloneLocation_csharp}
\end{figure}

\paragraph{\textbf{Overall location-based distribution of clones:}} Fig. \ref{fig:CloneLocation_python} shows the distribution of code clones (for the function granularity) in DL and traditional Python source code, regarding the locations of the clones. 
Based on the median values of the percentages of cloned lines of code for each location type in deep learning systems, we observe the following location-based distribution of clones: the percentages of cloned lines of code in `Same file' are higher than that in `same directory' which in turn are higher than the percentages of cloned lines of code in `different directories'. 
In traditional systems, we observe a location-based distribution similar to that of DL systems with the highest percentage of cloned lines being in the `same file' followed by the `same directory' category and finally the `different directories' category. The distributions of `same directory' and `different directories' categories overlap significantly for clones in traditional code.

\begin{table}[htpb]
\caption{Mann-Whitney Test and Cliff’s Delta Results Regarding the Distributions of Clones in  DL and Traditional (Trad) Python Projects.}
\label{tab:pvalueLocationCloneType_python}
\resizebox{\textwidth}{!}{%
\begin{tabular}{llllllll}
\cline{3-8}
 &
  \multicolumn{1}{l|}{} &
  \multicolumn{6}{c|}{\textbf{Location}} \\ \cline{3-8} 
 &
  \multicolumn{1}{l|}{} &
  \multicolumn{2}{c|}{\textbf{SF-SD}} &
  \multicolumn{2}{c|}{\textbf{SF-DD}} &
  \multicolumn{2}{c|}{\textbf{SD-DD}} \\ \hline
\multicolumn{1}{|l|}{\textbf{Clone Type}} &
  \multicolumn{1}{l|}{\textbf{Proj Type}} &
  \multicolumn{1}{l|}{\textbf{p-value}} &
  \multicolumn{1}{l|}{\textbf{CD}} &
  \multicolumn{1}{l|}{\textbf{p-value}} &
  \multicolumn{1}{l|}{\textbf{CD}} &
  \multicolumn{1}{l|}{\textbf{p-value}} &
  \multicolumn{1}{l|}{\textbf{CD}} \\ \hline
\multicolumn{1}{|l|}{\multirow{2}{*}{\textbf{ALL}}} &
  \multicolumn{1}{l|}{\textbf{DL}} &
  \multicolumn{1}{l|}{\textbf{2.911e-04}} &
  \multicolumn{1}{l|}{\textbf{0.4}} &
  \multicolumn{1}{l|}{\textbf{2.9e-10}} &
  \multicolumn{1}{l|}{\textbf{0.7}} &
  \multicolumn{1}{l|}{\textbf{4.79e-05}} &
  \multicolumn{1}{l|}{\textbf{0.46}} \\ \cline{2-8} 
\multicolumn{1}{|l|}{} &
  \multicolumn{1}{l|}{\textbf{Trad}} &
  \multicolumn{1}{l|}{\textbf{8.34e-13}} &
  \multicolumn{1}{l|}{\textbf{0.83}} &
  \multicolumn{1}{l|}{\textbf{2.05e-11}} &
  \multicolumn{1}{l|}{\textbf{0.81}} &
  \multicolumn{1}{l|}{0.15} &
  \multicolumn{1}{l|}{0.13} \\ \hline
\multicolumn{1}{|l|}{\multirow{2}{*}{\textbf{Type 1}}} &
  \multicolumn{1}{l|}{\textbf{DL}} &
  \multicolumn{1}{l|}{\textbf{1.94e-3}} &
  \multicolumn{1}{l|}{\textbf{0.41}} &
  \multicolumn{1}{l|}{\textbf{0.018}} &
  \multicolumn{1}{l|}{\textbf{0.32}} &
  \multicolumn{1}{l|}{\textbf{0.03}} &
  \multicolumn{1}{l|}{\textbf{0.28}} \\ \cline{2-8} 
\multicolumn{1}{|l|}{} &
  \multicolumn{1}{l|}{\textbf{Trad}} &
  \multicolumn{1}{l|}{0.16} &
  \multicolumn{1}{l|}{0.15} &
  \multicolumn{1}{l|}{\textbf{0.01}} &
  \multicolumn{1}{l|}{\textbf{0.42}} &
  \multicolumn{1}{l|}{\textbf{0.02}} &
  \multicolumn{1}{l|}{\textbf{0.35}} \\ \hline
\multicolumn{1}{|l|}{\multirow{2}{*}{\textbf{Type 2}}} &
  \multicolumn{1}{l|}{\textbf{DL}} &
  \multicolumn{1}{l|}{\textbf{7.64e-09}} &
  \multicolumn{1}{l|}{\textbf{0.76}} &
  \multicolumn{1}{l|}{\textbf{5.98e-08}} &
  \multicolumn{1}{l|}{\textbf{0.77}} &
  \multicolumn{1}{l|}{0.09} &
  \multicolumn{1}{l|}{0.21} \\ \cline{2-8} 
\multicolumn{1}{|l|}{} &
  \multicolumn{1}{l|}{\textbf{Trad}} &
  \multicolumn{1}{l|}{\textbf{3.4e-10}} &
  \multicolumn{1}{l|}{\textbf{0.86}} &
  \multicolumn{1}{l|}{\textbf{2.82e-10}} &
  \multicolumn{1}{l|}{\textbf{0.91}} &
  \multicolumn{1}{l|}{\textbf{6.36e-03}} &
  \multicolumn{1}{l|}{\textbf{0.41}} \\ \hline
\multicolumn{1}{|l|}{\multirow{2}{*}{\textbf{Type 3}}} &
  \multicolumn{1}{l|}{\textbf{DL}} &
  \multicolumn{1}{l|}{\textbf{6.98e-4}} &
  \multicolumn{1}{l|}{\textbf{0.36}} &
  \multicolumn{1}{l|}{\textbf{3.66e-10}} &
  \multicolumn{1}{l|}{\textbf{0.7}} &
  \multicolumn{1}{l|}{\textbf{2.57e-05}} &
  \multicolumn{1}{l|}{\textbf{0.48}} \\ \cline{2-8} 
\multicolumn{1}{|l|}{} &
  \multicolumn{1}{l|}{\textbf{Trad}} &
  \multicolumn{1}{l|}{\textbf{1.42e-12}} &
  \multicolumn{1}{l|}{\textbf{0.83}} &
  \multicolumn{1}{l|}{\textbf{1.59e-10}} &
  \multicolumn{1}{l|}{\textbf{0.77}} &
  \multicolumn{1}{l|}{0.29} &
  \multicolumn{1}{l|}{0.07} \\ \hline
\multicolumn{8}{c}{SF: Same File, SD: Same Directory, DD: Different Directories, CD: Cliff's Delta} \\ \hline
\end{tabular}%
}
\end{table}

\begin{table}[htpb]
\caption{Mann-Whitney Test and Cliff’s Delta Results Regarding the Distributions of Clones in  DL and Traditional (Trad) Java Projects.}
\label{tab:pvalueLocationCloneType_java}
\resizebox{\textwidth}{!}{%
\begin{tabular}{llllllll}
\cline{3-8}
 &
  \multicolumn{1}{l|}{} &
  \multicolumn{6}{c|}{\textbf{Location}} \\ \cline{3-8} 
 &
  \multicolumn{1}{l|}{} &
  \multicolumn{2}{c|}{\textbf{SF-SD}} &
  \multicolumn{2}{c|}{\textbf{SF-DD}} &
  \multicolumn{2}{c|}{\textbf{SD-DD}} \\ \hline
\multicolumn{1}{|l|}{\textbf{Clone Type}} &
  \multicolumn{1}{l|}{\textbf{Proj Type}} &
  \multicolumn{1}{l|}{\textbf{p-value}} &
  \multicolumn{1}{l|}{\textbf{CD}} &
  \multicolumn{1}{l|}{\textbf{p-value}} &
  \multicolumn{1}{l|}{\textbf{CD}} &
  \multicolumn{1}{l|}{\textbf{p-value}} &
  \multicolumn{1}{l|}{\textbf{CD}} \\ \hline
\multicolumn{1}{|l|}{\multirow{2}{*}{\textbf{ALL}}} &
  \multicolumn{1}{l|}{\textbf{DL}} &
  \multicolumn{1}{l|}{0.936} &
  \multicolumn{1}{l|}{0.05} &
  \multicolumn{1}{l|}{0.378} &
  \multicolumn{1}{l|}{0.33} &
  \multicolumn{1}{l|}{0.229} &
  \multicolumn{1}{l|}{0.44} \\ \cline{2-8} 
\multicolumn{1}{|l|}{} &
  \multicolumn{1}{l|}{\textbf{Trad}} &
  \multicolumn{1}{l|}{0.092} &
  \multicolumn{1}{l|}{0.61} &
  \multicolumn{1}{l|}{0.810} &
  \multicolumn{1}{l|}{0.11} &
  \multicolumn{1}{l|}{0.297} &
  \multicolumn{1}{l|}{0.38} \\ \hline
\multicolumn{1}{|l|}{\multirow{2}{*}{\textbf{Type 1}}} &
  \multicolumn{1}{l|}{\textbf{DL}} &
  \multicolumn{1}{l|}{0.155} &
  \multicolumn{1}{l|}{0.66} &
  \multicolumn{1}{l|}{0.155} &
  \multicolumn{1}{l|}{0.66} &
  \multicolumn{1}{l|}{0.810} &
  \multicolumn{1}{l|}{0.11} \\ \cline{2-8} 
\multicolumn{1}{|l|}{} &
  \multicolumn{1}{l|}{\textbf{Trad}} &
  \multicolumn{1}{l|}{\textbf{0.022}} &
  \multicolumn{1}{l|}{\textbf{0.86}} &
  \multicolumn{1}{l|}{\textbf{0.012}} &
  \multicolumn{1}{l|}{\textbf{1.0}} &
  \multicolumn{1}{l|}{0.082} &
  \multicolumn{1}{l|}{0.66} \\ \hline
\multicolumn{1}{|l|}{\multirow{2}{*}{\textbf{Type 2}}} &
  \multicolumn{1}{l|}{\textbf{DL}} &
  \multicolumn{1}{l|}{0.229} &
  \multicolumn{1}{l|}{0.44} &
  \multicolumn{1}{l|}{0.378} &
  \multicolumn{1}{l|}{0.33} &
  \multicolumn{1}{l|}{0.936} &
  \multicolumn{1}{l|}{0.05} \\ \cline{2-8} 
\multicolumn{1}{|l|}{} &
  \multicolumn{1}{l|}{\textbf{Trad}} &
  \multicolumn{1}{l|}{0.065} &
  \multicolumn{1}{l|}{0.66} &
  \multicolumn{1}{l|}{0.784} &
  \multicolumn{1}{l|}{0.13} &
  \multicolumn{1}{l|}{0.315} &
  \multicolumn{1}{l|}{0.4} \\ \hline
\multicolumn{1}{|l|}{\multirow{2}{*}{\textbf{Type 3}}} &
  \multicolumn{1}{l|}{\textbf{DL}} &
  \multicolumn{1}{l|}{0.810} &
  \multicolumn{1}{l|}{0.11} &
  \multicolumn{1}{l|}{0.297} &
  \multicolumn{1}{l|}{0.38} &
  \multicolumn{1}{l|}{0.297} &
  \multicolumn{1}{l|}{0.38} \\ \cline{2-8} 
\multicolumn{1}{|l|}{} &
  \multicolumn{1}{l|}{\textbf{Trad}} &
  \multicolumn{1}{l|}{0.170} &
  \multicolumn{1}{l|}{0.53} &
  \multicolumn{1}{l|}{0.936} &
  \multicolumn{1}{l|}{0.05} &
  \multicolumn{1}{l|}{0.522} &
  \multicolumn{1}{l|}{0.26} \\ \hline
\multicolumn{8}{c}{SF: Same File, SD: Same Directory, DD: Different Directories, CD: Cliff's Delta} \\ \hline
\end{tabular}%
}
\end{table}

\begin{table}[htpb]
\caption{Mann-Whitney Test and Cliff’s Delta Results Regarding the Distributions of Clones in  DL and Traditional (Trad) C\# Projects.}
\label{tab:pvalueLocationCloneType_csharp}
\resizebox{\textwidth}{!}{%
\begin{tabular}{llllllll}
\cline{3-8}
 &
  \multicolumn{1}{l|}{} &
  \multicolumn{6}{c|}{\textbf{Location}} \\ \cline{3-8} 
 &
  \multicolumn{1}{l|}{} &
  \multicolumn{2}{c|}{\textbf{SF-SD}} &
  \multicolumn{2}{c|}{\textbf{SF-DD}} &
  \multicolumn{2}{c|}{\textbf{SD-DD}} \\ \hline
\multicolumn{1}{|l|}{\textbf{Clone Type}} &
  \multicolumn{1}{l|}{\textbf{Proj Type}} &
  \multicolumn{1}{l|}{\textbf{p-value}} &
  \multicolumn{1}{l|}{\textbf{CD}} &
  \multicolumn{1}{l|}{\textbf{p-value}} &
  \multicolumn{1}{l|}{\textbf{CD}} &
  \multicolumn{1}{l|}{\textbf{p-value}} &
  \multicolumn{1}{l|}{\textbf{CD}} \\ \hline
\multicolumn{1}{|l|}{\multirow{2}{*}{\textbf{ALL}}} &
  \multicolumn{1}{l|}{\textbf{DL}} &
  \multicolumn{1}{l|}{0.368} &
  \multicolumn{1}{l|}{0.21} &
  \multicolumn{1}{l|}{0.752} &
  \multicolumn{1}{l|}{0.07} &
  \multicolumn{1}{l|}{0.56} &
  \multicolumn{1}{l|}{0.14} \\ \cline{2-8} 
\multicolumn{1}{|l|}{} &
  \multicolumn{1}{l|}{\textbf{Trad}} &
  \multicolumn{1}{l|}{\textbf{0.001}} &
  \multicolumn{1}{l|}{\textbf{0.71}} &
  \multicolumn{1}{l|}{\textbf{0.044}} &
  \multicolumn{1}{l|}{\textbf{0.46}} &
  \multicolumn{1}{l|}{0.544} &
  \multicolumn{1}{l|}{0.14} \\ \hline
\multicolumn{1}{|l|}{\multirow{2}{*}{\textbf{Type 1}}} &
  \multicolumn{1}{l|}{\textbf{DL}} &
  \multicolumn{1}{l|}{\textbf{0.039}} &
  \multicolumn{1}{l|}{\textbf{1.0}} &
  \multicolumn{1}{l|}{0.124} &
  \multicolumn{1}{l|}{0.77} &
  \multicolumn{1}{l|}{0.901} &
  \multicolumn{1}{l|}{0.04} \\ \cline{2-8} 
\multicolumn{1}{|l|}{} &
  \multicolumn{1}{l|}{\textbf{Trad}} &
  \multicolumn{1}{l|}{0.087} &
  \multicolumn{1}{l|}{0.51} &
  \multicolumn{1}{l|}{\textbf{0.0169}} &
  \multicolumn{1}{l|}{\textbf{0.72}} &
  \multicolumn{1}{l|}{0.221} &
  \multicolumn{1}{l|}{0.29} \\ \hline
\multicolumn{1}{|l|}{\multirow{2}{*}{\textbf{Type 2}}} &
  \multicolumn{1}{l|}{\textbf{DL}} &
  \multicolumn{1}{l|}{0.488} &
  \multicolumn{1}{l|}{0.18} &
  \multicolumn{1}{l|}{0.113} &
  \multicolumn{1}{l|}{0.43} &
  \multicolumn{1}{l|}{0.230} &
  \multicolumn{1}{l|}{0.35} \\ \cline{2-8} 
\multicolumn{1}{|l|}{} &
  \multicolumn{1}{l|}{\textbf{Trad}} &
  \multicolumn{1}{l|}{\textbf{0.0001}} &
  \multicolumn{1}{l|}{\textbf{0.86}} &
  \multicolumn{1}{l|}{\textbf{0.004}} &
  \multicolumn{1}{l|}{\textbf{0.66}} &
  \multicolumn{1}{l|}{0.85} &
  \multicolumn{1}{l|}{0.04} \\ \hline
\multicolumn{1}{|l|}{\multirow{2}{*}{\textbf{Type 3}}} &
  \multicolumn{1}{l|}{\textbf{DL}} &
  \multicolumn{1}{l|}{0.341} &
  \multicolumn{1}{l|}{0.22} &
  \multicolumn{1}{l|}{0.644} &
  \multicolumn{1}{l|}{0.11} &
  \multicolumn{1}{l|}{0.849} &
  \multicolumn{1}{l|}{0.05} \\ \cline{2-8} 
\multicolumn{1}{|l|}{} &
  \multicolumn{1}{l|}{\textbf{Trad}} &
  \multicolumn{1}{l|}{\textbf{0.0003}} &
  \multicolumn{1}{l|}{\textbf{0.80}} &
  \multicolumn{1}{l|}{\textbf{0.009}} &
  \multicolumn{1}{l|}{\textbf{0.59}} &
  \multicolumn{1}{l|}{0.576} &
  \multicolumn{1}{l|}{0.13} \\ \hline
\multicolumn{8}{c}{SF: Same File, SD: Same Directory, DD: Different Directories, CD: Cliff's Delta} \\ \hline
\end{tabular}%
}
\end{table}

\begin{figure}[htpb]
\centering
\includegraphics[width=.48\textwidth]{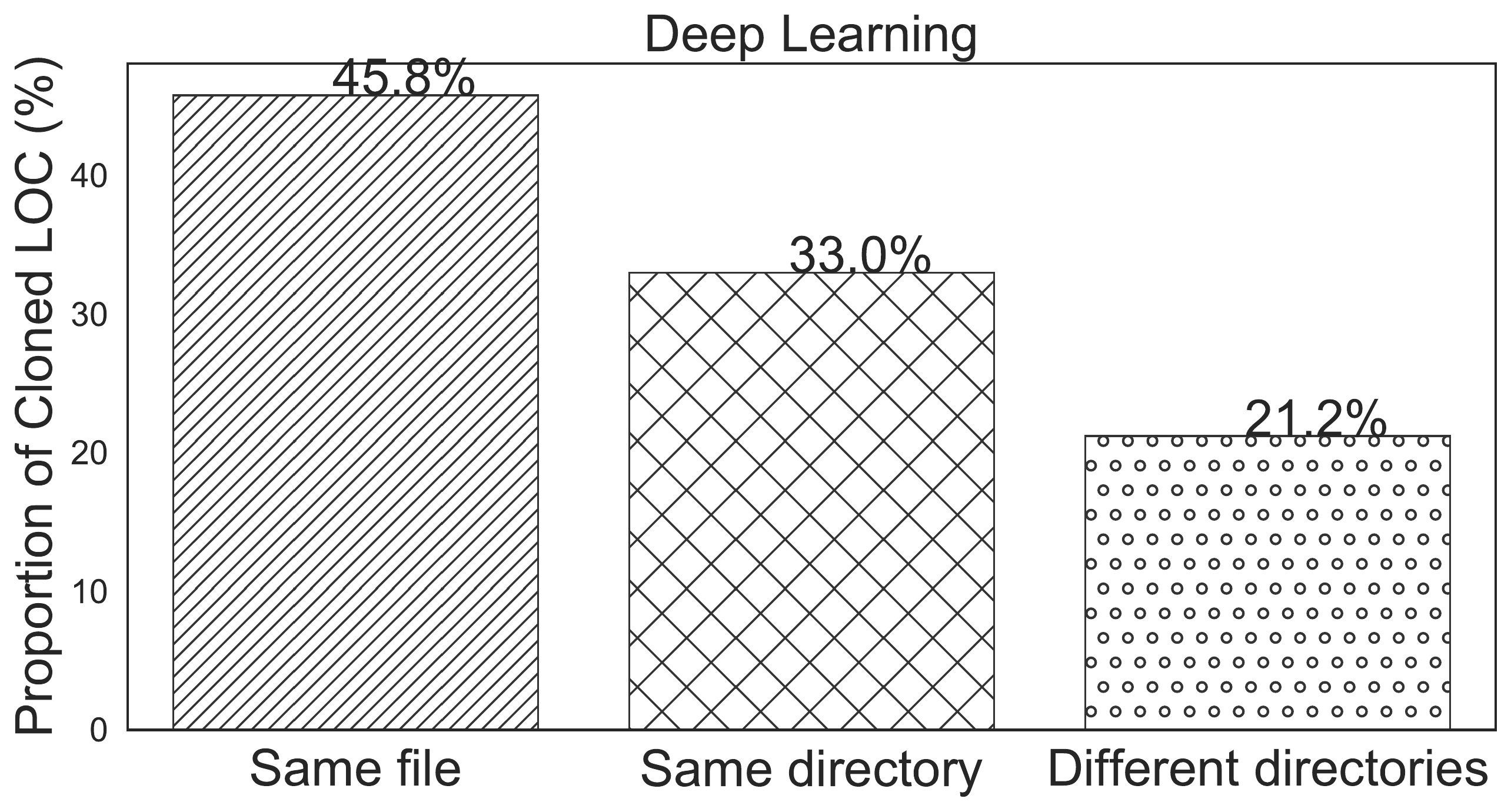}
\includegraphics[width=.48\textwidth]{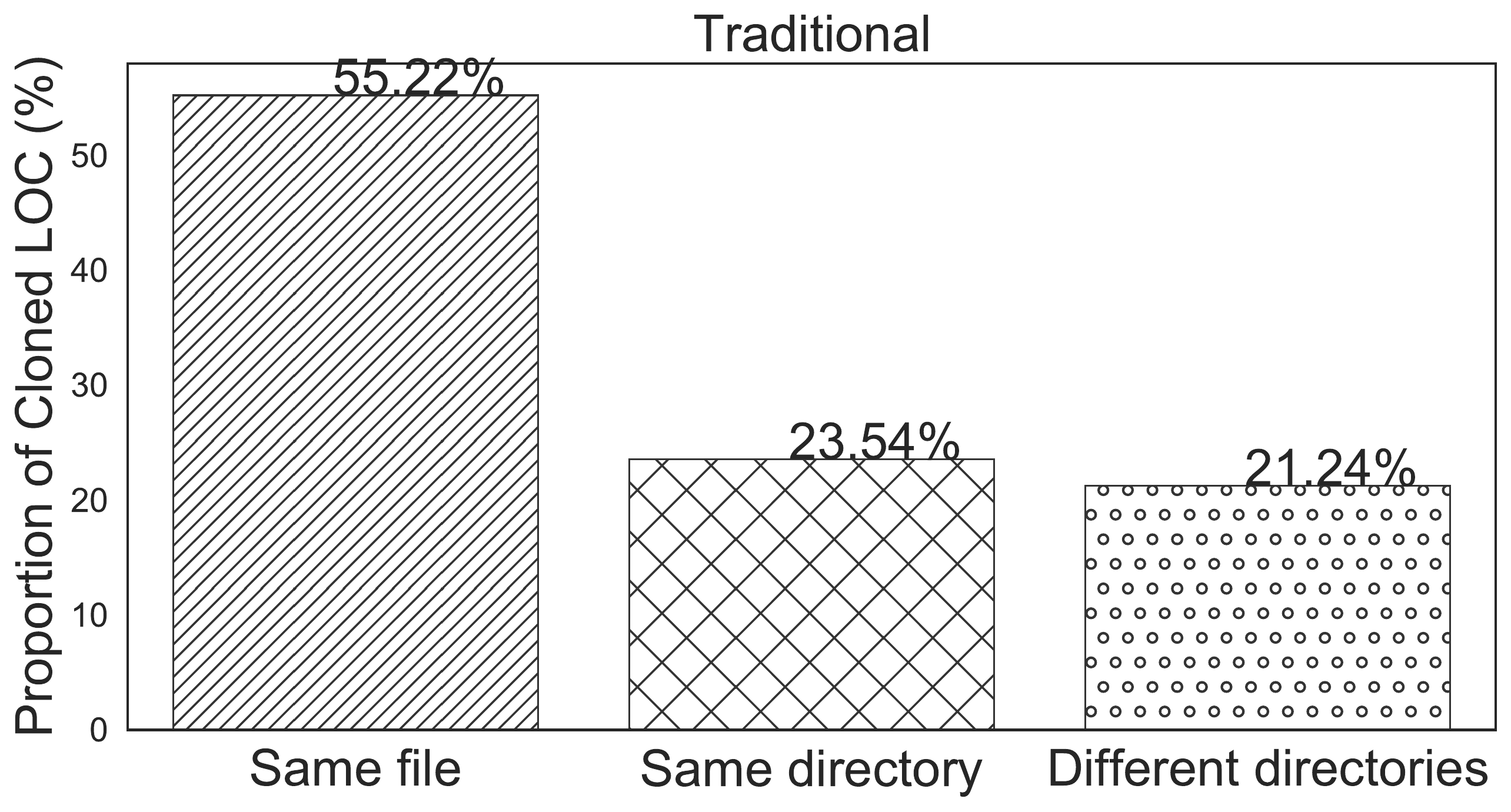}
\caption{Percentages of Lines of Code Clones by Location of Clones in both Deep Learning and Traditional Systems (Python)}
\label{fig:DLCloneLocation_python}
\end{figure} 

\begin{figure}[htpb]
\centering
\includegraphics[width=.48\textwidth]{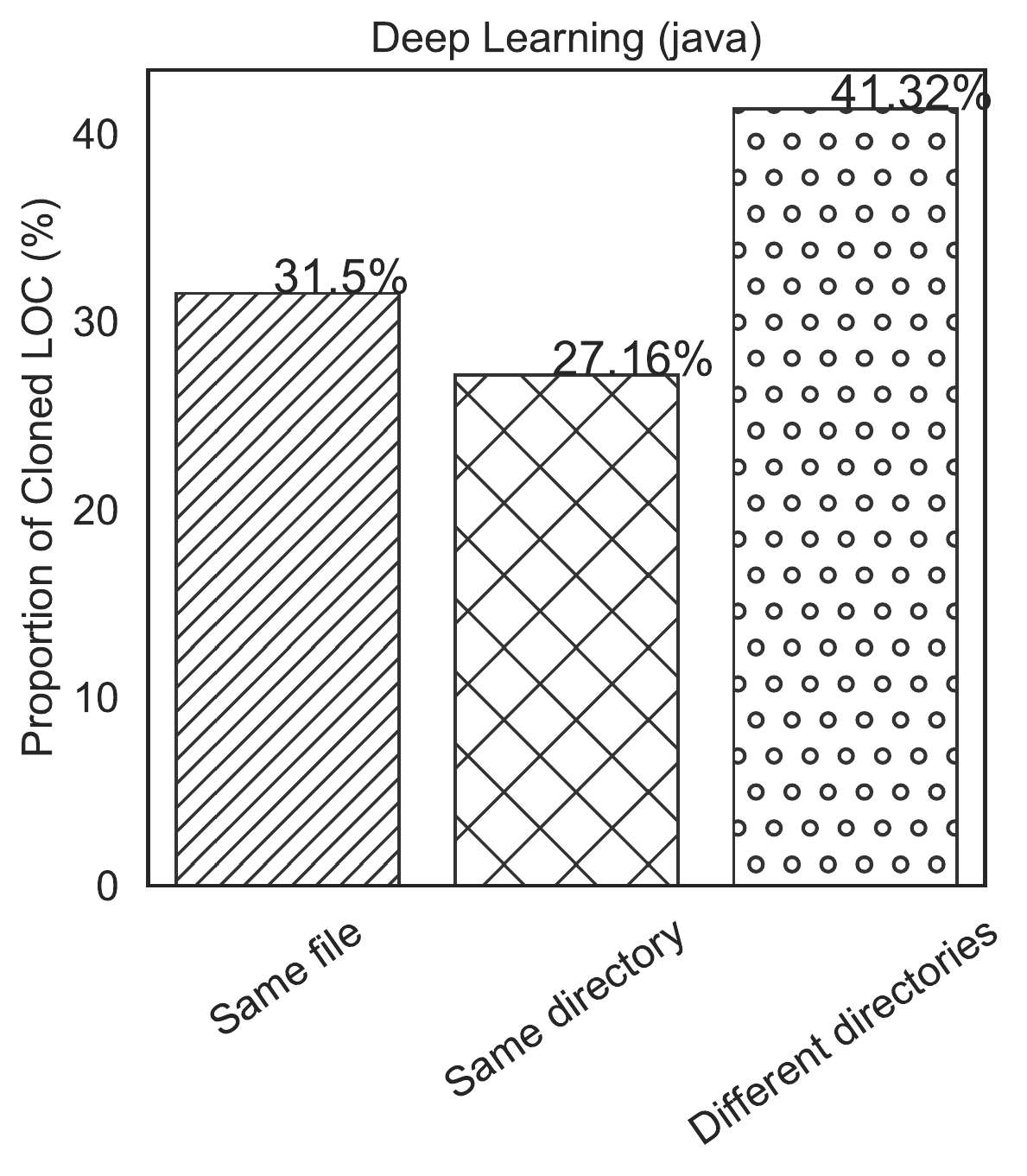}
\includegraphics[width=.48\textwidth]{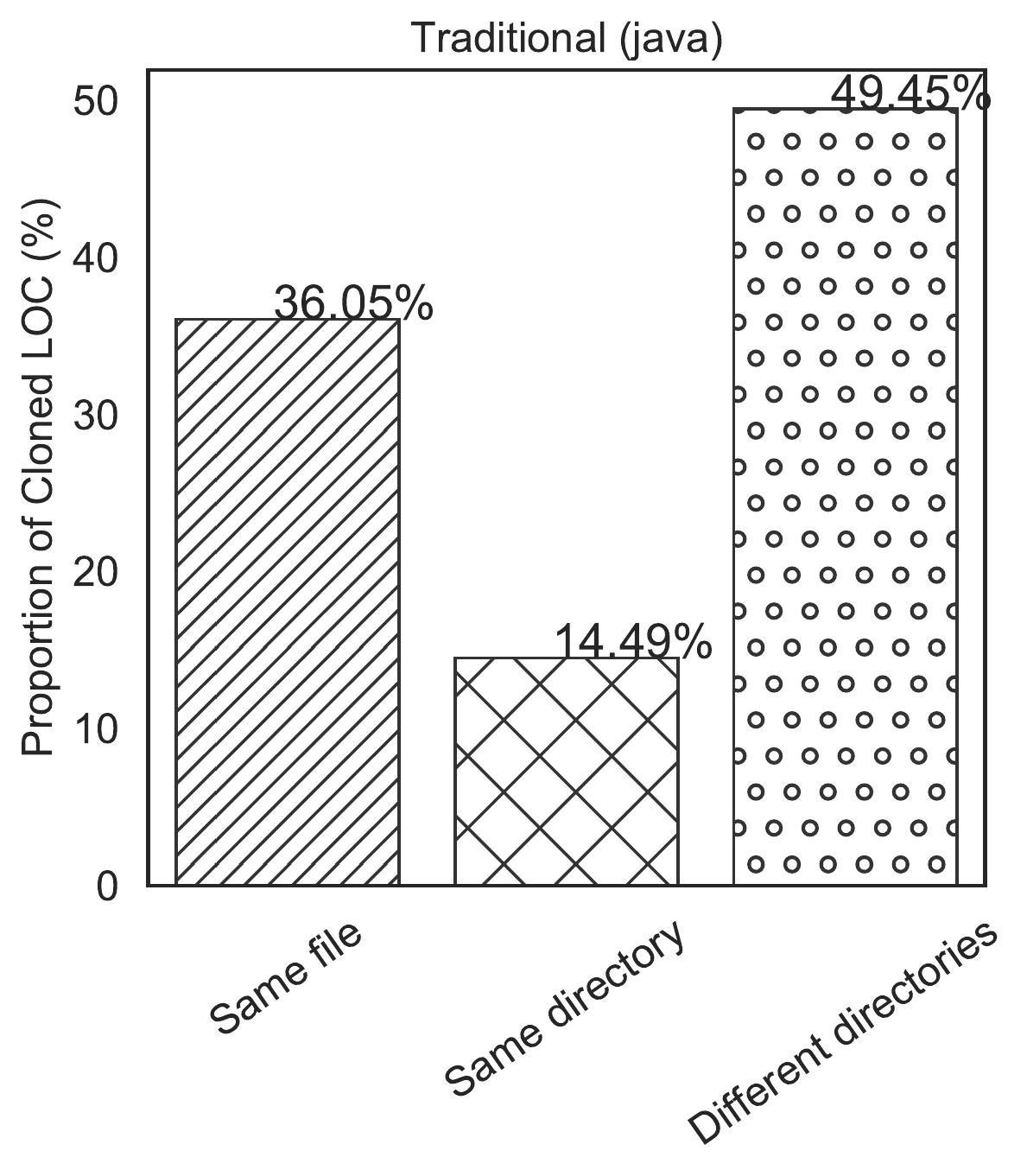}
\caption{Percentages of Lines of Code Clones by Location of Clones in both Deep Learning and Traditional Systems (Java)}
\label{fig:DLCloneLocation_java}
\end{figure} 

\begin{figure}[htpb]
\centering
\includegraphics[width=.48\textwidth]{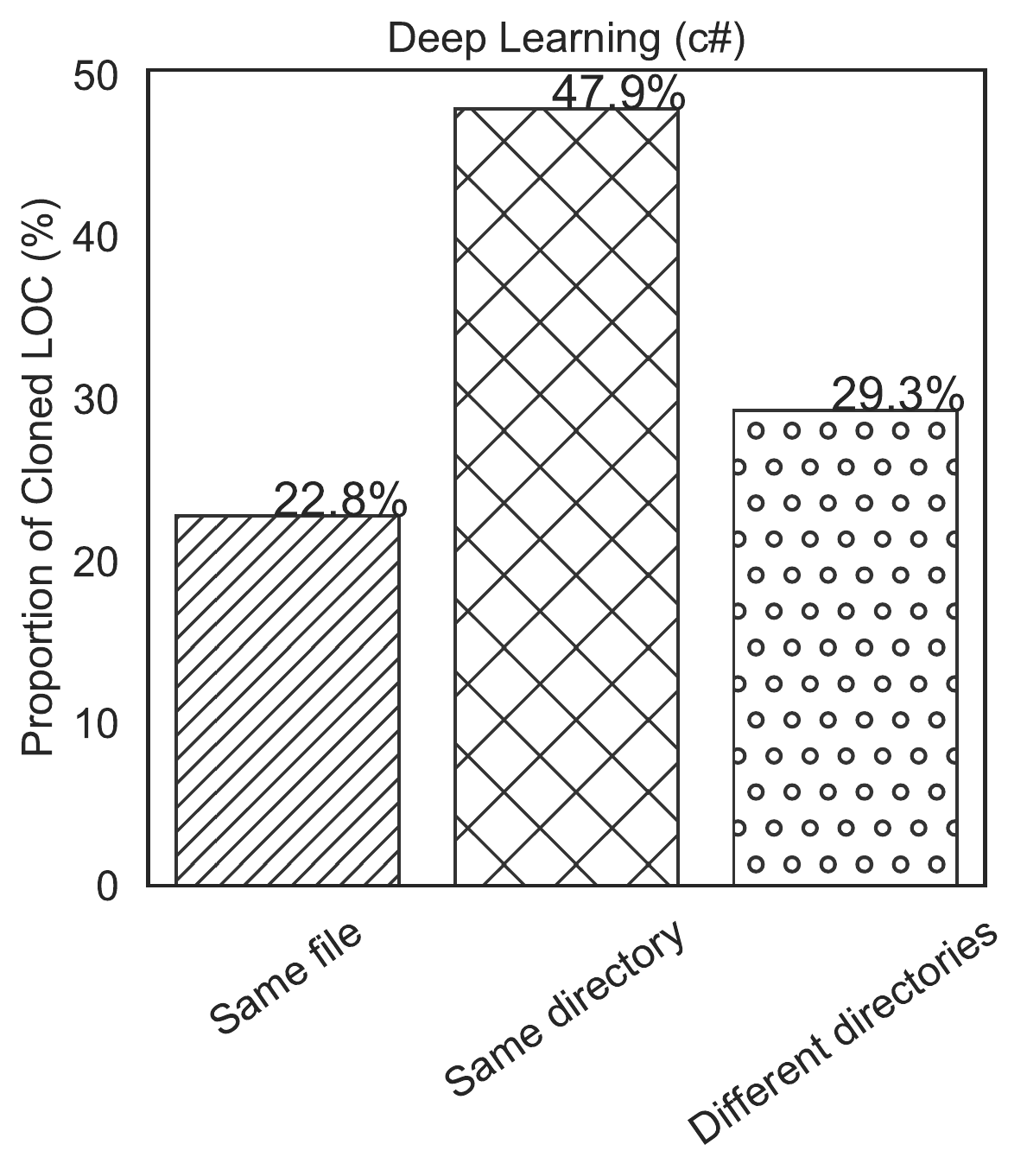}
\includegraphics[width=.48\textwidth]{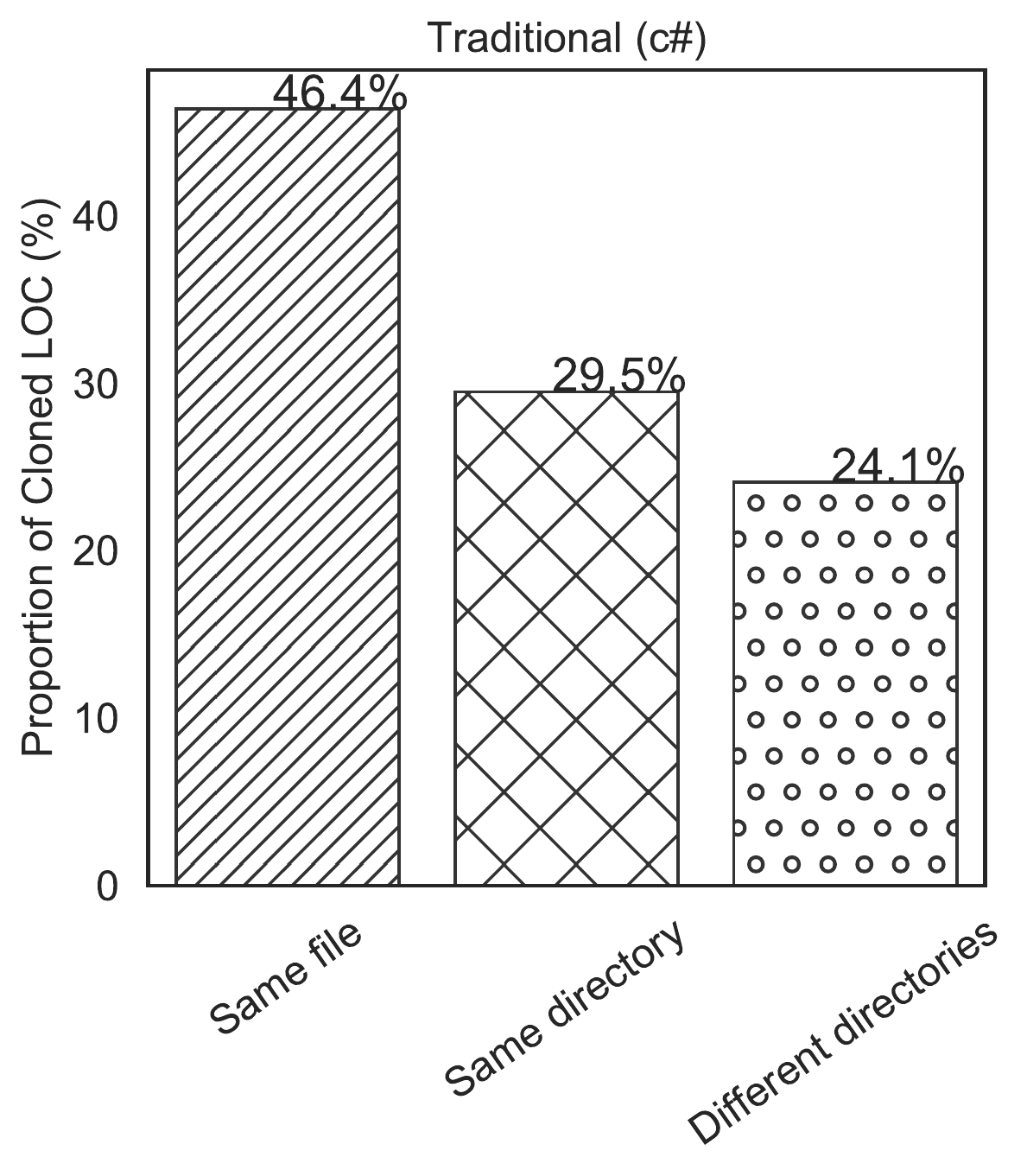}
\caption{Percentages of Lines of Code Clones by Location of Clones in both Deep Learning and Traditional Systems (C\#)}
\label{fig:DLCloneLocation_csharp}
\end{figure} 

Table \ref{tab:pvalueLocationCloneType_python} shows the p-values 
from the Mann-Whitney test and Cliff's delta values for different code clones location between the same type of systems (DL and traditional code) with respect to the relation between code clones locations and clone types. `ALL' in the Table \ref{tab:pvalueLocationCloneType_python} designates the unfiltered Type 3 (i.e, include all fragments from Type 1, Type 2, and Type 3). 
We found statistically significant differences between the `same file' category and both the `same directory' and the `different directories' categories in the DL code with p-values equal to 2.91e-04 and 2.9e-10 (which are $<0.05$) respectively and with medium effect size of 0.4 and 0.7, respectively. 
Thus, in DL systems, a significantly higher percentages of the cloned code reside in `same file' compared to the percentages of cloned code that reside in the `same directory' and in `different directories'. 
Similarly, the percentage of cloned lines in the `same directory' is significantly higher compared to the percentage of clones contained in `different directories' with p-value equals to 4.79e-05 ($<0.05$) and with a medium effect size. 
For traditional code, we observe that a higher percentage of cloned lines of code is located in the `same file' compared to the percentage of clones that are located in the `same directory' and in `different directories', with p-values $<0.05$ (8.34e-13 and 2.05e-11 respectively) and with large effect sizes (0.83 and 0.81 respectively). 
We found no statistically significant difference between the percentage of clones contained in the `same directory' category and the 'different directories' category, for traditional code. 

Fig. \ref{fig:CloneLocation_java} shows the proportion of cloned code (LOC \%) in different location category for Java systems. Based on the median values of the proportion of cloned code, we observe that the highest proportions of cloned code for Java DL systems reside in `different directory' followed by the proportion of cloned code in 'same directory' and in `same file' categories. Similar trend in the distribution of cloned code is observed in Java traditional systems (except for `same directory' category) with `different directory' category containing the highest median percentage of cloned LOC. This is an indication that clones in Java DL and traditional systems tend to be dispersed. However, from the results of the MWW test and Cliff's delta effect size for `All' types as shown in Table \ref{tab:pvalueLocationCloneType_java}, we do not observe the differences to be statistically significant.

For C\# systems as in Fig. \ref{fig:CloneLocation_csharp}, the median value for the distribution of clones in DL systems is the lowest for `same file' location category while it is the opposite for traditional systems. This shows that clones in C\# DL systems are likely to be dispersed compared to traditional C\# systems. However, from our statistical test results for C\# systems in Table \ref{tab:pvalueLocationCloneType_csharp}, we do not observe statistically significant differences in distribution of cloned code in different location categories. For traditional C\# systems, the differences in proportion of clones in `same file' are significantly high (p-values $<0.05$ with medium to large effect size) compared to `same directory' and `different directory' categories.

For further insights, we analyzed the average percentages of lines of cloned code by their locations in deep learning and traditional code. As shown in Fig. \ref{fig:DLCloneLocation_python}, we identify that 45.8\% of the DL-related clones are in the `same file', 33\% are in the `same directory' and 21.2\% are in `different directories'. Hence, DL clones are more dispersed, having fewer percentages of clones in the same file and more than 54\% (33\% + 21.2\%) in different files and directories. Code clones in traditional code, on the other hand, are more localized. More than the half of the code clones in non-DL code (55.22\%) are in the same file, 23.54\% are in the same directory and 21.24\% are in different directories. Therefore, according to our results, code clones in Python deep learning code are more dispersed than code clones in non-DL systems. For Java as in Fig. \ref{fig:DLCloneLocation_java}, the highest proportion of clones for both DL (41.32\%) and traditional (49.45\%) systems are in `different directories' contrasting to Python systems (Fig. \ref{fig:DLCloneLocation_python}). Although, traditional Java systems have higher proportion of clones in different directories compared to Java DL systems, DL systems have 68.48\% (41.32 + 27.16) of clones in `same directory' and `different directories' combined, compared to that of traditional Java code (49.45+14.49= 63.94\%). Thus, clones in Java DL systems are relatively more dispersed compared to clones in traditional systems. For C\# systems as in Fig. \ref{fig:DLCloneLocation_csharp}, 77.2\% (47.9 + 29.3) of cloned code is in `same directory' and `different directories' compared to 53.6\% (29.5 + 24.1) for traditional. This shows that clones in C\# DL systems are more dispersed compared to traditional cloned code.

For our location-based analysis of the distribution of clones, we observe that clones in DL systems are more dispersed compared to traditional clones. We also observe that clones in Java and C\# DL systems are more dispersed compared to Python DL systems. 
This dispersion of cloned code in the deep learning systems may harm the maintenance of duplicated code due to potential navigation and comprehension overhead. The degree of dispersion of clones may also vary depending on programming languages.

\begin{figure}[ht]
\centering
\includegraphics[width=.49\textwidth]{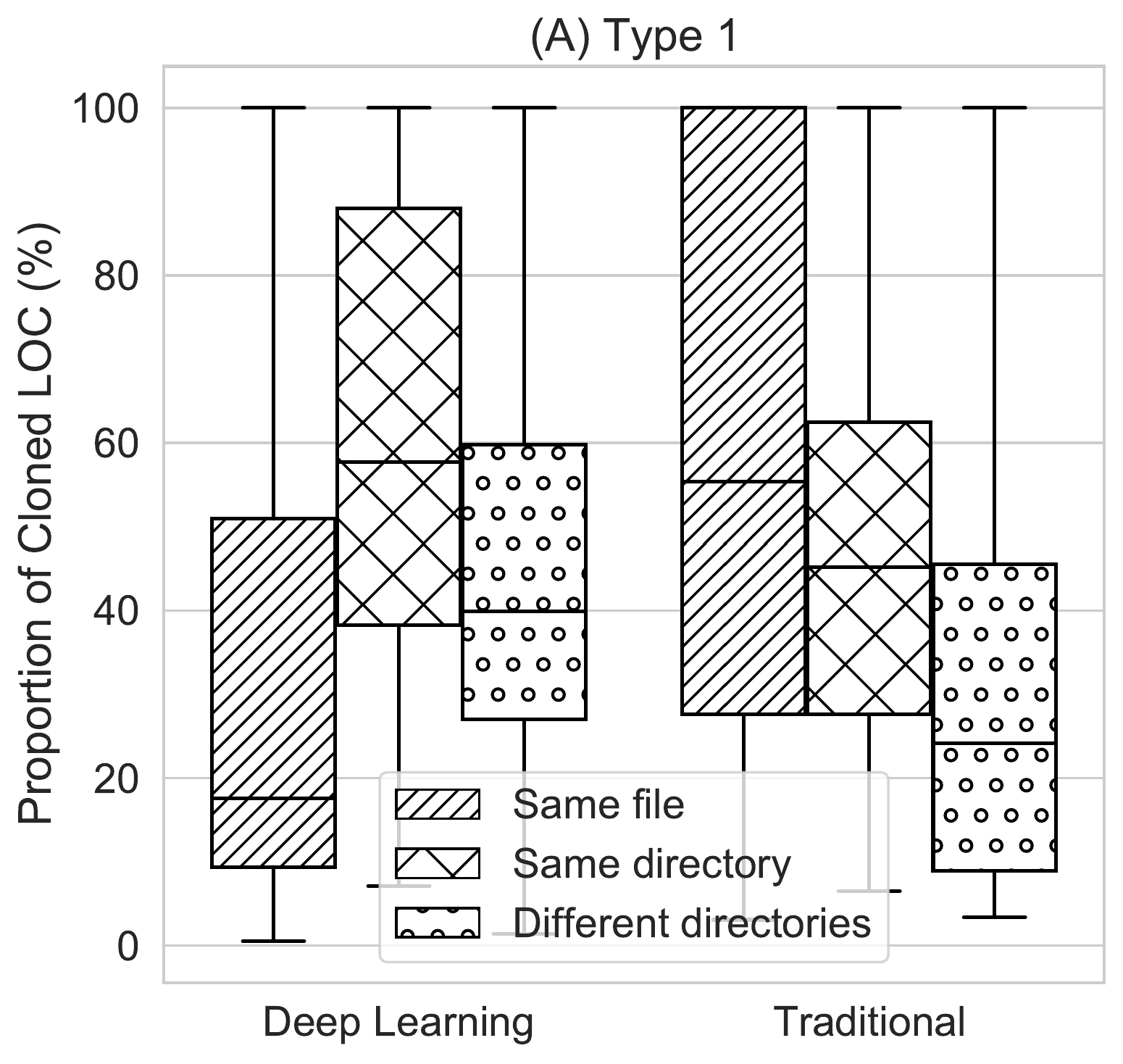}
\includegraphics[width=.49\textwidth]{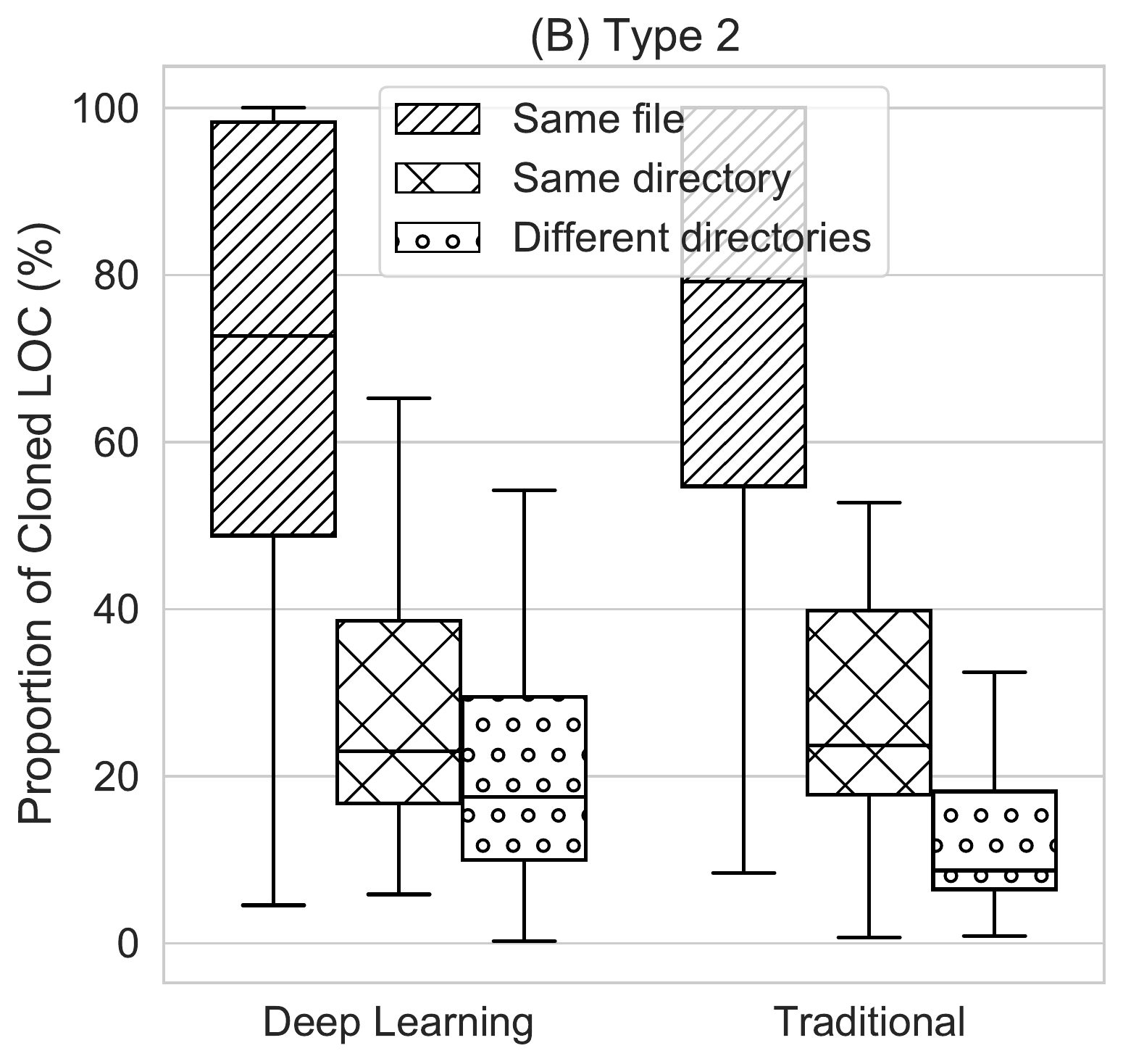}
\includegraphics[width=.49\textwidth]{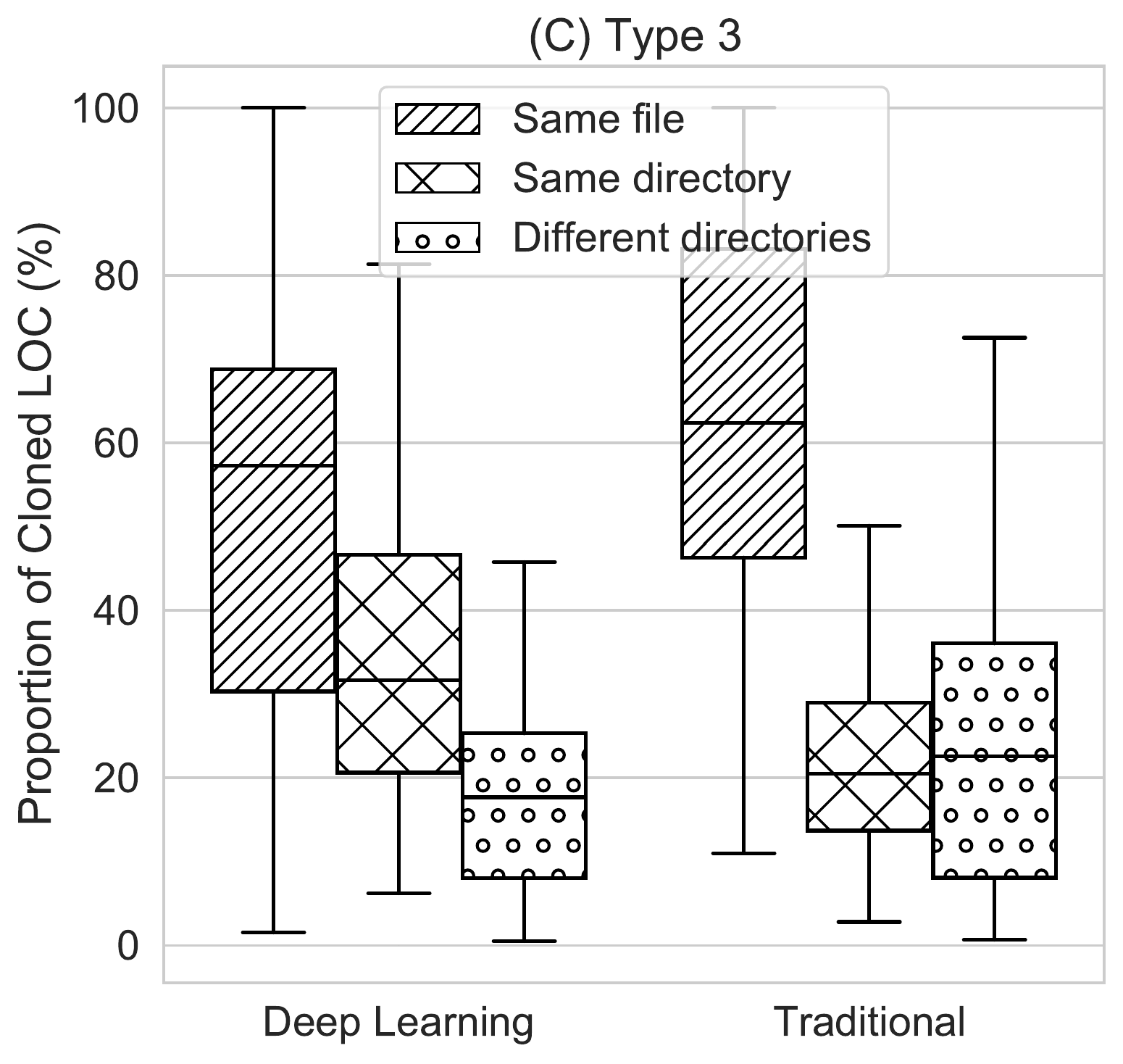}

\caption{Distribution of Different Types of Clones by Clone Location in DL and Traditional Code (Python)}
\label{fig:CloneLocationbyCloneType_python}
\end{figure}

\paragraph{\textbf{Location-based distribution of different types of clones:}} We analyze the location-based distribution of different types of clones as follows:

\paragraph{Type 1:} As shown in Fig. \ref{fig:CloneLocationbyCloneType_python}-A, the median of the distribution of the percentage of Type 1 cloned lines in `same file' in Python DL code is the lowest compared to the percentages of cloned lines in `same directory' and in `different directory'. This shows that Type 1 clones in DL code are dispersed in different files and directories. However, for traditional systems, we observe that majority of the Type 1 clones are in `same file' compared to `same directory' and `different directories'. This suggests that Type 1 clones in traditional systems reside in closer proximity, unlike the Type 1 clones in Python DL systems.   

To investigate whether the observed differences are statistically significant, we perform MWW tests (two-tailed, significance at 0.05) and measure Cliff's delta effect size. In Table \ref{tab:pvalueLocationCloneType_python}, we highlight in bold the statistically significant differences for the distributions of cloned lines in different clone locations for both DL and traditional Python code with respect to clone types where p-values are $<0.05$. Fig. \ref{fig:CloneLocationbyCloneType_python} represents these differences by showing the distribution of percentages of lines of code clones for each clone type and for both types of systems. 
Type 1 clones in deep learning Python code is less localized with a statistically significant difference between the same file location category and the others categories and with a small effect size. 

Whereas proportion of Type 1 clones in non-DL code shows a  statistically significant difference only between the `different directories' location and the others locations. The distribution of lines of code clones located in `different directories' is lower in comparison to the distribution of line of code clones in other locations (Same file (SF), and Same Directory (SD)).

The existence of exact clones (Type 1) in the same directory could shed lights on some implementation practices of deep learning developers in Python. For example, the occurrence of exact functions in the same directory may suggest that when DL developers have a working code that builds a model properly, they are inclined to copy-paste this same code in another file in the same directory to construct a similar model to try another configuration. One example is shown in Table \ref{tab:examplecodeSameDirectory}, where we found in the same directory 'inference', two models Movidius and Yolo that contains the exact function to calculate the accuracy of the model 'iou' which stands for Intersection Over Union for object detection. 
Building models may have the same common functions like computing accuracy or implementing the activation function. These functions could be exact for each model, which may explain the high occurrence of Type 1 clones in the same directory in deep learning projects, compared to traditional projects. 

\paragraph{Type 2:}  As shown in Fig. \ref{fig:CloneLocationbyCloneType_python}-B, the distribution of the percentage of the lines of Type 2 cloned code in `same file' in Python DL code is higher compared to the distributions of Type 2 clones in `same directory' and `different directories' categories. This implies that Type 2 clones in Python DL code reside in closer proximity. For traditional systems, we see a similar distribution of the percentage of Type 2 cloned lines in different location categories. However, the percentages of Type 2 clones in `different directories' in Python DL code tend to be slightly higher compared to that in traditional code.   

We found these differences between the percentages of cloned lines in `same file' and in both `same directory' and `different directories' categories for Type 2 clones in Python DL code statistically significant with p-values equal to 7.64e-09 and 5.98e-08, respectively (Table \ref{tab:pvalueLocationCloneType_python}). The values of effect size of these differences are large between `same file' and `same directory' (0.76) and `same file' and `different directories' (0.77). 

\begin{table}[htpb]
\caption{Clone Codes Example where the Location is in the Same Directory and Type 1 Clone }
\label{tab:examplecodeSameDirectory}
    \centering
    \begin{tabular}{|c|}
    \hline

\begin{minipage}{0.9\textwidth}
\vspace{0.1cm}
\begin{minted}[fontsize=\scriptsize]{python}
path: BerryNet/inference/movidius.py
def iou(box1, box2):
    tb = (min(box1[0] + 0.5 * box1[2], box2[0] + 0.5 * box2[2]) -
          max(box1[0] - 0.5 * box1[2], box2[0] - 0.5 * box2[2]))
    lr = (min(box1[1] + 0.5 * box1[3], box2[1] + 0.5 * box2[3]) -
          max(box1[1] - 0.5 * box1[3], box2[1] - 0.5 * box2[3]))
    if tb < 0 or lr < 0:
        intersection = 0
    else:
        intersection =  tb*lr
    return intersection / (box1[2] * box1[3] + box2[2] * box2[3] - intersection)
\end{minted}
\vspace{0.05cm}
\end{minipage}\\
\hline
\begin{minipage}{0.9\textwidth}
\vspace{0.1cm}
\begin{minted}[fontsize=\scriptsize]{python}
path: BerryNet/inference/yoloutils.py
def iou(box1, box2):
    tb = (min(box1[0] + 0.5 * box1[2], box2[0] + 0.5 * box2[2]) -
          max(box1[0] - 0.5 * box1[2], box2[0] - 0.5 * box2[2]))
    lr = (min(box1[1] + 0.5 * box1[3], box2[1] + 0.5 * box2[3]) -
          max(box1[1] - 0.5 * box1[3], box2[1] - 0.5 * box2[3]))
    if tb < 0 or lr < 0:
        intersection = 0
    else:
        intersection =  tb*lr
    return intersection / (box1[2] * box1[3] + box2[2] * box2[3] - intersection)
\end{minted}
\vspace{0.05cm}
\end{minipage}\\
\hline
    \end{tabular}
  
\end{table}

Hence, Type 2 clones are less dispersed in Python deep learning code than other types of clones. We notice a similar level of dispersion for Type 2 clones in non-DL Python code (as shown in Fig. \ref{fig:CloneLocationbyCloneType_python}-B).

\paragraph{Type 3:} As shown in Fig. \ref{fig:CloneLocationbyCloneType_python}-C, Type 3 clones are less dispersed compared to Type 1 clones but comparatively more dispersed than Type 2 clones in Python deep learning code. We present the results of the MWW tests in Table \ref{tab:pvalueLocationCloneType_python} where the p-values are $<0.05$ for deep learning code. The percentage of Type 3 clones located in 
the same file is the highest, in comparison to the percentages of Type 3 clones located in the `same directory' and `different directories'. In non-DL code, Type 3 clones are also located in the `same file' in high numbers. For traditional code, we found a higher percentage of lines of code clones located in the `same file' compared to the `same directory' and `different directories' categories. However, the difference is not 
statistically significant between the `same directory' and `different directories' categories, in traditional code as shown in Table \ref{tab:pvalueLocationCloneType_python}. 

From our clone-type centric analysis of the distribution of cloned code in Java  (Fig. \ref{fig:CloneLocationbyCloneType_java}) and C\# (Fig. \ref{fig:CloneLocationbyCloneType_csharp}) systems in the Appendix \ref{appRQ2}, Type 1 clones in DL systems are dispersed with major proportion of cloned code in 'same directory' and `different directories' location categories than the proportion of cloned code in `same file'. Type 2 clones on the other hand, tend to reside in `same file' having higher median value of the percentages of cloned code in both DL and traditional systems. The median values for the percentage of Type 2 cloned code is slightly higher in DL systems compared to the traditional code. For Type 3 clones in Traditional Java and C\# systems, the percentages of cloned code in `same file' category is higher compared to the percentages of clones in `same directory' and `different directories'. Java and C\# Traditional systems also have higher percentage of Type 3 clones in 'same file' compared to DL systems. This indicates that Type 3 clones in Java and C\# DL systems are relatively more dispersed compared to traditional systems. Although we observe clones tend to be relatively dispersed in DL systems compared to traditional code, we do not observe statistically significant differences in the location based distribution of clones in most cases for Java systems. For traditional C\# systems, we observe the differences between proportion of clones in 'same file' and in 'same directory' or 'different directories' to be statistically significant (except for Type 1).

From this analysis of the distribution of clones across the different files and directories of the studied projects, we can conclude that clones in deep learning code is more dispersed compared to clones in traditional code, although the differences in distribution may vary. Type 1 and Type 3 clones have relatively higher trends of being dispersed while Type 2 clones tend to be more localized (in the same file). However, the percentages of lines of cloned code may not always fully reflect their relative impacts on the systems. For example, a higher number of small sized cloned fragments scattered in distant locations may pose higher challenges in change propagation than a few larger cloned fragments located not too far apart. 
Therefore, we further analyze the distribution of the number of cloned fragments in different location categories.

\begin{figure}[htpb]
\centering
\includegraphics[width=.7\textwidth]{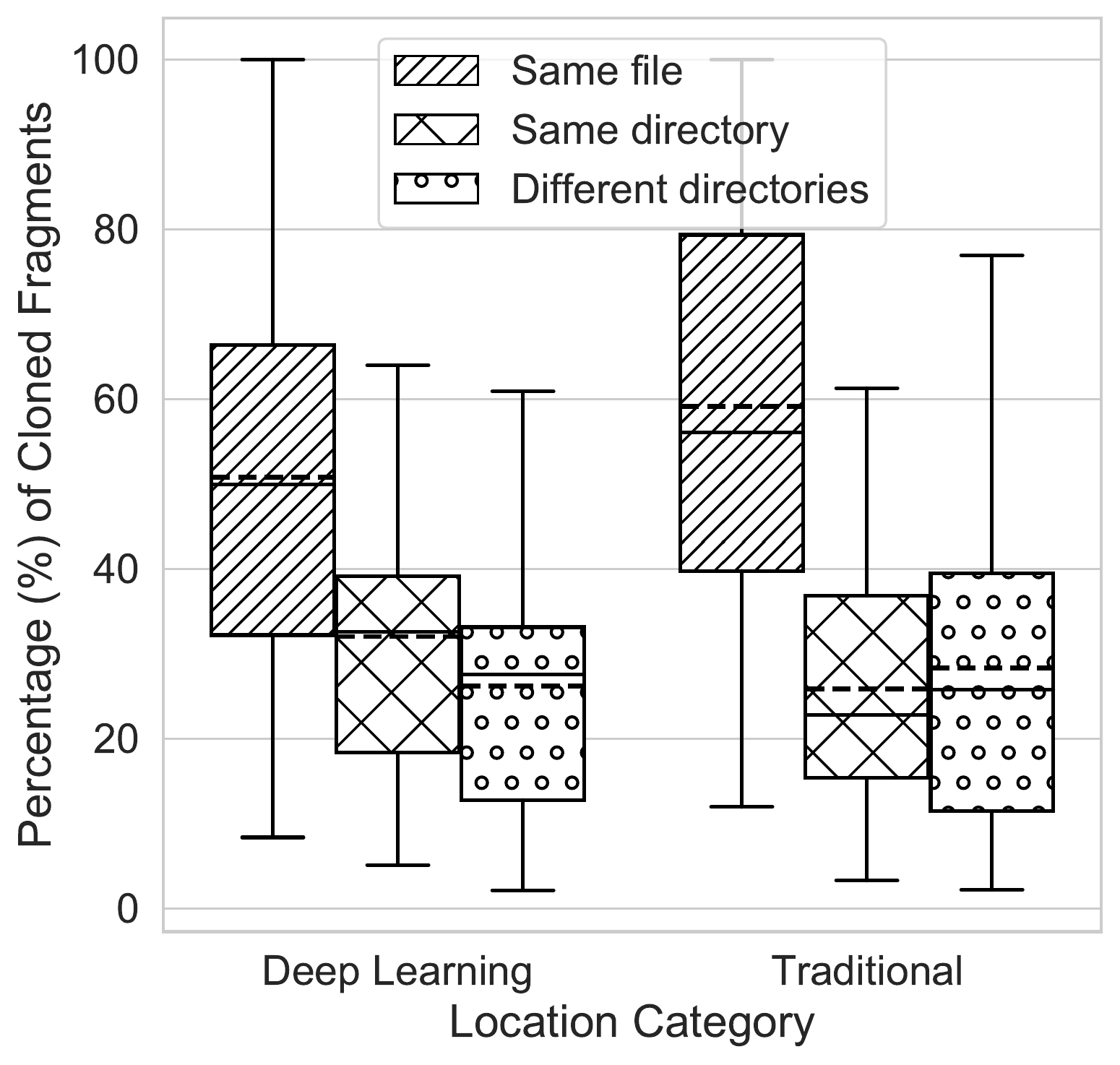}
\caption{Distribution of Percentage of Number of Fragments of Code Clones Classes per Clone Location in Python Systems.}
\label{fig:FragmentsPerLocation}
\end{figure}

\paragraph{\textbf{Location-based distribution of clone fragments:}}
Figure \ref{fig:FragmentsPerLocation} shows the percentages of number of fragments ((number of clone fragments in a location category/total number of clone fragments)x100) for each location category in Python systems. We observe that the proportion of clone fragments tend to be higher in the `same file' location category. We found a statistically significant difference between the distribution of the proportion of clone fragments located in the 'same file' and in both `same directory' and `different directories'. Their p-values are equal to 4.46e-05 and 1.94e-07 respectively with values of effect size of 0.44 (small) and 0.57 (medium), respectively in deep learning code.

\begin{figure}[htpb]
\centering
\includegraphics[width=.49\textwidth]{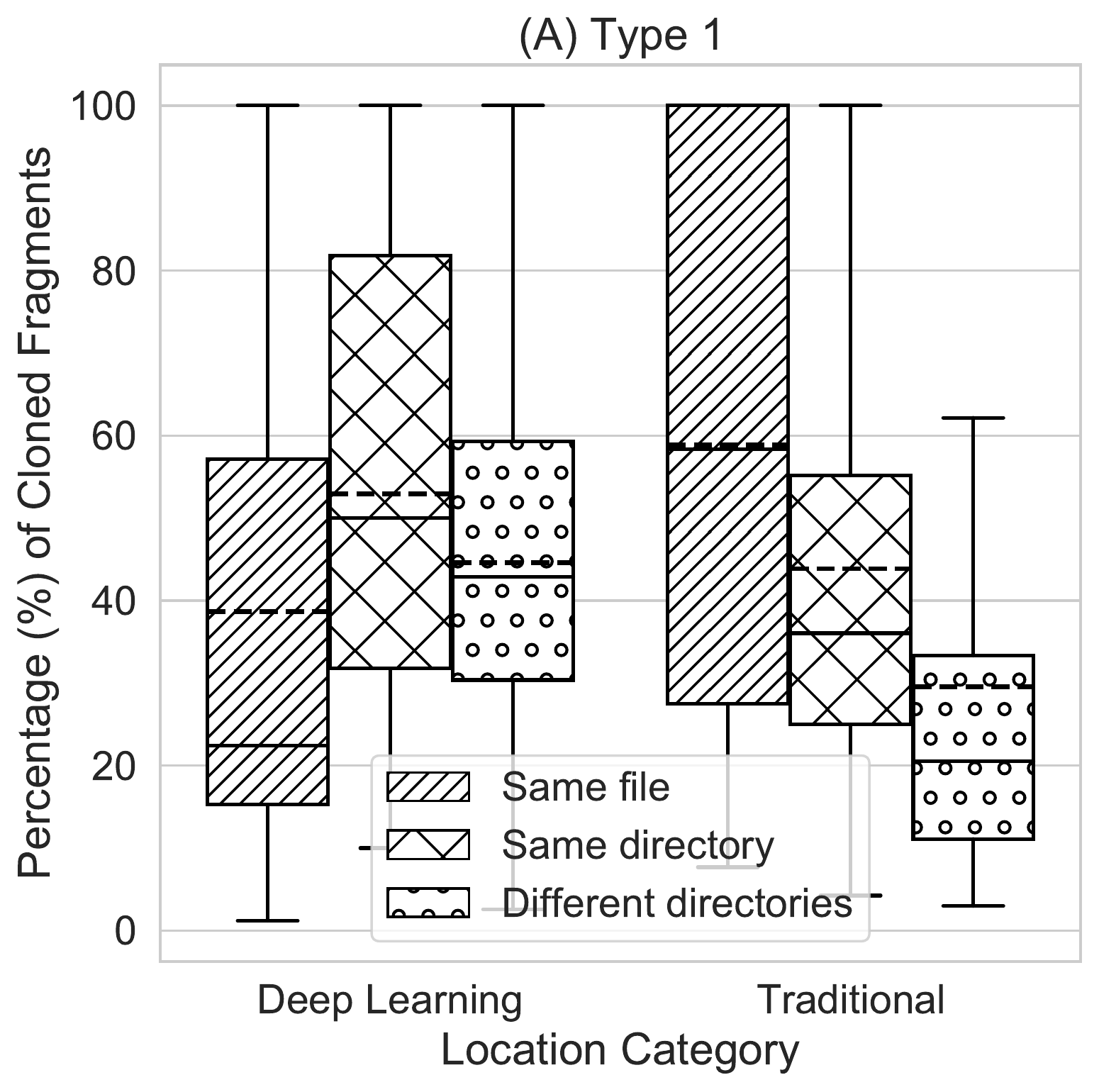}
\includegraphics[width=.49\textwidth]{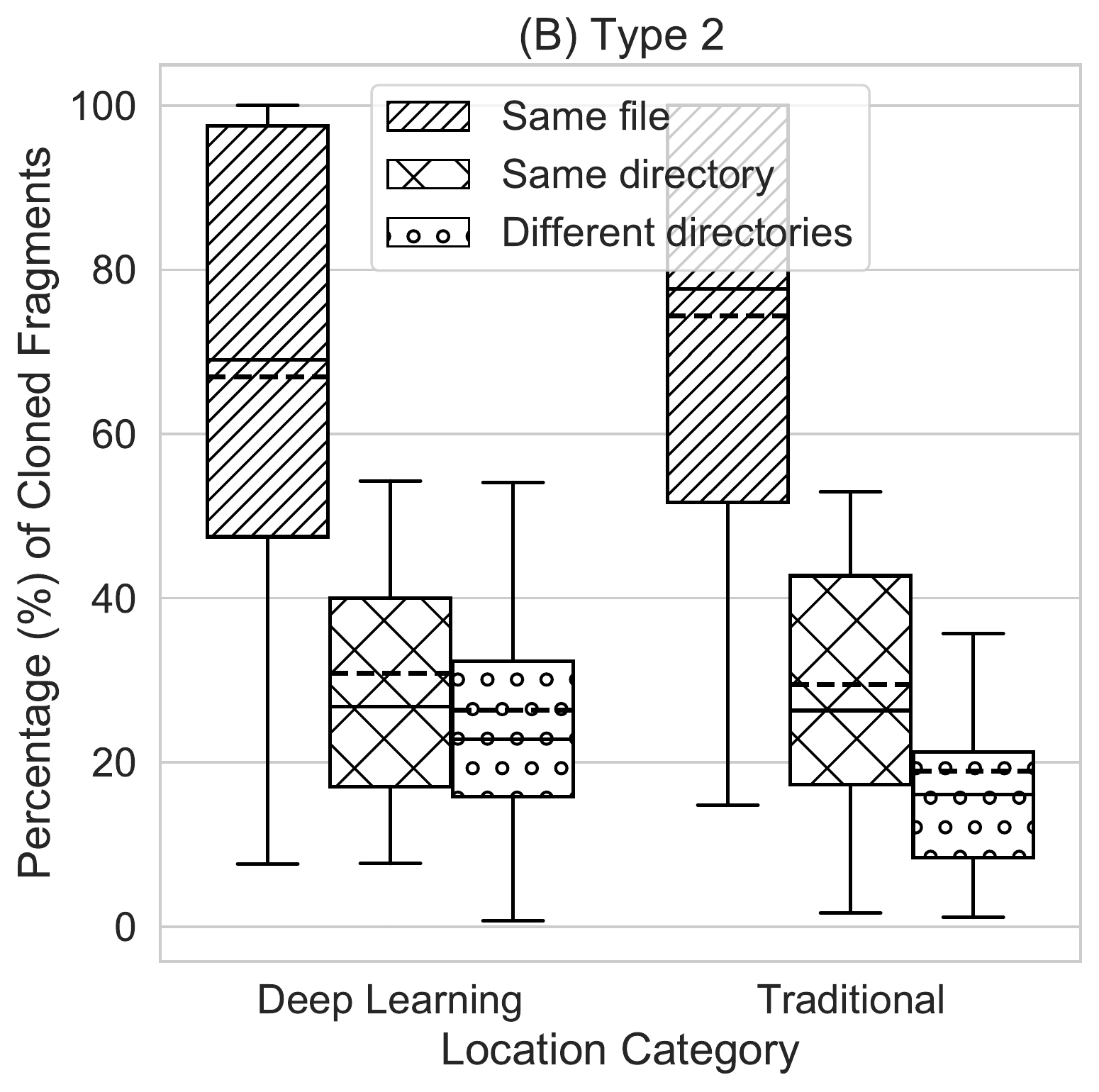}
\includegraphics[width=.49\textwidth]{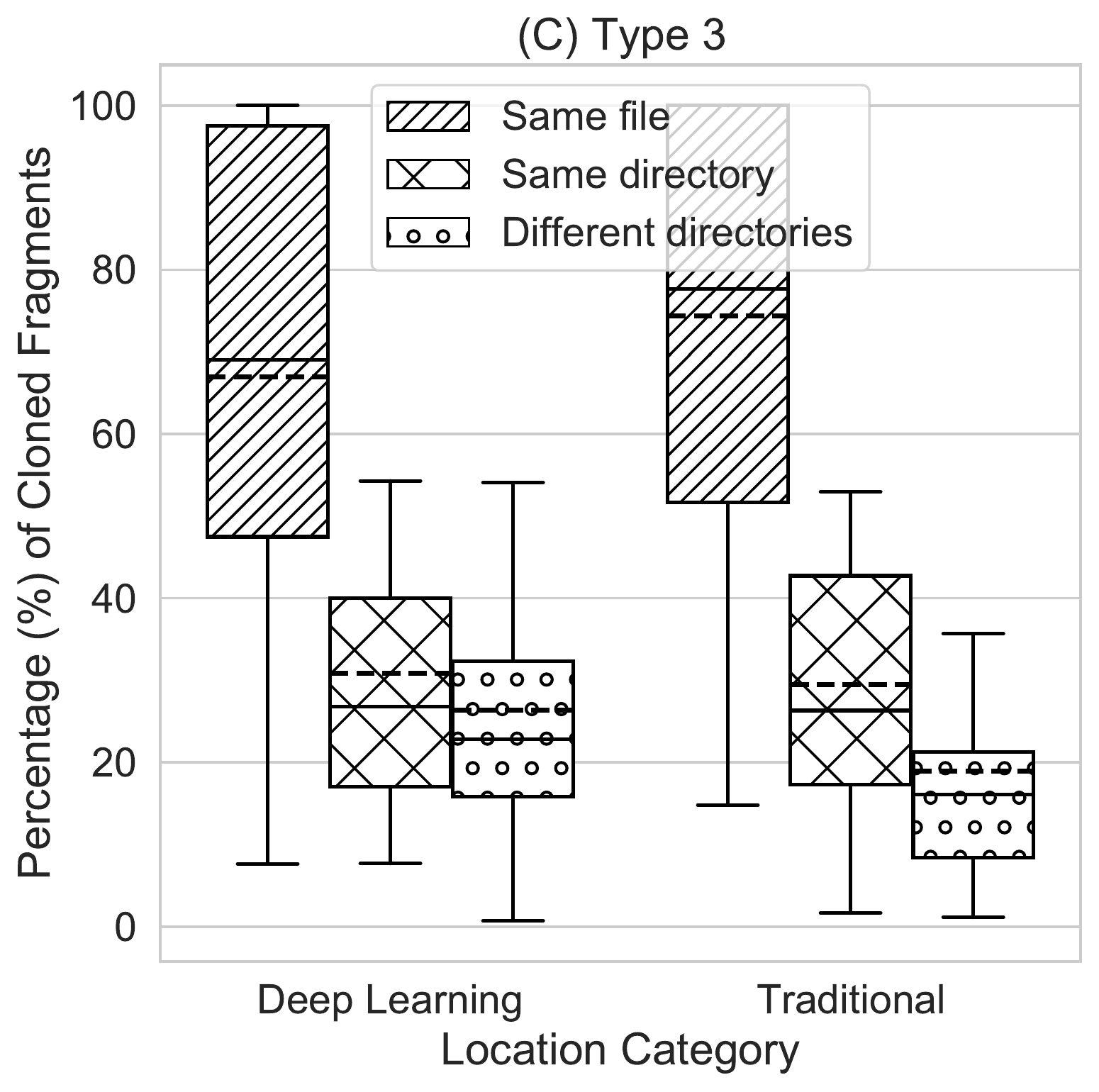}

\caption{Distribution of Percentage of Number of Fragments of Code Clones Classes per Clone Location.}
\label{fig:FragmentsPerLocationByCloneType}
\end{figure}

The mean value of the percentages of code fragments that belong to the same directory category in Python deep learning code is 32.04\% with a standard deviation (STD)
of 18.48. 
The number of fragments influences the degree to which the identified code clones from the same directory are difficult to maintain: the higher the number of fragments, the more troublesome their maintenance is likely to be. Hence, DL code may become more problematic with the spread of many duplicated code fragments that tend to be exact (Fig. \ref{fig:FragmentsPerLocationByCloneType}-A) and in different files, but in the same directory. We observe that the distributions of the percentages of clone fragments of different clone types in different locations (Fig. \ref{fig:FragmentsPerLocationByCloneType}) is similar to the distributions of the percentages of lines of cloned code (Fig. \ref{fig:CloneLocationbyCloneType_python}).  

\begin{figure}[htpb]
\centering
\includegraphics[width=.49\textwidth]{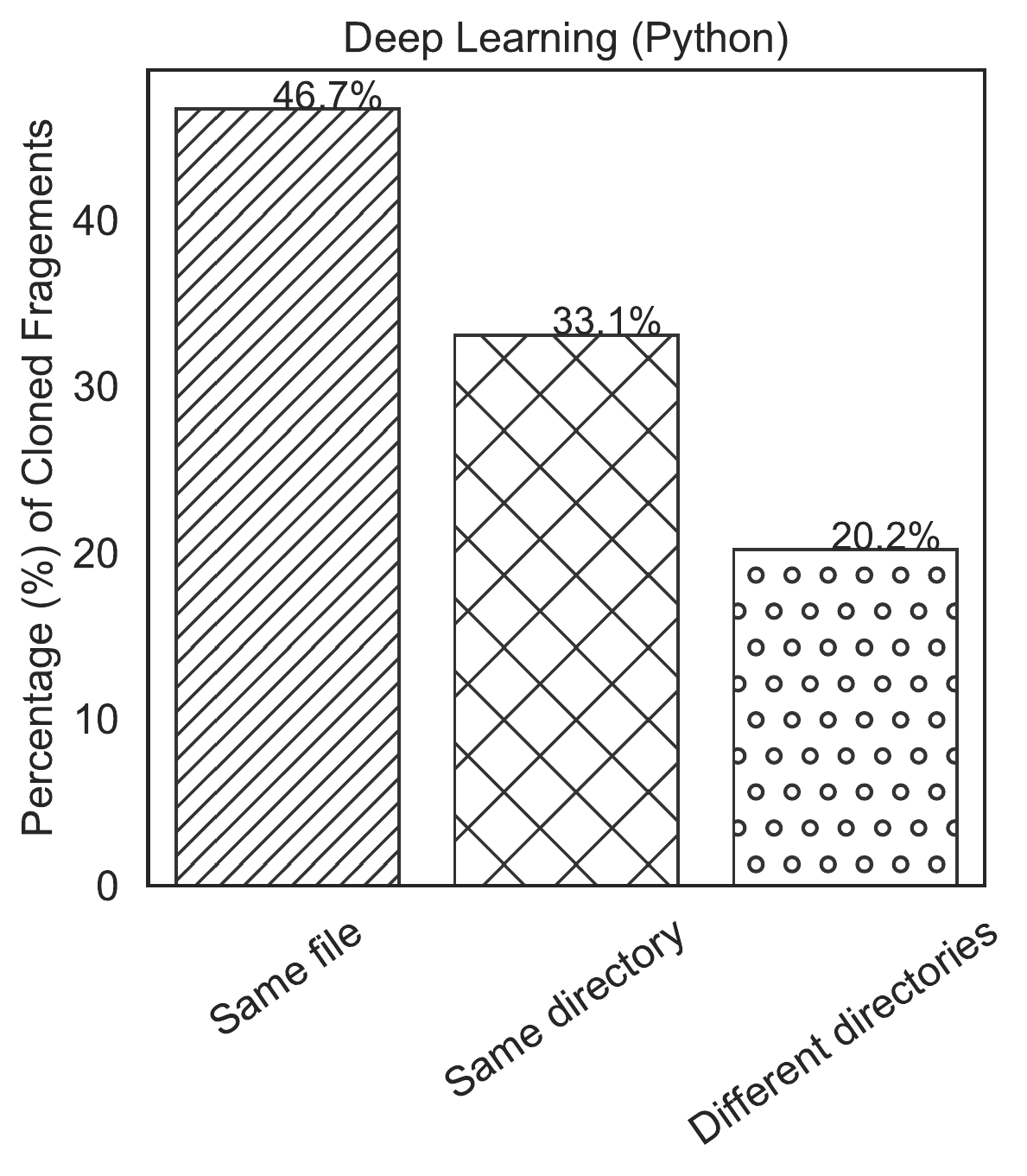}
\includegraphics[width=.49\textwidth]{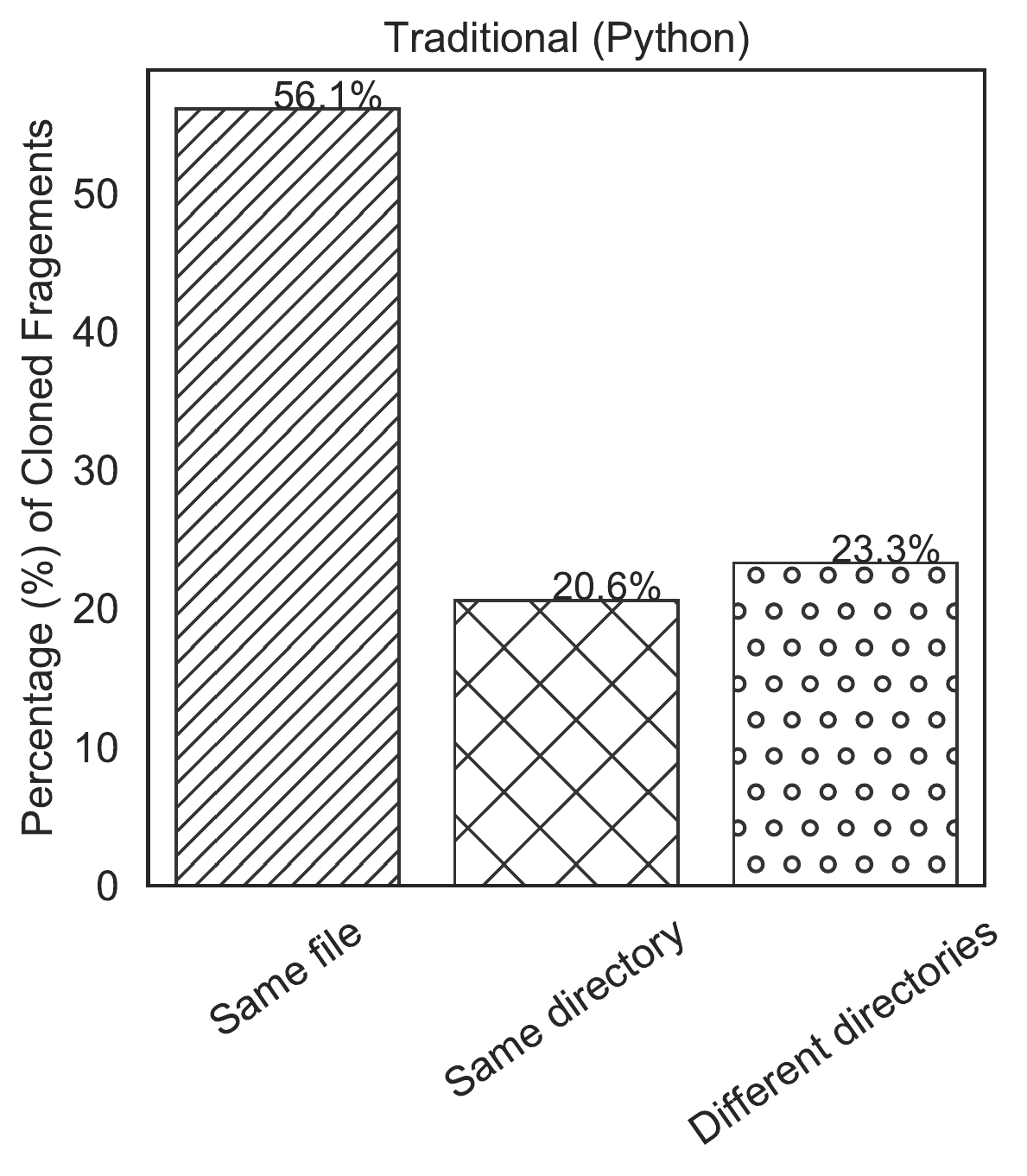}
\caption{Percentages of Average Number of Fragments of Code Clones by Location of Clones in both Deep Learning and Traditional Python Systems}
\label{fig:DLFragByCloneLocation_python}
\end{figure}

\begin{figure}[htpb]
\centering
\includegraphics[width=.49\textwidth]{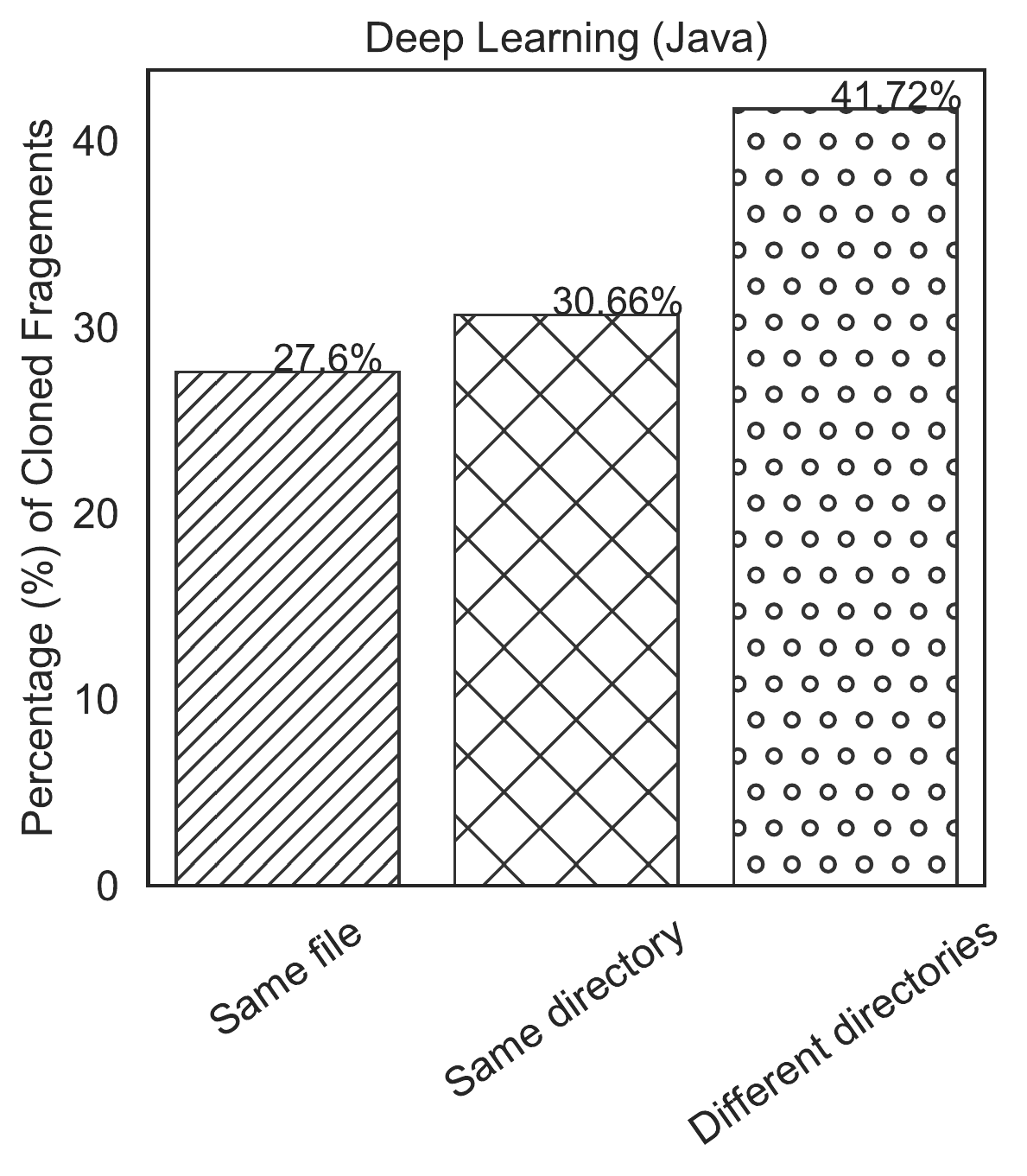}
\includegraphics[width=.49\textwidth]{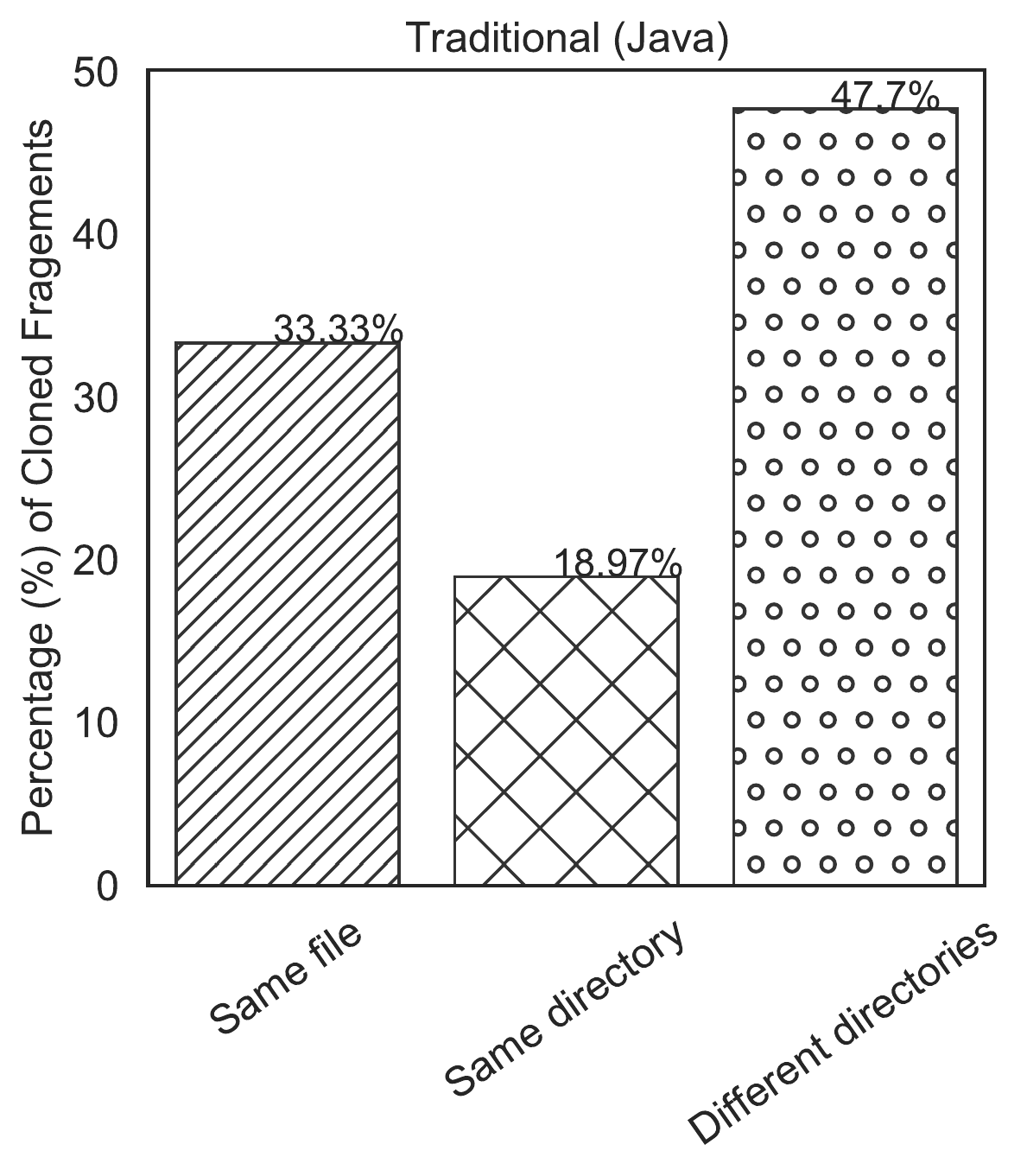}
\caption{Percentages of Average Number of Fragments of Code Clones by Location of Clones in both Deep Learning and Traditional Java Systems}

\label{fig:DLFragByCloneLocation_java}
\end{figure} 

\begin{figure}[htpb]
\centering
\includegraphics[width=.49\textwidth]{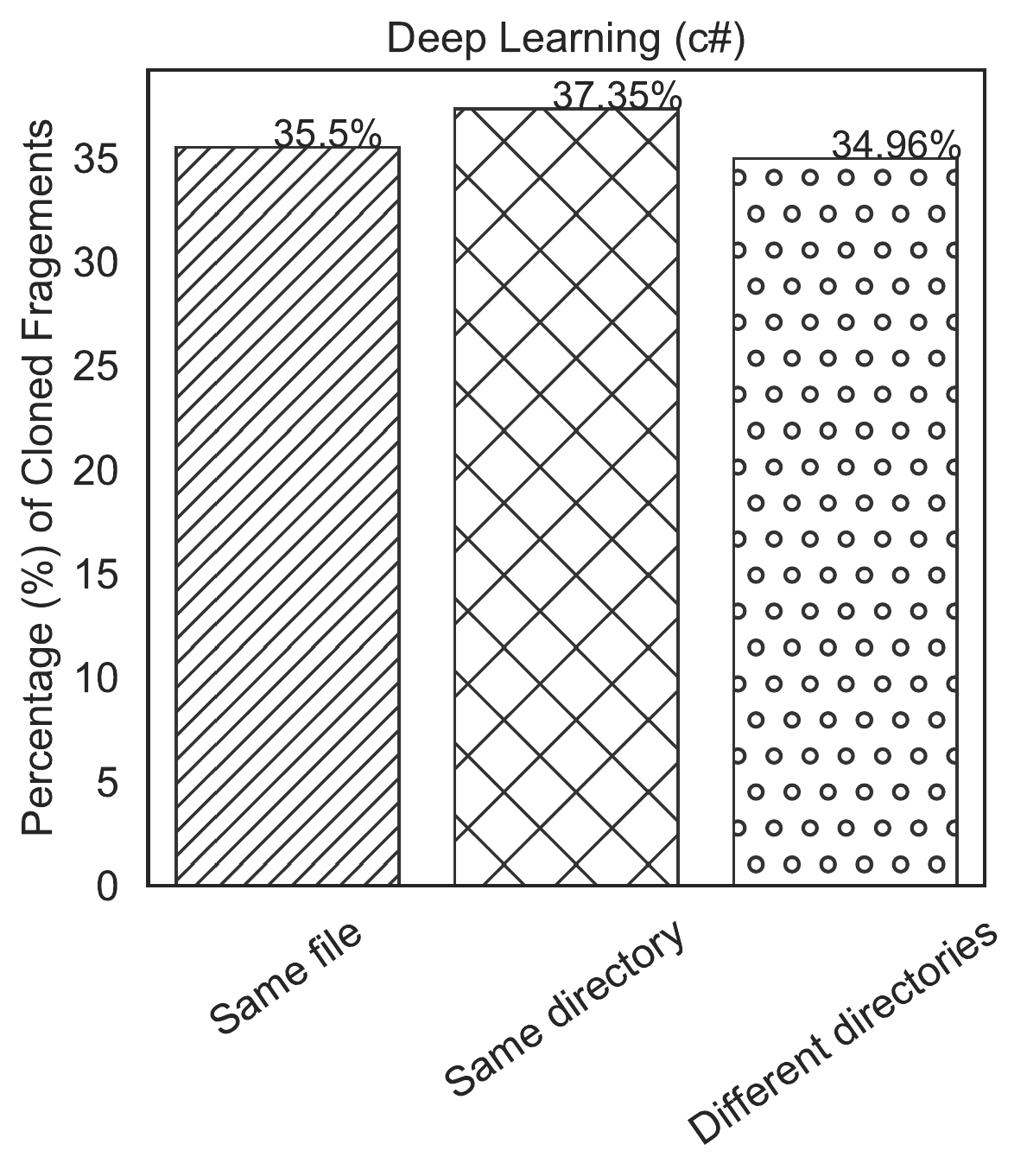}
\includegraphics[width=.49\textwidth]{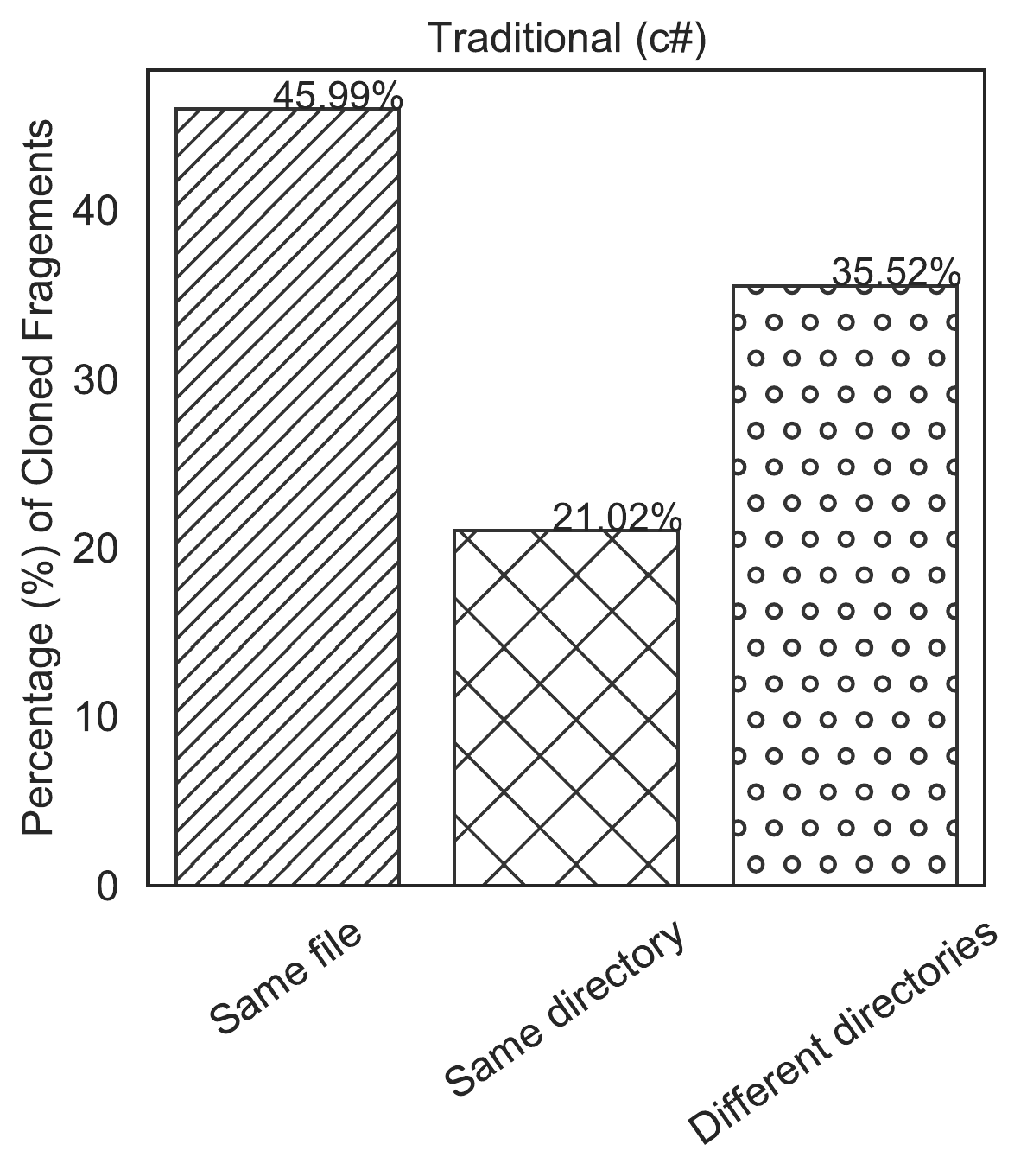}
\caption{Percentages of Average Number of Fragments of Code Clones by Location of Clones in both Deep Learning and Traditional C\# Systems}
\label{fig:DLFragByCloneLocation_csharp}
\end{figure} 

For further insights, we analyze the proportion
of the number of clones fragments in each clone class by their locations in deep learning and traditional code. As shown in Fig. \ref{fig:DLFragByCloneLocation_python} for Python systems, we identify that the distribution of the number of fragments in each clone class is of 46.7\% when clones are DL-related clones and are in the `same file', 33.1\% are in the `same directory' and 20.2\% are in `different directories'. Hence, based on the distribution of clone fragments, DL clones are found to be located in the  same file with a percentage equals to 46.7\%, which is nearly the same proportion as it is located in different files (i.e, in the same directory or in different directories (33.1\%+20.2\%)).
However, fragments of clones in Python deep learning code are relatively more dispersed compared to fragments of clones in traditional system as traditional systems have higher proportion of clones fragments in `same file' compared to deep learning code as shown in Fig. \ref{fig:DLFragByCloneLocation_python}. For Java systems (Fig. \ref{fig:DLFragByCloneLocation_java}), the major proportion of clone fragments in both DL and traditional code are in 'same directory' and 'different directories' containing 72.38\% (30.66 + 41.72) and 66.67\% (18.97 + 47.70) of cloned fragments respectively. So, clone fragments in Java DL systems are more dispersed compared to the clones in traditional Java systems. In Fig. \ref{fig:DLFragByCloneLocation_java} showing the distribution of cloned fragments in C\# systems, the propertion of clones in DL systems and traditional systems are 72.31\% (37.35+34.96) and 56.54\% (21.02 + 35.52) respectively. This also indicates that for C\#, clone fragments in DL code is more dispersed compared to traditional systems similar to Python and Java systems.
This higher dispersion clone fragments in the deep learning code likely to have negative consequences on maintenance due navigation and comprehension overhead for distant code. 

We also repeated the analysis for RQ2 with 20\% dissimilarity threshold for NiCad clone detector. However, the overall trends in the distribution of clones in different location categories remain the same as shown by our results as shown in Appendix \ref{appRQ2}.

\paragraph{\textbf{Qualitative analysis of the clones in different locations category: }}
We manually examined the clones contained in the different locations categories and observed that:

\textbf{Same file:} DL practitioners often duplicate functions in the same file when configuring different models. We found cloned functions in the same file with names representing the name of the model with slight modification in initializing (hyper)parameters of each model as shown in Table \ref{tab:examplecodesamefile}.
This type of code clones located in close proximity may be relatively less problematic. This is mainly due to the ease of navigation between code clones during maintenance as their locations are not too distant from each other. Consequently, they may be less prone to inconsistent updates, which is a key reason behind the introduction of faults in cloned code.
Duplicating code in close proximity is widely used to simplify the conception of the system \cite{kapser2003toward} by renaming functions to facilitate code reuse and to make the cloned functions' name more related to its purpose, which improves program comprehension. These type of cloned functions with structural similarity but with identifier naming and data type differences are Type 2 clones. The trends of such clones to be in closer proximity is also reflected in our results from Fig. \ref{fig:CloneLocationbyCloneType_python}-B.

\begin{table}[htpb]
\caption{Clone Codes Example where the Location is in the Same File (the Differences are Highlighted in Gray) }

\label{tab:examplecodesamefile}
    \centering
    \begin{tabular}{||c|c||}
    \hline

\begin{minipage}{0.45\textwidth}
\vspace{0.1cm}
\begin{minted}[fontsize=\scriptsize, highlightlines={1,3,5,7-8,12},highlightcolor=mygray]{python}
def mobile_imagenet_config():
  return tf.contrib.training.HParams(
      stem_multiplier=1.0,
      dense_dropout_keep_prob=0.5,
      num_cells=12,
      filter_scaling_rate=2.0,
      drop_path_keep_prob=1.0,
      num_conv_filters=44,
      use_aux_head=1,
      num_reduction_layers=2,
      data_format='NHWC',
      skip_reduction_layer_input=0,
      total_training_steps=250000,
      use_bounded_activation=False,
  )
\end{minted}
\vspace{0.05cm}
\end{minipage}
& 
\begin{minipage}{0.45\textwidth}
\vspace{0.1cm}

\begin{minted}[fontsize=\scriptsize,highlightlines={1,3,5,7-8,12},highlightcolor=mygray]{python}
def large_imagenet_config():
  return tf.contrib.training.HParams(
      stem_multiplier=3.0,
      dense_dropout_keep_prob=0.5,
      num_cells=18,
      filter_scaling_rate=2.0,
      num_conv_filters=168,
      drop_path_keep_prob=0.7,
      use_aux_head=1,
      num_reduction_layers=2,
      data_format='NHWC',
      skip_reduction_layer_input=1,
      total_training_steps=250000,
      use_bounded_activation=False,
  )
  \end{minted}
  \vspace{0.05cm}
  \end{minipage} \\
    \hline
    \end{tabular}
\end{table}

Table \ref{tab:examplecodesamefile} represents examples of cloned code where the location is in the same file. 
Through a manual analysis, we tried to understand where DL developers duplicate code regarding DL phases perspectives. It was found that 40 code clones classes were in the same file and were corresponding to the DL phase. We found 27 of them to be similar functions with a common purpose. These functions were cloned to perform the same or closely similar tasks (to build the model with some modifications). The modifications were achieved by renaming functions (giving them names that are more meaningful and relevant to the task context) and parameterizing the code with different values that are specific to each model. As shown in Table \ref{tab:examplecodesamefile}, we have two functions that parameterize two different models. This is performed by calling a library routine capable of setting the hyperparameters of the model. It is configured as key-value pairs to build the model. Each function is renamed to be relevant to the model and we see slight differences between the values of each hyperparameter. An important number of this type of duplication is cloning similar functions with different names and parameter types, leading to Type 2 clones. This type of code duplication may explain the high percentage of Type 2 clones that are found to reside in the same file (Fig. \ref{fig:CloneLocationbyCloneType_python}-B). Our findings also confirm the results of previous work \cite{kapser2003toward}. 

\textbf{Same directory:} Regarding the second category where 
clones exist in different files but in the same directory, it is common to find duplicated functions without or with minor changes \cite{kapser2003toward}. In our case and as shown in Figure \ref{fig:CloneLocationbyCloneType_python}-A, for deep learning code, Type 1 clones are the type of clones that are frequently located in the same directory but in different files. 
 From a manual examination of all of the code clones (85) that exist in the same directory but in different files,
we found in 49 code clones classes (57\%) a file named utility containing useful functions needed to perform the construction of deep learning models.
This suggests that DL developers either refactor their code without deleting the old functions, or different developers working on the same project are not conscious about the existence of such files.

\textbf{Different directories:} Regarding clones located in different directories, we manually analyzed all the code clones fragments (102) that were located in different directories
that 63\% of them are not related to deep learning code. We often detect this type of duplication when it comes to verifying libraries' versions to choose the right routine call, deallocate memory, or getting model metadata (logging).

\vspace{4mm}
\fbox{\begin{minipage}{35em}
 \hspace{4mm} 
\textit{\textbf{Summary of findings (RQ2):}} According to our results, code clones in deep learning code are more dispersed than in traditional code in terms of the distribution of code clones. While for clone types, Type 1 clones are more dispersed in DL systems while Type 2 clones tend to be localized in the same file. Type 3 clones are spread in different locations but with a high percentage of lines of code clones residing in the same file. Regarding the distribution of the number of clone fragments, clones in deep learning code tend be more dispersed compared to clones in traditional software systems. 

\end{minipage}}

\subsection{\textbf{RQ\(3\): Do cloned and non-cloned code suffer similarly from bug-proneness in deep learning projects?}}

Since deep learning systems are relatively fast to develop and deploy, code quality is often overlooked and it is frequently the case that the code is re-used and rarely refactored \cite{wan2019does}. Several previous studies in traditional code highlighted the negative impacts of code clones on the maintenance and comprehension of code. Barbour et al. \cite{barbour2018investigation} found clones to be related to a high risk of bugs. Given the complexity of deep learning systems, it is likely that clones can have a similar averse effect on maintenance and bug-proneness. Hence, in this work, we analyze the bug-proneness of clones in deep learning code from two perspectives: (1) we examine correlations between the co-occurrences of clones and bugs in deep learning code, and (2) examine whether clones affect the time required to fix bugs in deep learning systems. 
We perform these investigations first on all clones and then for individual clone types (Type 1, Type 2, Type 3). We analyze all the commit history to identify all buggy commits (details are presented in Section \ref{subsubsec:cloneBugMethod}).  
In order to determine the co-existence between bugs and code clones in deep learning code, we match code changes in bug-fixing commits with code clones by finding the intersection between the lines changed to fix bugs and the cloned lines of code. 

We consider that a bug fix commit is related to code clones when the buggy lines belong to any duplicated code, otherwise it is considered as related to non-cloned code. Then, we calculate the percentage of bug-fix commits related to cloned and non-cloned code for each project. Finally, we compute the average percentage of bug-fix commits related to cloned and non-cloned code, to comparatively evaluate their bug-proneness in the context of deep learning code. 

Our results show that \textbf{75.85\%} of bug-fix commits in Python DL systems are related to clones, i.e., in other words, more than three-quarters of the bugs in deep learning code are related to clones.
When we do similar analysis on bug-fix commits in C\# and Java DL systems, we found that \textbf{27.54\%} are related to clones in C\# DL systems and \textbf{45.31\%} are related to clones in Java DL systems.
We perform MWW tests for the distributions of code clones and non-clone code in the Python DL bug-fix commits. We found a statistically significant difference between the distribution of the number of commits that fix bugs on cloned lines and the distribution of bug-fix commits on non-cloned code, with a p-value equals to 0.026 and an effect size of 0.55 (medium).
Thus, the bug-proneness of cloned code in deep learning code is higher compared to that of non-clone code. However, the degree of bug-proneness may vary with systems of different programming languages, as we observe in our results.

\begin{figure}[htpb]
\centering
\includegraphics[width=.8\textwidth]{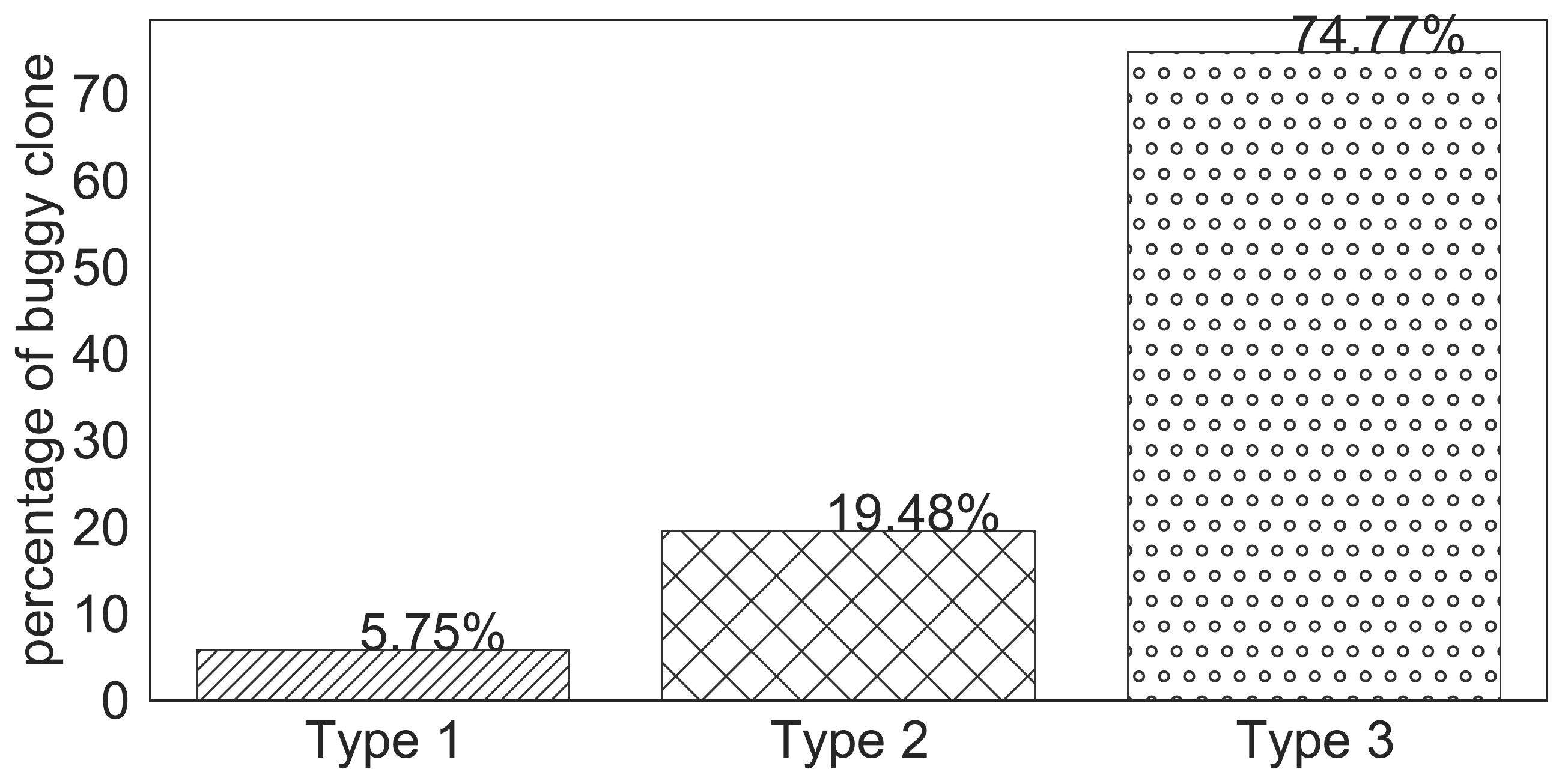}
\caption{Buggy Code Clones Occurrences by Clone Type for Python DL systems}
\label{fig:buggy_clone_type}
\end{figure}

\begin{figure}[htpb]
\centering
\includegraphics[width=.8\textwidth]{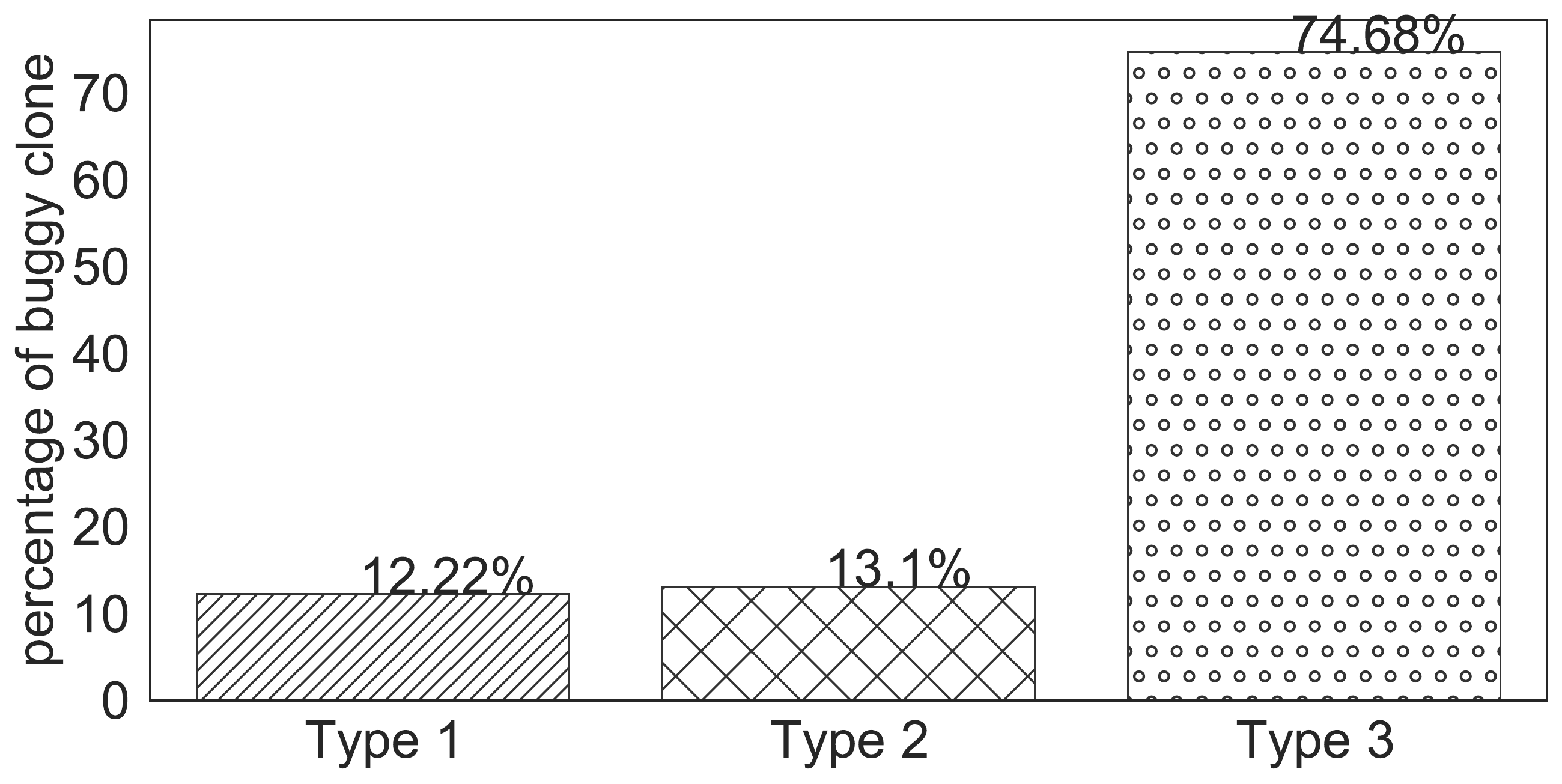}
\caption{Buggy Code Clones Occurrences by Clone Type for C\# DL systems}
\label{fig:buggy_clone_type_cs}
\end{figure}

\begin{figure}[htpb]
\centering
\includegraphics[width=.8\textwidth]{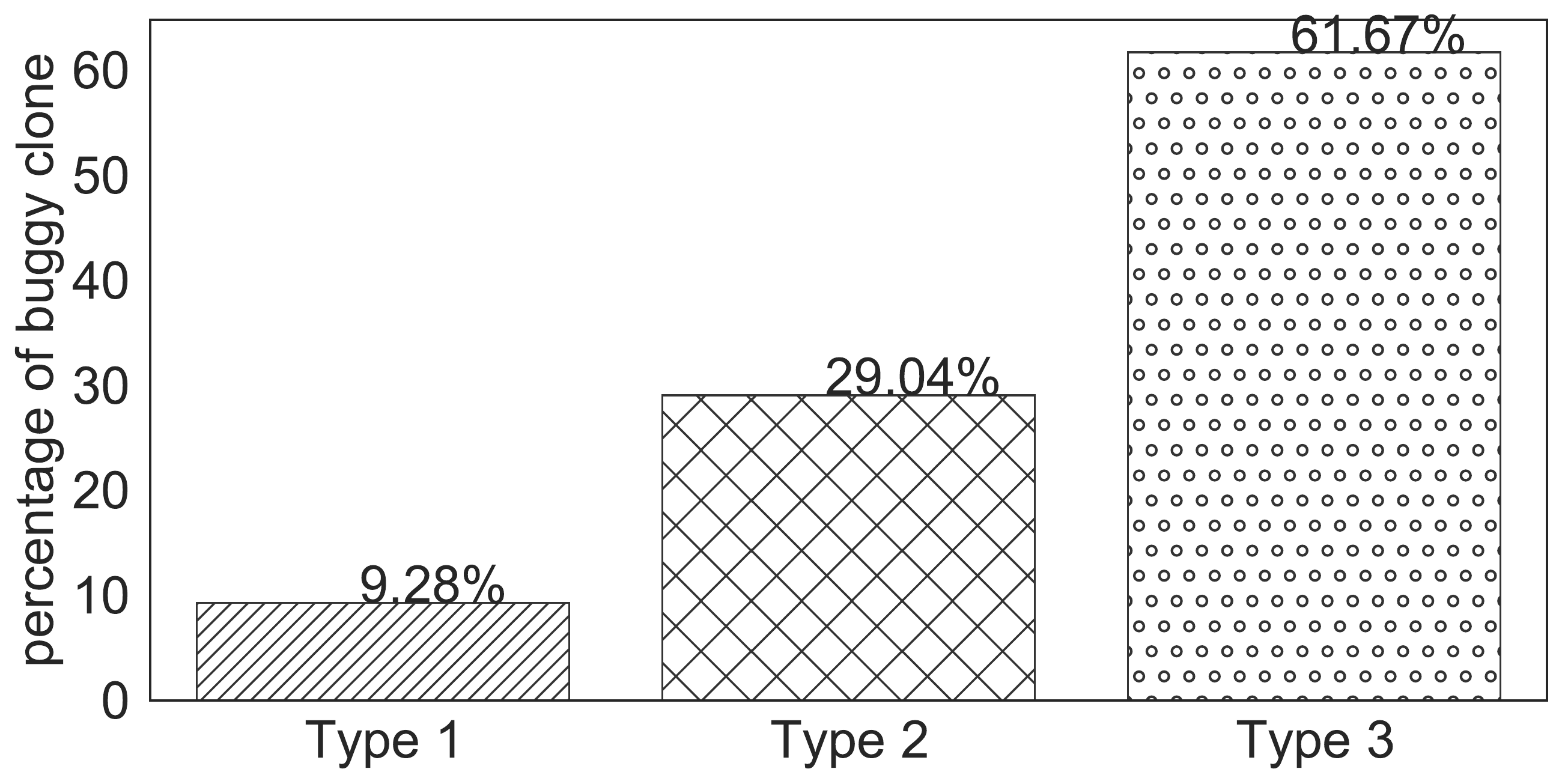}
\caption{Buggy Code Clones Occurrences by Clone Type for Java DL systems}
\label{fig:buggy_clone_type_java}
\end{figure}

Now, we further investigate the bug-proneness of different types of clones in DL code to gain deeper insights on the types of clones that are likely to be more risky (in terms of bug occurrence). 
This information would help deep learning developers to carefully prioritize clones for refactoring and tracking. 
Figure \ref{fig:buggy_clone_type}, \ref{fig:buggy_clone_type_cs} and \ref{fig:buggy_clone_type_java} show the percentages of clones from different types (Type 1, Type 2, and Type 3) that are related to bugs for Python, C\# and Java DL systems respectively. 


We find that Type 3 clones are the most likely to be buggy in all types of DL systems as \textbf{74.77\%, 74.68\%} and \textbf{61.67\%} of clone related bugs are Type 3 clones for Python, C\# and Java DL projects. Then, we have Type 2 clones with a percentage of \textbf{19.48\%, 13.1\%} and \textbf{29.04\%} for Python, C\# and Java respectively, and finally Type 1 clones with a percentage of \textbf{5.75\%, 12.22\%} and \textbf{9.28\%} clones related to bugs. These results obtained from deep learning code are similar to the findings of previous works comparing the bug-proneness of Type 1, Type 2, and Type 3 clones in traditional software systems \cite{Mondal_ICSME_2015}.

\begin{figure}[htpb]
\centering
\includegraphics[width=.8\textwidth]{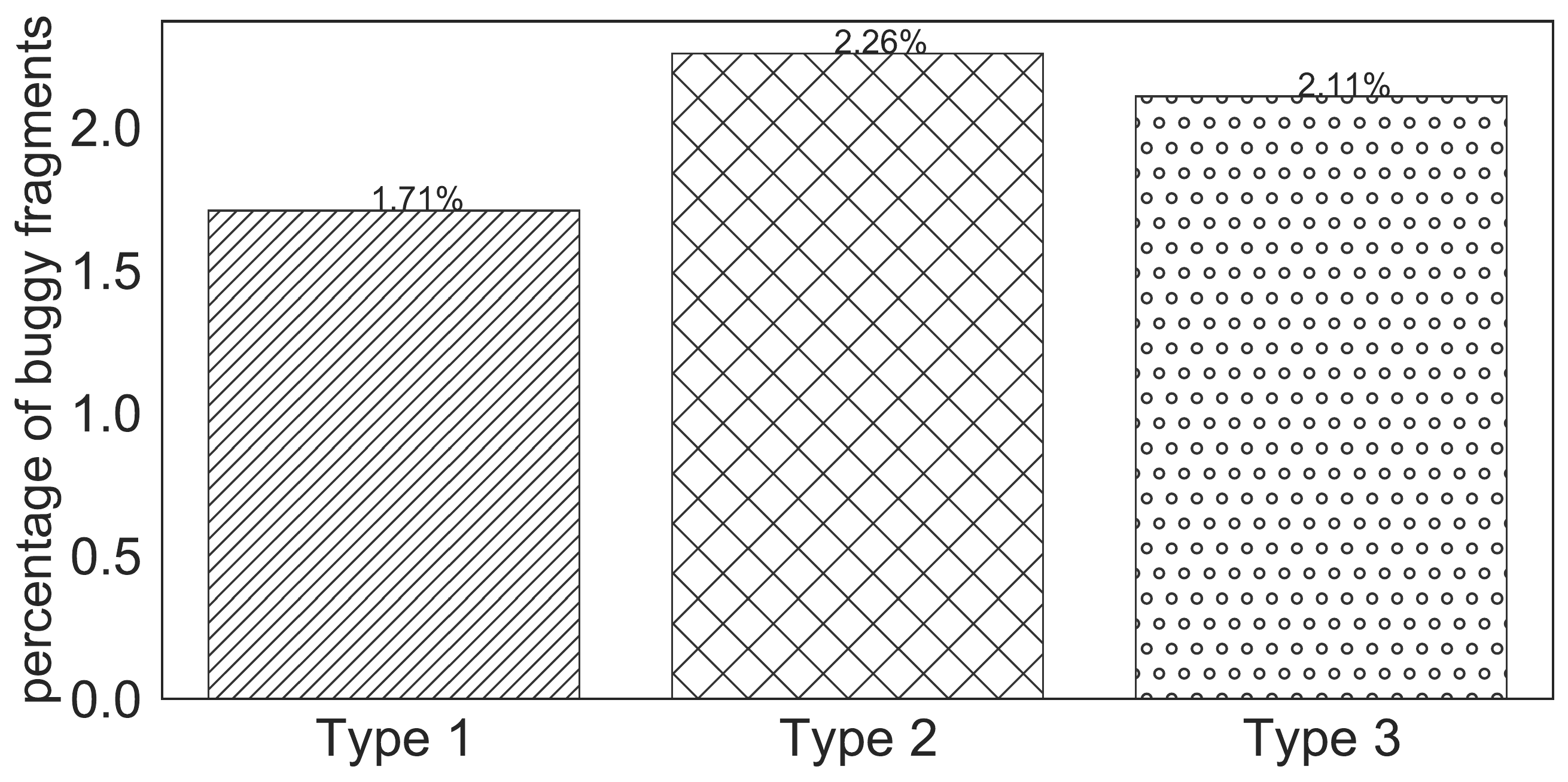}
\caption{Percentages of Buggy Code Fragments by Clone Type for python DL systems}
\label{fig:buggy_fragments_clone_type}
\end{figure}

Since Type 3 clones are higher in density (Fig. \ref{fig:dl_trad_clone_type_python}) and prevalent according to the distribution of the percentages of lines of code in different types of clones, they are possibly being associated with higher percentages of bugs too. Thus, we investigate further their bug-proneness by studying the percentage of clone fragments in each clone types (Type 1, Type 2, Type 3) that are related to bugs. As shown in Figure \ref{fig:buggy_fragments_clone_type}, we find that 1.71\% of clone fragments are buggy in Type 1 clones, 2.26\% of clone fragments are buggy in Type 2 clones and 2.11\% of clone fragments are buggy in Type 3 clones. 
This shows that a higher percentage of Type 2 clone fragments are related to bugs, followed by Type 3 clones and then Type 1 clones. However, there are more Type 3 clones in the deep learning code (Fig. \ref{fig:dl_trad_clone_type_python}) compared to Type 1 and Type 2 clones. This explains the observation that Type 3 clones contains the highest fractions of clone related bugs (Fig. \ref{fig:buggy_clone_type}) despite the percentages of buggy clone fragments in Type 3 not being higher than that of Type 2 clones.

\begin{figure}[htpb]
\centering
\includegraphics[width=.8\textwidth]{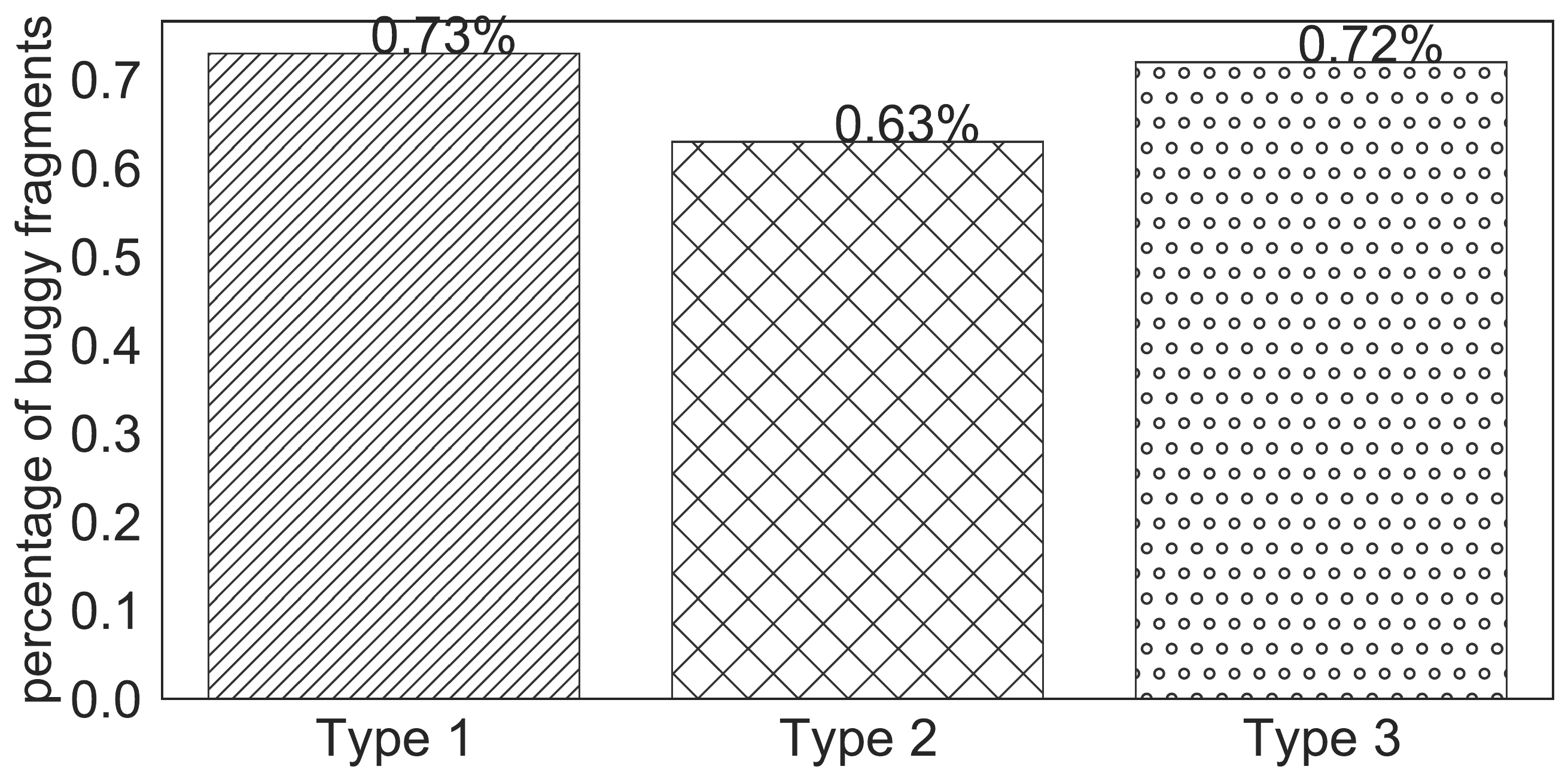}
\caption{Percentages of Buggy Code Fragments by Clone Type for C\# DL systems}
\label{fig:buggy_fragments_clone_type_cs}
\end{figure}

\begin{figure}[htpb]
\centering
\includegraphics[width=.8\textwidth]{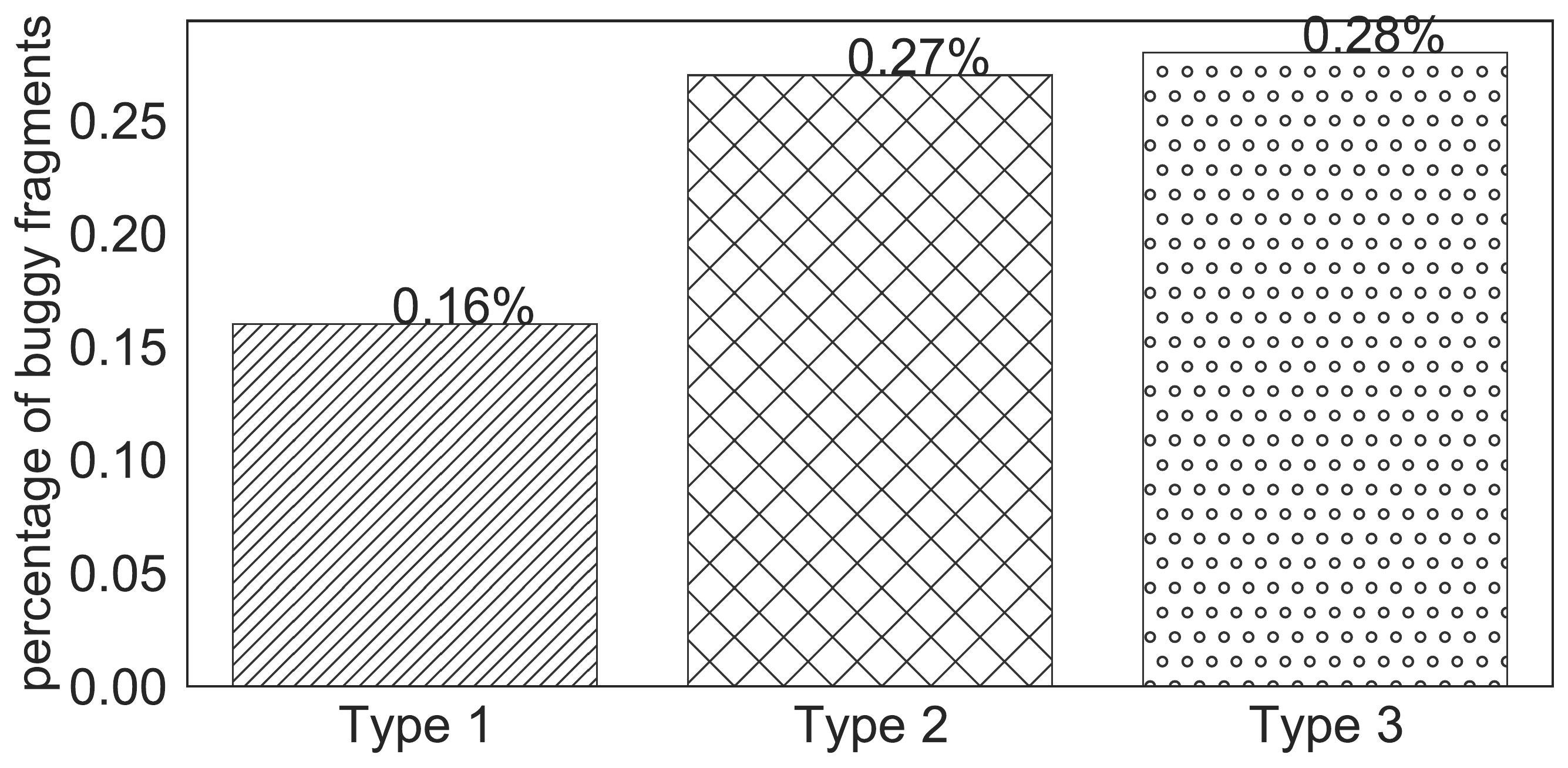}
\caption{Percentages of Buggy Code Fragments by Clone Type for Java DL systems}
\label{fig:buggy_fragments_clone_type_java}
\end{figure}


Figure \ref{fig:buggy_fragments_clone_type_cs} and \ref{fig:buggy_fragments_clone_type_java} shows the trends for C\# and Java DL systems. We find that 0.73\% and 0.16\% clone fragments are buggy in type 1 clones, 0.63\% and 0.27\% are buggy in Type 2 clones and 0.72\% and 0.28\% are buggy in Type 3 clones for C\# and Java DL systems respectively. The highest percentage is from Type 3 for Java and Type 1 for C\#.


Our results show that clones in deep learning systems are more related to bugs compared to non-cloned code. This observed bug-proneness of clones in DL code is likely to be related to different confounding factors such as relative size and distribution of cloned and non-cloned code, clone types, programming language, clone detection tool and different configuration settings, etc. We carefully considered these confounding factors to avoid potential threats. As size of the clones and non-cloned is known to be an important confounding factor for empirical analysis of clones \cite{SainiEMSE2018}, we investigated the relative size of the cloned and non-cloned code. We observed that on an average 27.45\% of code in the DL systems are cloned code. We also compared the relative size of the cloned and non-cloned functions. NiCad extracts all functions in the system as candidate clone list for clone detection. We used those extracted functions to separate cloned and non-cloned functions for our analysis by comparing to the clone detection results. We then analyzed the comparative size distribution of cloned and non-cloned functions. As shown in the Appendix \ref{appRQ3} (Fig. \ref{fig:FunctionSize}, Fig. \ref{fig:MeanFunctionSize_sys}, Fig. \ref{fig:MeanFunctionSize_lang}), we do not observe any statistically significant difference between the size of the cloned and non-cloned functions. However, non-cloned functions tend to be larger in size compared to cloned functions as can be seen on the median values of their distributions. The differences between the sizes of cloned and non-cloned functions are not statistically significant in DL and traditional code, and we observe a similar trend for programming language variations (Fig. \ref{fig:MeanFunctionSize_lang}).

Now, we study how clones in deep learning code affect the time to fix bugs in deep learning code. We investigate whether the time required to fix bugs in code clones is more than when bugs are not in code clones. Thus, we calculated the time between bug-fixing commit and the corresponding bug-inducing commit. Then, we compare the average time spent to fix bugs when bugs are related to clones (bug-fix lines intersect with code clones) and when the bug is not clone related. 

\begin{figure}[htpb]
\centering
\includegraphics[width=.7\textwidth]{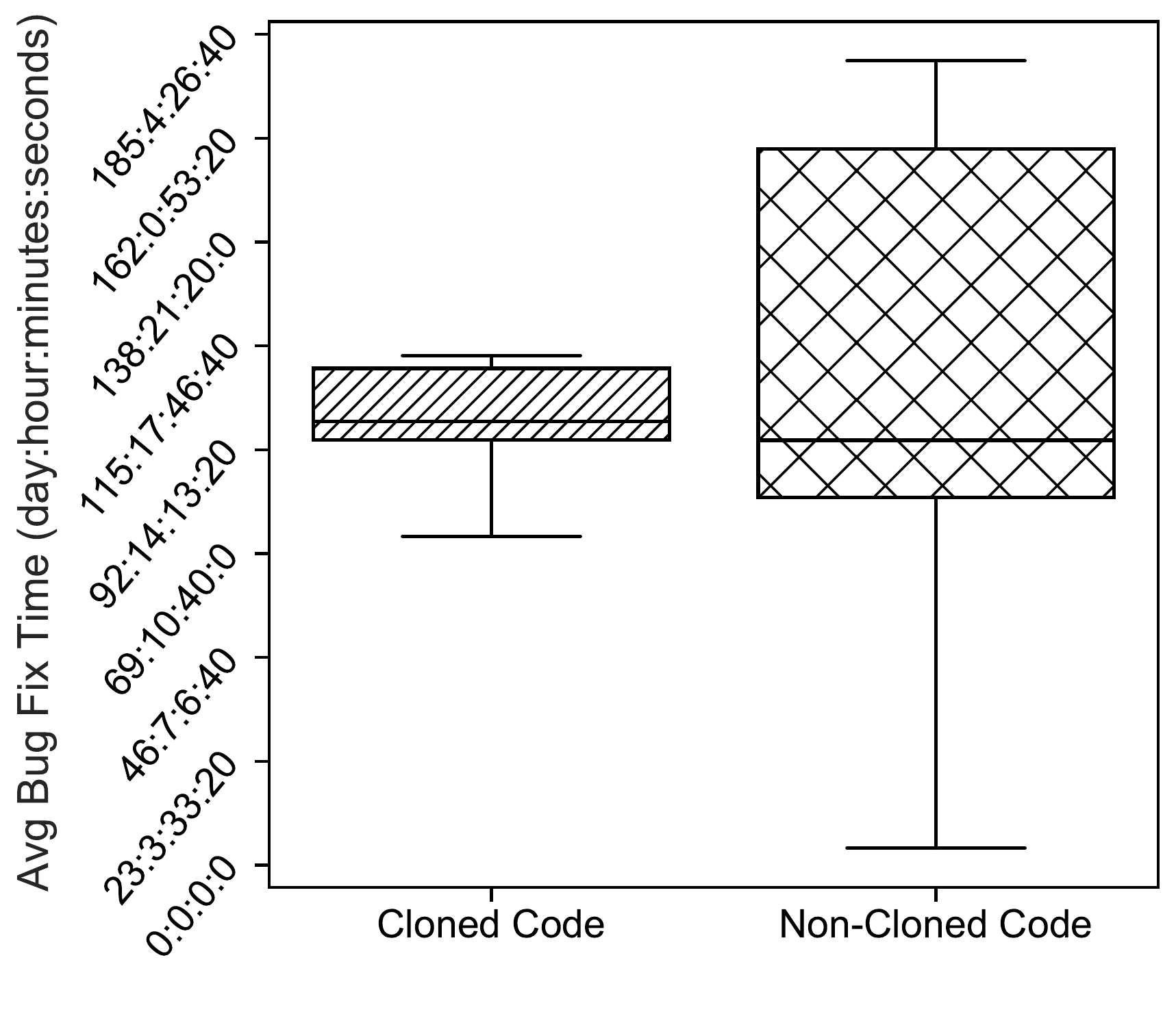}
\caption{Comparative Bug-Fix Times for Cloned and non-Cloned Code in python DL Systems.}
\label{fig:time_fix_bugs}
\end{figure}

Figure \ref{fig:time_fix_bugs} shows the distribution of the average times to fix bugs for python DL systems when the bug is located in a clone and when it is not.  
Comparing the median between the two distributions of time, we can see on Figure \ref{fig:time_fix_bugs} that the time required to fix bugs when there is cloned code is slightly higher than the time required to fix a bug when the bug is not related to a code clone. The mean value of the average time to fix bugs in clones in DL code is 134.0 days, 6.0 hours, 13.0 minutes and 20.0 seconds with a STD of 95.0 days, 22.0 hours, 46.0 minutes and 40.0 seconds, and when it comes to non cloned code, the mean time to fix a bug is 122.0 days, 16.0 hours, 26.0 minutes and 40.0 seconds with a STD of 96.0 days 15.0 hours 26.0 minutes and 40.0 seconds. Buggy cloned code seems to be taking comparatively more time to get fixed in deep learning code. 
We perform a Man-Whitney test comparing the distribution of times. However, we found no statistically significant difference between the time to fix bugs in cloned and non-cloned code (p-value =0.34, effect size 0.2).

Similarly, the average bug fix time (mean=77 days) for cloned-codes in C\# DL systems is higher than non-cloned code (mean=51 days). ConvNetSharp DL system has average bug fix time of 132 days for cloned code and 89 days for non cloned code while NeuralNetwork.NET DL system has average bug fix time of 22 days for cloned code and 13 days for non cloned code. On the other hand, the average bug fix time of cloned-codes in Java DL systems (mean=63 days) is slightly lower than that of non-cloned codes (mean=64 days). The knime-deeplearning Java DL system has average bug fix time of 105 days for clone-code and 102 days for non-clone code while Neuralnetworks DL system has average bug-fix time of 21 for cloned code and 26 for non-cloned code.


We further check what percentages of DL systems have higher bug-fix time for clones and what percentages of systems have the opposite. Among 10 DL subject systems, we observe in seven DL systems that the bug-fix time for clones is higher and for the rest (i.e., three systems), the bug-fix time for clones is lower. Overall, we can conclude that in a majority of cases (70\%), bugs related to clones take a longer time to get fixed, suggesting that bugs occurring in cloned code may be more challenging to fix.

The observation that bugs in cloned code take comparatively longer time to fix means that the existence of code clones in deep learning code may hinder the maintenance of this type of system. 

\vspace{4mm}
\fbox{\begin{minipage}{35em}
\hspace{4mm}
\textit{\textbf{Summary of findings (RQ3):}} According to our results, we find that cloned code is likely to be more bug-prone than non-cloned code in deep learning systems. In addition, Type 3 clones have a relatively higher odd to be involved in bugs in the deep learning code, followed by Type 2 and Type 1 clones respectively. Also, bugs related to clones in DL code tend to take more time to get fixed compared to other bugs. 

\end{minipage}}

\subsection{\textbf{RQ\(4\): Why do deep learning developers clone code?}}

In this research question, we examine the reasons behind the practice of code cloning in deep learning systems. We manually analyzed the detected code clones for a selected subset of six deep learning projects. We labeled each detected code clone class by the functionality it serves (task). Then, we assign each labeled clone class to its corresponding DL phase making sure that the relation between the labeled task and the DL phase is `to perform' (more details in section \ref{subsubsec:DLtaxonomy}). 

Table \ref{tab:dlCodeCloneTaxonomy} shows the taxonomy of code clones that resulted from our manual analysis of the selected six DL projects. We show only the related DL phases that co-occur with the detected code clones. Therefore, we have neither all the DL phases nor all of its subcategories presented in Table \ref{tab:dlCodeCloneTaxonomy}. Also, the DL phase subcategories are not exclusive, since functions may be used in tasks related to different phases.

In this research question, we study clones at the granularity of the function. A function can implement one or more tasks 
that are involved in a DL phase. In the following, we discuss the characteristics of clones in deep learning systems that are associated with different functionalities and development phases of deep learning systems. The following phases are listed based on the clones occurrences ratio from highest to lowest. 

\begin{table}[htp]
\caption{Percentages of Occurrence of Code Clones in DL Phases}
\label{tab:dlCodeCloneTaxonomy}
\resizebox{\textwidth}{!}{%
\setlength{\tabcolsep}{1.2pt} 
\begin{tabular}{|l|l|c|c|c|c|c|}
\hline
\textbf{\begin{tabular}[c]{@{}l@{}}dl\_phase\\ category\end{tabular}} &
\textbf{\begin{tabular}[c]{@{}l@{}}dl\_phase\\ subcategory\end{tabular}} &
  \textbf{\begin{tabular}[c]{@{}l@{}}Type 1\\ \%occs\end{tabular}} &
  \textbf{\begin{tabular}[c]{@{}l@{}}Type 2\\ \% occs\end{tabular}} &
  \textbf{\begin{tabular}[c]{@{}l@{}}Type 3\\ \% occs\end{tabular}} &
  \textbf{\begin{tabular}[c]{@{}l@{}}\% occs 
 \\ in subcat
  \end{tabular}} &
  \textbf{\begin{tabular}[c]{@{}l@{}}\% occs\\  in total\end{tabular}} \\
  \hline
\begin{tabular}[c]{@{}l@{}}Preliminary\\ preparation\end{tabular}                        & hardware requirements            & 100   & 0     & 0     & 100.0 & 1.03                    \\ \hline
\multirow{2}{*}{\begin{tabular}[c]{@{}l@{}} Data\\ collection\end{tabular} }     & load data                        & 20    & 20    & 40    & 80.0  & \multirow{2}{*}{5.15}   \\ \cline{2-6}
                                     & load label                       & 20    & 0     & 0     & 20.0  &                         \\ \hline
\multirow{4}{*}{\begin{tabular}[c]{@{}l@{}} Data\\ postprocessing\end{tabular} } & compute output shape             & 0     & 0     & 12.5  & 12.5  & \multirow{4}{*}{8.25}   \\ \cline{2-6}
                                     & object localization              & 25    & 0     & 12.5  & 37.5  &                         \\ \cline{2-6}
                                     & process output                   & 25    & 0     & 12.5  & 37.5  &                         \\ \cline{2-6}
                                     & set shape of output data         & 12.5  & 0     & 0     & 12.5  &                         \\ \hline
\multirow{12}{*}{\begin{tabular}[c]{@{}l@{}} Data\\ preprocessing\end{tabular} } & apply data augmentation          & 5.55  & 0     & 0     & 5.55  & \multirow{12}{*}{18.56} \\ \cline{2-6}
                                     & data normalization               & 0     & 0     & 11.11 & 11.11 &                         \\ \cline{2-6}
                                     & get batches of data              & 0     & 5.55  & 0     & 5.55  &                         \\ \cline{2-6}
                                     & get numerical feature columns    & 5.55  & 0     & 5.55  & 11.11 &                         \\ \cline{2-6}
                                     & parse arguments                  & 0     & 0     & 5.55  & 5.55  &                         \\ \cline{2-6}
                                     & prepare tensor                   & 11.11 & 0     & 0     & 11.11 &                         \\ \cline{2-6}
                                     & process input                    & 0     & 0     & 16.66 & 16.66 &                         \\ \cline{2-6}
                                     & resize image                     & 5.55  & 0     & 0     & 5.55  &                         \\ \cline{2-6}
                                     & set shape of input data          & 0     & 0     & 11.11 & 11.11 &                         \\ \cline{2-6}
                                     & set type of input data           & 0     & 0     & 5.55  & 5.55  &                         \\ \cline{2-6}
                                     & setting format input data        & 0     & 0     & 5.55  & 5.55  &                         \\ \cline{2-6}
                                     & split data                       & 0     & 0     & 5.55  & 5.55  &                         \\ \hline
\begin{tabular}[c]{@{}l@{}} Model\\ prediction\end{tabular}                              & inference                        & 100   & 0     & 0     & 100.0 & 2.06                    \\ \hline
\multirow{11}{*}{\begin{tabular}[c]{@{}l@{}} Model\\ construction\end{tabular}  } & model component format verif.    & 2.86  & 0     & 0     & 2.86  & \multirow{11}{*}{36.08} \\ \cline{2-6}
                                     & activation function call         & 0     & 0     & 2.86 & 2.86  &                         \\ \cline{2-6}
                                     & build model                      & 2.86  & 0     & 0     & 2.86  &                         \\ \cline{2-6}
                                     & build one subnetwork             & 0     & 0     & 2.86  & 2.86  &                         \\ \cline{2-6}
                                     & compute model outputs            & 0     & 0     & 2.86  & 2.86  &                         \\ \cline{2-6}
                                     & init evaluation metrics          & 0     & 0     & 2.86  & 2.86  &                         \\ \cline{2-6}
                                     & initialize model graph           & 0     & 0     & 2.86  & 2.86  &                         \\ \cline{2-6}
                                     & initialize model output          & 2.86  & 0     & 0     & 2.86  &                         \\ \cline{2-6}
                                     & layer construction               & 0     & 2.86  & 2.86  & 5.71  &                         \\ \cline{2-6}
                                     & model architecture instantiation & 0     & 0     & 5.71  & 5.71  &                         \\ \cline{2-6}
                                     & model (hyper)parameters init     & 14.29 & 20    & 28.75 & 62.86 &                         \\ \hline
\multirow{2}{*}{\begin{tabular}[c]{@{}l@{}} Model\\ evaluation\end{tabular}  }    & performance metric computation   & 0     & 22.22 & 66.66 & 88.89 & \multirow{2}{*}{9.28}   \\ \cline{2-6}
                                     & test data prediction             & 0     & 0     & 11.11 & 11.11 &                         \\ \hline
\multirow{7}{*}{\begin{tabular}[c]{@{}l@{}} Model\\ training\end{tabular}  }      & compute loss                     & 27.77 & 0     & 0     & 27.78 & \multirow{7}{*}{18.56}  \\ \cline{2-6}
                                     & get pooling info                 & 0     & 0     & 5.56  & 5.56  &                         \\ \cline{2-6}
                                     & measure model accuracy           & 5.56  & 0     & 0     & 5.56  &                         \\ \cline{2-6}
                                     & model training                   & 5.56  & 0     & 11.11 & 16.67 &                         \\ \cline{2-6}
                                     & one model step training          & 11.11 & 5.56  & 11.11 & 27.78 &                         \\ \cline{2-6}
                                     & training procedure               & 0     & 0     & 5.56  & 5.56  &                         \\ \cline{2-6}
                                     & weight normalization             & 0     & 0     & 11.11 & 11.11 &                         \\ \hline
\begin{tabular}[c]{@{}l@{}} Model\\ tuning\end{tabular}                  & Minibatch size            & 0     & 0     & 100   & 100   & 1.03                    \\ \hline
\end{tabular}%
}
\end{table}

\textbf{Model construction:} Our manual analysis shows that the most frequent DL-phase category that co-exists with code clones is the model construction phase with 36.08\% of DL-related code clones. This is an indication that DL practitioners duplicate code frequently when building the model and specifically when initializing hyperparameters/parameters with 62.86\% of code clones classes being related to the DL-phase subcategory of the model construction. 
Table \ref{tab:dlCodeCloneTaxonomy} shows that the majority of clones created by developers when initializing (hyper)parameters during the construction of the model are Type 3 clones (they represent 28.75\% overall).  The \emph{data preprocessing} and \emph{model training} phases contained 18.56\% of all the clones that we manually analyzed. 

\textbf{Model training:} Computing loss and training in each step of the model are the most frequent activities performed during model training. These activities are associated with 27.78\% of clones from the model training subcategory. According to our manual analysis, computing loss functions are often copied/pasted from other functions located in the same location
as the model implementation, or written from
scratch. i.e., by calling DL libraries routines to perform loss computation. Some developers may also reuse the corresponding code from the online sources.
The duplication of code for loss function is illustrated in the example in the table \ref{tab:ExampleCodeComputeLoss}. The first line in the table corresponds to calculating the loss of RankingLoss. The second line corresponds to computing the loss of Softmax. The two functions are Type 3 clones to each other.  Ranking and Softmax are two types of loss functions in deep learning. 
In fact, the loss computation is often common between deep learning models. Even when they are different, some of them have similar implementation logic. Hence, the prevalence of duplicated code that computes loss functions.

\begin{table}[htpb]
\caption{Example of Model Training (Compute Loss) Type 3 Clone }

\label{tab:ExampleCodeComputeLoss}
    \centering
    
    \begin{tabular}{|c|}
    \hline

\begin{minipage}{0.9\textwidth}
\vspace{0.1cm}
\begin{minted}[fontsize=\scriptsize,highlightlines={2-19,20,21},highlightcolor=mygray]{python}
    def compute(self, labels, logits, weights, reduction):
    """Computes the reduced loss for tf.estimator (not tf.keras).

    Note that this function is not compatible with keras.

    Args:
      labels: A `Tensor` of the same shape as `logits` representing graded
        relevance.
      logits: A `Tensor` with shape [batch_size, list_size]. Each value is the
        ranking score of the corresponding item.
      weights: A scalar, a `Tensor` with shape [batch_size, 1] for list-wise
        weights, or a `Tensor` with shape [batch_size, list_size] for item-wise
        weights.
      reduction: One of `tf.losses.Reduction` except `NONE`. Describes how to
        reduce training loss over batch.

    Returns:
      Reduced loss for training and eval.
    """
    losses, loss_weights = self.compute_unreduced_loss(labels, logits)
    weights = tf.multiply(self.normalize_weights(labels, weights), loss_weights)
    return tf.compat.v1.losses.compute_weighted_loss(
        losses, weights, reduction=reduction) 

\end{minted}
\vspace{0.05cm}
\end{minipage}
\\
\hline
\begin{minipage}{0.9\textwidth}
\vspace{0.1cm}

\begin{minted}[fontsize=\scriptsize,highlightlines={2,3,4},highlightcolor=mygray]{python}
def compute(self, labels, logits, weights, reduction):
    """See `_RankingLoss`."""
    labels, logits = self.precompute(labels, logits, weights)
    losses, weights = self.compute_unreduced_loss(labels, logits)
    return tf.compat.v1.losses.compute_weighted_loss(
        losses, weights, reduction=reduction)
  \end{minted}
  \vspace{0.05cm}
  \end{minipage} \\
    \hline
   
    \end{tabular}
    
\end{table}

\textbf{Data preprocessing}
Processing input is related to 16.66\% of clones associated with the `data preprocessing' phase of the DL development workflow. Input processing includes all the transformation needed to apply on the input data to prepare data for model training (e.g., processing input of model inception v3 by normalizing each pixel of input image). 

\textbf{Model evaluation:} Each DL model is evaluated to improve its performance. Overall, 9.28\% of the DL-related cloned code corresponds to model evaluation of which 89\% of clones are related to performance metric computation and 11.11\% to test data prediction. Measurement metrics used to evaluate the models tend to be duplicated frequently for each model and for each metric. One example of cloning metric computation code can be seen in Table \ref{tab:ExampleCodeComputeMetric} \footnote{\url{https://github.com/tensorflow/ranking}}, where we have the implementation of two measures: Mean Reciprocal Rank and Mean Average Precision. The two functions are clones of each other. The clone is of Type 3. The differences are in the renaming and function calls that corresponds to each metric computation. 

\begin{table}[t]
\caption{Example of Model Evaluation (Compute Metrics) Type 3 Clone}

\label{tab:ExampleCodeComputeMetric}
    \centering
    
    \begin{tabular}{|c|}
    \hline

\begin{minipage}{0.9\textwidth}
\vspace{0.1cm}
\begin{minted}[fontsize=\scriptsize,highlightlines={1,6,14,15,19,21,22,24,25},highlightcolor=mygray]{python}
    
def mean_reciprocal_rank(labels,
                         predictions,
                         weights=None,
                         topn=None,
                         name=None):
  """Computes mean reciprocal rank (MRR).
  Args:
    labels: A `Tensor` of the same shape as `predictions`. A value >= 1 means a
      relevant example.
    predictions: A `Tensor` with shape [batch_size, list_size]. Each value is
      the ranking score of the corresponding example.
    weights: A `Tensor` of the same shape of predictions or [batch_size, 1]. The
      former case is per-example and the latter case is per-list.
    topn: An integer cutoff specifying how many examples to consider for this
      metric. If None, the whole list is considered.
    name: A string used as the name for this metric.

  Returns:
    A metric for the weighted mean reciprocal rank of the batch.
  """
  metric = metrics_impl.MRRMetric(name, topn)
  with tf.compat.v1.name_scope(metric.name, 'mean_reciprocal_rank',
                               (labels, predictions, weights)):
    mrr, per_list_weights = metric.compute(labels, predictions, weights)
    return tf.compat.v1.metrics.mean(mrr, per_list_weights)

\end{minted}
\vspace{0.05cm}
\end{minipage}
\\
\hline
\begin{minipage}{0.9\textwidth}
\vspace{0.1cm}

\begin{minted}[fontsize=\scriptsize,highlightlines={1,6,14,18,20,21,23,24,25},highlightcolor=mygray]{python}
def mean_average_precision(labels,
                           predictions,
                           weights=None,
                           topn=None,
                           name=None):
  """Computes mean average precision (MAP).
  Args:
    labels: A `Tensor` of the same shape as `predictions`. A value >= 1 means a
      relevant example.
    predictions: A `Tensor` with shape [batch_size, list_size]. Each value is
      the ranking score of the corresponding example.
    weights: A `Tensor` of the same shape of predictions or [batch_size, 1]. The
      former case is per-example and the latter case is per-list.
    topn: A cutoff for how many examples to consider for this metric.
    name: A string used as the name for this metric.

  Returns:
    A metric for the mean average precision.
  """
  metric = metrics_impl.MeanAveragePrecisionMetric(name, topn)
  with tf.compat.v1.name_scope(metric.name, 'mean_average_precision',
                               (labels, predictions, weights)):
    per_list_map, per_list_weights = metric.compute(labels, predictions,
                                                    weights)
  return tf.compat.v1.metrics.mean(per_list_map, per_list_weights)

  \end{minted}
  \vspace{0.05cm}
  \end{minipage} \\
    \hline
   
    \end{tabular}
    
\end{table}

\textbf{Data post-processing:} Data is often post-processed after an inductive process and converted into a format recommended by the stakeholders of the model and to satisfy the requirement of the application using the model.
Our findings show that 8.25\% of cloned code are related to DL functions used in the data post-processing phase. There are various techniques to perform this phase. We found that object localization functions are duplicated functions with 37.5\% from the total cloned functions to perform data post-processing phase. For example, in object detection the object localization is used to interpret the output by assigning each object to a class with a higher probability or by drawing bounding boxes on an image from inference results.
37.5\% of cloned functions are classified as processing output and the rest are found duplicated for computing shape of output data.

\textbf{Data collection:} Data collection operations are frequently cloned.
5.15\% of our manually analyzed clones were related to data collection. Among them, we found 80\% of clones to be related to loading data either from files or from an URL or using a library to get data. The rest are derived from load class labels data (20\%).

\textbf{Model prediction:}  2.06\% of our manual analysis code clones are related to inference. All of them are Type 1 clones. Subsequently, according to our manual analysis, we can say 
that DL developers often duplicate the same code to create an inference process from a trained model.

\textbf{Model tuning:} We found clones in the code used to tune different model configurations (e.g., best batch size to train the model) or hyperparameters. 
All the clones found to be related to this category were Type 3 clones.
Meaning that DL developers often duplicate the hyperparameter tuning code of other models and apply some modifications to it (adding few extra lines), for example to adjust the batch size. 
Clones in hyperparameter tuning code represents 1.03\% of the analyzed clones.

\textbf{Preliminary preparation:} The code used to prepare the environment for model training appears to also contain clones. 
To optimise the model training time, developers write code to manage the hardware, e.g., CPU and GPU management. Among the manually analyzed clones, 1.03\% of them belong to the environment configuration category. All these clones were Type 1 clones, suggesting that developers often duplicate these configuration codes without modifications. 

Our analysis show that code duplication is a common practice among DL developers. They duplicate code during almost all the phases of the development process of deep learning models, 
in addition to duplicating traditional methods like test and logging. Since duplicating code may lead to bug propagation and inconsistency in the program, we recommend that DL developers pay a close attention to these clones during the maintenance and evolution of their systems.

\vspace{4mm}
\fbox{\begin{minipage}{35em}
\hspace{4mm}
\textit{\textbf{Summary of findings (RQ4):}} According to our findings, code duplication is more prevalent during the model construction phase of deep learning development. Code related to the initialization of model hyperparameters are cloned most frequently, followed by code related to model training and data preprocessing. 

\end{minipage}}

\subsection{\textbf{RQ\(5\): In which phases of deep learning development code cloning is more prone to faults?}}

After applying our taxonomy to the selected code clones from the analyzed subset of six systems, it is of interest to identify deep learning activities during which cloning has the highest risk of bugs. 
Our results of RQ3 show that code cloning can lead to bugs. A better understanding of the circumstances in which bugs frequently occur on cloned code will help raise the awareness of the developers to understand and mitigate the risks associated with their code cloning activities.

\begin{figure}[htpb]
\centering
\fbox{\includegraphics[width=1\textwidth]{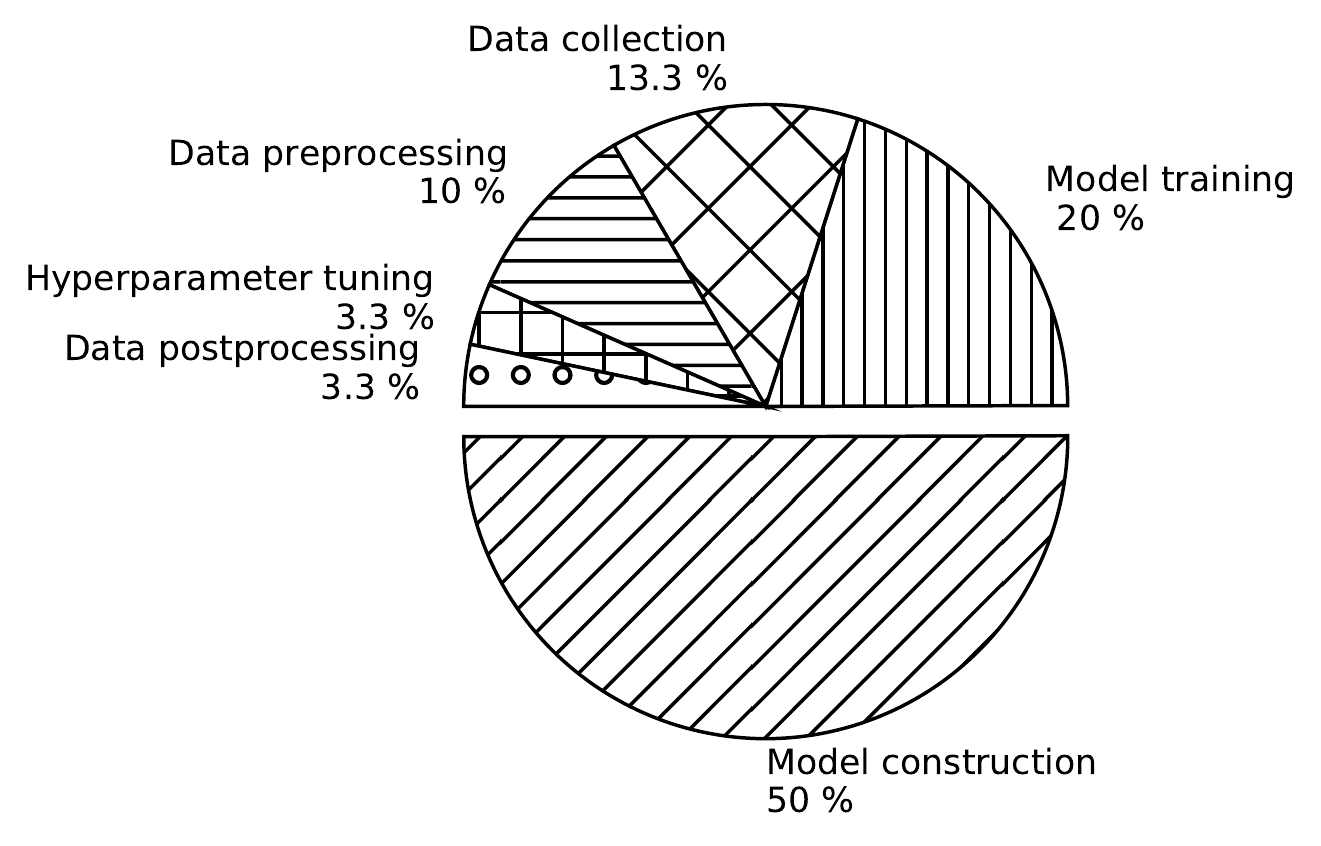}}
\caption{Percentage of Bug-Fix Occurred with Cloned Functions with respect to Deep Learning Phases}
\label{fig:DLPhaseBugPercentage}
\end{figure}

To carry out this investigation, we consider the relation between bugs and code clones and determine which part of the DL code is more prone to bugs when it is duplicated. 
We consider a clone fragment as 'buggy' when cloned lines intersect with lines modified by bug-fixing commit. We use the labelling of clones as in RQ4 to group the clones regarding the DL phases the clones are related. We computed the percentages of bugs related cloned functions for each DL phase. Our result shows that code cloned during the model construction phase are related to bugs in higher numbers; 50\% of them are buggy as shown in Figure \ref{fig:DLPhaseBugPercentage}.

Table \ref{tab:taxonomy_sub_class_dl_bugs} shows which DL-related cloned functions (tasks) are the most involved with bugs. The corresponding DL phases are also provided in the Table. We display only DL-related tasks and phases where clones are involved in bugs (i.e., other phases of the DL workflow were the phenomenon is not observed are omitted). As mentioned earlier, the model construction phase contains the highest proportion of buggy clones (i.e., 50\% of all buggy clones in our manually analyzed clone data).
The majority of them are related to model (hyper)parameters initialization (46.66\%). 
Table \ref{tab:ExampleCodeBugFix} shows an example of bug fix in a buggy clone. The presented cloned code fragments are from a bug-fix commit with the message `Minor optimizer consistency fixes'\footnote{ \url{https://github.com/keras-team/keras/commit/2d8739d}}. The optimizers in deep learning are capable of reducing the losses by changing the attributes of the neural network (i.e., learning rate). Those optimizers have a common implementation of the hyperparameter initialization function. In the example of commit bug-fix from Table \ref{tab:ExampleCodeBugFix}, to fix the instantiation of the number of iteration by adding a data type, the deep learning developer had to propagate the same change to several optimizer initializations. In this example, we have seven optimizers, i.e, SGD, RMSProp, Adagrad, Adadelta, Adam, Adamax, and Nadam that share the same initialization implementation and the developer needed to propagate the fixing change seven times.

\begin{table}[t]
\caption{Example of Bug Fix Commit Code Change }

\label{tab:ExampleCodeBugFix}
    \centering
    
    \begin{tabular}{|c|}
    \hline

\begin{minipage}{0.9\textwidth}
\vspace{0.1cm}
\begin{minted}[fontsize=\scriptsize,highlightlines={6},highlightcolor=mygray]{python}
    
def __init__(self, lr=0.01, epsilon=None, decay=0., **kwargs):
        super(Adagrad, self).__init__(**kwargs)
        with K.name_scope(self.__class__.__name__):
            self.lr = K.variable(lr, name='lr')
            self.decay = K.variable(decay, name='decay')
            self.iterations = K.variable(0, name='iterations')
        if epsilon is None:
            epsilon = K.epsilon()
        self.epsilon = epsilon
        self.initial_decay = decay

\end{minted}
\vspace{0.05cm}
\end{minipage}
\\
\hline
\begin{minipage}{0.9\textwidth}
\vspace{0.1cm}

\begin{minted}[fontsize=\scriptsize,highlightlines={6},highlightcolor=mygray]{python}
def __init__(self, lr=0.01, epsilon=None, decay=0., **kwargs):
        super(Adagrad, self).__init__(**kwargs)
        with K.name_scope(self.__class__.__name__):
            self.lr = K.variable(lr, name='lr')
            self.decay = K.variable(decay, name='decay')
            self.iterations = K.variable(0, dtype='int64', name='iterations')
        if epsilon is None:
            epsilon = K.epsilon()
        self.epsilon = epsilon
        self.initial_decay = decay

  \end{minted}
  \vspace{0.05cm}
  \end{minipage} \\
    \hline
   
    \end{tabular}
    
\end{table}

The DL phase with the second highest proportion of buggy clones is 
`Model training' with a proportion of 20\%. The phase with the third highest proportion of buggy clones is the data collection phase with 13.3\%. 10\% of buggy clones are related to data pre-processing; the majority of them are in code related to tensor operations 
(66.66\%) and code for setting the shape of input data. Only a small amount of the analyzed buggy clones were found to be related to 
data post-processing (3.3\%) and tuning of the hyperparameters of the model (3.3\%).
In light of these results, we recommend that DL developers pay particular attention when duplicating code during the model construction phase. Although it may seems like a good idea to copy the code of an existing model, to speed up the model construction phase, there are perils to this practice. 

\begin{table}[htpb]
\caption{Percentage of DL-related Cloned Functions with Bugs}
\label{tab:taxonomy_sub_class_dl_bugs}
\resizebox{\textwidth}{!}{%
\begin{tabular}{|l|l|l|l|}
\hline
\textbf{Taxonomy}                    & \textbf{Task}                       & \textbf{\% occs in DL step} & \textbf{\% occs from total} \\ \hline
Data collection       & Load data                              & 100   & 13.3 \\ \hline
Data post-processing  & Set shape of output data               & 100   & 3.3  \\ \hline
\multirow{2}{*}{Data Pre-processing} & Prepare tensor                      & 66.66                       & \multirow{2}{*}{10}         \\ \cline{2-3}
                      & Set shape of input data                & 33.33 &      \\ \hline
Hyperparameter Tuning & Hyperparameter tuning                  & 100   & 3.3  \\ \hline
\multirow{6}{*}{Model Construction}  & Model component format verification & 6.66                        & \multirow{6}{*}{50}         \\ \cline{2-3}
                      & Initialize model graph                 & 6.66  &      \\ \cline{2-3}
                      & Initialize model output                & 6.66  &      \\ \cline{2-3}
                      & Layer construction                     & 13.33 &      \\ \cline{2-3}
                      & Model architecture instantiation       & 20    &      \\ \cline{2-3}
                      & Model (hyper)parameters initialization & 46.66 &      \\ \hline
\multirow{2}{*}{Model training}      & Model training                      & 33.33                       & \multirow{2}{*}{20}         \\ \cline{2-3}
                      & One model step training                & 66.66 &      \\ \hline
\end{tabular}%
}
\end{table}

\vspace{4mm}
\fbox{\begin{minipage}{35em}
\hspace{4mm}
\textit{\textbf{Summary of findings (RQ5):}} Code clones that are related to model construction are the most bug-prone among deep learning clones followed by the clones related to model training and data collection. 

\end{minipage}}

\section{Research Implications}
\label{sec:5}
In this section, we discuss the implications of our findings with regard to cloning activity in the DL code.
\paragraph{\textbf{Code clones are prevalent in deep learning code:}}
In light of the higher density of code-clones identified in deep learning code, it appears that DL developers
prefer to reuse existing solutions instead of creating new ones from scratch. 
As they need to experiment with different configurations to find the best DL model, duplicating a code that works often seem a good idea to save time and effort. Our results show that developers often copy-paste exact code (frequently for loss computation) in the same location as the calling statement. We assume that this proximity aims to ease the maintenance of the resulting code. However, maintaining multiple clone copies always increase the risk of failing to propagate changes consistently; leading to bugs. 
Our results show that clones are more prevalent in deep learning code in comparison to traditional code. We attribute this phenomenon to the fact that a same decision logic can be used several times in a deep learning model and also across models.
For example, when creating a convolutional neural network \cite{lecun1990handwritten,lecun1998gradient} model that consists of a set of layers. Each layer is initialized according to its type and its parameter values needed, and blocks of codes are stacked to create
the architecture. These blocks are exact or similar copies of each other. Hence, the prevalence of code clones in the code of this model.

\paragraph{\textbf{Deep learning code clones are dispersed:}}
Considering the code clones being distant as found in deep learning code, such dispersion of code clones is problematic \cite{koschke2007survey} from a maintenance point of view. 
When changing code, it is most likely easier to change the code in the same file than in different files or folders. Fragments of related code in distant locations may add navigation and comprehension overhead during code change. Thus, the maintenance will be difficult to handle. Due to the high percentages of distant code clones (same directory and different directories), deep learning practitioners should be aware of the potential negative impacts of such cloning practices. 
In addition, our analysis of the percentages of clone fragments in 
different location categories show that clones in deep learning code are more dispersed in the code, which is also problematic.
Because of the negative impact of code clones on maintenance, developers should consider refactoring them. 
We noticed some signs of refactoring in some of the studied deep learning systems. Specifically, we observed the use of files with a name ending with `\_utility' that contains all the useful functions and functions likely to be used in different parts of the system \cite{BerrynetGithub,RankingGithub,AdanetGithub}.

\paragraph{\textbf{Code clones in DL code are related to bugs:}}

Our results show that cloned code may be more bug-prone than non-cloned code in deep learning systems. In addition, Type 3 clones have a relatively higher odd to be involved in bugs in the deep learning code than Type 2 and Type1 clones. Also, code clones that are related to model construction phase are the most bug-prone. In particular those related to model (hyper)parameters initialization.   
Since the main challenge of DL developers is to provide a model with high accuracy, setting model (hyper)parameters is an important step to implement an efficient model. Therefore bugs occurring in the code responsible for this critical task is likely to have a severe impact on the quality of the deep learning system. 

Due to the data-driven nature of deep learning, data collection is a crucial task \cite{munappy2019data} and any bug occurring in the code responsible for this phase is also likely to significantly impact the quality of the deep learning system. 

Our findings regarding the prevalence and distribution of clones in deep learning code, their bug-proneness and insights on the characteristics and impacts of the clones related to different DL phases are thus important for deep learning practitioners. These findings can help researchers to further investigate the characteristics and evolution of clones in deep learning code and also guide practitioners to adapt the best software development practices to the maintenance and evolution of the deep learning systems.  

\section{Threats to Validity}
\label{sec:6}
In this section, we discuss the potential threats to the validity of our research methodology and findings.

In terms of \textbf{Internal validity}, we manually labeled each detected code clones class to its corresponding DL phase. Then, we also manually assigned them to one of the steps of the DL code process. However, this relies on the subjective judgment of the persons who performed the manual classification. This is a threat to the internal validity of our experiment. To mitigate this threat, the manual classification for creating the taxonomy was done by two authors having academic and industry background. The results were then cross-validated, and disagreements were resolved by group discussion. We believe this process decreased the chances of incorrect tagging. However, future research may further improve our approach and provide additional perspectives about our results by surveying deep learning practitioners.  

In terms of \textbf{construct validity} threats, which concern the relationship between theory and observation. We followed the approach proposed by Rosen et al. \cite{rosen2015commit} to detect bug-fix commits by employing a set of keywords that are bug fixing related. If the commit message contains one of the keywords, it will be labeled as a bug-fix commit. To reduce the imprecision in the bug-fix commit detection process, we reviewed a sample of the labeled commits (102) and confirmed that they corresponded to bug-fixes with high accuracy (100\%).

For detecting clones, we used the NiCad clone detector \cite{cordy2011nicad}. Since different settings can have different effects, that we call a confounding configuration problem \cite{wang2013searching}, we have carefully set the parameters of NiCad by employing a standard configuration \cite{roy2008nicad} and with these settings, NiCad is reported to be very accurate in clone detection \cite{roy2008nicad,roy2009mutation}.  Thus, we believe that our findings on code clones in deep learning code have relevance significance. We also repeated our analysis for varying dissimilarity thresholds (20\% and 30\%) for clone detection.

With regard to \textbf{external validity}, our analysis is primarily focused on deep learning repositories that are written in Python with small number of Java and C\# systems. As Python is the most popular programming language in the machine learning field, we can assume that the small data used brings a lot of knowledge. In addition, we selected only six DL repositories to be manually analyzed for the creation of the taxonomy. Therefore, this may threaten the generalizability of our results. We believe that our small scale but detail analysis of deep learning repositories will provide a comprehensive overview of how and why deep learning practitioners had to duplicate code. 
We assume that, for each deep learning project, we can have different percentages of code clones occurrences alternating between the different phases of deep learning workflow. The fact that they exist may challenge the development of this type of system. Future studies should validate the generalizability of our findings with other DL systems that are written in other programming languages.

In terms of \textbf{threats to reliability}, we investigate in our study open-source deep learning and traditional projects that are available on GitHub. And we provide a replication package that contains needed data and scripts to replicate our study \cite{ReplicationPackage}.

And with respect to \textbf{threats to conclusion validity}, we use non-parametric statistical tests to analyze the difference between distributions. Non-parametric tests are adequate because they make no assumption on the nature of the data distribution. 

\section{Related Work}
\label{sec:7}
In this section, we discuss relevant studies from the literature that tackle the challenges faced by developers when building AI/ML/DL systems. We also review previous studies that examine the quality assurance of deep learning systems. In addition, we examine the impact of code clones on traditional systems quality.

\subsection{Software Engineering for AI-based system}

Thanks to the democratization of powerful open-source AI/ML/DL libraries/ frameworks, complex prediction systems are built quickly. Due to this rapid release of this type of system, the software quality is often sacrificed. Thus, it becomes challenging and expensive to maintain them over time because of technical debt. Sculley et al. \cite{techDebt} discuss the challenges in designing ML systems and explain how poor engineering choices can be very expensive. The challenges discussed include: hidden feedback loops, data dependencies, configuration debt, common ML code smells, etc. 
Amershi et al \cite{amershi2019software} reported the best practices used by Microsoft software engineers while developing projects that are related to Artificial Intelligence and Machine learning.
They mainly focused on the differences between ML-based software projects and traditional projects, and the challenges of adapting Agile principles to ML-based systems. Their study was conducted via interviews with selected Microsoft developers and a large-scale survey within the company. 
They report that maintaining and versioning data is a crucial task for ML-based systems. They also remark that data is harder to version than code. And that in addition to being a software engineer, ML skills are needed to build ML-based systems. Furthermore, it is more challenging to handle distant modules in ML-based systems.


\textbf{Testing and Monitoring:} One of the important strategies to reduce technical debt and lower long-term maintenance costs is testing and monitoring. ML-based systems are more difficult to test than traditional software systems \cite{BRAIEK2020}.
This is a consequence of the heavy dependence of ML on data and models.
Breck et al. \cite{breck2017ml} have outlined specific testing and monitoring needs based on practical experience at Google. They provide 28 actionable tests that can be used to measure the production readiness of a ML-based system and reduce technical debt.
Breck et al.'s study focuses more on model quality rather than the infrastructure quality of machine learning systems. 
Zhang et al. \cite{zhang2020machine} provide a comprehensive survey of ML testing covering 138 papers. The study of Zhang et al. \cite{zhang2020machine} presents definitions and research status of many testing properties such as correctness, robustness, and fairness. In addition, they discuss the need to test the different components involved in the ML model building (data, learning program, and framework). Since ML testing remains at an early stage in its development, they present many challenges. Among these challenges, they found challenges in test input generation, test assessment criteria, the oracle problem, and testing cost reduction. Furthermore, Zhang et al. \cite{zhang2020machine} analyze some research directions to benefit ML developers and the research community. They suggest testing more application scenarios since most of previous studies focus on image classification. It will be worth investigating many other areas such as speech recognition or natural language processing.
They also mentioned uncovered testing opportunities like testing unsupervised and reinforcement learning systems.

\textbf{Software Engineering Practices and Challenges:} Amershi et al. \cite{amershi2019software} performed a survey of Software Engineering practices for ML-based systems at Microsoft.
They interviewed developers from Microsoft to understand their development practices and the benefits of these practices. 
Another study related to software engineering practices for DL applications was conduced by Zhang et al. \cite{zhang2019software}. Zhang et al. also surveyed DL practitioners about their software engineering practices. 
They proposed recommendations to improve the development process of DL applications.
Wan et al.\cite{wan2019does} studied the features and impacts of machine learning on software development. They compare various aspects of software engineering and work characteristics in both the machine learning systems and non-machine learning software systems. 
A recent study by Chen et al. \cite{chen2020understanding} examined challenges in deploying DL software by analyzing Stack Overflow posts and posts from other popular Q\&A website for developers. They proposed a taxonomy of the challenges faced by developers when deploying DL software.

\textbf{Bugs in Deep Learning Code:} Islam et al. \cite{islam2019comprehensive} analyzed stack overflow posts and bug-fix commits from popular deep learning frameworks to understand the characteristics of DL systems’ bugs (their types, root causes, and effects). Zhang et al. \cite{zhang2018empirical} studied deep learning applications built on top of TensorFlow \cite{rampasek2016tensorflow} by collecting their program bugs from GitHub and Stack Overflow. They identified the root causes and symptoms of the collected bugs. They also studied the detection and localization challenges of these bugs. 

\textbf{Code Smells in Deep Learning Applications:} Jebnoun et al. \cite{jebnounscent} studied the prevalence, evolution, and bug proneness of code smells in deep learning applications. They identify three types of smells that occur more frequently: long lambda functions, long ternary conditional expression, and complex container comprehension. They report that the number of code smells increases across releases. They focused on the relationship between code smells and bugs reported the overlap of nearly 63\% of the changed file with files that contain code smells, and that frequent smells co-exist with software bugs more frequently than the others. 
Jiakun Liu et al \cite{liu2020using} investigated technical debt in deep learning frameworks by mining the self-admitted technical debt comments provided by developers. Among the types of design debt, Jiakun Liu et al \cite{liu2020using} report that DL developers consider code duplication to be a contributing factor to technical debt and increased maintenance costs.

\textbf{Computational Notebooks:} Nowadays, we are witnessing a proliferation of computational notebooks in data science studies, thanks to
their strengths in presenting data stories and their flexibility.
However, using these notebooks in a real project may induce technical debt, because of their lack of abstraction, modularisation, and automated tests. 
Recently, a fair amount of research works has been conducted on computational notebooks, mostly focusing on the challenges that they pose to data scientists and the poor software engineering practices observed in these notebooks. 

One common bad practice that is frequently observed in computational notebooks is the copying and pasting of code, by data scientists in order to save time and effort.
Kery et al. \cite{kery2018story} conducted two case studies where they interviewed 21 data scientists and surveyed 45 data scientists to understand the use of literate programming tools. They studied Jupyter Notebook as it is the most popular literate tool \cite{shen2014interactive}. They \cite{kery2018story} identified the good and bad practices employed by data scientists. 
One practical limit of Jupyter Notebook is its size and performance. The limited size often leads data scientists to copy-paste code into a new Notebook when the maximum size is reached.  
In addition, they copy-paste code to ensure that the code dependencies are properly located next to the new code, instead of extracting the code to a function. 
Pimentel et al. \cite{pimentel2019large} conducted an empirical study of 
1.4 million notebooks from GitHub; examining reproducibility issues, and 
challenges related to the implementation of projects within Jupyter notebook. They also provide a set of best practices to  
improve reproducibility. Additionally, they identified and reported
good and bad practices followed by developers of 
computational notebooks. One best practice that is reported is
the use of markdown and visualization, which are two key features of literate notebooks. The use of a convenient and comprehensive filename is also reported to be a frequent good practice in computational notebooks. The bad practices identified include the lack of testing code, as well as poor programming practices that make reasoning and reproducing results more difficult, such as non-executed code cells and hidden states.
Psallidas et al. \cite{psallidas2019data} also examined the quality of notebooks (through an analysis of 6 million python notebooks from GitHub and 2 million enterprise DS pipelines developed within COMPANYX). They also performed an analysis of 12 popular deep learning libraries over 900 releases. They report that the majority of notebooks use only a few libraries and that commonly used tools are mature and popular.
Koenzen et al. \cite{koenzen2020code} examined how code is cloned in Jupyter notebooks and 
found that 7.6\% of code clones are self-duplication. They also performed an observational lab study and found that frequently reuse code is copied from web tutorial sites, API documentation, and Stack Overflow. 

\subsection{Impacts of Code Clones}
Since we investigate code clones in the deep learning code, we are interested in reviewing the existing literature on the impact of code clones in traditional software systems.

Roy and Cordy \cite{RoyCK2010} report that code clones represents between 
7\%-23\% of the code of traditional software systems. Multiple studies from the literature have examined 
the impacts of clones on traditional software systems from different software quality perspectives, e.g., change-proneness, bug-proneness, challenges in consistent update, and overall maintenance efforts and costs. 

Sajnani et al. \cite{sajnani2014comparative} showed that, contrary to intuition, the cloned code contains less problematic patterns than non-cloned code. Along the same line, Rahman et al. \cite{rahman2012clones} reports no correlation between bug-proneness and code clones. 
However, these conclusions about the lack of harmfulness of clones are contradicted by Islam et al. \cite{islam2016bug} who found that code cloning activities contribute to replicating bugs. Islam et al. suggest to prioritize refactoring and tracking for clone fragments containing method calls and/or if-conditions, to prevent bugs being replicated. 
Another study by Islam et al. \cite{islam2017comparative} examined bugs that were reported during the evolution of a software system for two different programming languages (Java and C) and found that 
clone code tends to be more bug-prone than non-clone code. 

Aversano et al. \cite{aversano2007clones}  investigated how clones are maintained considering the inconsistency that may induce code clones when fixing a bug in just one fragment or when
evolving code fragments. They found that the majority of clone classes are always maintained consistently. Similar work was also performed 
by Göde et al. \cite{gode2011clone}, confirming the findings of Aversano et al. Göde et al. also found 
cloned code to be even more stable than non cloned code. They also report that near-miss clones (Type 2 and Type 3) are more stable than exact clones (Type 1). Krinke \cite{krinke2011cloned} conducted a comparative study (in terms of the average age) between cloned code and non-cloned code. They observed that cloned code is usually older than non-cloned code and that cloned code in a file is usually older than the non-cloned code in the same file. This confirms previous observations that code clones are more stable than non-cloned code. Therefore, maintaining code clones is not necessarily more expensive than maintaining a non-cloned code. 

Jiang et al. \cite{jiang2007context} examined the bug-proneness of cloned code and confirmed that code cloning can be error-prone and directly related to inconsistencies in the code. They proposed an algorithm able to locate clone related bugs by detecting such inconsistencies. When a code fragment contains bugs and is reused by duplicating it with some adjustments w.r.t to the need, it may increase the spread of bugs in the system. Several other previous studies from the literature  \cite{barbour2011late,Barbour_JSEP:2013,juergens2009code,Li2006cpminer,Li_ICSE_2012,Rahman_SCAM_2017,Wagner_SANER_2016} report a similar conclusion about code clones, i.e., that they 
make the code bug-prone and increase maintenance costs. 
Juergens et al. \cite{juergens2009code}  examined the root cause of faults in cloned code and report 
that one of the major sources of faults is inconsistent code clones.
They provided an open-source algorithm for the detection of inconsistent clones. G\"ode and Koschke. \cite{gode2011frequency} provide empirical evidences showing that 
unintentional inconsistencies of code clones leads to faults.

Barbour et al. \cite{barbour2011late} examined late propagation evolutionary patterns of clones and 
identified 8 types of late propagation. They further examined the risk of faults in these evolutionary patterns and found that late propagation in which 
a clone is modified and then re-synchronized without any modification to the other clone in the clone pair is the riskiest pattern.
Researchers have also examined the 
maintenance efforts that result from duplicating code. 
Hotta et al. \cite{hotta2010duplicate} conducted an empirical study on 15 open-source systems, comparing the modification frequency of code clones and non-clones code. They concluded that the existence of clones does not impact software evolution negatively. 
Kapser and Godfrey \cite{kapser2008cloning} performed an empirical study of code cloning patterns, reporting the reasons behind the different patterns. They also report that the majority of clones have a positive impact on software maintainability. However, their claim is contradicted by 
Kim et al.  \cite{kim2005empirical} who suggest that refactoring techniques cannot tackle consistently changing code clones. 
Li et al \cite{Li2006cpminer} advise that maintaining duplicate code would be very beneficial for developers, as this would avoid introducing
hard to detect bugs.  Lozano and Wermelinger \cite{lozano2010tracking} have also shown the negative impacts of code clones in terms of maintenance cost and system stability. They found that code clones have a higher density of changes than non-cloned code. Lozano and Wermelinger  \cite{Lozano:2008} also show that the existence of code clones may increase the change effort.

A recent study by Mondal et al. \cite{Mondal_2018_emse} shows that cloned code are more unstable than non-cloned code in general. However according to Selim et al. \cite{selim2010studying}, the bug-proneness of code clones might be system dependent. Mondal et al. empirical study \cite{mondal2017does}  shows that cloned code tends to require more effort in maintenance than non-cloned code and that Type 2 and Type 3 clones often need a special attention when making management decisions since they require more effort.

While previous works examined the prevalence and impacts of code clones in traditional software systems, we investigate the distribution and impacts (bug-proneness) of code clones in the deep learning code. We manually investigate clones in deep learning systems with the aim to derive insights on 'what' functions deep learning practitioners clone and 'why' they clone. 
Previous studies on duplicated codes in data scientists' projects have almost exclusively focused on analyzing computational notebooks. A number of studies' results have shown that copying and pasting of cells within the same notebook is a widespread practice. 
There are some works \cite{kery2018story,koenzen2020code} discussing the common code duplication practices of deep learning practitioners. However, these works were limited to interviews with practitioners and focuses only on computational notebooks. 
In this paper, we examine the distribution of code clones deep learning systems, in terms of occurrences and location, and propose a taxonomy of code clones in deep learning systems. We also study the relationship between code clones and bug fixes, and examine the model construction phases in which cloning has the highest risk of bugs.

\section{Conclusion}
\label{sec:8}

This paper presents an empirical study of code clones in deep learning systems. Through quantitative and qualitative analyses, we have examined the characteristics, distribution, and impacts of clones in deep learning code. We have shown that code clones are prevalent and dispersed in deep learning systems (which may add navigation and comprehension overhead). In addition, our results show a higher association between code clones and bug occurrences. Furthermore, cloning code responsible for model (hyper)parameters initialisation appeared to be a very risky activity, since a large proportion of clones in this part of the code were found to be buggy. Although duplicating code may lead to short term productivity gains, deep learning practitioners should be aware of the perils of such practice.

As future work, we need further studies on the evolution patterns and the impacts of clones on different aspects of the quality of deep learning code, to guide the practitioners to better manage clones in deep learning systems. Thus, we plan our future research towards the investigation of the clone genealogy in deep learning code, to have deeper insights into how clones in deep learning code evolve, which in turn can help practitioners adopt safer code reuse practices, leverage existing libraries and open-source resources in the rapidly growing domain of deep learning and other machine learning based system development. 

\begin{acknowledgements}
This work is supported by Fonds de Recherche du Quebec (FRQ) and
the Natural Sciences and Engineering Research Council of Canada
(NSERC).
We would like to thank Dr. Amin Nikanjam for his valuable comments on the manuscript.
\end{acknowledgements}

%
%
\bibliographystyle{spmpsci}      
\bibliography{Mybibfile}   


\begin{appendices}
\section{Study Design}
\label{appendixStudyDesign}
Table \ref{tab:6DLReposDesc} shows the name, url, number of lines of code (SLOC), number of commits and the size of each selected 6 DL repository. 
\begin{table}[htpb]
\caption{The 6 analyzed DL repositories details}
\label{tab:6DLReposDesc}
\resizebox{\textwidth}{!}{%
\begin{tabular}{|l|l|l|l|l|}
\hline
\textbf{Repository} & \textbf{URL}                                         & \textbf{SLOC} & \textbf{\#commits} & \textbf{Size} \\ \hline
keras-applications  & https://github.com/keras-team/keras-applications.git & 4263          & 89                 & small         \\ \hline
DeepCTR  & https://github.com/shenweichen/DeepCTR.git & 4886  & 91  & small  \\ \hline
nn-wtf   & https://github.com/lene/nn-wtf.git         & 2688  & 119 & small  \\ \hline
ranking  & https://github.com/tensorflow/ranking.git  & 10778 & 145 & medium \\ \hline
BerryNet & https://github.com/DT42/BerryNet.git       & 12963 & 396 & medium \\ \hline
adanet   & https://github.com/tensorflow/adanet.git   & 25165 & 432 & large  \\ \hline
\end{tabular}%
}
\end{table}
\section{RQ1 Additional Results}\label{appRQ1}
\subsection{Results of Clone Detection Using Threshold of 20\% }
\begin{figure}[htpb]
\centering
\includegraphics[width=.45\textwidth]{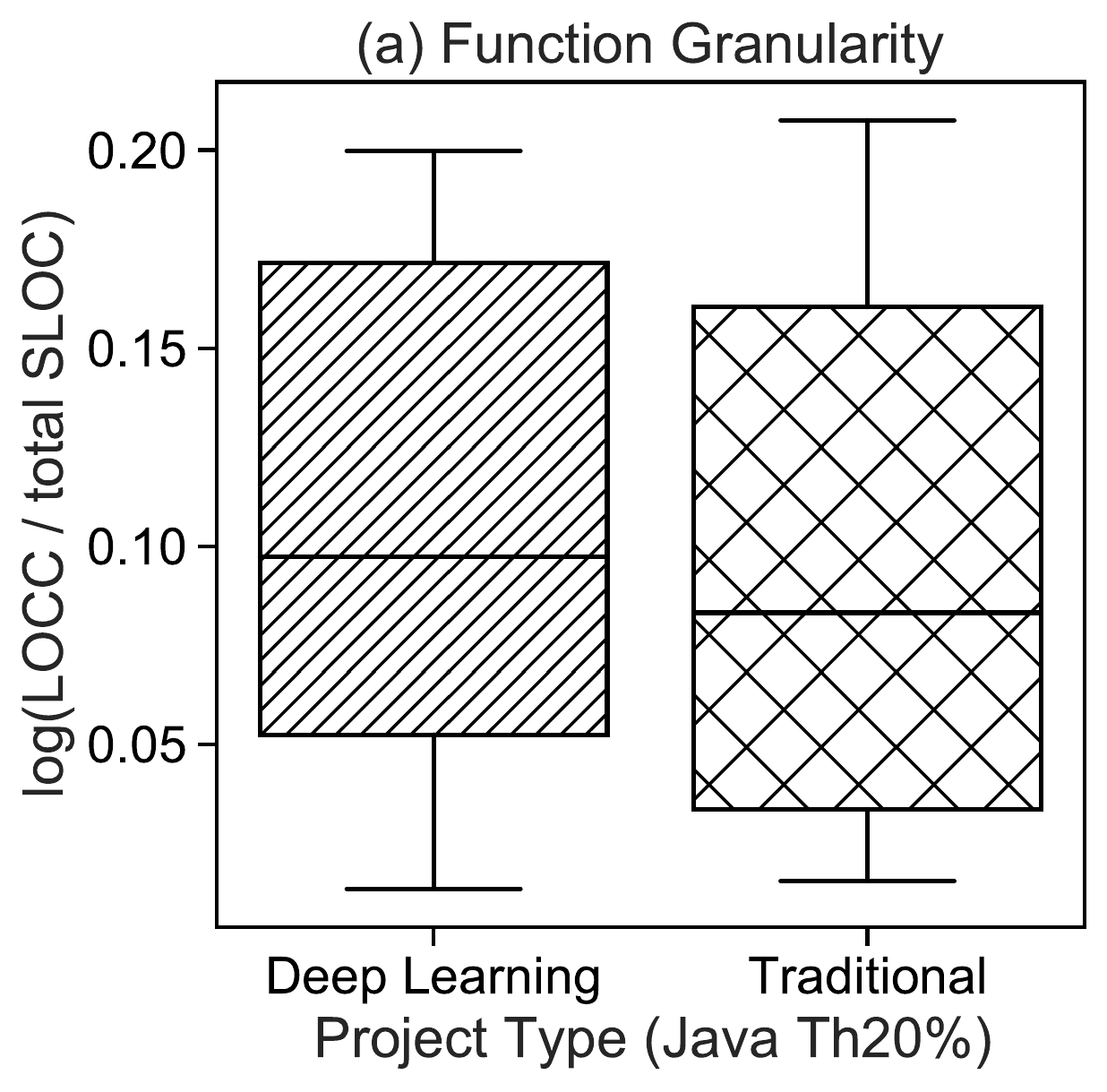}
\includegraphics[width=.45\textwidth]{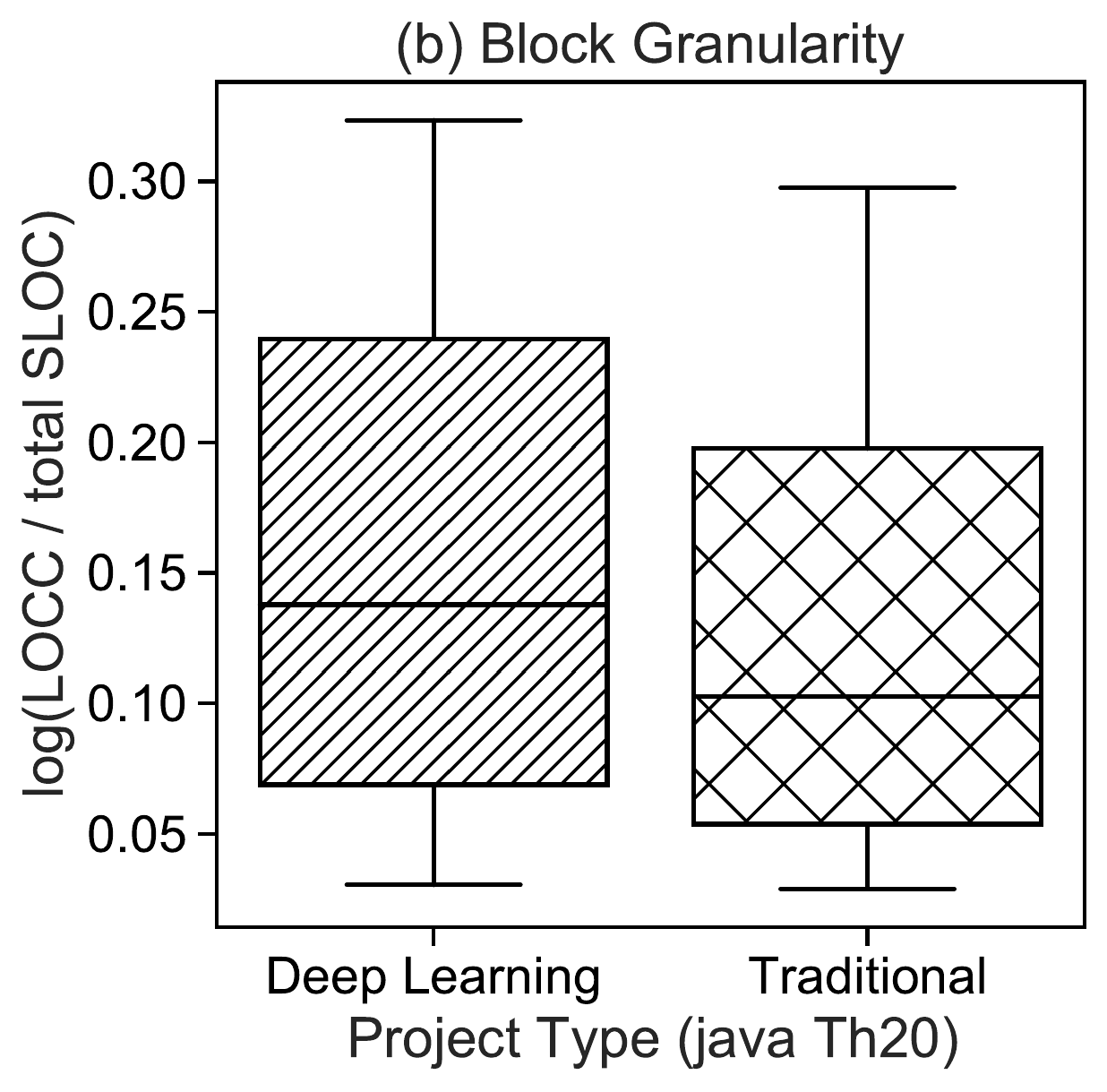}
\caption{Code Clones Occurrences in DL and Traditional Java Projects Using threshold as 20\% for Both Code Clones Granularities: (a) Function, (b) Block. \textbf{\textit{LOCC}}: Lines Of Code Clones, \textbf{\textit{SLOC}}: Source Lines Of Code.}
\label{fig:dl_trad_java_20}
\end{figure}

\begin{figure}[htpb]
\centering
\includegraphics[width=.48\textwidth]{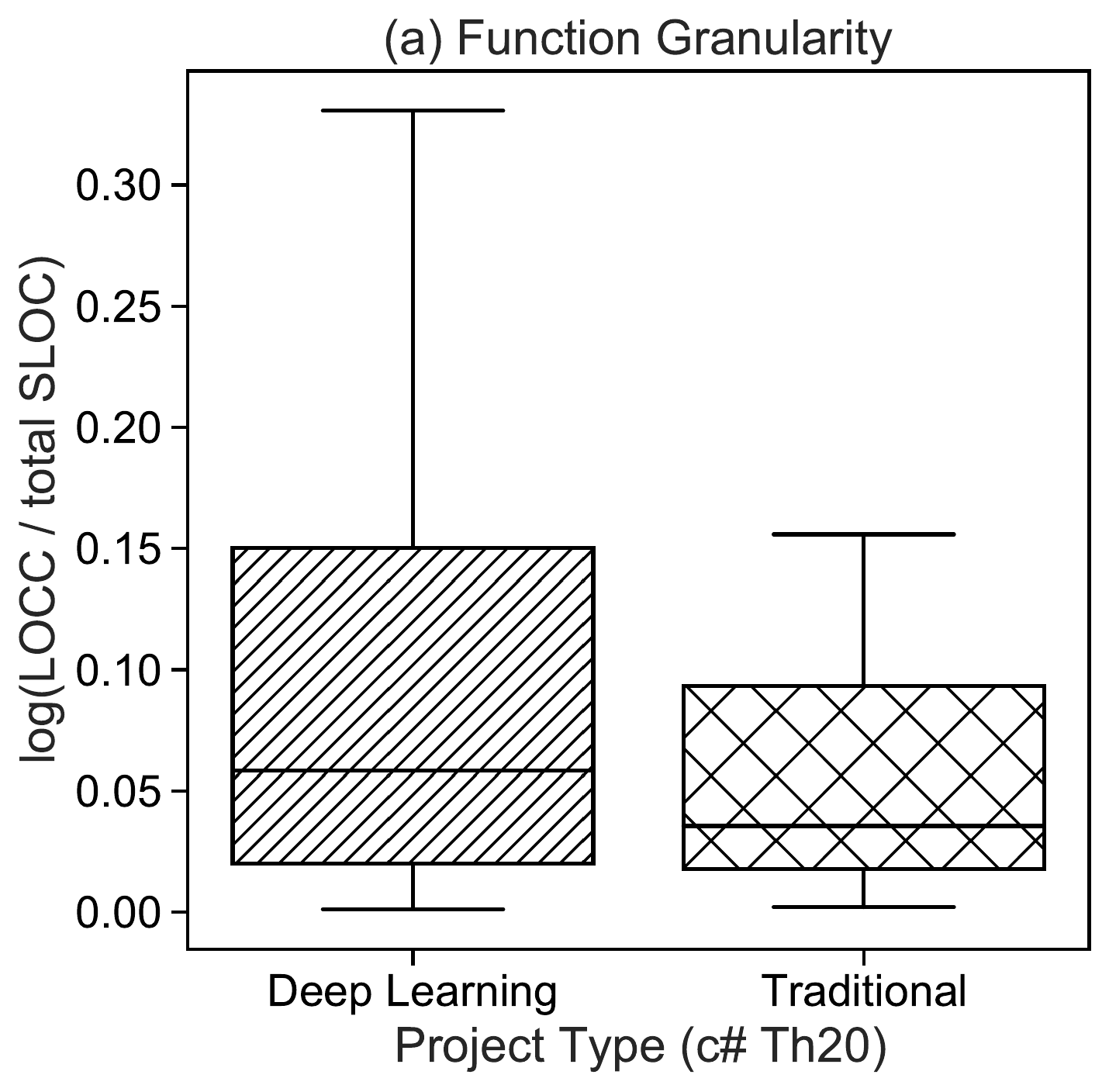}
\includegraphics[width=.48\textwidth]{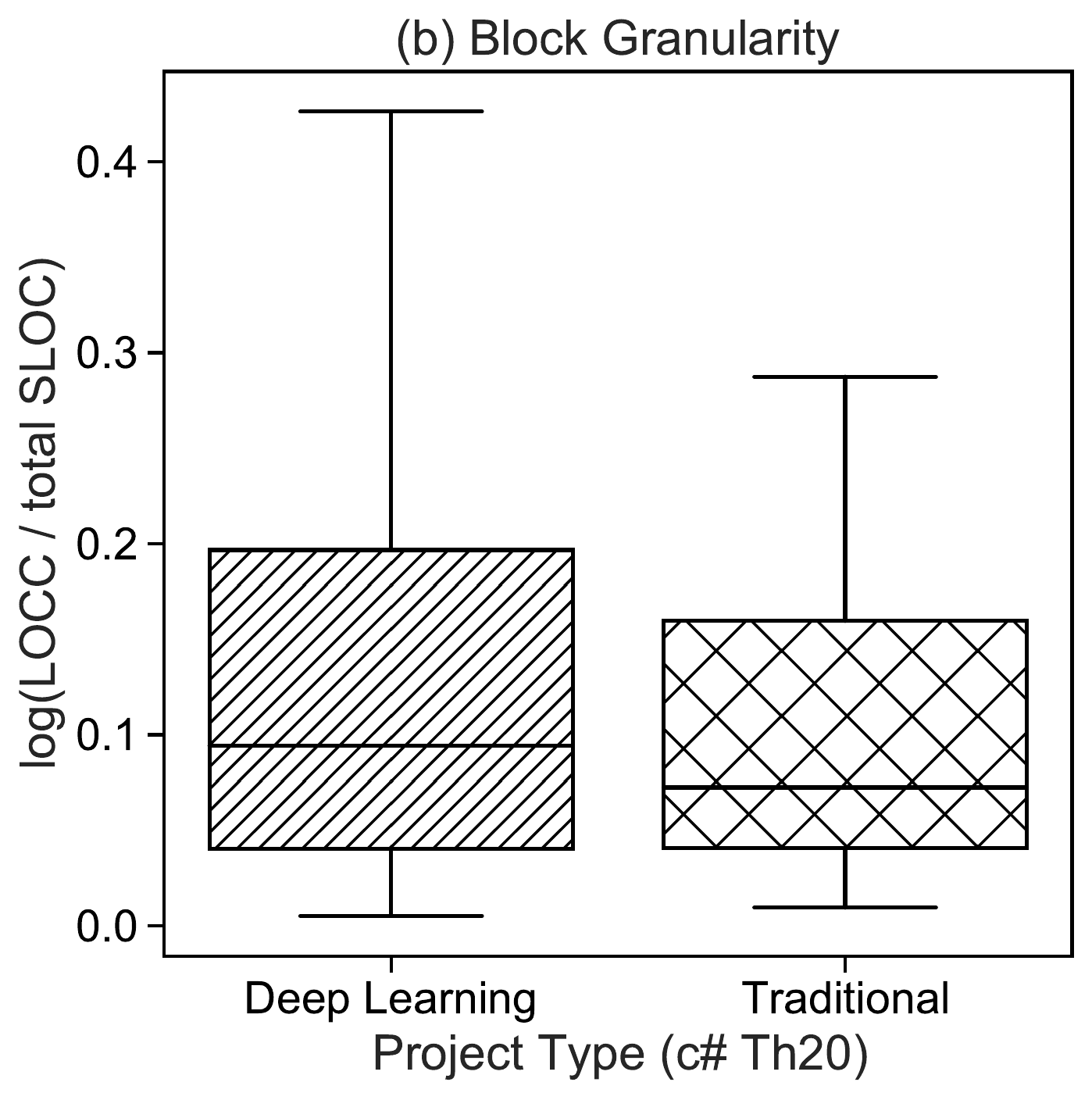}
\caption{Code Clones Occurrences in DL and Traditional C\# Projects for Both Code Clones Granularities: (a) Function, (b) Block and using 20\% as threshold. \textbf{\textit{LOCC}}: Lines Of Code Clones, \textbf{\textit{SLOC}}: Source Lines Of Code.}
\label{fig:dl_trad_csharp_20}
\end{figure}

In this section, we provide additional results for both programming languages (Java and C\#) when using a dissimilarity threshold 20\% in order to explore the impact of threshold on clone detection as we use in our analysis 30\% as threshold. Figure \ref{fig:dl_trad_java_20} shows the code clones occurrences in DL and Traditional Java projects for both code clones granularities. Figure \ref{fig:dl_trad_csharp_20} shows the same analysis but for C\# projects.

\begin{figure}[htpb]
\centering
\includegraphics[width=.45\textwidth]{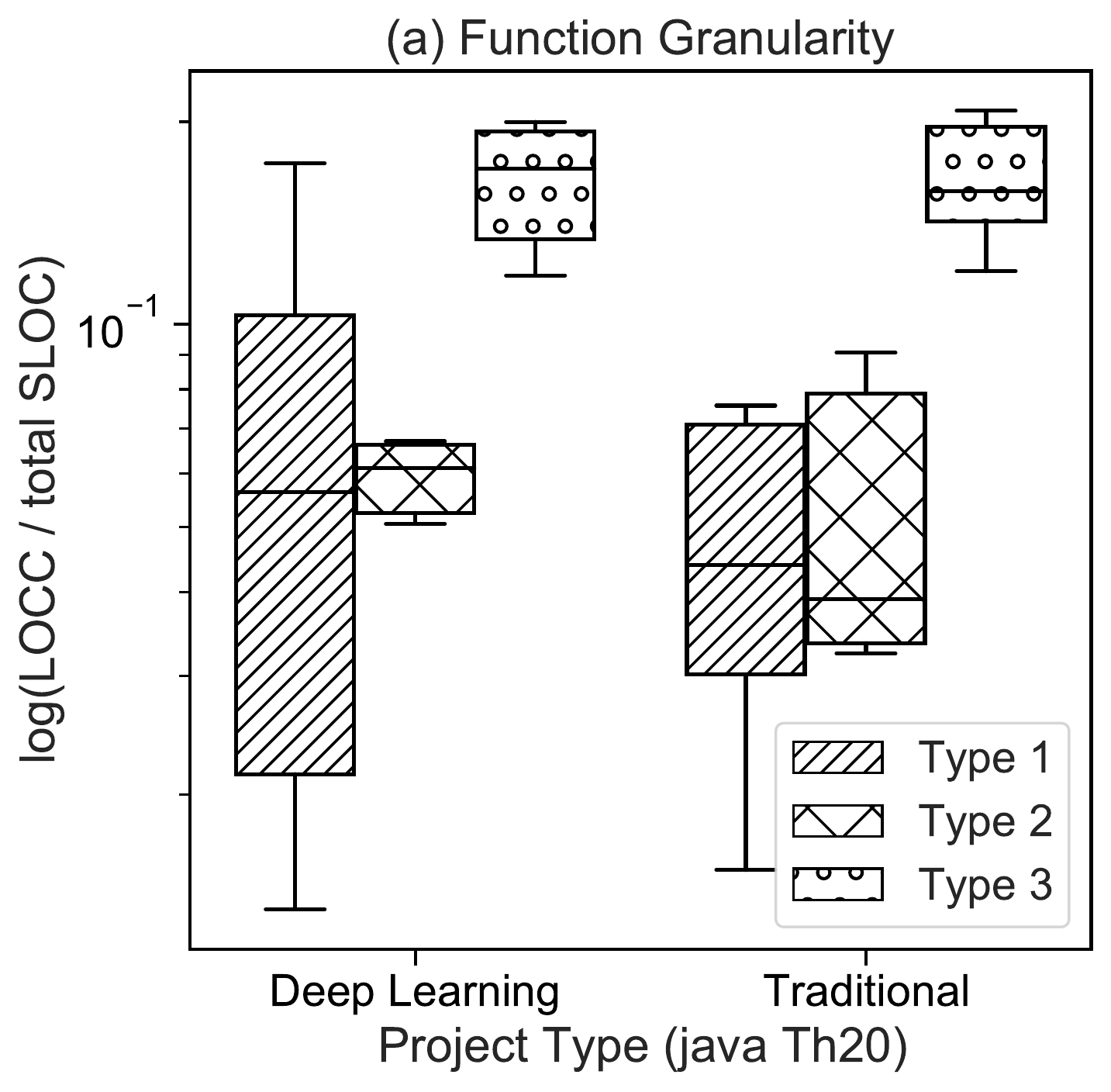}
\includegraphics[width=.45\textwidth]{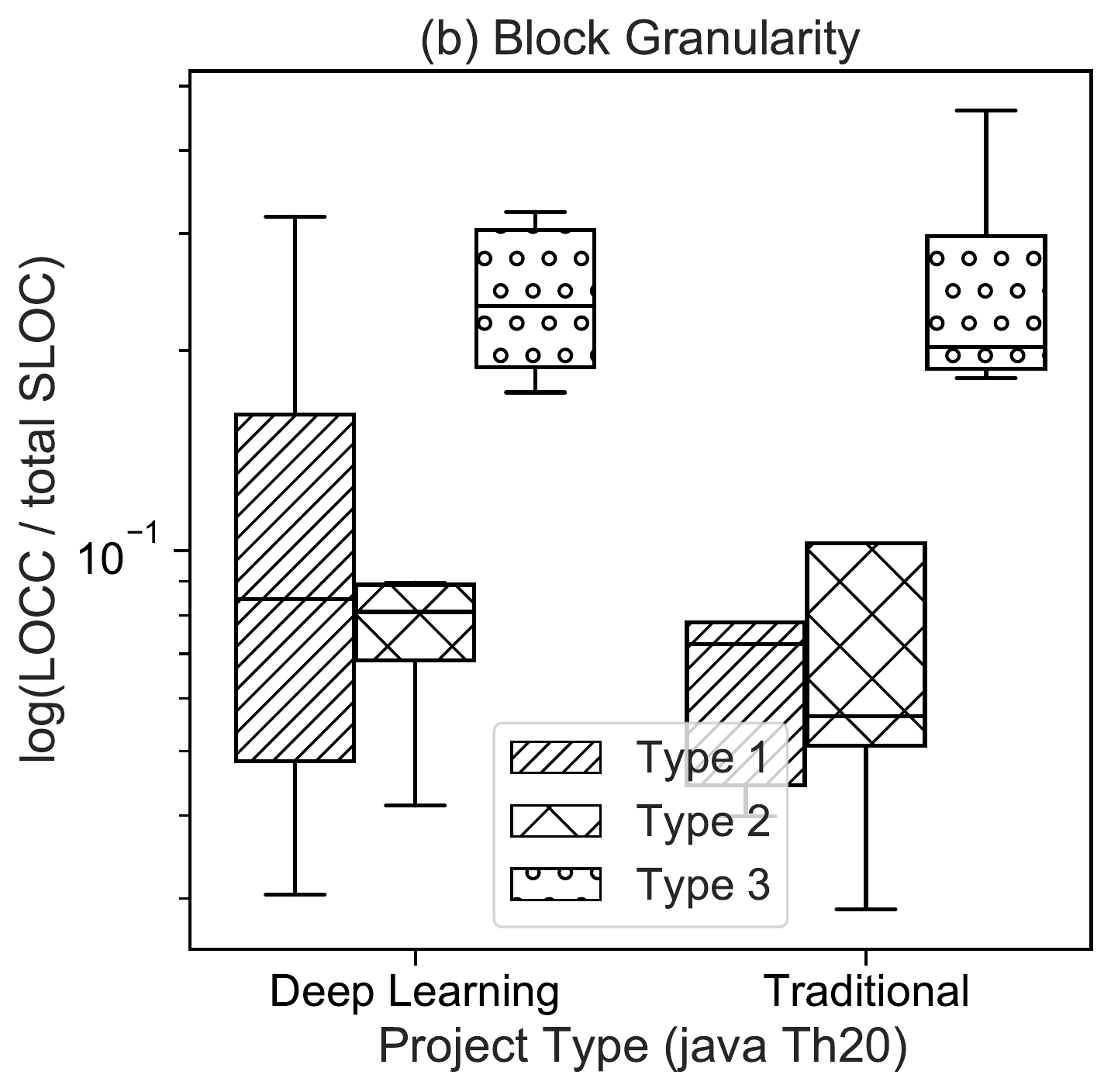}
\caption{Clone Density in DL and Traditional Java Projects for Clone Types and Granularity Using threshold as 20\% \textit{\textbf{LOCC: } Lines Of Code Clones}}
\label{fig:dl_trad_clone_type_java_20}
\end{figure}

\begin{figure}[htpb]
\centering
\includegraphics[width=.45\textwidth]{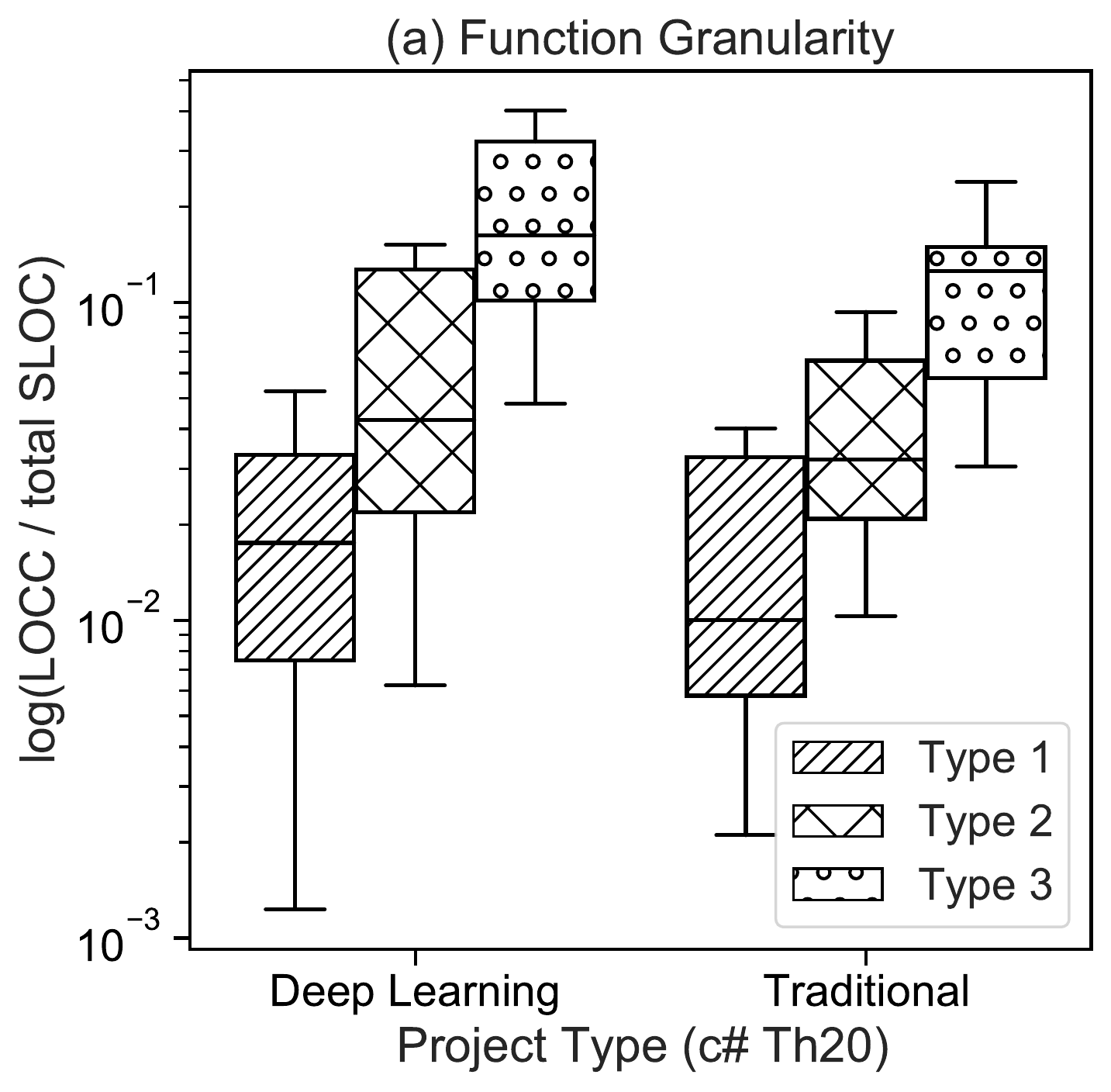}
\includegraphics[width=.45\textwidth]{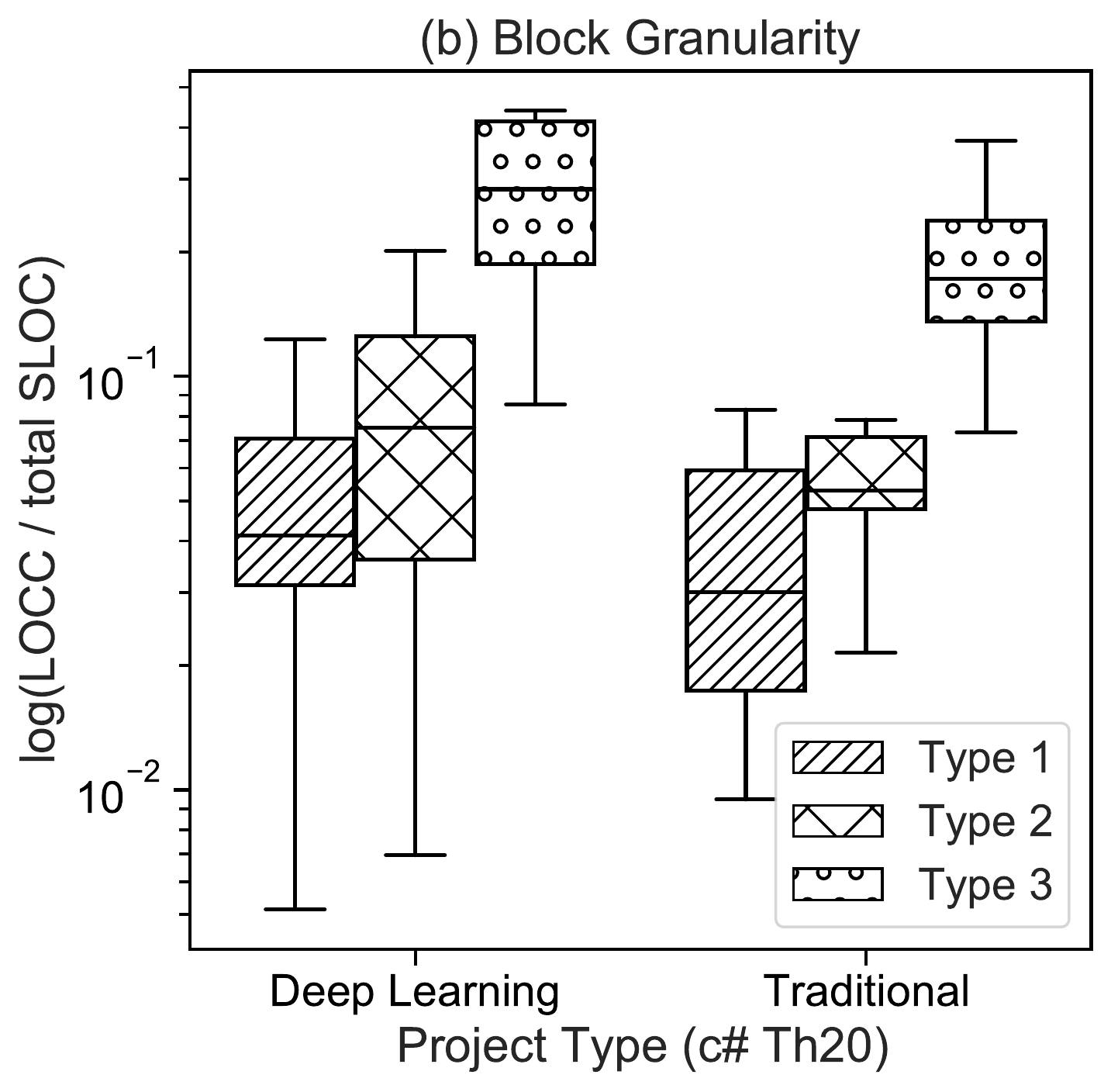}
\caption{Clone Density in DL and Traditional C\# Projects for Clone Types and Granularity and using 20\% as threshold. \textit{\textbf{LOCC: } Lines Of Code Clones}}
\label{fig:dl_trad_clone_type_csharp_20}
\end{figure}

\newpage
We further extend our analysis by comparing clones types. Figure \ref{fig:dl_trad_clone_type_java_20} and Figure \ref{fig:dl_trad_clone_type_csharp_20} illustrate the clone density in DL and traditional projects for clone types and granularity for the two programming languages (Java and C\# respectively).

\section{RQ2 Additional Results}
\label{appRQ2}
\subsection{Other Programming Languages Analysis Results}
\begin{figure}[htpb]
\centering
\includegraphics[width=.49\textwidth]{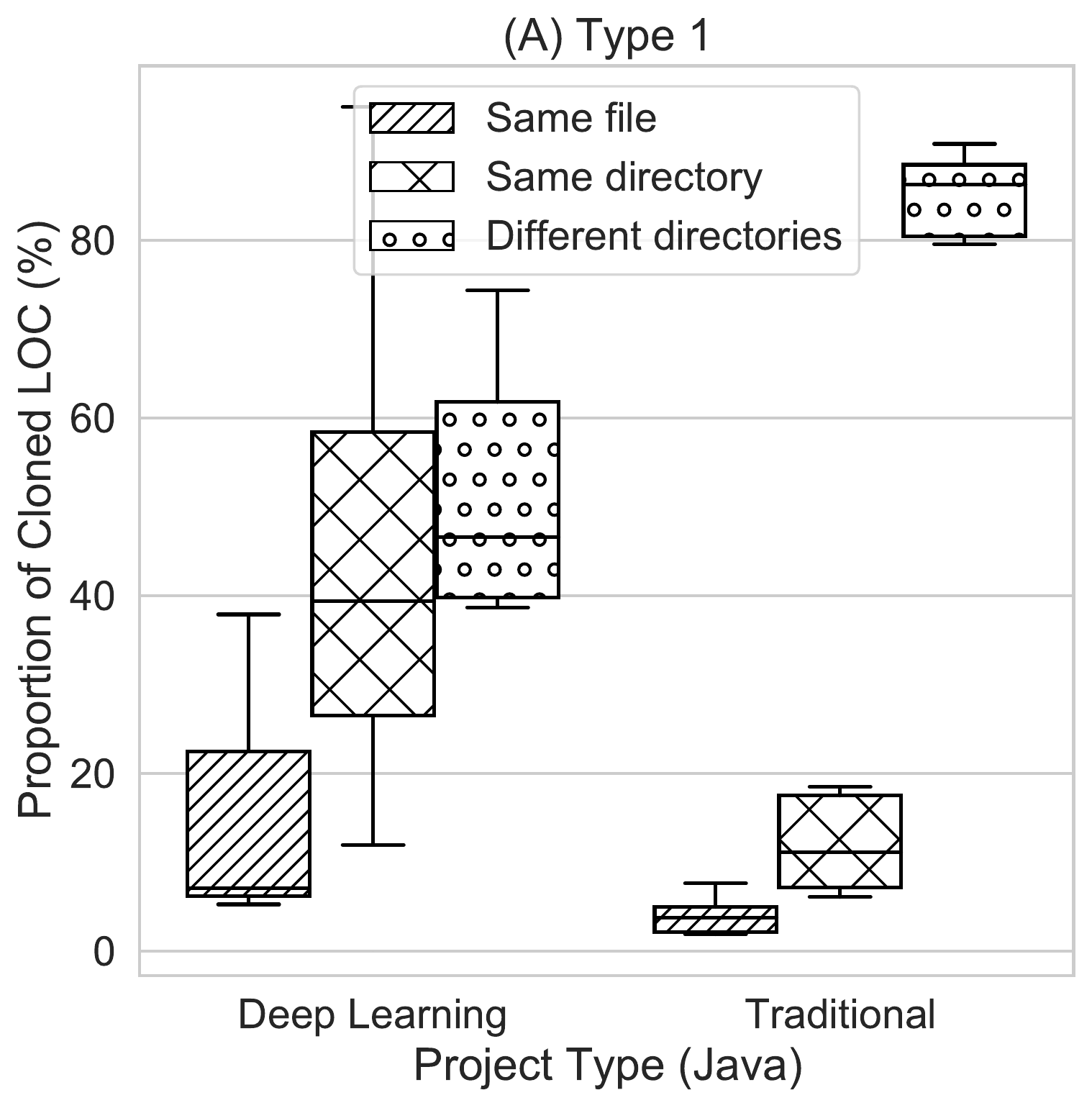}
\includegraphics[width=.49\textwidth]{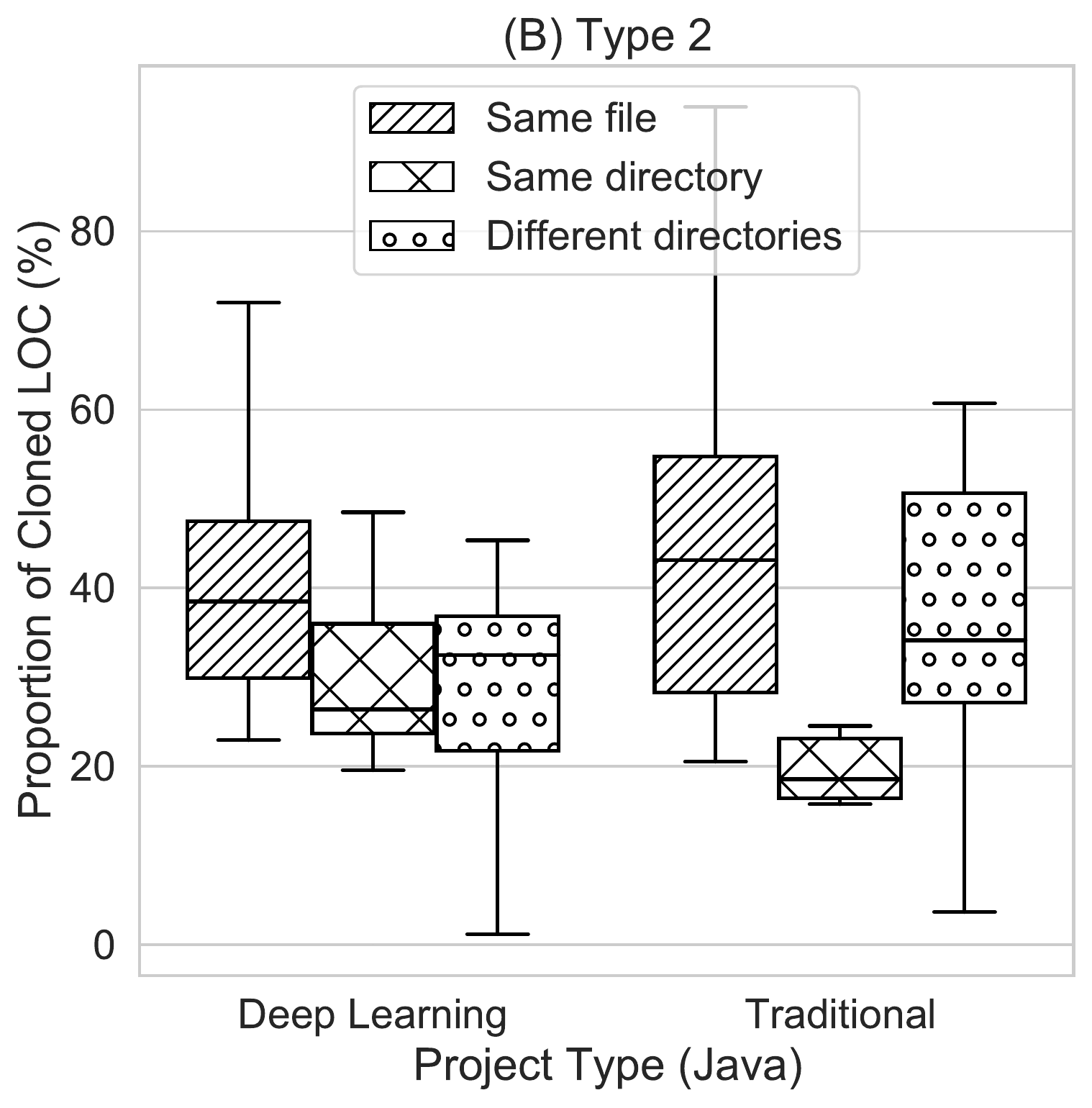}
\includegraphics[width=.49\textwidth]{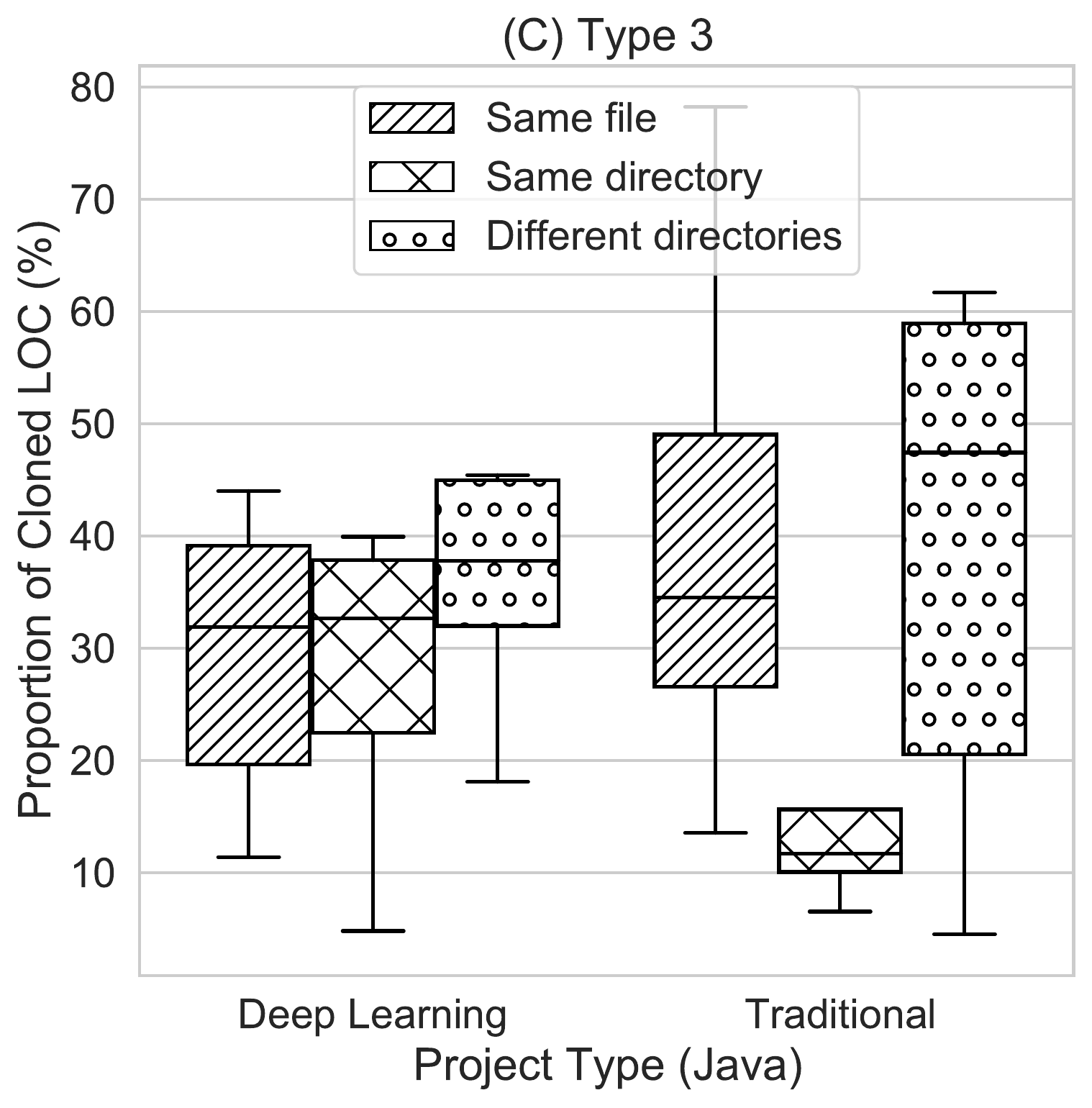}

\caption{Distribution of Different Types of Clones by Clone Location in DL and Traditional Code (Java)}
\label{fig:CloneLocationbyCloneType_java}
\end{figure}

\begin{figure}[htpb]
\centering
\includegraphics[width=.49\textwidth]{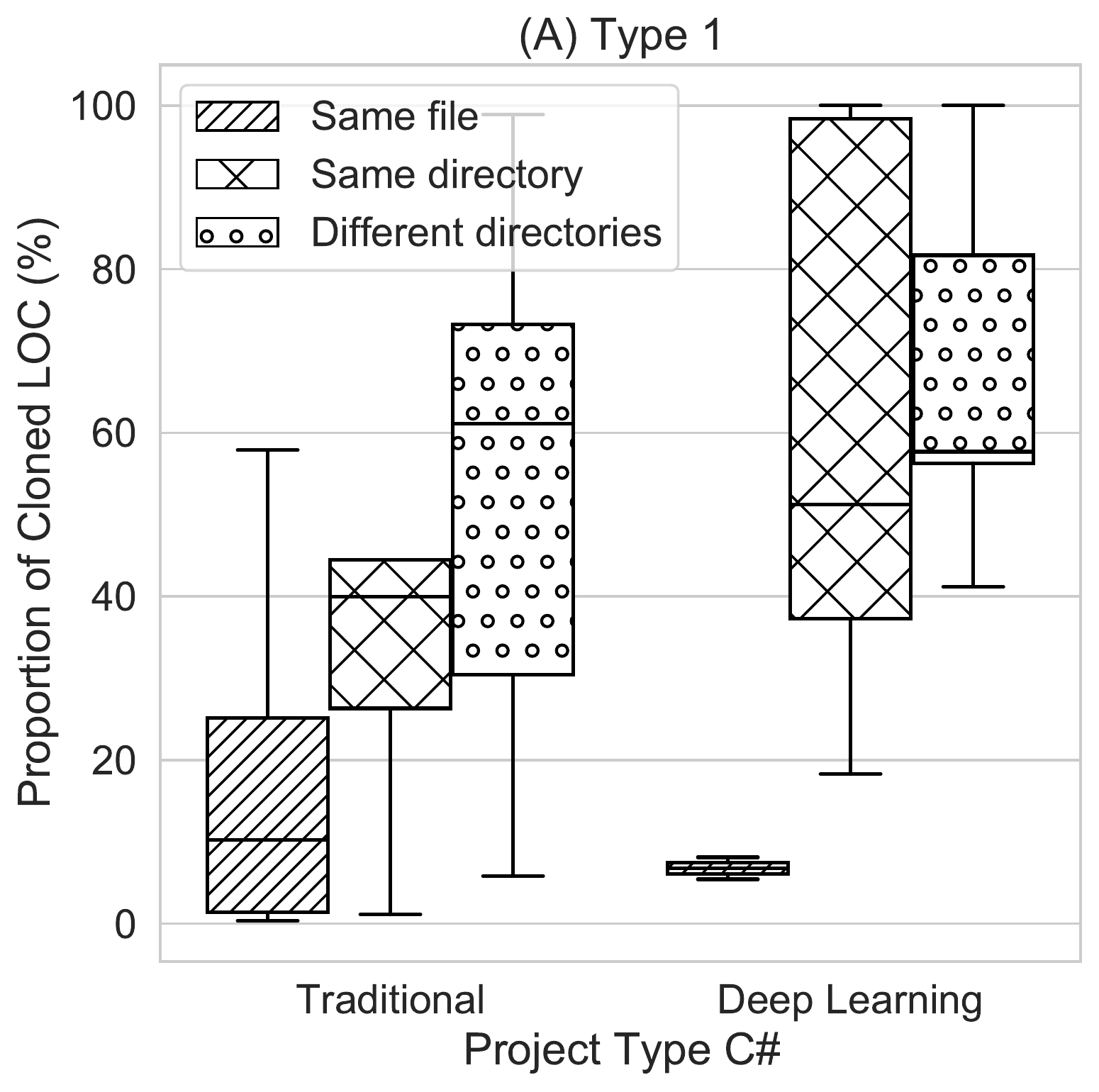}
\includegraphics[width=.49\textwidth]{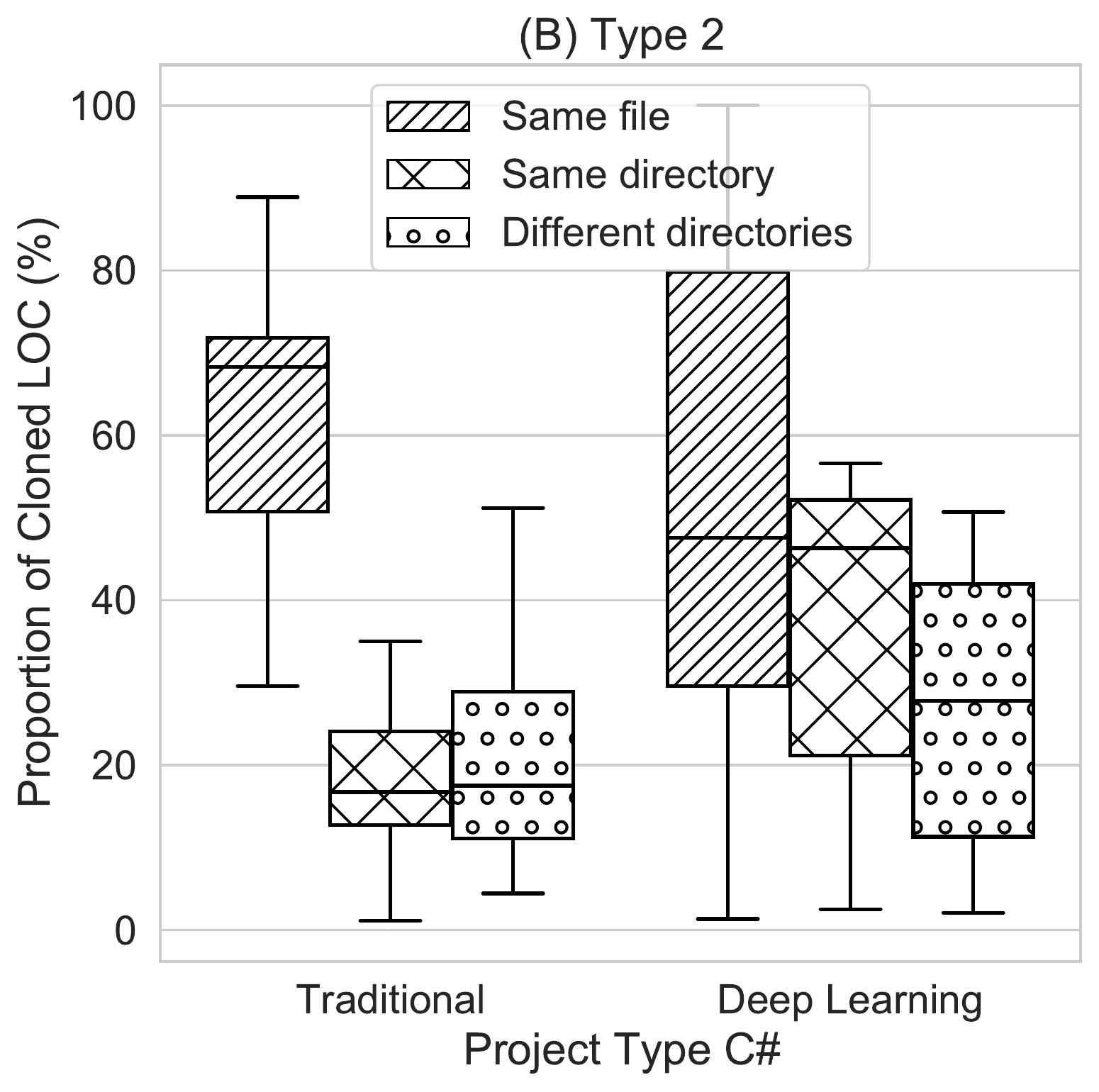}
\includegraphics[width=.49\textwidth]{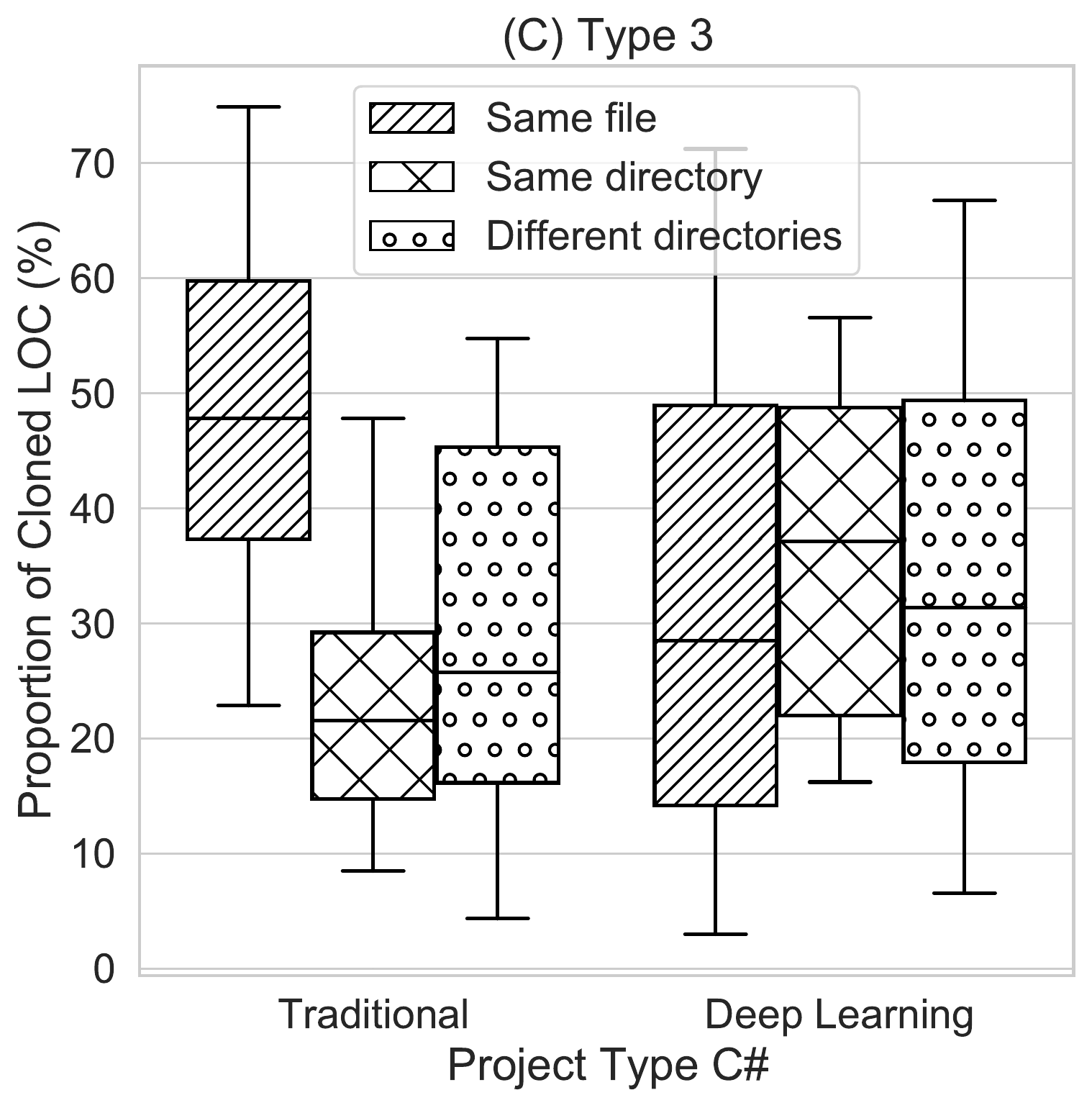}

\caption{Distribution of Different Types of Clones by Clone Location in DL and Traditional Code (C\#)}
\label{fig:CloneLocationbyCloneType_csharp}
\end{figure}

We study the distribution of different clones types by clone location in DL and traditional code in java projects (Figure \ref{fig:CloneLocationbyCloneType_java}) and in C\# projects (Figure \ref{fig:CloneLocationbyCloneType_csharp})

\subsection{Results of Clone Detection Using Threshold of 20\% }
\begin{figure}[htpb]
\centering
\includegraphics[scale=0.4]{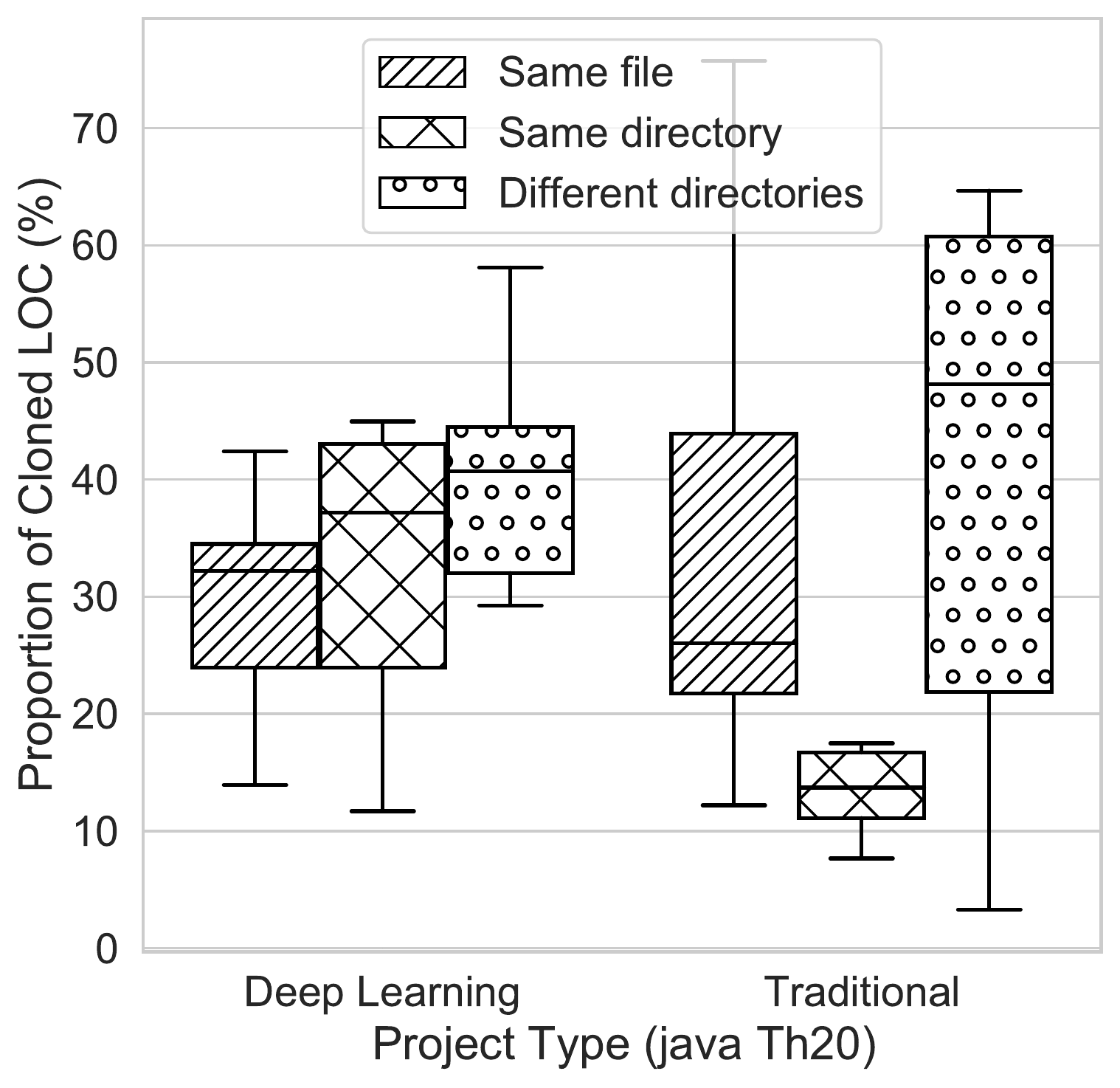}
\caption{Code Clones Distribution by Location in DL and Traditional java Systems using 20\% as Threshold Regarding Percentage of Lines of Code Clones (LOCC). i.e, (LOCC/total LOCC)x 100}
\label{fig:CloneLocation_java_20}
\end{figure}

\begin{figure}[htpb]
\centering
\includegraphics[scale=0.4]{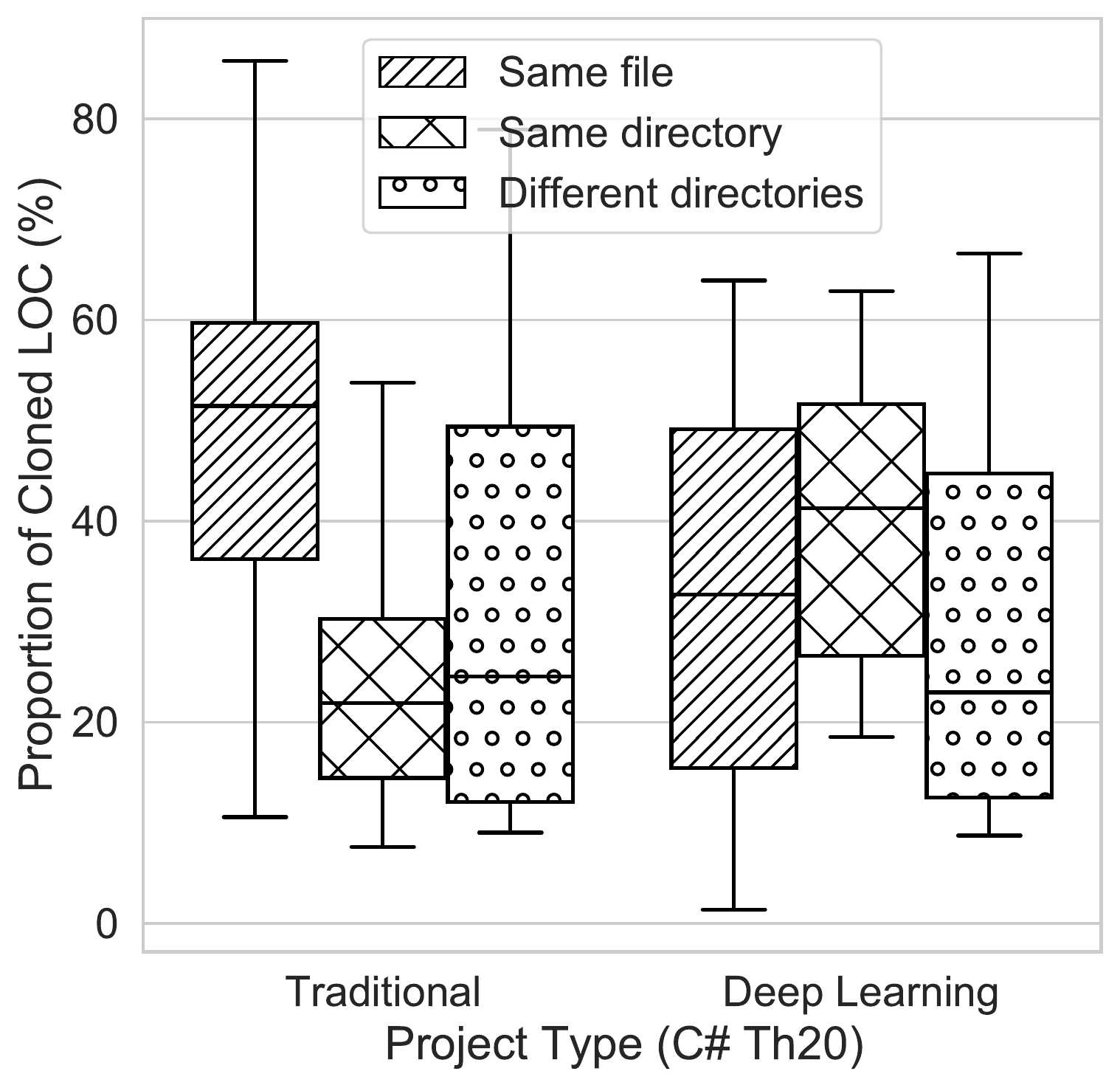}
\caption{Code Clones Distribution by Location in DL and Traditional C\# Systems using 20\% as Threshold Regarding Percentage of Lines of Code Clones (LOCC). i.e, (LOCC/total LOCC)x 100}
\label{fig:CloneLocation_csharp_20}
\end{figure}

In this section, we present the additional analysis we performed to address RQ2. We examine the code clone distribution by location in DL and traditional java (Figure \ref{fig:CloneLocation_java_20}) and C\# (Figure \ref{fig:CloneLocation_csharp_20}) systems using 20\% as dissimilarity threshold.

\begin{figure}[htpb]
\centering
\includegraphics[width=.45\textwidth]{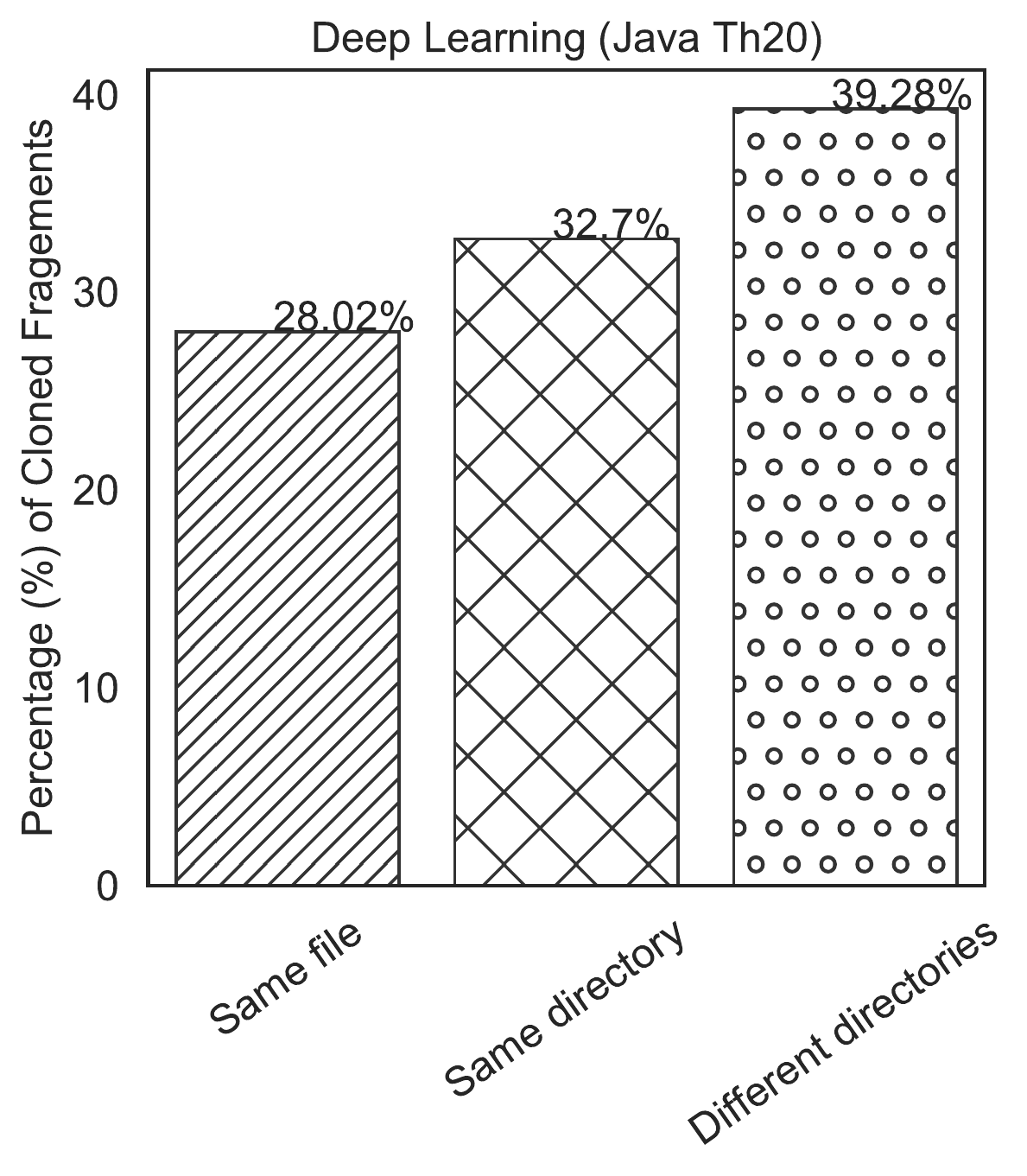}
\includegraphics[width=.45\textwidth]{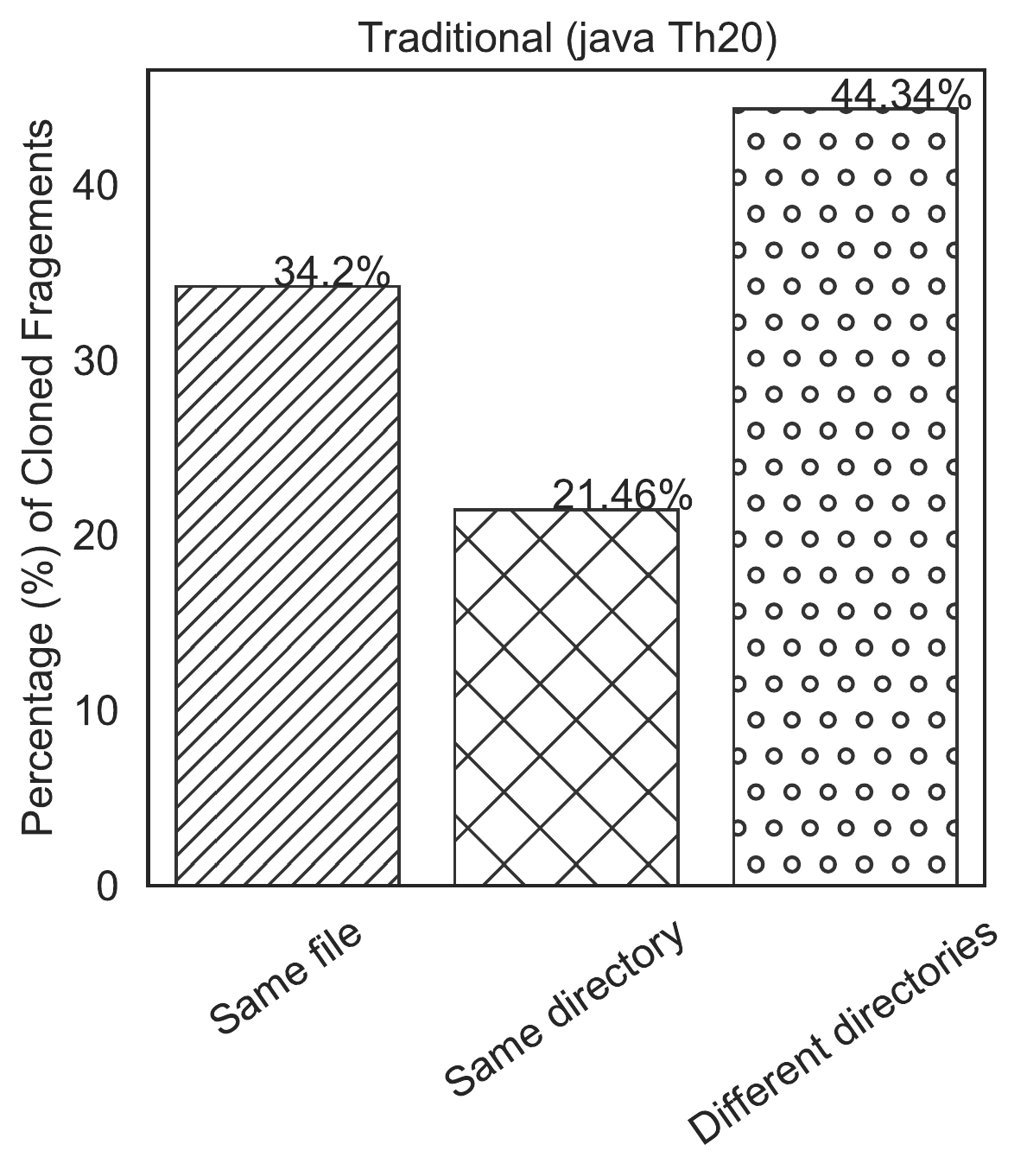}
\caption{ Percentages of Average Number of Fragments of Code Clones by Location of Clones in both Deep Learning and Traditional Java Systems using 20\% as threshold}

\label{fig:DLFragByCloneLocation_java_20}
\end{figure}

\begin{figure}[htpb]
\centering
\includegraphics[width=.45\textwidth]{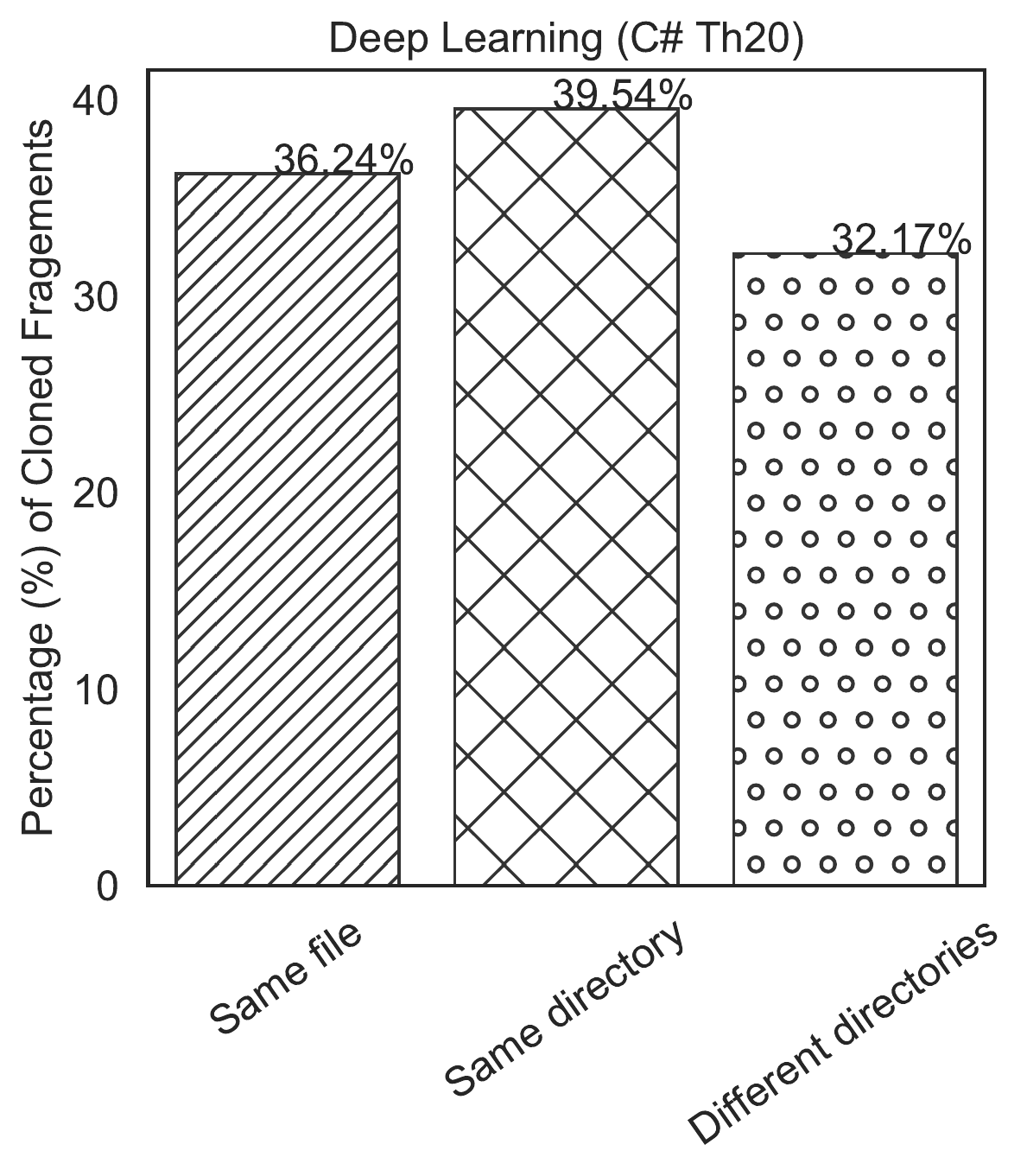}
\includegraphics[width=.45\textwidth]{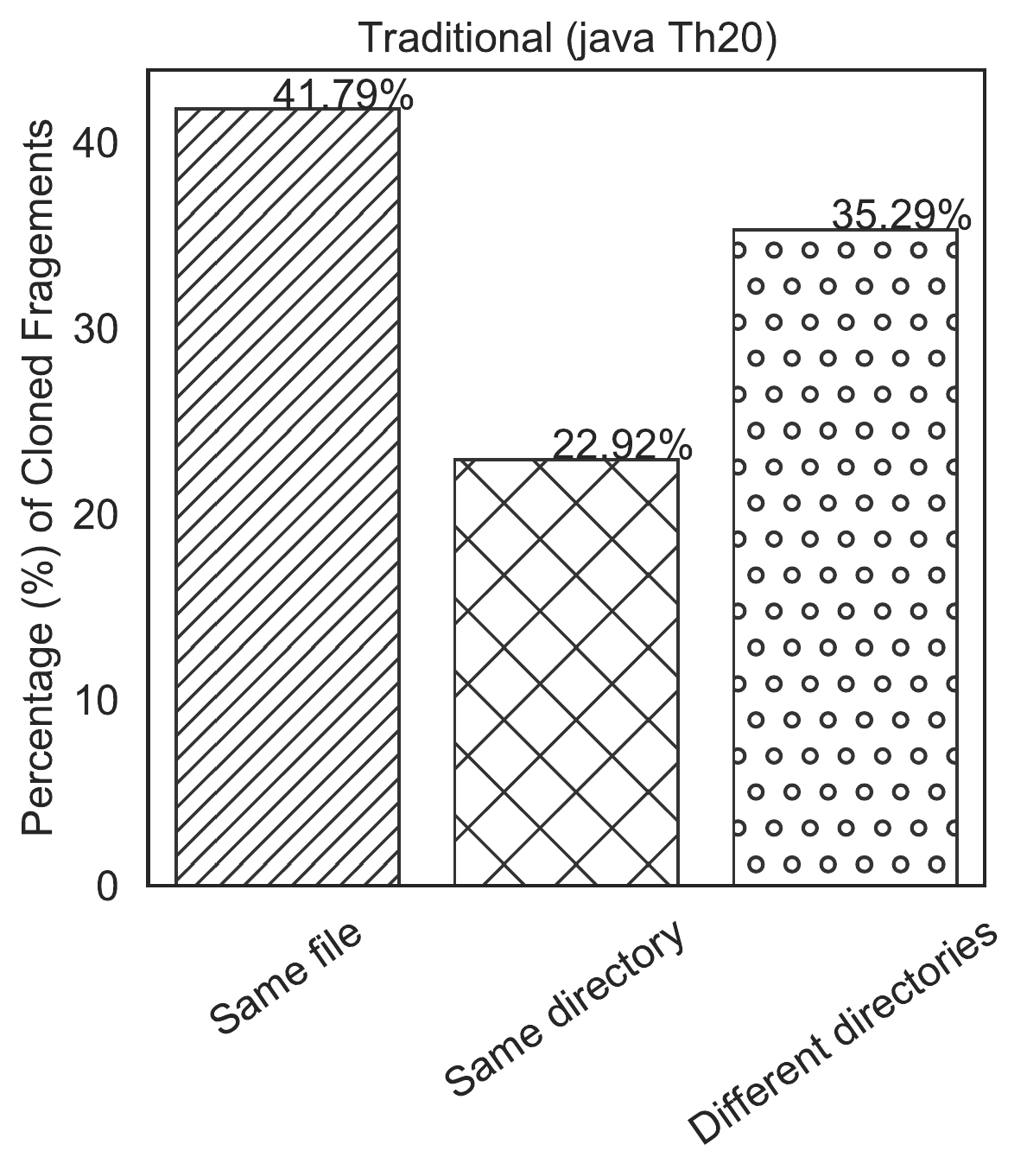}
\caption{ Percentages of Average Number of Fragments of Code Clones by Location of Clones in both Deep Learning and Traditional C\# Systems using 20\% as threshold}
\label{fig:DLFragByCloneLocation_csharp_20}
\end{figure}

We further study the percentages of average number of fragments of code clones by location of clones in both deep learning and traditional using 20\% dissimilarity threshold for the two programming language Java (Figure \ref{fig:DLFragByCloneLocation_java_20}) and C\# (Figure \ref{fig:DLFragByCloneLocation_csharp_20}).

\begin{figure}[htpb]
\centering
\includegraphics[width=.49\textwidth]{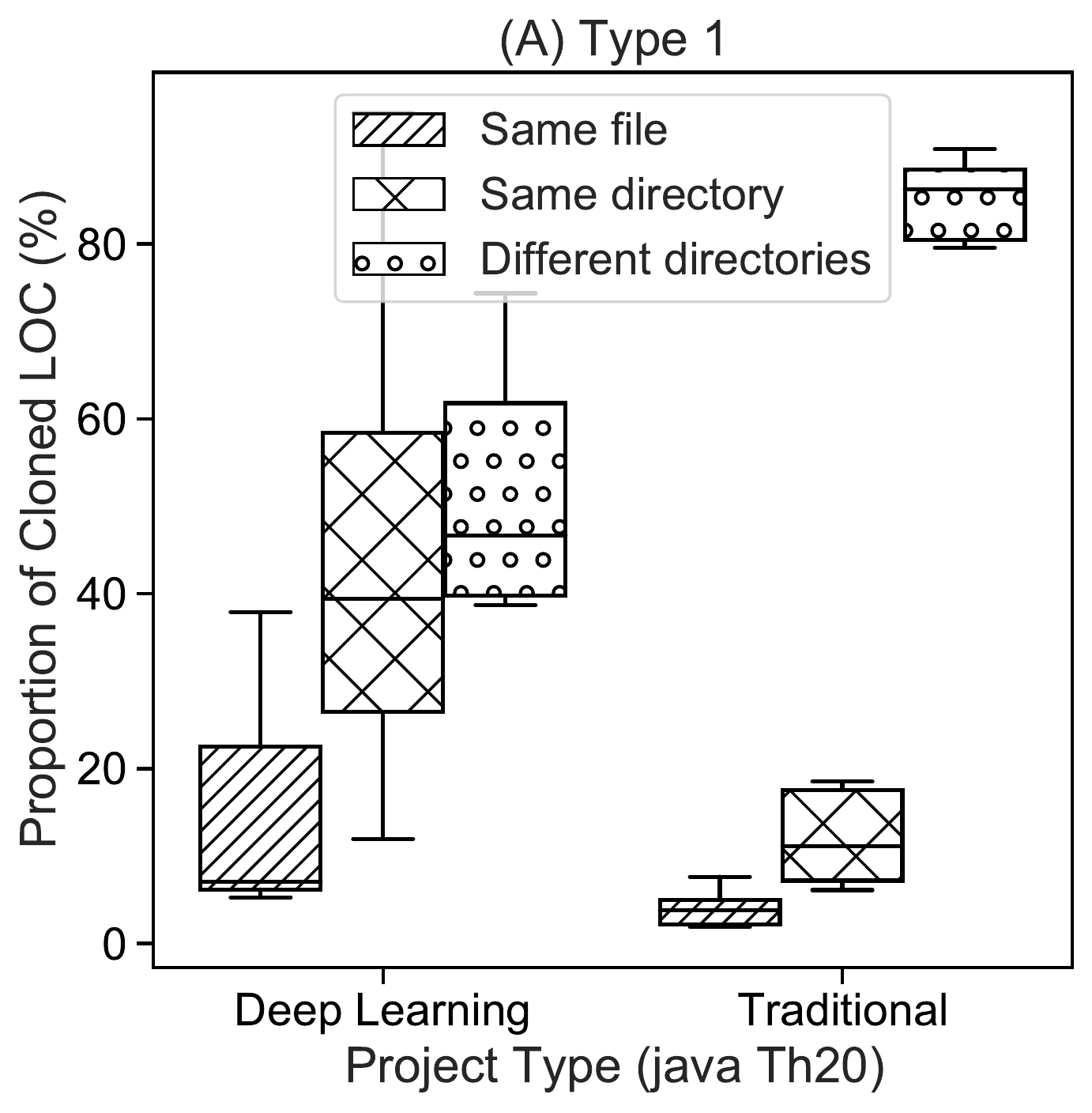}
\includegraphics[width=.49\textwidth]{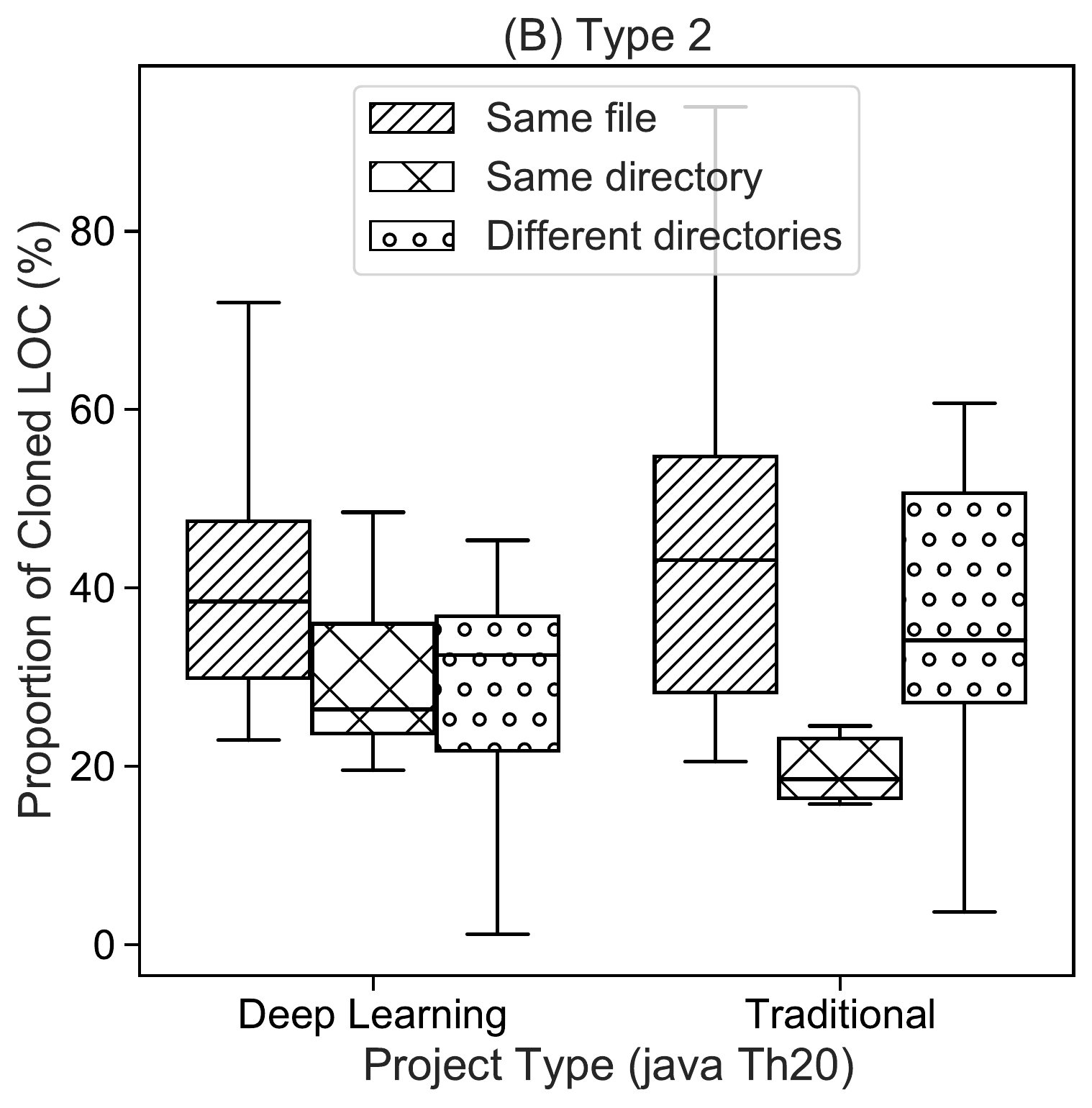}
\includegraphics[width=.49\textwidth]{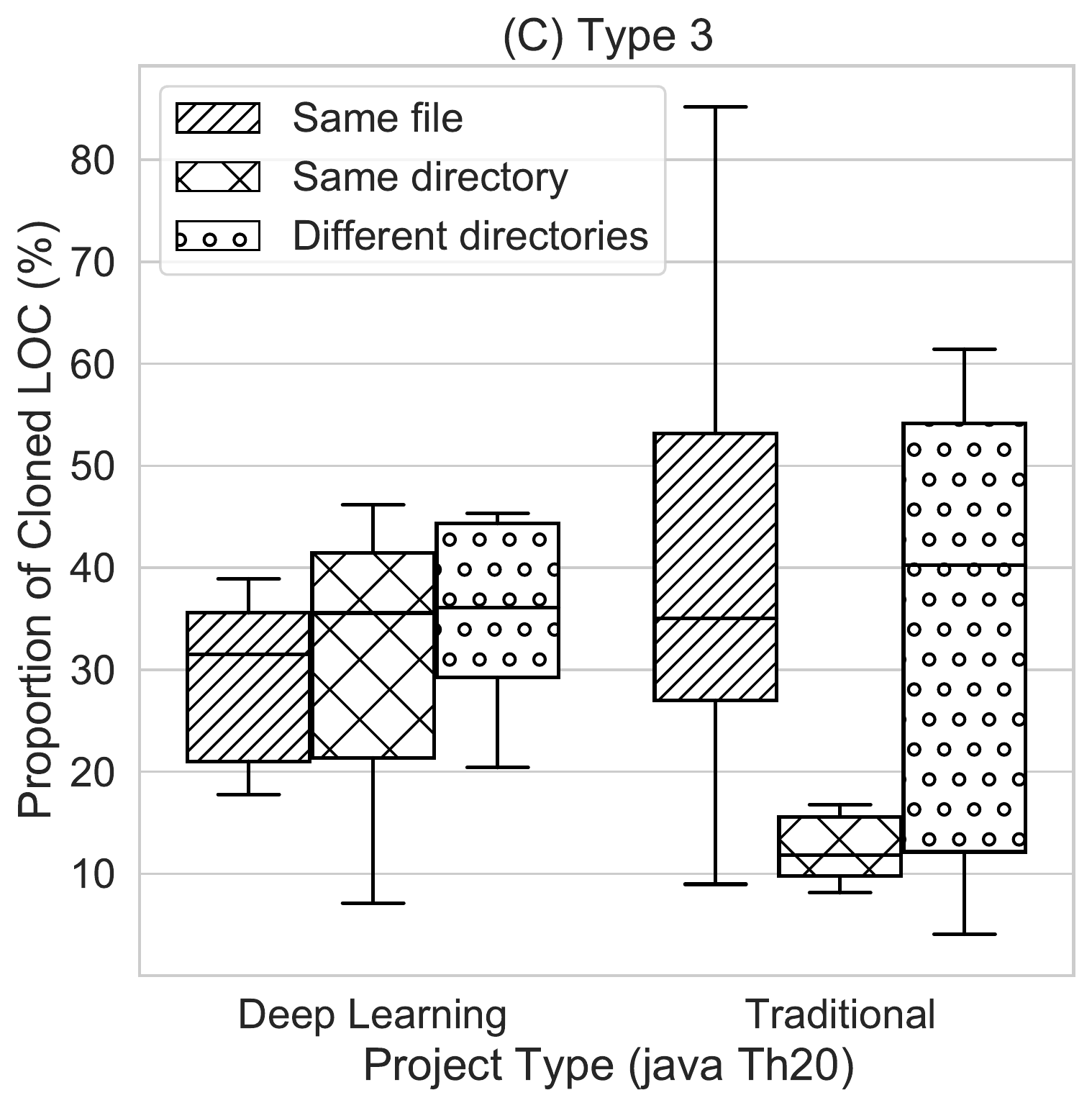}

\caption{Distribution of Different Types of Clones by Clone Location in DL and Traditional Code (Java) with 20\% as threshold}
\label{fig:CloneLocationbyCloneType_java_20}
\end{figure}

\begin{figure}[htpb]
\centering
\includegraphics[width=.49\textwidth]{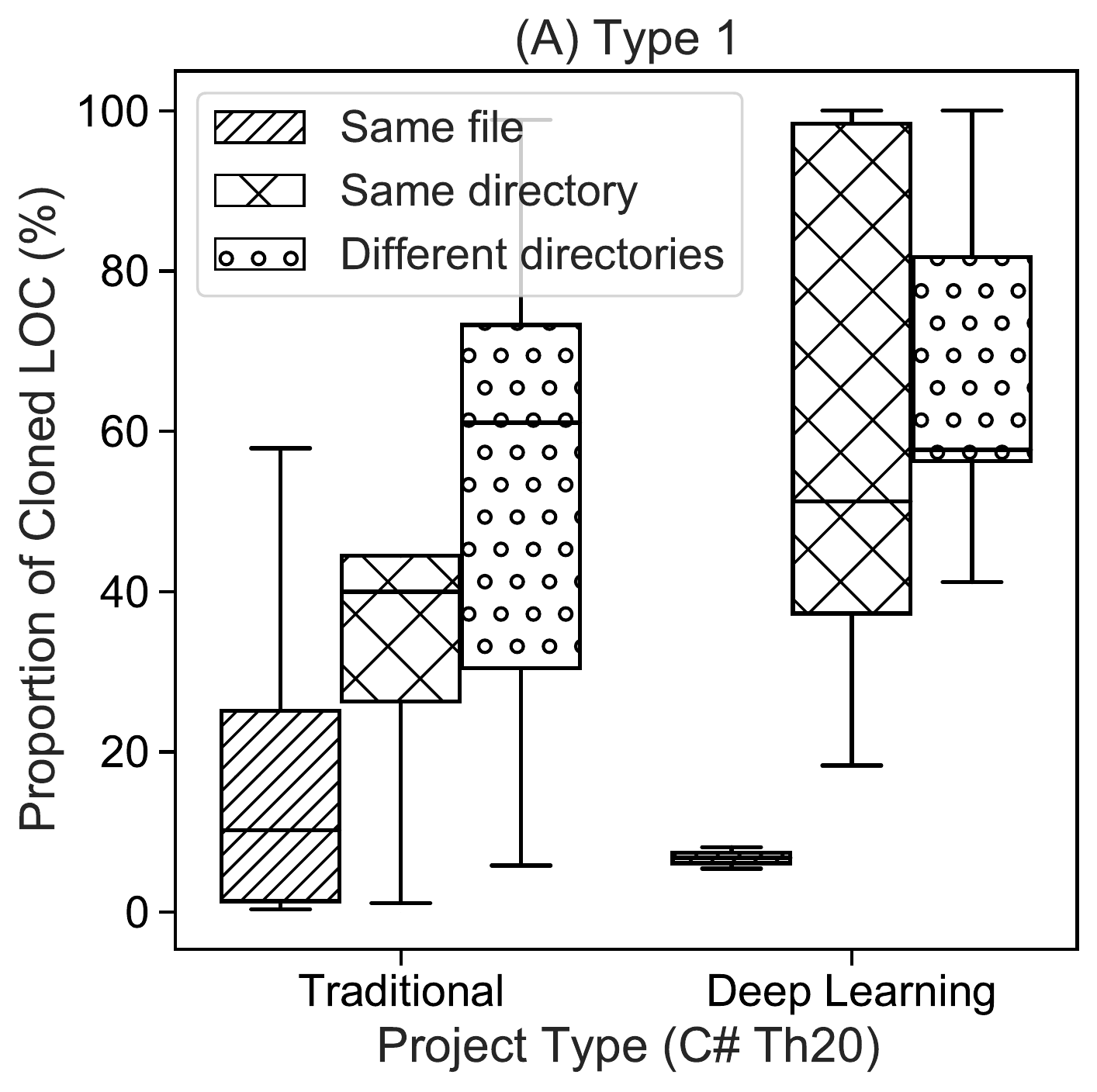}
\includegraphics[width=.49\textwidth]{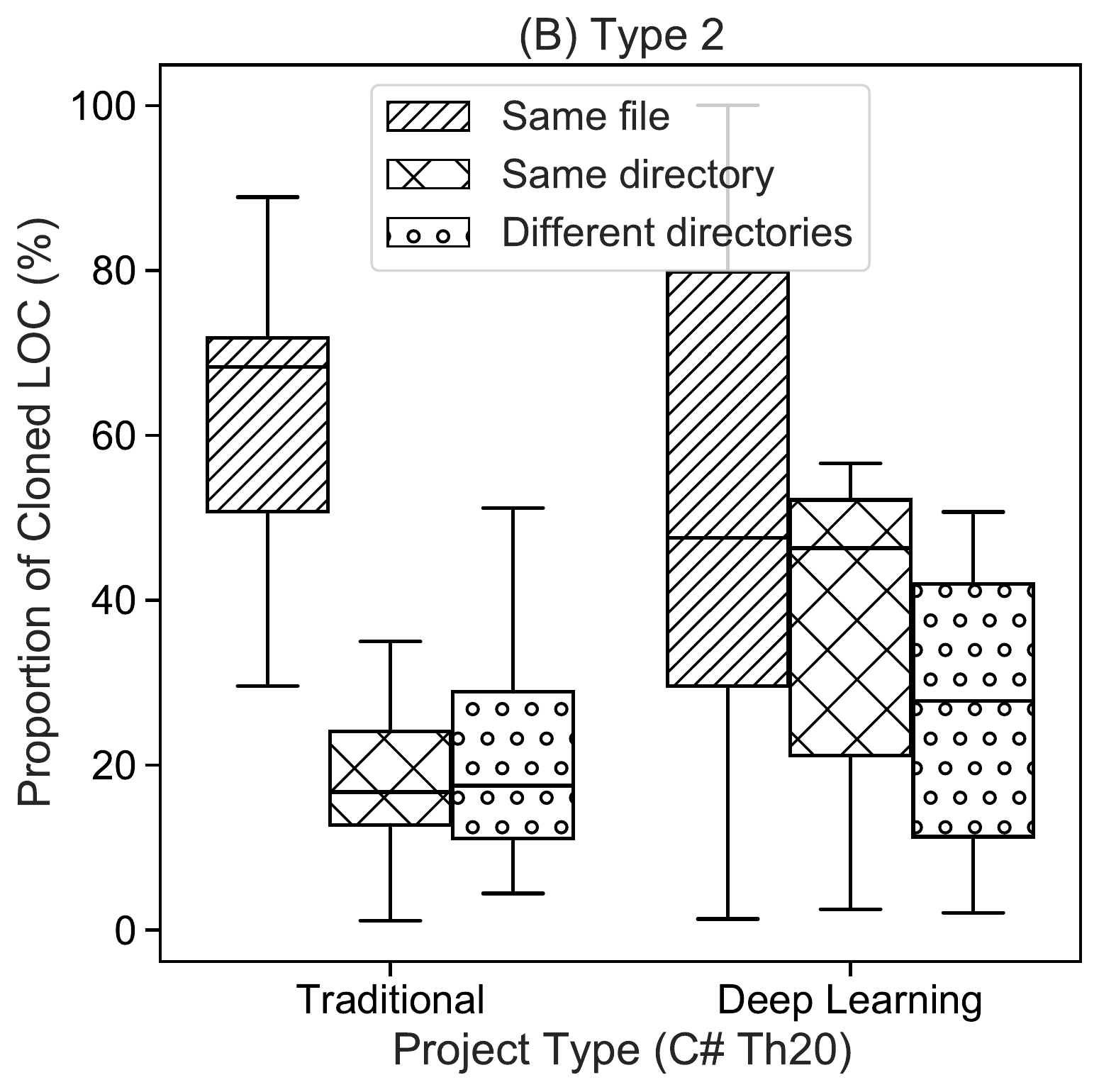}
\includegraphics[width=.49\textwidth]{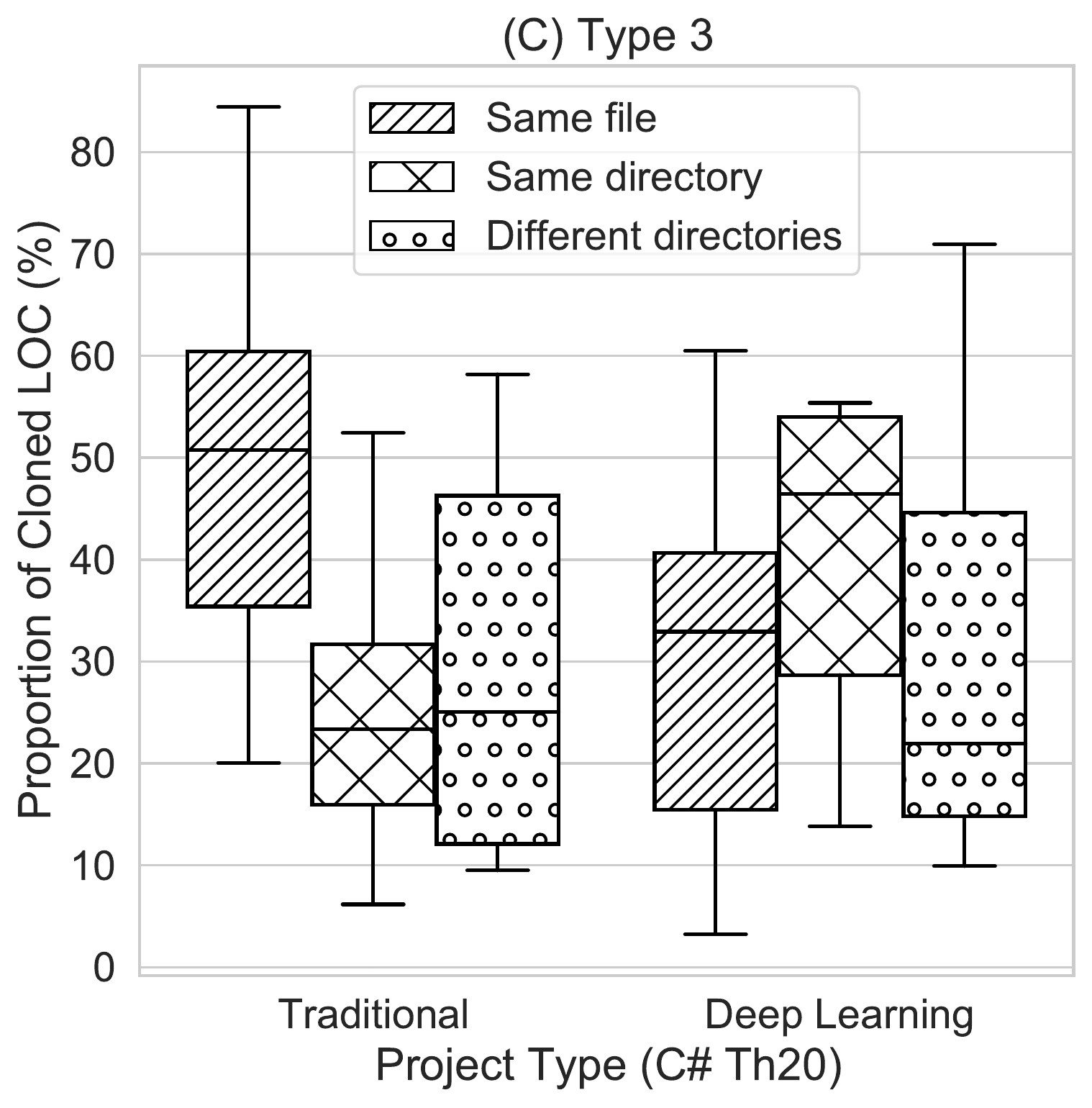}

\caption{Distribution of Different Types of Clones by Clone Location in DL and Traditional Code (C\#) with 20\% as threshold}
\label{fig:CloneLocationbyCloneType_csharp_20}
\end{figure}

We then study the distribution of different types of clones in the different clones location (Same file, Same directory, and different directories) using 20\% dissimilarity threshold for the two programming languages Java (Figure \ref{fig:CloneLocationbyCloneType_java_20}) and C\# (Figure \ref{fig:CloneLocationbyCloneType_csharp_20}).

\section{RQ3 Additional Results} \label{appRQ3}
\begin{figure}[htpb]
\centering
\includegraphics[width=.45\textwidth]{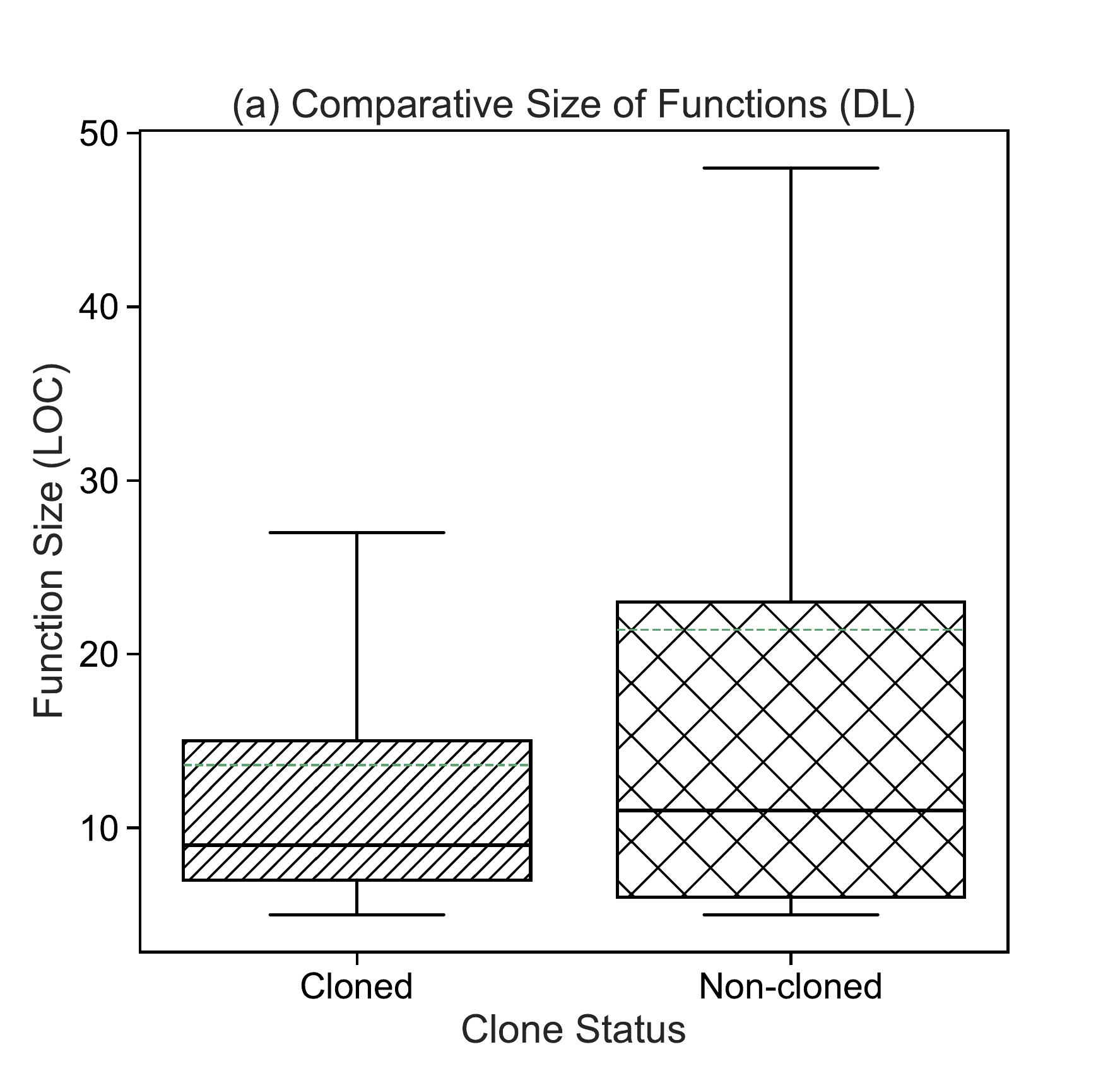}
\includegraphics[width=.45\textwidth]{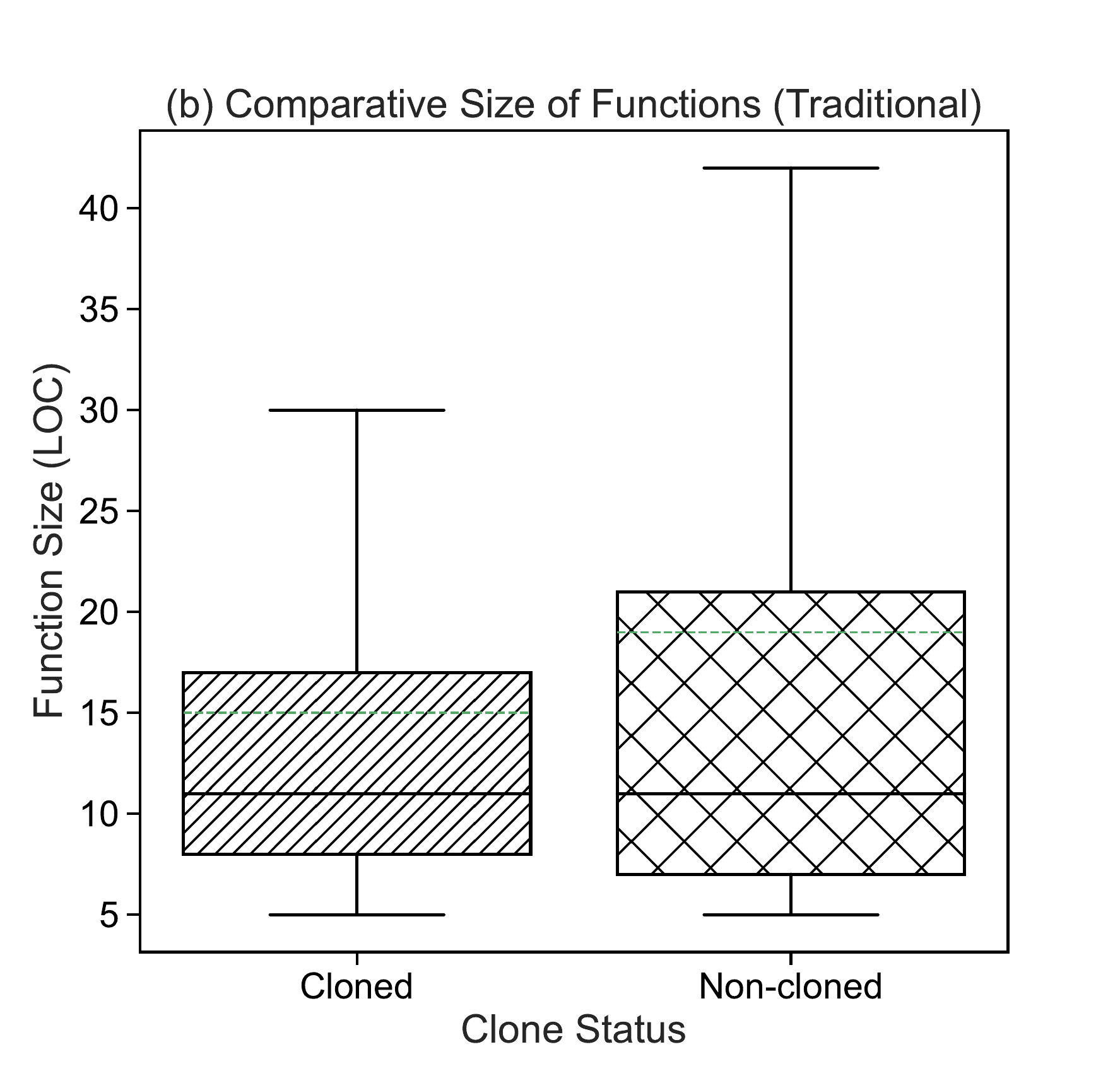}
\caption{Distribution of the Size of Cloned and Non-cloned Functions in DL and Traditional Systems}
\label{fig:FunctionSize}
\end{figure}

\begin{figure}[htpb]
\centering
\includegraphics[width=.45\textwidth]{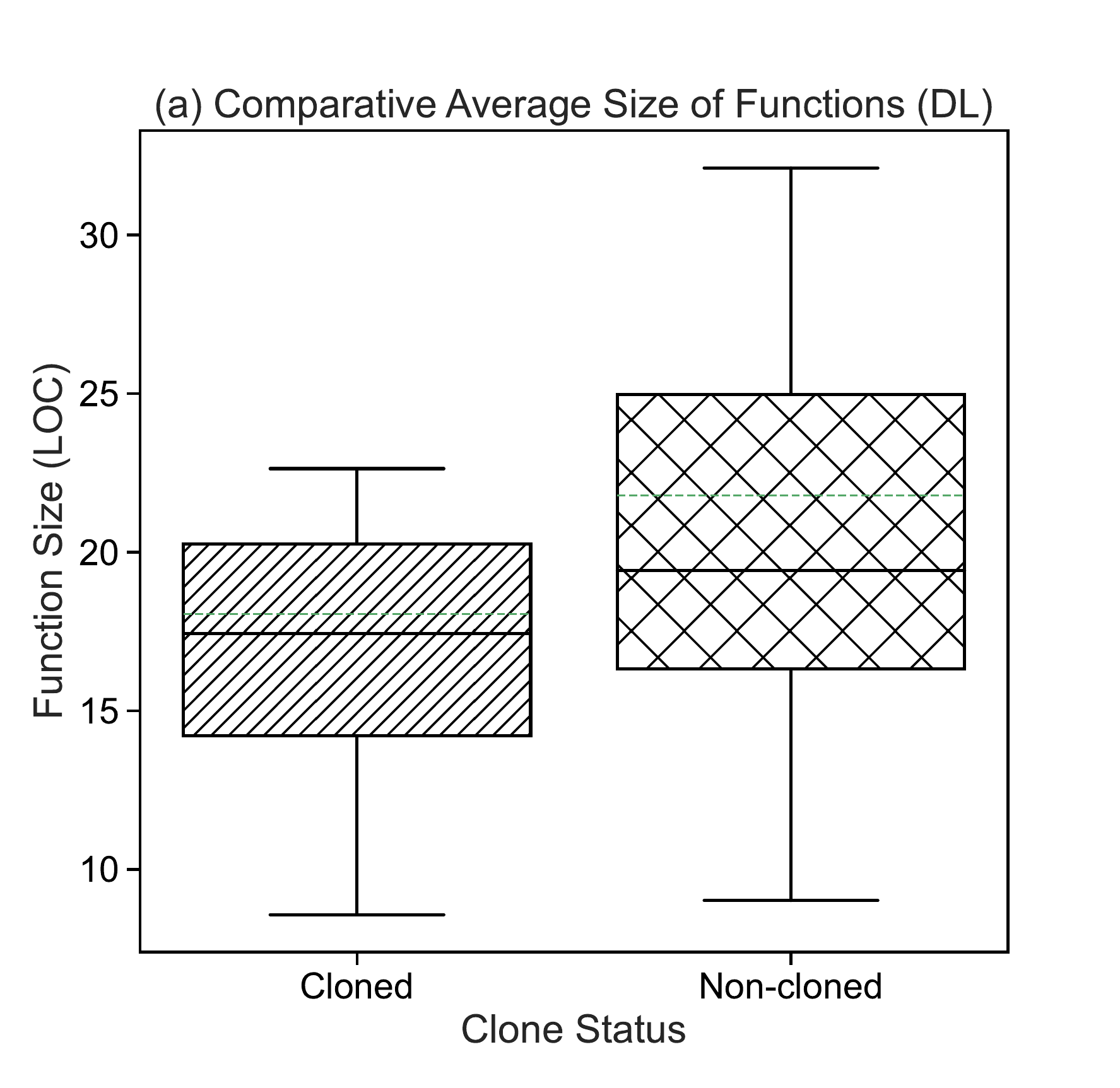}
\includegraphics[width=.45\textwidth]{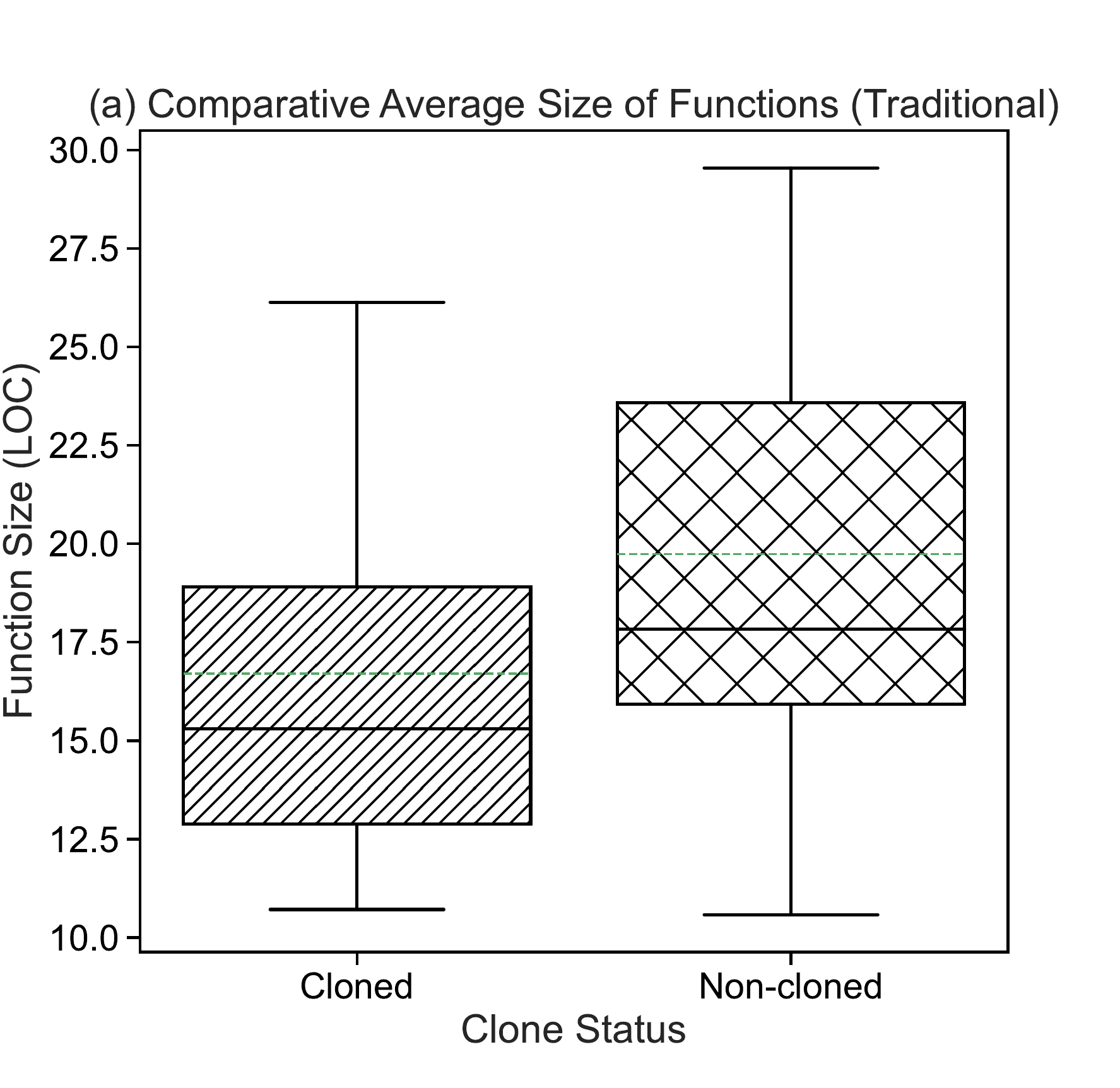}
\caption{Distribution of Mean Size of Cloned and Non-cloned Functions Per Systems in DL and Traditional Systems}
\label{fig:MeanFunctionSize_sys}
\end{figure}

\begin{figure}[htpb]
\centering
\includegraphics[width=.45\textwidth]{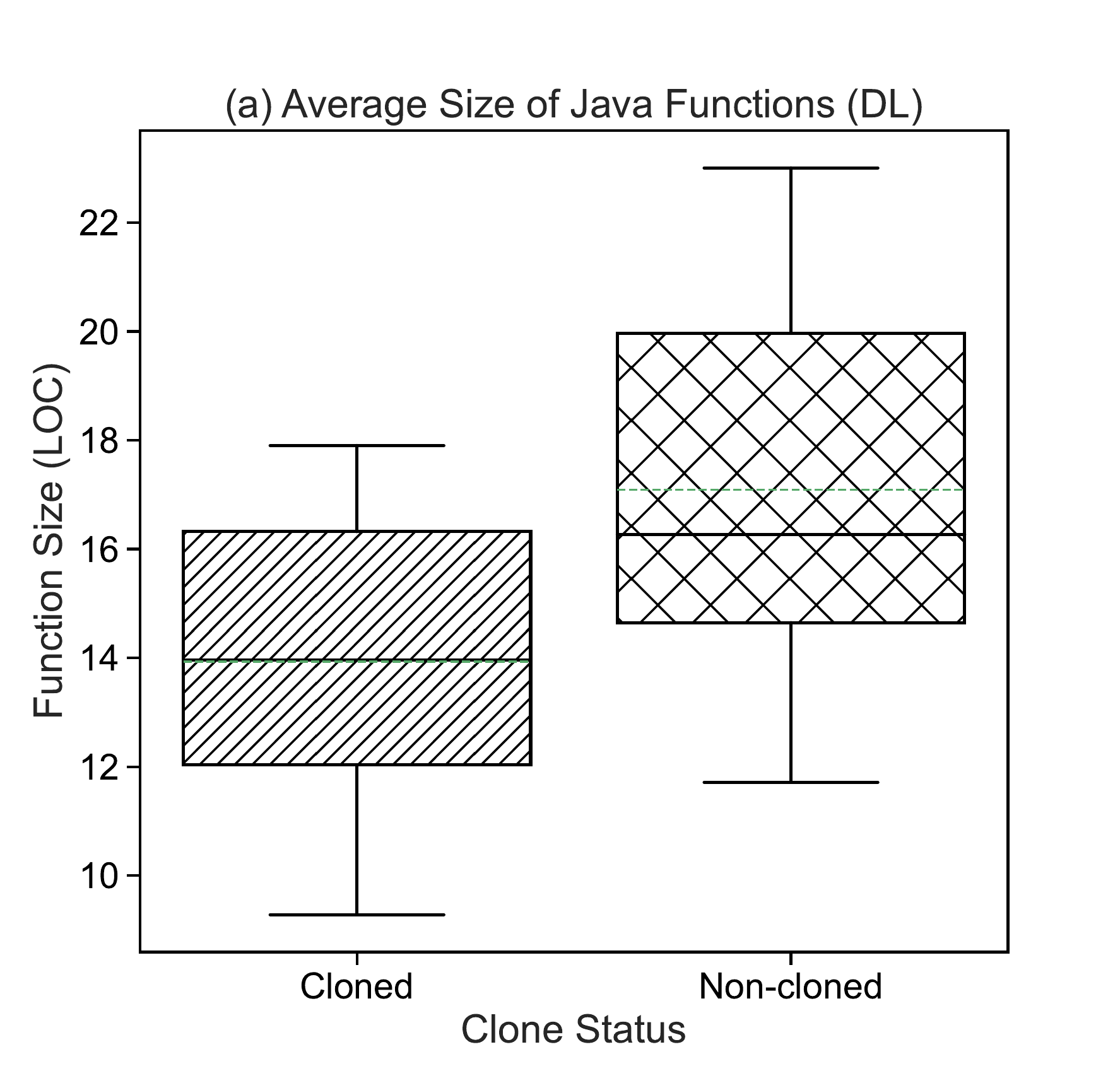}
\includegraphics[width=.45\textwidth]{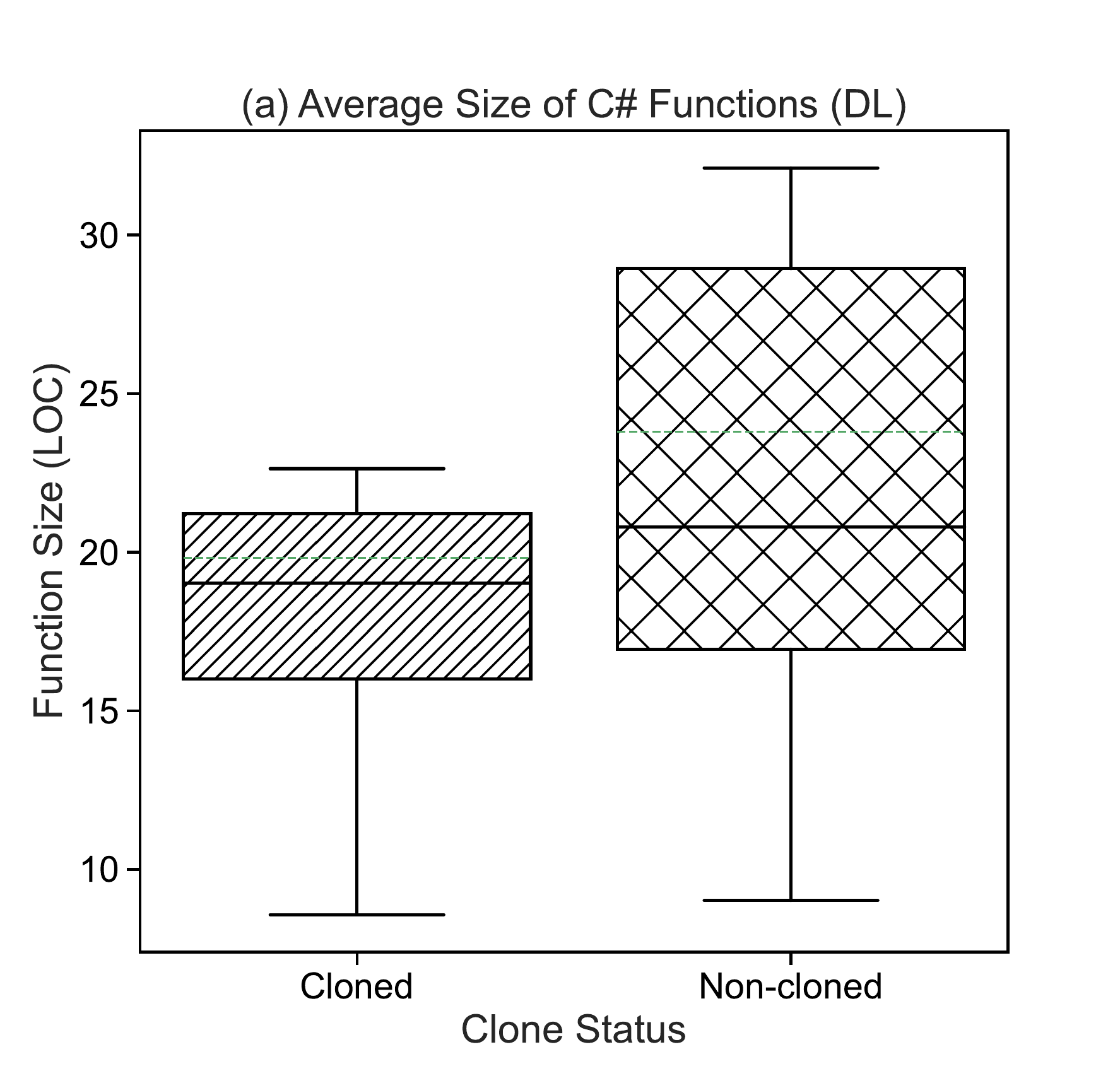}
\caption{Distribution of Mean Size of Cloned and Non-cloned Functions Per Systems in Java and C\# DL Systems}
\label{fig:MeanFunctionSize_lang}
\end{figure}

In this section, we provide additional analysis on the distribution of the size of cloned and non-cloned functions in DL and traditional systems (Figure \ref{fig:FunctionSize}). This is done to understand if size is playing an important confounding role in identifying bug fixing commits that are related to clones. We study the distribution of the mean size of cloned and non-cloned functions per systems in DL and traditional systems in Python projects (Figure \ref{fig:MeanFunctionSize_sys}) and for Java and C\# (Figure \ref{fig:MeanFunctionSize_lang}).

\section{RQ4 Additional Results}
As explained in RQ4 but in percentages, Table \ref{tab:dlCodeCloneTaxonomy_number} shows the total number of code clones attributed to the DL phases. The total number of code clones manually analyzed is 595.

\begin{table}[htp]
\caption{Total number of Occurrence of Code Clones in DL Phases}
\label{tab:dlCodeCloneTaxonomy_number}
\resizebox{\textwidth}{!}{%
\setlength{\tabcolsep}{1.2pt} 
\begin{tabular}{|l|l|c|c|c|c|c|}
\hline
\textbf{\begin{tabular}[c]{@{}l@{}}dl\_phase\\ category\end{tabular}} &
\textbf{\begin{tabular}[c]{@{}l@{}}dl\_phase\\ subcategory\end{tabular}} &
  \textbf{\begin{tabular}[c]{@{}l@{}}Type 1\\ occs\end{tabular}} &
  \textbf{\begin{tabular}[c]{@{}l@{}}Type 2\\  occs\end{tabular}} &
  \textbf{\begin{tabular}[c]{@{}l@{}}Type 3\\  occs\end{tabular}} &
  \textbf{\begin{tabular}[c]{@{}l@{}} occs 
 \\ in subcat
  \end{tabular}} &
  \textbf{\begin{tabular}[c]{@{}l@{}} occs\\  in total\end{tabular}} \\
  \hline
\begin{tabular}[c]{@{}l@{}}Preliminary\\ preparation\end{tabular}                        & hardware requirements            & 6   & 0     & 0     & 6 & 6                  \\ \hline
\multirow{2}{*}{\begin{tabular}[c]{@{}l@{}} Data\\ collection\end{tabular} }     & load data                        & 6    & 6    & 12    & 24  & \multirow{2}{*}{30}   \\ \cline{2-6}
                                     & load label                       & 6    & 0     & 0     & 6  &                         \\ \hline
\multirow{4}{*}{\begin{tabular}[c]{@{}l@{}} Data\\ postprocessing\end{tabular} } & compute output shape             & 0     & 0     & 6  & 6 & \multirow{4}{*}{50}   \\ \cline{2-6}
                                     & object localization              & 12   & 0     & 7  & 19  &                         \\ \cline{2-6}
                                     & process output                   & 12    & 0     & 7  & 19  &                         \\ \cline{2-6}
                                     & set shape of output data         & 6  & 0     & 0     & 6  &                         \\ \hline
\multirow{12}{*}{\begin{tabular}[c]{@{}l@{}} Data\\ preprocessing\end{tabular} } & apply data augmentation          & 6  & 0     & 0     & 6  & \multirow{12}{*}{110} \\ \cline{2-6}
                                     & data normalization               & 0     & 0     & 12 & 12 &                         \\ \cline{2-6}
                                     & get batches of data              & 0     & 6  & 0     & 6  &                         \\ \cline{2-6}
                                     & get numerical feature columns    & 6  & 0     & 6  & 12 &                         \\ \cline{2-6}
                                     & parse arguments                  & 0     & 0     & 6  & 6  &                         \\ \cline{2-6}
                                     & prepare tensor                   & 12 & 0     & 0     & 12 &                         \\ \cline{2-6}
                                     & process input                    & 0     & 0     & 20 & 20 &                         \\ \cline{2-6}
                                     & resize image                     & 6  & 0     & 0     & 6  &                         \\ \cline{2-6}
                                     & set shape of input data          & 0     & 0     & 12 & 12 &                         \\ \cline{2-6}
                                     & set type of input data           & 0     & 0     & 6  & 6  &                         \\ \cline{2-6}
                                     & setting format input data        & 0     & 0     & 6  & 6  &                         \\ \cline{2-6}
                                     & split data                       & 0     & 0     & 6  & 6  &                         \\ \hline
\begin{tabular}[c]{@{}l@{}} Model\\ prediction\end{tabular}                              & inference                        & 12   & 0     & 0     & 12 & 12                    \\ \hline
\multirow{11}{*}{\begin{tabular}[c]{@{}l@{}} Model\\ construction\end{tabular}  } & model component format verif.    & 6  & 0     & 0     & 6  & \multirow{11}{*}{216} \\ \cline{2-6}
                                     & activation function call         & 0     & 0     & 6 & 6  &                         \\ \cline{2-6}
                                     & build model                      & 6  & 0     & 0     & 6  &                         \\ \cline{2-6}
                                     & build one subnetwork             & 0     & 0     & 6  & 6  &                         \\ \cline{2-6}
                                     & compute model outputs            & 0     & 0     & 6  & 6  &                         \\ \cline{2-6}
                                     & init evaluation metrics          & 0     & 0     & 6  & 6  &                         \\ \cline{2-6}
                                     & initialize model graph           & 0     & 0     & 6  & 6  &                         \\ \cline{2-6}
                                     & initialize model output          & 6  & 0     & 0     & 6  &                         \\ \cline{2-6}
                                     & layer construction               & 0     & 6  & 6  & 12  &                         \\ \cline{2-6}
                                     & model architecture instantiation & 0     & 0     & 12  & 12  &                         \\ \cline{2-6}
                                     & model (hyper)parameters init     & 33 & 46    & 65 & 144 &                         \\ \hline
\multirow{2}{*}{\begin{tabular}[c]{@{}l@{}} Model\\ evaluation\end{tabular}  }    & performance metric computation   & 0     & 12 & 36 & 48 & \multirow{2}{*}{55}   \\ \cline{2-6}
                                     & test data prediction             & 0     & 0     & 7 & 7 &                         \\ \hline
\multirow{7}{*}{\begin{tabular}[c]{@{}l@{}} Model\\ training\end{tabular}  }      & compute loss                     & 30 & 0     & 0     & 30 & \multirow{7}{*}{110}  \\ \cline{2-6}
                                     & get pooling info                 & 0     & 0     & 6  & 6  &                         \\ \cline{2-6}
                                     & measure model accuracy           & 7  & 0     & 0     & 7  &                         \\ \cline{2-6}
                                     & model training                   & 6  & 0     & 12 & 18 &                         \\ \cline{2-6}
                                     & one model step training          & 12 & 6  & 12 & 30 &                         \\ \cline{2-6}
                                     & training procedure               & 0     & 0     & 7  & 7  &                         \\ \cline{2-6}
                                     & weight normalization             & 0     & 0     & 12 & 12 &                         \\ \hline
\begin{tabular}[c]{@{}l@{}} Model\\ tuning\end{tabular}                  & Minibatch size            & 0     & 0     & 6   & 6   & 6                    \\ \hline
\end{tabular}%
}
\end{table}

\end{appendices}

\end{document}